\newcommand{\figures}{./}
\documentclass[aps,showpacs,twocolumn,superscriptaddress,letterpaper,longbibliography]{revtex4-1}
\usepackage{graphicx}
\usepackage{dcolumn}
\usepackage{bm}
\usepackage{amsmath, amssymb}
\usepackage{epstopdf}
\usepackage{color}
\usepackage[normalem]{ulem}
\usepackage{multirow}
\usepackage{enumitem}
\usepackage{comment}
\usepackage{hyperref}
\usepackage{cleveref}
\usepackage{lineno}
\usepackage{ragged2e}
\usepackage{float}
\usepackage{soul}
\usepackage{adjustbox}
\usepackage{makecell}
\usepackage[maxfloats=256]{morefloats}
\hypersetup{breaklinks = true, colorlinks = true, citecolor = blue, linkcolor = blue, urlcolor = blue}

\newcommand{\uB}{MicroBooNE}
\newcommand{\CCIpOpi}{CC1p0$\pi$}

\begin{document}


\title{Multi-Differential Cross Section Measurements of \texorpdfstring{$\nu_{\mu}$}{numu}-Argon\texorpdfstring{\\}{}Quasielastic-like Reactions with the MicroBooNE Detector}
\date{\today}


\newcommand{\ANL}{Argonne National Laboratory (ANL), Lemont, IL, 60439, USA}
\newcommand{\Bern}{Universit{\"a}t Bern, Bern CH-3012, Switzerland}
\newcommand{\BNL}{Brookhaven National Laboratory (BNL), Upton, NY, 11973, USA}
\newcommand{\UCSB}{University of California, Santa Barbara, CA, 93106, USA}
\newcommand{\Cambridge}{University of Cambridge, Cambridge CB3 0HE, United Kingdom}
\newcommand{\CIEMAT}{Centro de Investigaciones Energ\'{e}ticas, Medioambientales y Tecnol\'{o}gicas (CIEMAT), Madrid E-28040, Spain}
\newcommand{\Chicago}{University of Chicago, Chicago, IL, 60637, USA}
\newcommand{\Cincinnati}{University of Cincinnati, Cincinnati, OH, 45221, USA}
\newcommand{\CSU}{Colorado State University, Fort Collins, CO, 80523, USA}
\newcommand{\Columbia}{Columbia University, New York, NY, 10027, USA}
\newcommand{\Edinburgh}{University of Edinburgh, Edinburgh EH9 3FD, United Kingdom}
\newcommand{\FNAL}{Fermi National Accelerator Laboratory (FNAL), Batavia, IL 60510, USA}
\newcommand{\Granada}{Universidad de Granada, Granada E-18071, Spain}
\newcommand{\Harvard}{Harvard University, Cambridge, MA 02138, USA}
\newcommand{\IIT}{Illinois Institute of Technology (IIT), Chicago, IL 60616, USA}
\newcommand{\KSU}{Kansas State University (KSU), Manhattan, KS, 66506, USA}
\newcommand{\Lancaster}{Lancaster University, Lancaster LA1 4YW, United Kingdom}
\newcommand{\LANL}{Los Alamos National Laboratory (LANL), Los Alamos, NM, 87545, USA}
\newcommand{\Louisiana}{Louisiana State University, Baton Rouge, LA, 70803, USA}
\newcommand{\Manchester}{The University of Manchester, Manchester M13 9PL, United Kingdom}
\newcommand{\MIT}{Massachusetts Institute of Technology (MIT), Cambridge, MA, 02139, USA}
\newcommand{\Michigan}{University of Michigan, Ann Arbor, MI, 48109, USA}
\newcommand{\Minnesota}{University of Minnesota, Minneapolis, MN, 55455, USA}
\newcommand{\NMSU}{New Mexico State University (NMSU), Las Cruces, NM, 88003, USA}
\newcommand{\Oxford}{University of Oxford, Oxford OX1 3RH, United Kingdom}
\newcommand{\Pitt}{University of Pittsburgh, Pittsburgh, PA, 15260, USA}
\newcommand{\Rutgers}{Rutgers University, Piscataway, NJ, 08854, USA}
\newcommand{\SLAC}{SLAC National Accelerator Laboratory, Menlo Park, CA, 94025, USA}
\newcommand{\SDSMT}{South Dakota School of Mines and Technology (SDSMT), Rapid City, SD, 57701, USA}
\newcommand{\Maine}{University of Southern Maine, Portland, ME, 04104, USA}
\newcommand{\Syracuse}{Syracuse University, Syracuse, NY, 13244, USA}
\newcommand{\TelAviv}{Tel Aviv University, Tel Aviv, Israel, 69978}
\newcommand{\Tennessee}{University of Tennessee, Knoxville, TN, 37996, USA}
\newcommand{\UTA}{University of Texas, Arlington, TX, 76019, USA}
\newcommand{\Tufts}{Tufts University, Medford, MA, 02155, USA}
\newcommand{\UCL}{University College London, London WC1E 6BT, United Kingdom}
\newcommand{\VTech}{Center for Neutrino Physics, Virginia Tech, Blacksburg, VA, 24061, USA}
\newcommand{\Warwick}{University of Warwick, Coventry CV4 7AL, United Kingdom}
\newcommand{\Yale}{Wright Laboratory, Department of Physics, Yale University, New Haven, CT, 06520, USA}

\affiliation{\ANL}
\affiliation{\Bern}
\affiliation{\BNL}
\affiliation{\UCSB}
\affiliation{\Cambridge}
\affiliation{\CIEMAT}
\affiliation{\Chicago}
\affiliation{\Cincinnati}
\affiliation{\CSU}
\affiliation{\Columbia}
\affiliation{\Edinburgh}
\affiliation{\FNAL}
\affiliation{\Granada}
\affiliation{\Harvard}
\affiliation{\IIT}
\affiliation{\KSU}
\affiliation{\Lancaster}
\affiliation{\LANL}
\affiliation{\Louisiana}
\affiliation{\Manchester}
\affiliation{\MIT}
\affiliation{\Michigan}
\affiliation{\Minnesota}
\affiliation{\NMSU}
\affiliation{\Oxford}
\affiliation{\Pitt}
\affiliation{\Rutgers}
\affiliation{\SLAC}
\affiliation{\SDSMT}
\affiliation{\Maine}
\affiliation{\Syracuse}
\affiliation{\TelAviv}
\affiliation{\Tennessee}
\affiliation{\UTA}
\affiliation{\Tufts}
\affiliation{\UCL}
\affiliation{\VTech}
\affiliation{\Warwick}
\affiliation{\Yale}

\author{P.~Abratenko} \affiliation{\Tufts}
\author{O.~Alterkait} \affiliation{\Tufts}
\author{D.~Andrade~Aldana} \affiliation{\IIT}
\author{J.~Anthony} \affiliation{\Cambridge}
\author{L.~Arellano} \affiliation{\Manchester}
\author{J.~Asaadi} \affiliation{\UTA}
\author{A.~Ashkenazi}\affiliation{\TelAviv}
\author{S.~Balasubramanian}\affiliation{\FNAL}
\author{B.~Baller} \affiliation{\FNAL}
\author{G.~Barr} \affiliation{\Oxford}
% ...

\collaboration{The MicroBooNE Collaboration}
\thanks{microboone\_info@fnal.gov}\noaffiliation


\begin{abstract}
We report on a flux-integrated multi-differential measurement of charged-current muon neutrino scattering on argon with one muon and one proton in the final state using the Booster Neutrino Beam and \uB\ detector at Fermi National Accelerator Laboratory.
The data are studied as a function of various kinematic imbalance variables and of a neutrino energy estimator, and are compared to a number of event generator predictions.
We find that the measured cross sections in different phase-space regions are sensitive to nuclear effects.
Our results provide precision data to test and improve the neutrino-nucleus interaction models needed to perform high-accuracy oscillation analyses. 
Specific regions of phase-space are identified where further model refinements are most needed.
\end{abstract}

\maketitle


\section{Introduction}\label{intro}

High-precision measurements of the neutrino mixing angles, mass differences, and charge-parity violating phase, and the search for physics beyond the Standard Model are the primary physics goals of many currently operating as well as next-generation neutrino experiments~\cite{pdg2018,T2KNature20,DUNE1:2016oaz,DUNE2:2016oaz,DUNE3:2016oaz,HK}.
These measurements require reliable comparisons of measured and theoretically-expected neutrino interaction rates in the corresponding detectors. 
Thus, understanding the neutrino-nucleus scattering processes in detail is a prerequisite for these experiments to reach their discovery potential.
A number of neutrino oscillation experiments employ liquid argon time projection chambers (LArTPCs)~\cite{DUNE1:2016oaz,DUNE2:2016oaz,DUNE3:2016oaz,Antonello:2015lea,Tortorici:2018yns,Abi:2018dnh} to detect the particles produced in neutrino interactions.
The ultimate goal of these efforts is both to reconstruct the energy of the neutrino based on the kinematics of the outgoing particles and to enable few-percent-level modeling of neutrino-argon interaction rates~\cite{Abi:2020evt}.
Therefore, high-accuracy modeling of neutrino-argon interactions is of the utmost importance~\cite{Dolan:2018sbb,PhysRevLett.116.192501,Rocco2020}.

This work presents the first measurement of flux-integrated single- and double-differential cross sections for muon-neutrino-argon ($\nu_{\mu}$-Ar) charged-current (CC) quasielastic (QE)-like scattering reactions as a function of kinematic imbalance variables~\cite{PhysRevC.94.015503,Abe:2018pwo,PhysRevLett.121.022504,PhysRevD.101.092001,Bathe-Peters:2022kkj}.
Double-differential measurements as a function of a neutrino energy estimator are further reported for the first time in kinematic imbalance bins on argon.
Motivated by a previous analysis with a similar signal event topology~\cite{PhysRevLett.125.201803}, we focus on reactions where a single muon-proton pair is reconstructed with no additional detected particles.
The results reported here use the \uB\ detector~\cite{Acciarri:2016smi} with an exposure of $6.79 \times 10^{20}$ protons on target from the Booster Neutrino Beam (BNB)~\cite{AguilarArevalo:2008yp} at Fermi National Accelerator Laboratory.

The experimental setup is presented in Sec.~\ref{exp}, followed by the signal definition and event selection in Sec.~\ref{sig}.
The observables of interest are defined in Sec.~\ref{obs}. 
Section~\ref{xsec} describes the cross section extraction and systematics procedure and Sec.~\ref{model} outlines the modeling configurations used for comparison to the data. 
The results are reported in Sec.~\ref{results} and the conclusions are discussed in Sec.~\ref{concl}.


\section{Experimental Setup}\label{exp}

The \uB\ LArTPC has an active volume that contains 85 tonnes of argon.
It is exposed to BNB neutrinos, with an energy spectrum that peaks around 0.8\,GeV and extends to 2\,GeV.

Charged particles are produced after the primary neutrino interaction with the argon nuclei in the LArTPC active volume.
Scintillation light and electron ionization trails are produced while these charged particles travel through the liquid argon.
In the presence of an electric field of 273\,V/cm, the ionization electrons drift towards a system of three anode wire planes.
Photomultiplier tubes (PMTs) are used to measure the scintillation light.

If the PMT signals are in time coincidence with the beam arrival time, then events are recorded.
Trigger hardware and software selection criteria are designed to minimize the contribution from background events, which are primarily cosmic muons.
After these are applied, enriched data samples are obtained in which a neutrino interaction occurs in $\approx$ 15\% of selected beam spills~\cite{Kaleko:2013eda}. 

Individual particle tracks are reconstructed with $\texttt{Pandora}$ pattern recognition algorithms based on the measured ionization signals in the enriched data samples~\cite{Acciarri:2017hat}.
Particles are identified based on the measured track energy deposition profile, while the particle momenta are obtained based on the track length~\cite{PDG_spline_table,stoppingpowersource}.


\section{Signal Definition \& Event Selection}\label{sig}

The QE-like signal definition used in this analysis includes all $\nu_{\mu}$-Ar scattering events with a final-state muon with momentum 0.1 $< p_{\mu}<$ 1.2\,GeV/$c$, and exactly one proton with 0.3 $< p_{p} <$ 1\,GeV/$c$.
Events with final-state neutral pions at any momentum are excluded.
Signal events may contain any number of protons below 300 MeV/$c$ or above 1\,GeV/$c$, neutrons at any momentum, and charged pions with momentum lower than 70 MeV/$c$.
We refer to the events passing this definition as CC1p0$\pi$.
The aforementioned momentum ranges are driven by considering resolution effects, as well as regions of the phase space with non-zero efficiencies and systematic uncertainties that are well understood.
This signal consists predominantly of QE events.
More complex interactions as labeled at a generator level, namely meson exchange currents (MEC), resonance interactions (RES) and deep inelastic scattering events (DIS), can still yield CC1p0$\pi$ events.
That can be the case due to final-state interactions (FSI), such as pion absorption, or due to the presence of particles outside the momentum range of interest in the CC1p0$\pi$ signal definition as defined above.

Events that satisfy the CC1p0$\pi$ signal definition at a reconstruction level but not at truth level are treated as  background events.
We refer to these events as non-CC1p0$\pi$.
Based on simulation predictions, we find that the dominant background contribution originates from events with two protons in the momentum range of interest, where the second proton was not reconstructed.
These events are referred to as CC1$\mu$2p0$\pi$ and are the focus of a dedicated MicroBooNE cross section analysis that demonstrated good data-simulation agreement~\cite{MicroBooNE:2022emb}.

\begin{figure}[htb!]
\centering 
\includegraphics[width=\linewidth]{\figures 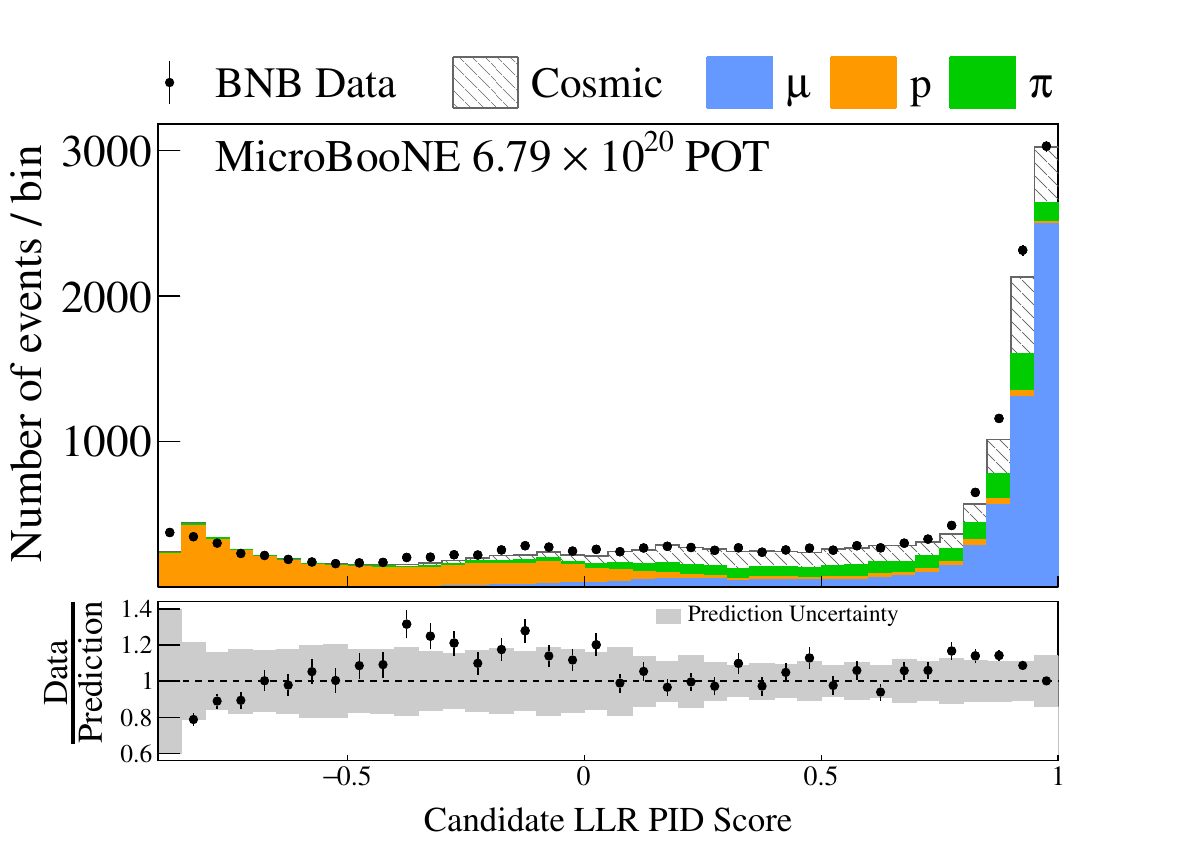}
\caption{
The log-likelihood ratio (LLR) particle identification (PID) score distribution used to tag the muon and proton candidates.
}
\label{LLRPlot}
\end{figure}

Candidate muon-proton pairs are isolated by requiring the existence of precisely two track-like and no shower-like objects, as classified by $\texttt{Pandora}$ using a track-score variable~\cite{VanDePontseele:2020tqz,PhysRevD.105.112004}.
The log-likelihood ratio (LLR) particle identification (PID) score~\cite{NicoloPID} is used to identify the muon and proton candidates.
Figure~\ref{LLRPlot} shows the particle composition breakdown of the sample as a function of the LLR PID score.

Muons tend to have higher LLR PID score values than protons, thus the track with the highest score is tagged as the candidate muon.
Meanwhile, the track with the lower score is treated as the candidate proton.

\begin{figure}[htb!]
\centering  
\includegraphics[width=\linewidth]{\figures 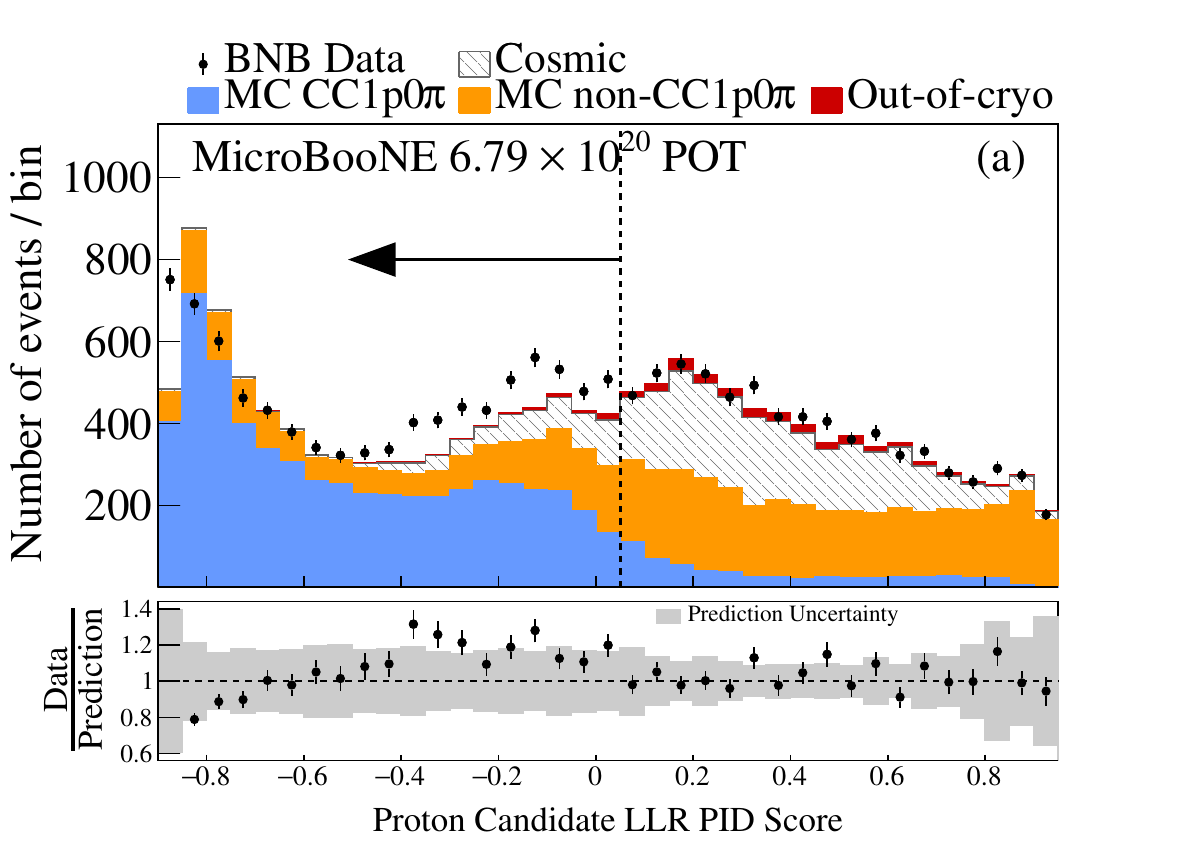}
\includegraphics[width=\linewidth]{\figures 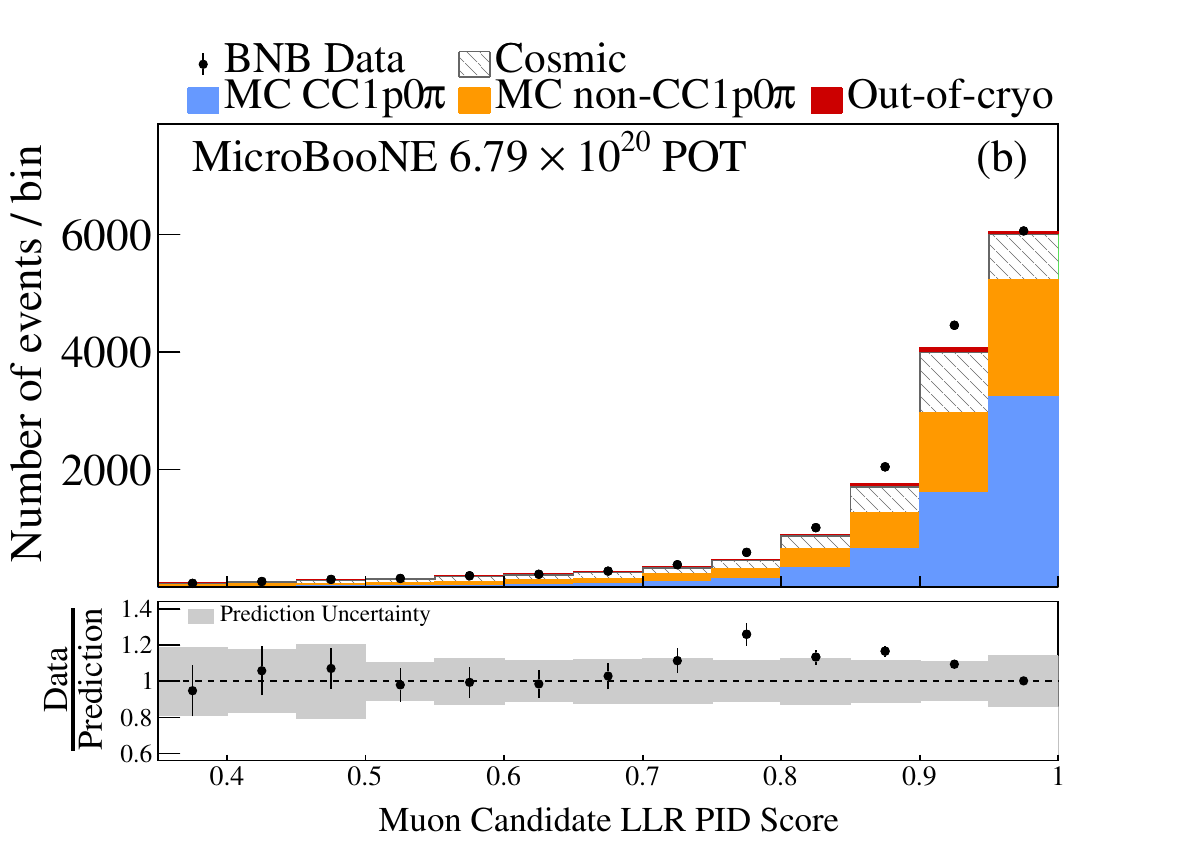}
\caption{
(Top) the proton candidate LLR PID score distribution, illustrating the fitness of a cut at LLR PID $<$ 0.05 to reject cosmic and non-CC1p0$\pi$ background events.
(Bottom) the muon candidate LLR PID score distribution, illustrating a peak close to one.
Only statistical uncertainties are shown on the data.
The bottom panel shows the ratio of data to prediction.
}
\label{picpid}
\end{figure}

Cosmic muon and non-\CCIpOpi\ contamination backgrounds were significantly reduced by applying a requirement on the candidate proton LLR PID score.
We studied the effect of cutting on different values of this quantity, which has a strong discrimination power for rejecting MC non-CC1p0$\pi$ background, out-of-cryostat and cosmic events. 
That yielded an optimal cut on the proton candidate LLR score of $<$ 0.05, as shown in Fig.~\ref{picpid}a.
Figure~\ref{picpid}b shows the corresponding muon candidate LLR score, which is peaked at values close to one.
The uncertainty bands account for potential data-MC discrepancies observed for both particle scores.
The particle composition of the panels included in Fig.~\ref{picpid} is shown in the Supplemental Material. 

\begin{figure}[htb!]
\centering  
\includegraphics[width=\linewidth]{\figures 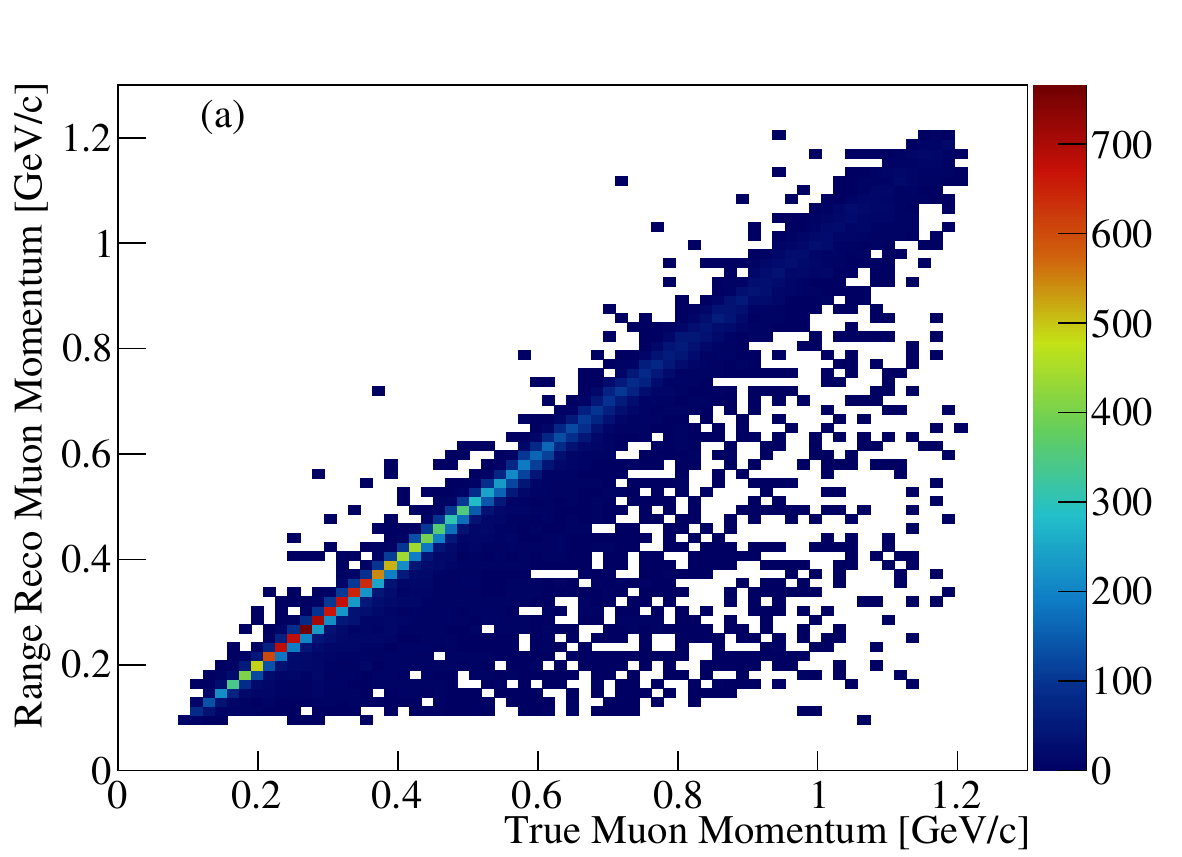}
\includegraphics[width=\linewidth]{\figures 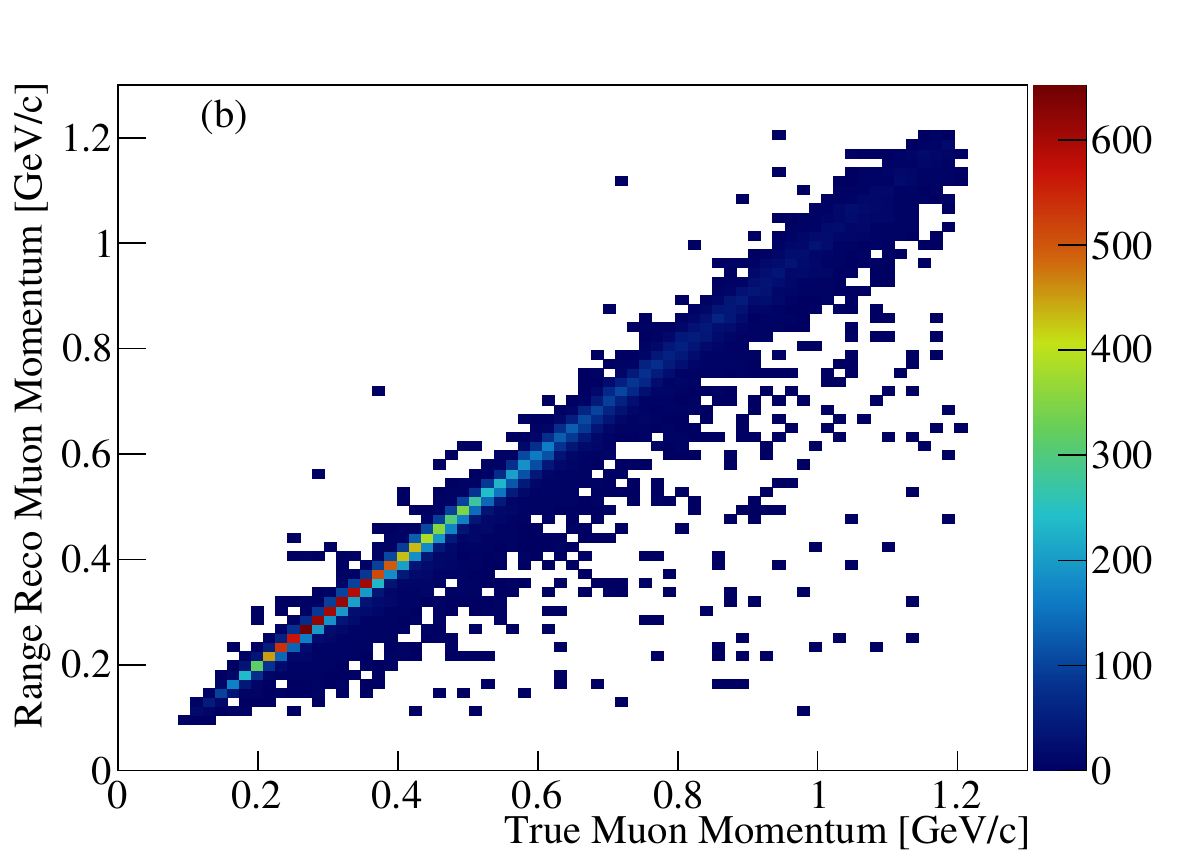}
\caption{Muon momentum reconstruction (top) before and (bottom) after the application of the muon momentum quality cut using contained muon tracks.}
\label{TwoDMuonMomentumQualityCut}
\end{figure}

To further minimize the contribution of mis-reconstructed track directions, we took advantage of two muon momentum reconstruction methods available for contained tracks, namely the momentum from range~\cite{osti_139791} and the momentum from Multiple Coulomb Scattering (MCS)~\cite{Abratenko:2017nki}. 
The range and MCS muon momenta needed to be in agreement within 25\% and the improvement in the muon momentum reconstruction can be seen in Fig.~\ref{TwoDMuonMomentumQualityCut}.
We required that the distance between the track start points and the vertex is smaller than the corresponding distance between the track end points and the vertex.
We also demanded that the distance between the start points of the two candidate tracks is smaller than the distance between the two end points. 

Further reduction of the cosmic tracks and minimization of bin-migration effects is achieved by considering only fully contained candidate muon-proton pairs within a fiducial volume of 10\,cm inside the edge of the detector active volume.
We retain 9051 data events that satisfy all event selection criteria.

In order to provide an accurate description of the dominant cosmic backgrounds pertinent to surface detectors, the full Monte Carlo (MC) simulation consists of a combination of simulated neutrino interactions overlaid on top of beam-off background data.
This approach has been extensively used by MicroBooNE~\cite{Adams:2018lzd,PhysRevLett.125.201803,PhysRevLett.128.151801,PhysRevD.105.L051102}.
The $\texttt{GENIE v3.0.6}$ event generator is used to simulate neutrino interactions with the $\texttt{G18\_10a\_02\_11a}$ configuration~\cite{Andreopoulos:2009rq,Andreopoulos:2015wxa}.
The CCQE and CCMEC predictions have been additionally tuned to T2K $\nu_{\mu}$-carbon CC0$\pi$ data with any number of protons in the final state~\cite{PhysRevD.93.112012,GENIEKnobs}.
The different target nuclei across T2K and MicroBooNE might result in particle reinteraction differences that can affect the reconstructed final state topologies, such as different absorption effects.
Yet, the T2K data sets used for tuning are dominated by CCQE and CCMEC interaction processes, which are the main contributors to the CC1p0$\pi$ topology presented in this work.
Predictions for more complex interactions, such as RES, remain unaltered and no additional MC constraints are applied.
We refer to the corresponding tuned prediction as $\texttt{G18}$.
All the final state particles following the primary neutrino interaction are generated by $\texttt{GENIE}$.
They are further propagated in $\texttt{GENIE}$ through the nucleus to account for FSI.
The propagation of the particles outside the nucleus is simulated using $\texttt{GEANT4}$~\cite{Geant4}.
The \uB\ detector response is modeled using the $\texttt{LArSoft}$ framework~\cite{Pordes:2016ycs,Snider:2017wjd}.
Based on this MC prediction, we obtain a purity of $\approx$ 70\% and an efficiency for selecting \CCIpOpi\ events of $\approx$ 10\%.
The final efficiency is primarily driven by the demand for exactly two fully contained track-like candidates.


\section{Observables}\label{obs}

\begin{figure}[htb!]
\centering 
\includegraphics[width=\linewidth]{\figures 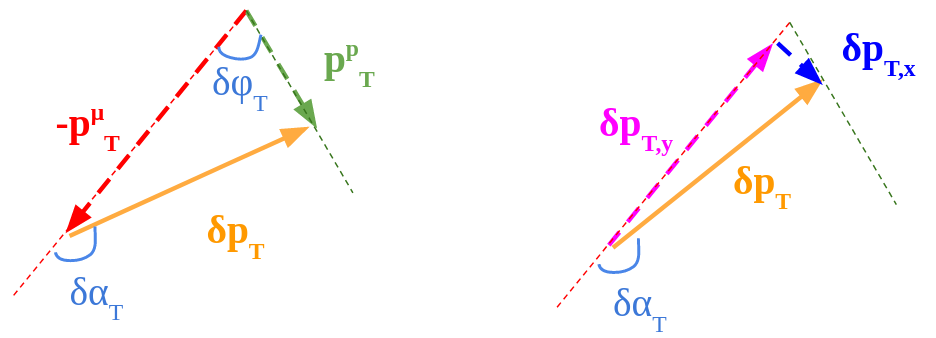}
\caption{
Schematic representation of the kinematic imbalance variables on the plane transverse to the beam direction using CC1p0$\pi$ events.
}
\label{TransVar}
\end{figure}

In neutrino-nucleus scattering events, there is an imbalance between the true initial neutrino momentum and the true sum of final-state lepton and hadron momenta as a result of nuclear effects~\cite{PhysRevC.94.015503}.
A schematic representation of the kinematic imbalance variables of interest in this work is shown in Fig.~\ref{TransVar}.

\begin{figure}[htb!]
\centering 
\includegraphics[width=\linewidth]{\figures 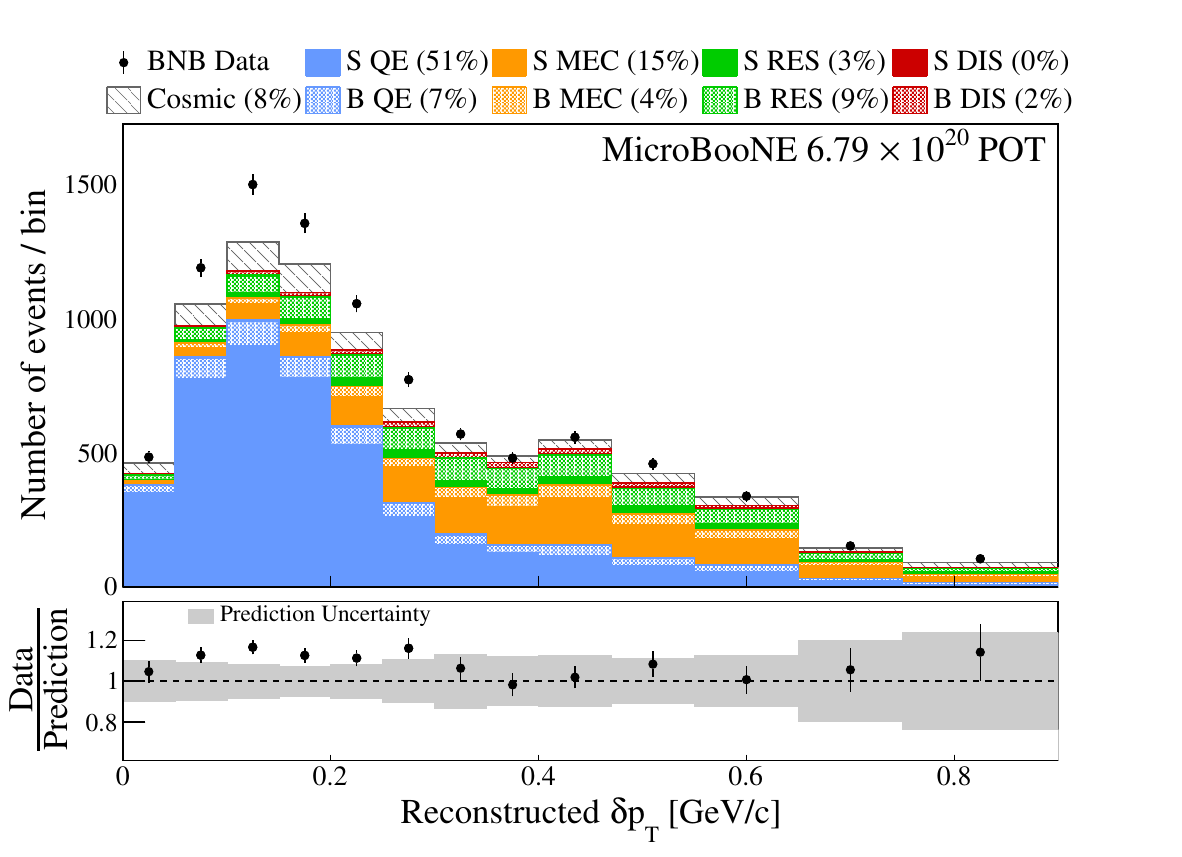}
\caption{
Distribution of the selected CC1p0$\pi$ events as a function of the transverse missing momentum $\delta p_{T}$.
Only statistical uncertainties are shown on the data.
The interaction contributions are obtained from simulation and their separation in signal (S) and background (B) events is presented.
The bottom panel shows the ratio of data to prediction.
}
\label{DeltaPTEvents}
\end{figure}

Using the CC1p0$\pi$ candidate muon-proton pair kinematics, the missing momentum in the plane transverse to the beam direction is defined as 

\begin{equation}
\label{deltapt}
\begin{array}{c}
\delta p_{T} = |\vec{p}_{T}\,^{\mu} + \vec{p}_{T}\,^{p}|,
\end{array}
\end{equation}        
where $\vec{p}_{T}\,^{\mu}$ and $\vec{p}_{T}\,^{p}$ are the projections of the momenta  of  the  outgoing  lepton  and  proton  on  the transverse plane, respectively.
In the absence of nuclear effects, purely QE interactions would yield $\delta p_{T}$ = 0.
In the presence of the dense nuclear medium, this variable encapsulates information related to the Fermi motion, but it is smeared due to FSI and non-QE interactions, as can be seen in Fig.~\ref{DeltaPTEvents}.
Further discussion on the FSI smearing effects can be found in Sec.~\ref{results}.

\begin{figure}[htb!]
\centering 
\includegraphics[width=\linewidth]{\figures 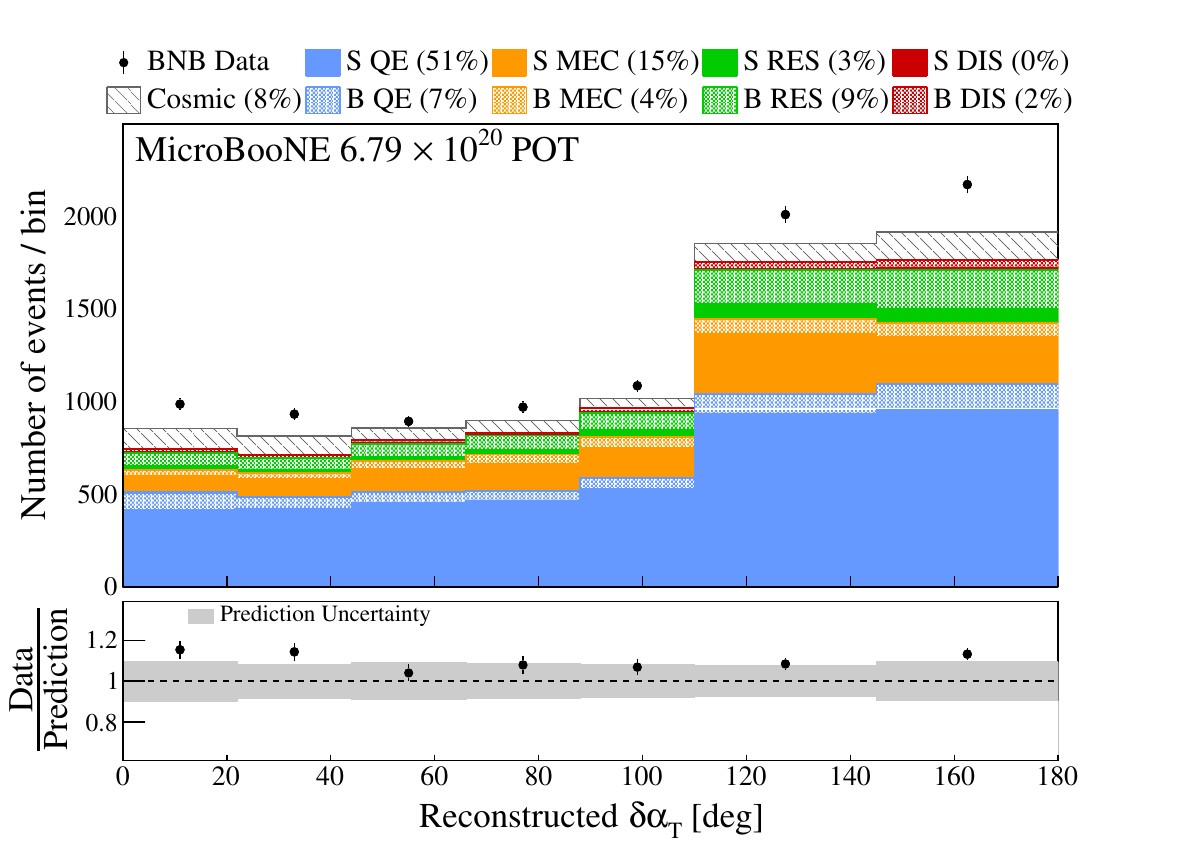}
\caption{
Distribution of the selected CC1p0$\pi$ events as a function of the transverse missing momentum direction $\delta\alpha_{T}$.
Only statistical uncertainties are shown on the data.
The interaction contributions are obtained from simulation and their separation in signal (S) and background (B) events is presented.
The bottom panel shows the ratio of data to prediction.
}
\label{DeltaAlphaTEvents}
\end{figure}

The direction of the transverse momentum imbalance $\delta p_{T}$ is described by the angle

\begin{equation}
\label{deltaalphat}
\begin{array}{c}
\delta \alpha_{T} = \arccos\left(\cfrac{- \vec{p}_{T}\,^{\mu} \cdot \delta \vec{p}_{T}}{p_{T}\,^{\mu} \,\, \delta p_{T}}\right),
\end{array}
\end{equation} 
which is uniformly distributed in the absence of FSI due to the isotropic nature of the Fermi motion.
In the presence of FSI, the proton momentum is generally reduced and the $\delta \alpha_{T}$ distribution becomes weighted towards 180$^{\circ}$, as can be seen in Fig.~\ref{DeltaAlphaTEvents}.

\begin{figure}[htb!]
\centering 
\includegraphics[width=\linewidth]{\figures 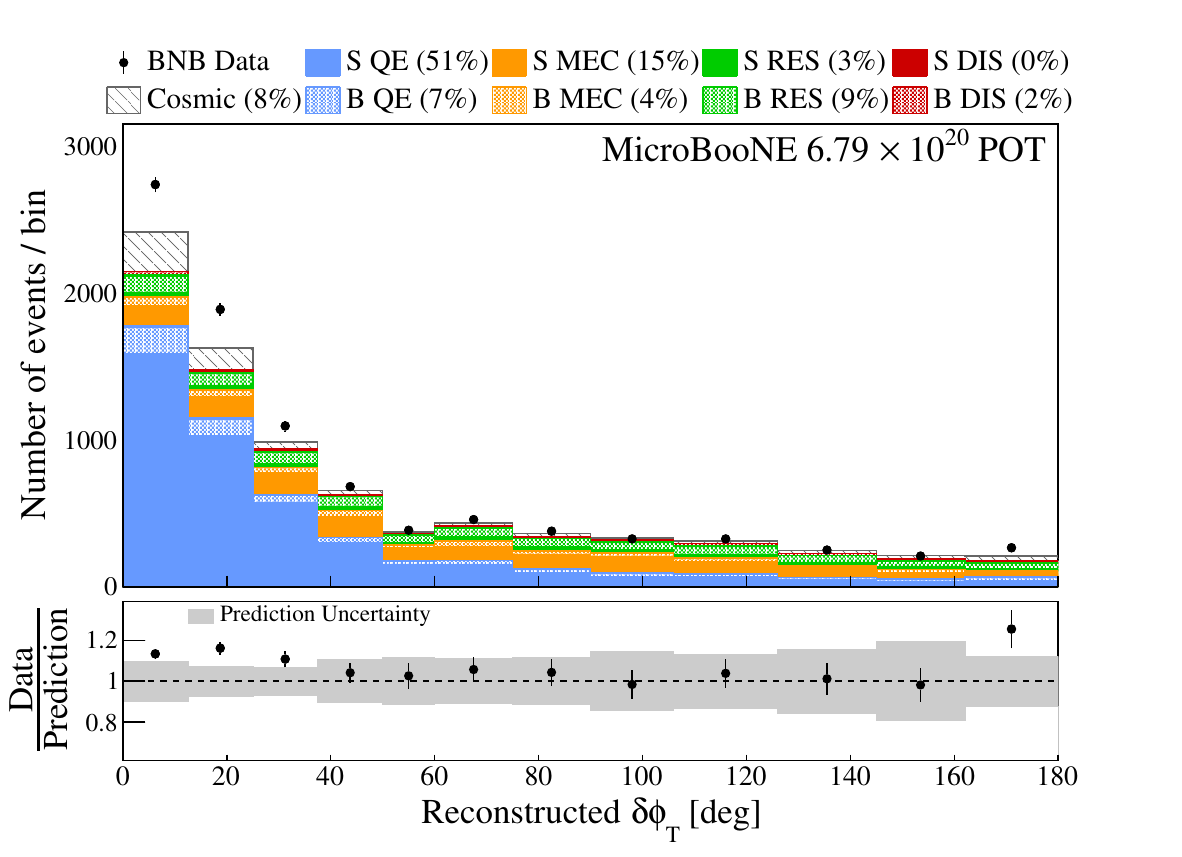}
\caption{
Distribution of the selected CC1p0$\pi$ events as a function of the muon-proton transverse opening angle $\delta\phi_{T}$.
Only statistical uncertainties are shown on the data.
The interaction contributions are obtained from simulation and their separation in signal (S) and background (B) events is presented.
The bottom panel shows the ratio of data to prediction.
}
\label{DeltaPhiTEvents}
\end{figure}

The opening angle $\delta \phi_{T}$ between the correlated candidate muon-proton pair on the transverse plane is given by

\begin{equation}
\label{deltaphit}
\begin{array}{c}
\delta \phi_{T} = \arccos\left(\cfrac{- \vec{p}_{T}\,^{\mu} \cdot \vec{p}_{T}\,^{p}}{p_{T}\,^{\mu} \,\, p_{T}\,^{p}}\right).
\end{array}
\end{equation}
In the absence of nuclear effects, QE events would be concentrated at $\delta\phi_{T}$ = 0.
When nuclear effects are present, QE events can occupy wider angles.
At the same time, non-QE events are dominant in the high $\delta\phi_{T}$ part of the tail and their contribution is fairly flat across all angles, as can be seen in Fig.~\ref{DeltaPhiTEvents}.

\begin{figure}[htb!]
\centering 
\includegraphics[width=\linewidth]{\figures 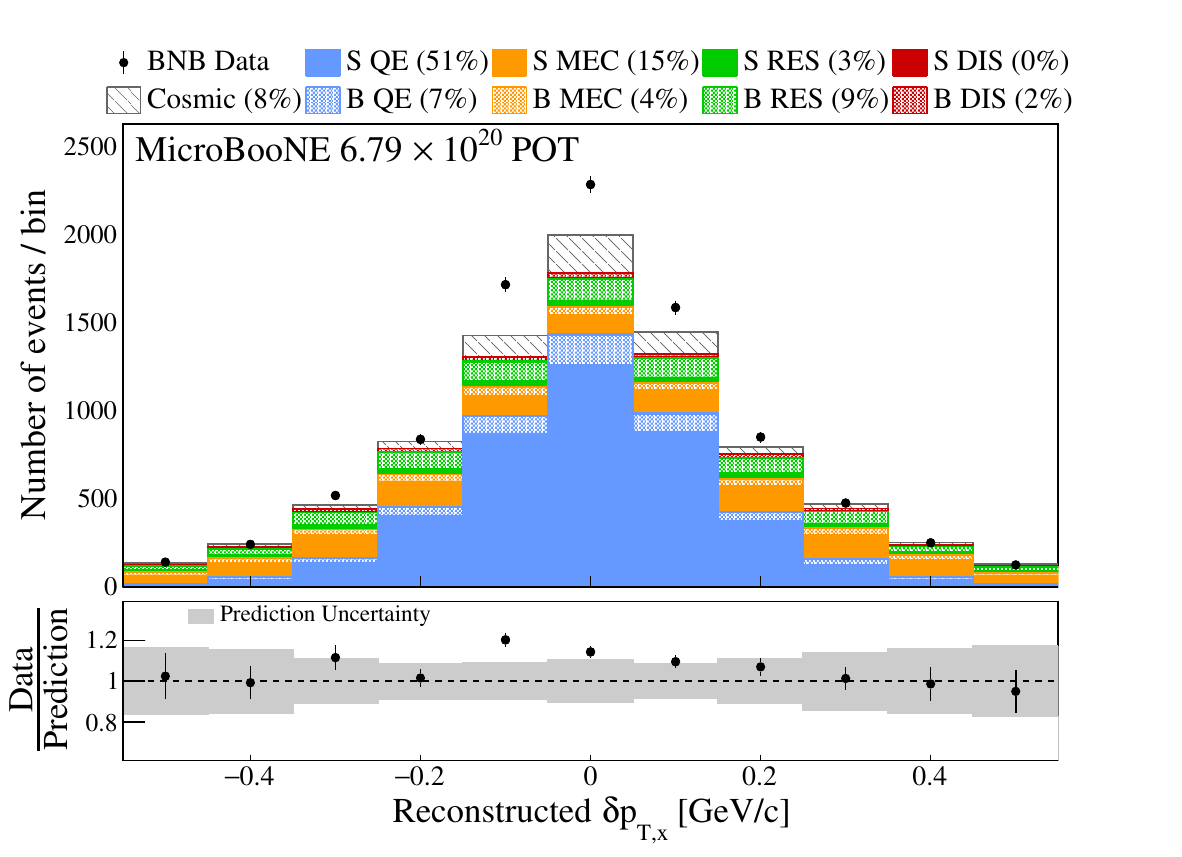}
\caption{
Distribution of the selected CC1p0$\pi$ events as a function of the perpendicular component of the transverse missing momentum $\delta p_{T,x}$.
Only statistical uncertainties are shown on the data.
The interaction contributions are obtained from simulation and their separation in signal (S) and background (B) events is presented.
The bottom panel shows the ratio of data to prediction.
}
\label{DeltaPtxEvents}
\end{figure}

\begin{figure}[htb!]
\centering 
\includegraphics[width=\linewidth]{\figures 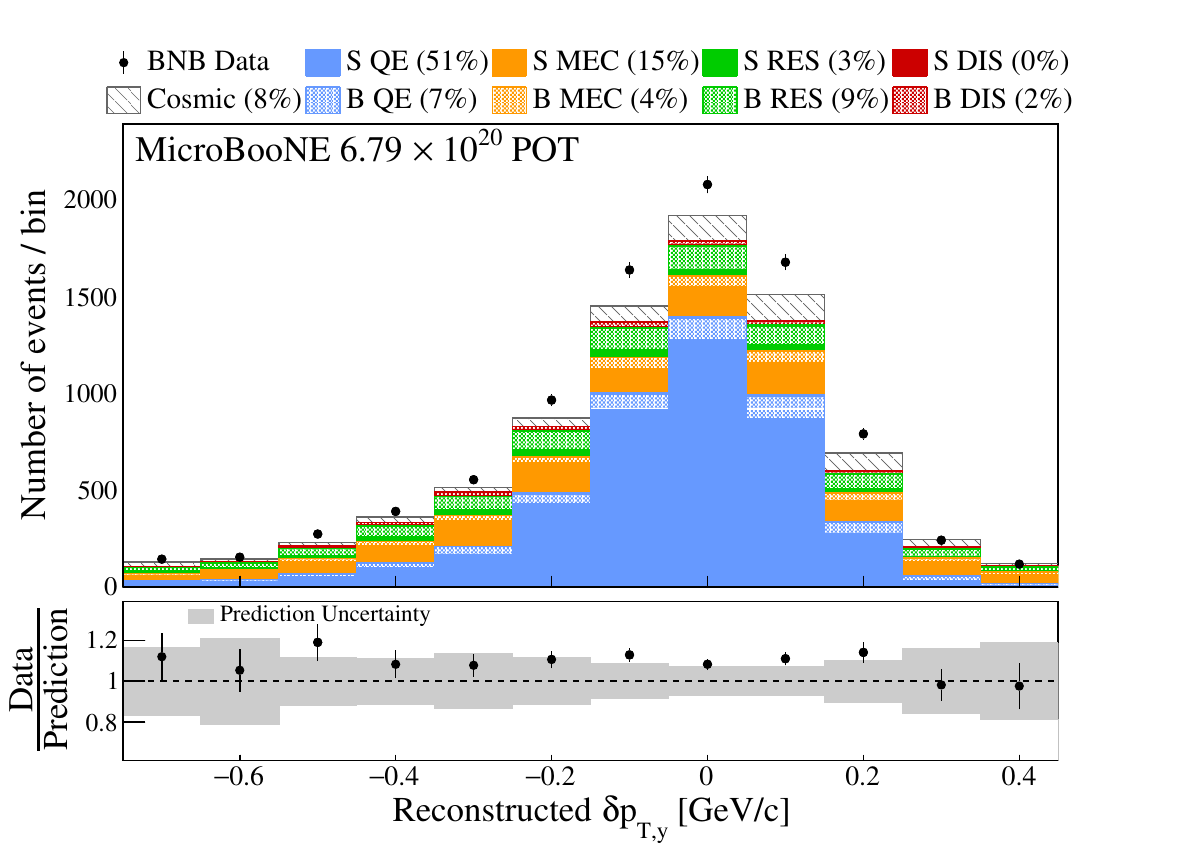}
\caption{
Distribution of the selected CC1p0$\pi$ events as a function of the longitudinal component of the transverse missing momentum $\delta p_{T,y}$.
Only statistical uncertainties are shown on the data.
The interaction contributions are obtained from simulation and their separation in signal (S) and background (B) events is presented.
The bottom panel shows the ratio of data to prediction.
}
\label{DeltaPtyEvents}
\end{figure}

The muon-proton momentum imbalances transverse and longitudinal to the transverse lepton momentum~\cite{PhysRevD.101.092001} are defined as
 	
\begin{equation}
\begin{split}
\delta p_{T,x} = (\hat{p}_{\nu} \times \hat{p}_{T}^{\mu}) \cdot \delta\vec{p}_{T}\\
\delta p_{T,y} = - \hat{p}_{T}^{\mu} \cdot \delta\vec{p}_{T},
\end{split}
\label{imbalanceVect}
\end{equation}	
and can also be written as
 	
\begin{equation}
\begin{split}
\delta p_{T,x} = \delta p_{T} \cdot \sin\delta\alpha_{T}\\
\delta p_{T,y} = \delta p_{T} \cdot \cos\delta\alpha_{T}.
\end{split}
\label{imbalance}
\end{equation}
These distributions can be seen in Fig.~\ref{DeltaPtxEvents} and Fig.~\ref{DeltaPtyEvents}, respectively.
The $\delta p_{T,x}$ distribution is symmetric around 0\,GeV/$c$ due to the presence of the $\sin\delta\alpha_{T}$ factor in Eq.~\ref{imbalance} and the fact that $\delta\alpha_{T}$ ranges from 0$^{\mathrm{o}}$ to 180$^{\mathrm{o}}$.
The width of the distribution is driven by the Fermi motion that affects the $\delta p_{T}$ magnitude.
Unlike $\delta p_{T,x}$, the $\delta p_{T,y}$ distribution is asymmetric with an enhanced contribution from negative values.
The asymmetry is driven by the presence of the $\cos\delta\alpha_{T}$ factor in Eq.~\ref{imbalance} and the fact that $\delta\alpha_{T}$ is mainly peaked around 180$^{\mathrm{o}}$.
Given that the forward $\delta\alpha_{T}$ peak is driven by FSI, the size of the $\delta p_{T,y}$ asymmetry is also sensitive to the FSI strength.

\begin{figure}[htb!]
\centering 
\includegraphics[width=\linewidth]{\figures 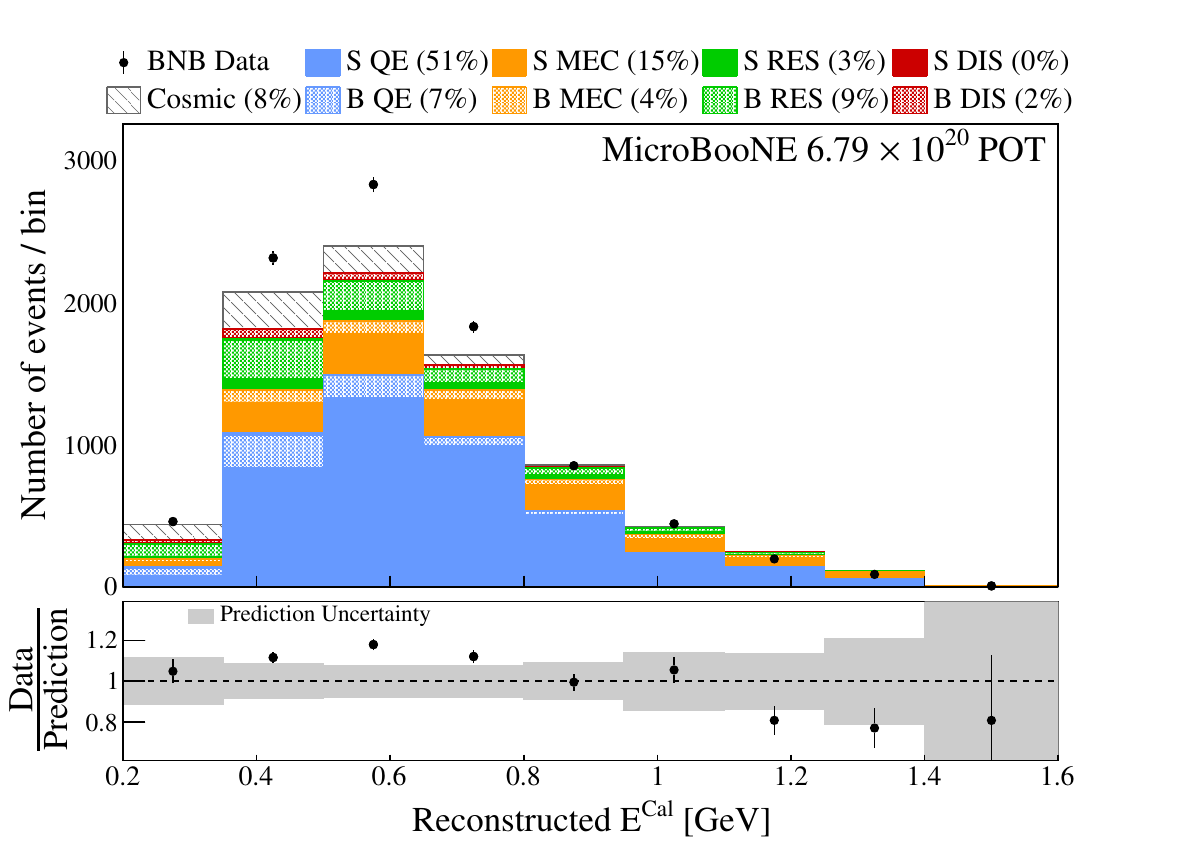}
\caption{
Distribution of the selected CC1p0$\pi$ events as a function of the calorimetric energy reconstruction $E^{Cal}$.
Only statistical uncertainties are shown on the data.
The interaction contributions are obtained from simulation and their separation in signal (S) and background (B) events is presented.
The bottom panel shows the ratio of data to prediction.
}
\label{ECalEvents}
\end{figure}

Finally, the calorimetric energy reconstruction 

\begin{equation}
\begin{split}
E^{Cal} = E_{\mu} + T_{p} + BE
\end{split}
\label{ecaleq}
\end{equation}
is investigated, where $E_{\mu}$ is the muon energy, $T_{p}$ is the proton kinetic energy and $BE$ = 0.04\,GeV is the average binding energy for argon~\cite{BindE}.
This energy estimator, shown in Fig.~\ref{ECalEvents}, is an approximation for the true energy of the incoming neutrino and is used in oscillation searches.

\section{Cross Section Extraction \& Systematics}\label{xsec}

The flux-averaged differential event rate as a function of a given variable $x$ in bin $i$ is obtained by
\begin{equation}
\frac{dR}{dx_{i}} = \frac{N_{i} - B_{i}}{T \cdot \Phi_{\nu} \cdot \Delta_{i}}
\label{fluxnormrate}
\end{equation}
where $N_{i}$ and $B_{i}$ are the number of measured events and the expected background events, respectively.
$T$ is the number of target argon nuclei in the fiducial volume of interest.
$\Phi_{\nu}$ corresponds to the integrated BNB flux and $\Delta_{i}$ corresponds to the $i$-th bin width or area for the single- and double-differential results, respectively.

We report the extracted cross sections for CC1p0$\pi$ interactions using the Wiener singular value decomposition (Wiener-SVD) unfolding technique as a function of unfolded kinematic variables~\cite{Tang_2017}.
This unfolding procedure corrects a measured event rate for inefficiency and resolution effects.
This is achieved by performing a minimization of a $\chi^{2}$ score that compares data to a prediction and allows for a regularization term. 
A Wiener filter determines the level of regularization that is required to minimize the mean square error between the variance and bias of the result. 
In addition to the measured event rate, the method uses a covariance matrix calculated from simulated events accounting for the statistical and systematic uncertainties on the measurement as input.
It also requires the construction of a response matrix describing the expected detector smearing and reconstruction efficiency. 

The output of the method is an unfolded differential cross section, a covariance matrix describing the total uncertainty on the unfolded result, and an additional smearing matrix that we refer to as $A_{C}$. 
The latter contains information about the regularization and bias of the measurement. 
The corresponding $A_{C}$ matrices have been applied to all the cross section predictions included in this work when a comparison to the unfolded data is performed.
The $A_{C}$ matrix should be applied to any independent theoretical prediction when a comparison is performed to the data reported in this paper.
The data release, the unfolded covariance matrices, and the additional matrices $A_{C}$ can be found in the Supplemental Material.

The total covariance matrix $E_{ij}$ = $E^{\mathrm{stat}}_{ij}$ + $E^{\mathrm{syst}}_{ij}$ includes the statistical and systematic uncertainties on the differential event rate associated with our measurement.
$E^{\mathrm{stat}}_{ij}$ is a diagonal covariance matrix with the statistical uncertainties and $E^{\mathrm{syst}}_{ij}$ is a covariance matrix that incorporates the total systematic uncertainties both on the CC1p0$\pi$ signal and on the non-CC1p0$\pi$ background events as detailed below.

The neutrino flux is predicted using the flux simulation of the MiniBooNE collaboration that used the same beam line~\cite{Aguilar-Arevalo:2013dva}. 
Neutrino cross section modeling uncertainties were estimated using the $\texttt{GENIE}$ framework of event reweighting~\cite{Andreopoulos:2009rq,Andreopoulos:2015wxa,GENIEKnobs}.  
The rescattering uncertainties were obtained using $\texttt{GEANT4}$ and the relevant reweighting package~\cite{Calcutt_2021}.
For each of these sources of uncertainty, we use a multisim technique~\cite{Roe:2007hw}, which consists of generating a large number of MC replicas, each one called a ``universe''\kern-.2em, where model parameters are varied within their uncertainties. 
The simultaneous varying of many model parameters provides a correct treatment of their correlations.
A total of $n$ such universes are used to construct a covariance matrix corresponding to each source of uncertainty,
\begin{equation}
E_{ij} = \frac{1}{n} \sum_{k = 1}^{k = n} \left(R^{k}_{i} - R^{CV}_{i}\right) \cdot \left(R^{k}_{j} - R^{CV}_{j}\right)
\label{multicovform}
\end{equation}
where $R^{CV}_{i}$ ($R^{CV}_{j}$) and $R^{k}_{i}$ ($R^{k}_{j}$) are the flux-averaged event rates for the central value and systematic universe $k$ in a measured bin $i$ ($j$), respectively.
The resulting covariance matrices are summed together to estimate the relevant uncertainty from each source.

In order to account for potential biases due to the nominal MC modeling prediction used in the unfolding procedure and presented in Sec.~\ref{model}, an additional cross section uncertainty using the $\texttt{NuWro v19.02.2}$ event generator prediction~\cite{GOLAN2012499} as an alternative universe has been added. 
The relevant $\texttt{NuWro}$ modeling is significantly different when compared to the nominal MC one, as detailed in Sect.~\ref{model}.
The flux-integrated $\texttt{NuWro}$ cross sections are obtained using Eq.~\ref{fluxnormrate} and the corresponding covariance matrices are constructed using Eq.~\ref{multicovform} and a single universe ($n$ = 1).

For detector model systematic uncertainties, one detector parameter is varied each time by $1\sigma$ and is referred to as a ``unisim''.  
These include variations in the light yield, the ionization electron recombination model, space-charge effects, and waveform deconvolution~\cite{WireMod}.
We then examine the impact of each parameter variation on the MC event rates by obtaining the differences with respect to the central value on a bin-by-bin basis.
We define the total detector $1\sigma$ systematic uncertainty by summing in quadrature the effect of $m$ detector variations using the formalism introduced in Eq.~\ref{multicovform},
\begin{equation}
E_{ij} = \sum_{k = 1}^{k = m} \left(R^{k}_{i} - R^{CV}_{i}\right) \cdot \left(R^{k}_{j} - R^{CV}_{j}\right).
\label{unicovform}
\end{equation}

The full fractional uncertainty on the integrated total cross section is 11\% and includes contributions from the neutrino flux prediction (7.3\%), neutrino interaction cross section modeling (6\%), detector response modeling (4.9\%), beam exposure (2.3\%), statistics (1.5\%), number-of-scattering-targets (1.2\%), reinteractions (1\%), and out-of-cryostat (dirt) interaction modeling (0.2\%).
The Supplemental Material includes tables detailing all the cross section uncertainties used in this work.
The main contributors are found to be the strength of the RPA correction and CCMEC cross section shape.
The signal related cross section uncertainties are found to be 8.6\%, while the background ones account for 6.3\%.
Note that the individual contributions are higher than the total cross section uncertainty of 6\% due to correlations between the signal and background events, since the same interaction processes can contribute both as signal and background.

In the results presented below, the inner error bars on the reported cross sections correspond to the statistical uncertainties.
The systematic uncertainties were decomposed into shape- and normalization-related sources following the procedure outlined in~\cite{MatrixDecomv}.
The cross-term uncertainties were incorporated in the normalization part.
The outer error bars on the reported cross sections correspond to statistical and shape uncertainties added in quadrature.
The normalization uncertainties are presented with the gray band at the bottom of each plot.
Overflow (underflow) values are included in the last (first) bin.
The relevant $A_{C}$ matrices have been applied to the theoretical predictions to account for regularization effects.


\section{Modeling Configurations}\label{model}

The nominal MC neutrino interaction prediction ($\texttt{G18}$) uses the local Fermi gas (LFG) model~\cite{Carrasco:1989vq}, the Nieves CCQE scattering prescription~\cite{Nieves:2012yz} which includes Coulomb corrections for the outgoing muon~\cite{Engel:1997fy} and random phase approximation (RPA) corrections~\cite{RPA}.
Additionally, it uses the Nieves MEC model~\cite{Schwehr:2016pvn}, the KLN-BS RES~\cite{Nowak:2009se,Kuzmin:2003ji,Berger:2007rq,Graczyk:2007bc} and Berger-Sehgal coherent (COH)~\cite{Berger:2008xs} scattering models, the hA2018 FSI model~\cite{Ashery:1981tq}, and MicroBooNE-specific tuning of model parameters~\cite{GENIEKnobs}.

Our results are also compared to a number of alternative event generators.
$\texttt{GiBUU 2021 (GiBUU)}$ uses similar models, but they are implemented in a coherent way by solving the Boltzmann-Uehling-Uhlenbeck transport equation~\cite{Mosel:2019vhx}. 
The modeling includes the LFG model~\cite{Carrasco:1989vq}, a standard CCQE expression~\cite{Leitner:2006ww}, an empirical MEC model and a dedicated spin dependent resonance amplitude calculation following the $\texttt{MAID}$ analysis~\cite{Mosel:2019vhx}. 
The DIS model is from $\texttt{PYTHIA}$~\cite{Sjostrand:2006za}.
$\texttt{GiBUU}$'s FSI treatment propagates the hadrons through the residual nucleus in a nuclear potential which is consistent with the initial state.
$\texttt{NuWro v19.02.2 (NuWro)}$ uses the LFG model~\cite{Carrasco:1989vq}, the Llewellyn Smith  model for QE events~\cite{LlewellynSmith:1971uhs}, the Nieves model for MEC events~\cite{ValenciaModel}, the Adler-Rarita-Schwinger formalism to calculate the $\Delta$ resonance explicitly~\cite{Graczyk:2007bc}, the BS COH~\cite{Berger:2008xs} scattering model and an intranuclear cascade model for FSI~\cite{ValenciaModel}.
$\texttt{NEUT v5.4.0 (NEUT)}$ uses the LFG model~\cite{Carrasco:1989vq}, the Nieves CCQE scattering prescription~\cite{Nieves:2012yz}, the Nieves MEC model~\cite{Schwehr:2016pvn}, the BS RES~\cite{Nowak:2009se,Kuzmin:2003ji,Berger:2007rq,Graczyk:2007bc} and BS COH~\cite{Berger:2008xs} scattering models, and FSI with Oset medium corrections for pions~\cite{Andreopoulos:2009rq,Andreopoulos:2015wxa}.

In addition to the alternative event generators, our results are compared to a number of different $\texttt{GENIE}$ configurations.
These include an older version, $\texttt{GENIE v2.12.10 (Gv2)}$~\cite{Andreopoulos:2009rq,Andreopoulos:2015wxa}, which uses the Bodek-Ritchie Fermi Gas model, the Llewellyn Smith CCQE scattering prescription~\cite{LlewellynSmith:1971uhs}, the empirical MEC model~\cite{Katori:2013eoa}, a Rein-Sehgal RES and COH scattering model~\cite{Rein:1980wg}, and a data driven FSI model denoted as ``hA''~\cite{Mashnik:2005ay}.
Another model, ``$\texttt{Untuned}$'', uses the $\texttt{GENIE v3.0.6 G18\_10a\_02\_11a}$ configuration without additional MicroBooNE-specific tuning.
Finally, the newly added theory-driven $\texttt{GENIE v3.2.0 G21\_11b\_00\_000}$ configuration ($\texttt{G21}$) is shown. 
This includes the SuSAv2 prediction for the QE and MEC scattering parts~\cite{PhysRevD.101.033003} and the hN2018 FSI model~\cite{hN2018}.	
The modeling options for RES, DIS, and COH interactions are the same as for $\texttt{G18}$.

To quantify the data-simulation agreement, the $\chi^{2}$/bins ratio data comparison for each generator is shown on all the figures and is calculated by taking into account the total covariance matrix.
Ratios close to unity are indicative of a sufficiently accurate modeling performance.
Theoretical uncertainties on the models themselves are not included.


\section{Results}\label{results}



\begin{figure*}[htb!]
\centering
\includegraphics[width=0.49\linewidth]{\figures 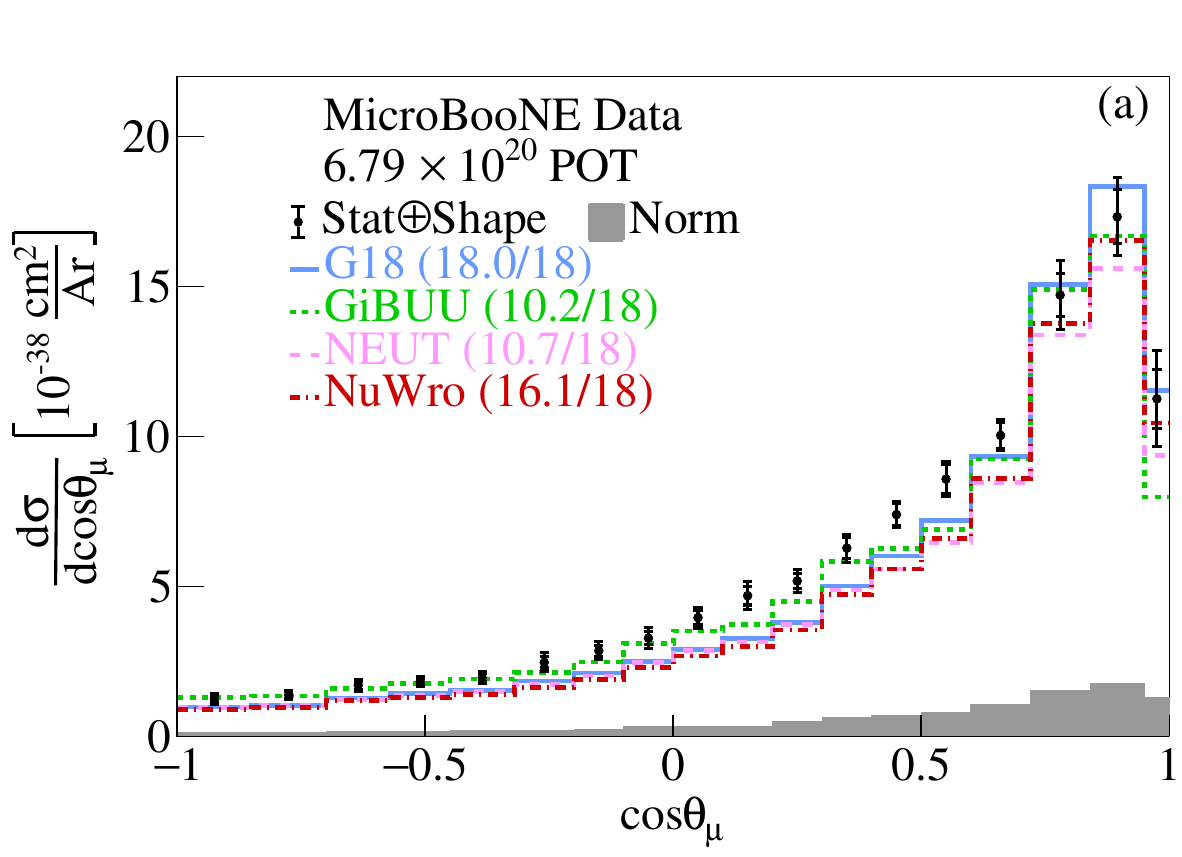}
\includegraphics[width=0.49\linewidth]{\figures 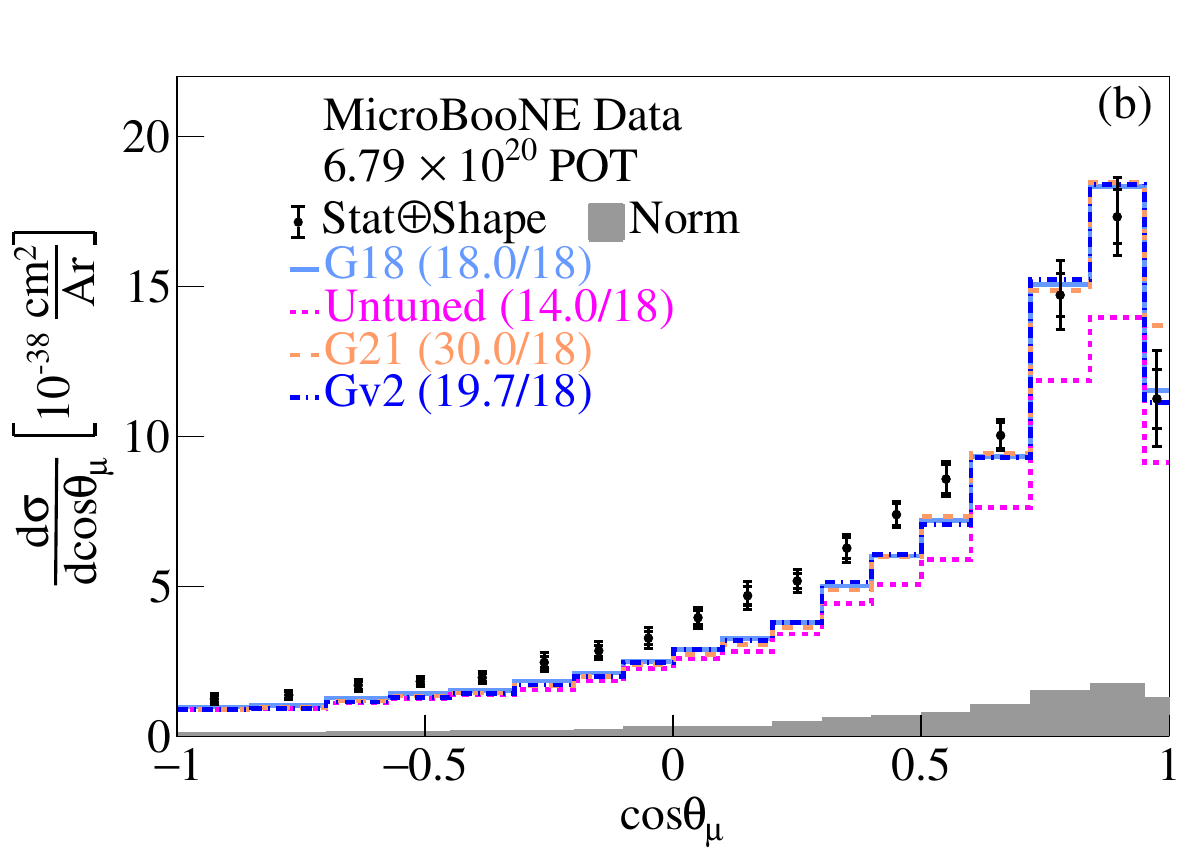}\\
\caption{
The flux-integrated single-differential cross sections as a function of cos$\theta_{\mu}$. 
(a) Generator and (b) $\texttt{GENIE}$ configuration predictions are compared to data.
Inner and outer error bars show the statistical and total (statistical and shape systematic) uncertainty at the 1$\sigma$, or 68\%, confidence level. 
The gray band shows the normalization systematic uncertainty.
The numbers in parentheses show the $\chi^{2}$/bins calculation for each one of the predictions.
}
\label{CosThetaMuGen}
\end{figure*}

Along with the aforementioned kinematic imbalance and energy estimator results, the data are also presented as a function of the lepton angular orientation (Fig.~\ref{CosThetaMuGen}). 
Previous MicroBooNE measurements using different signal definitions~\cite{PhysRevLett.125.201803,PhysRevLett.123.131801,PhysRevD.102.112013} showed discrepancies in that quantity, primarily in the forward direction.
These analyses used an older simulation prediction, namely $\texttt{GENIE v2.12.2}$, to account for the efficiency corrections and beam-induced backgrounds. 
This work illustrates that all generator (Fig.~\ref{CosThetaMuGen}a) and $\texttt{GENIE}$ configuration (Fig.~\ref{CosThetaMuGen}b) predictions are in good agreement with the data when reported as a function of cos$\theta_{\mu}$.



\begin{figure*}[htb!]
\centering
\includegraphics[width=0.49\linewidth]{\figures 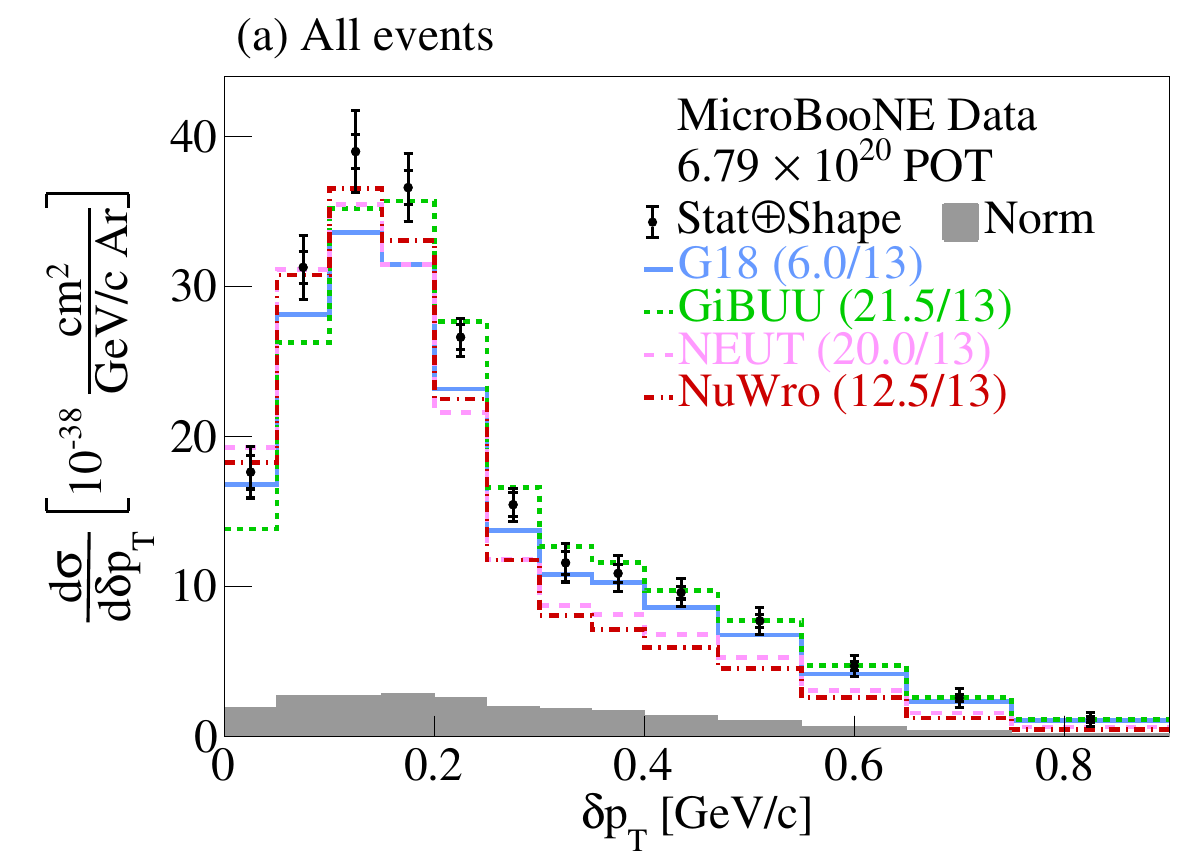}\\
\includegraphics[width=0.49\linewidth]{\figures 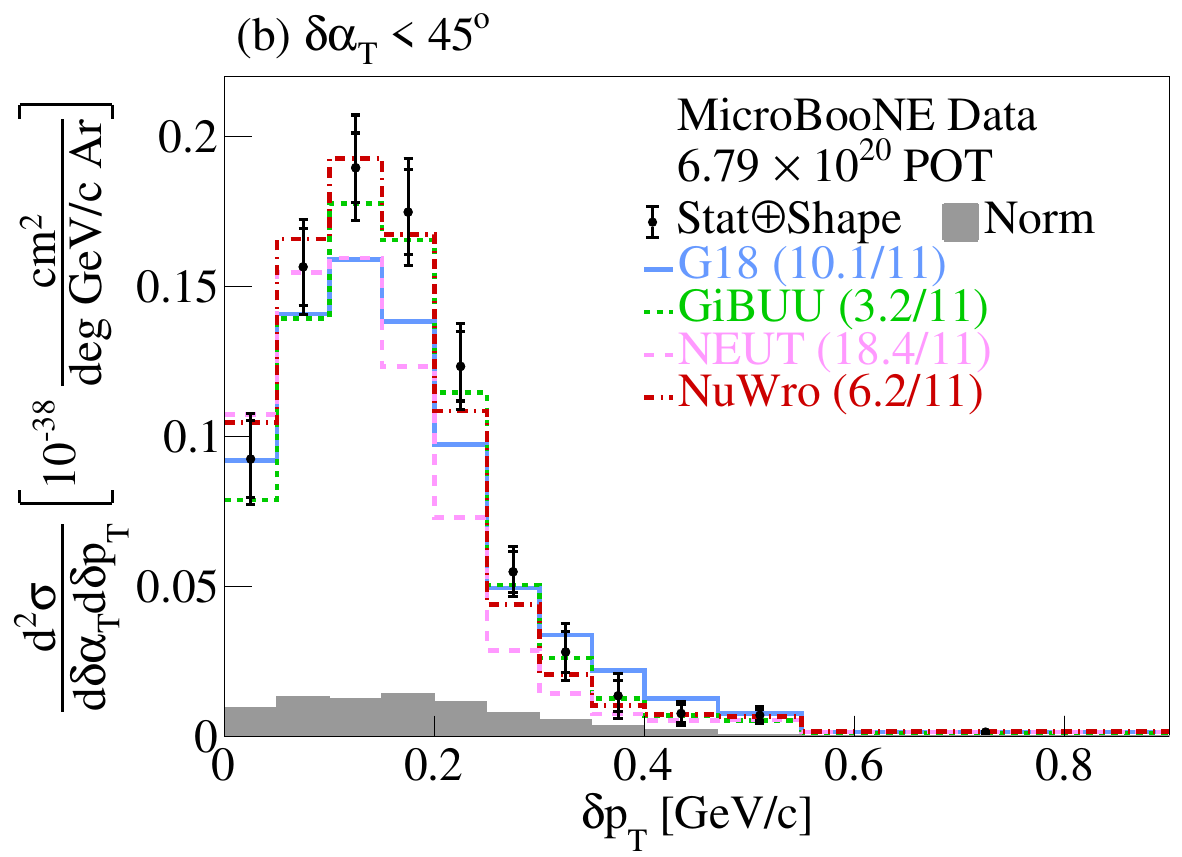}
\includegraphics[width=0.49\linewidth]{\figures 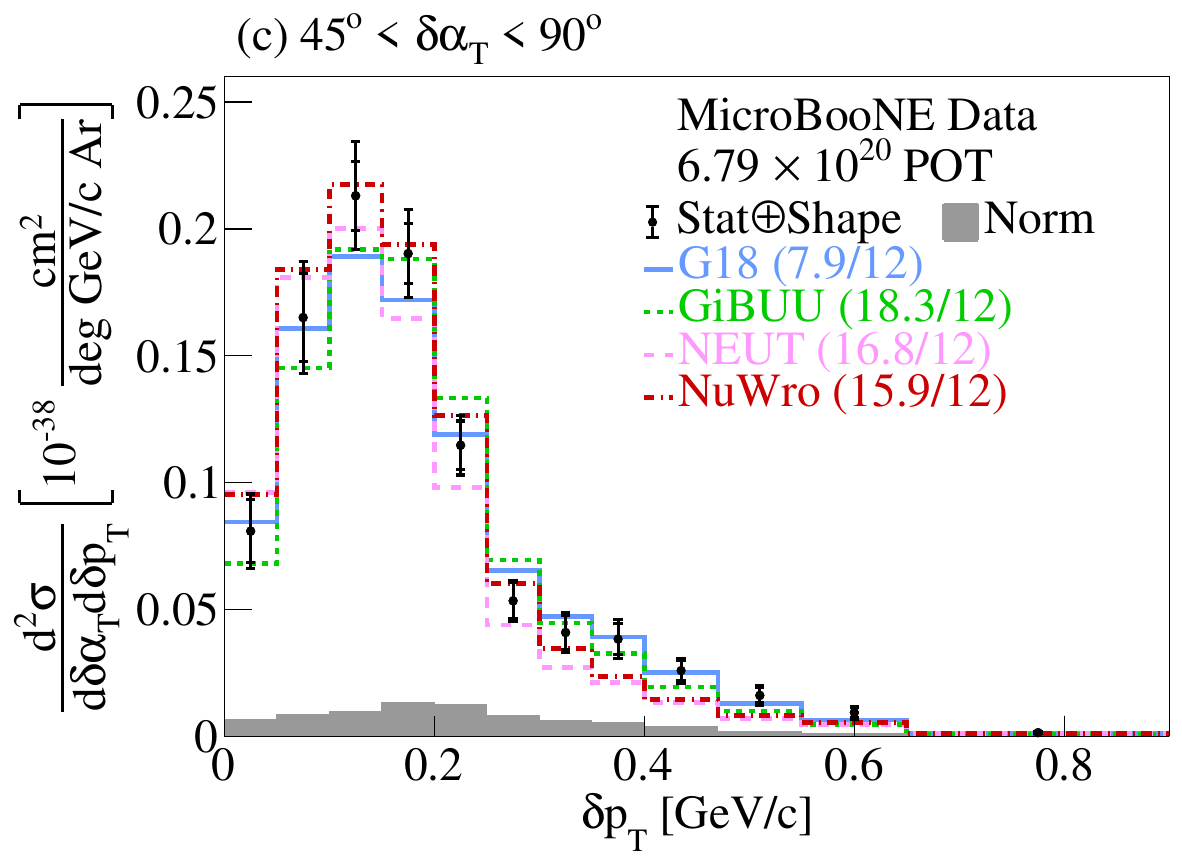}\\
\includegraphics[width=0.49\linewidth]{\figures 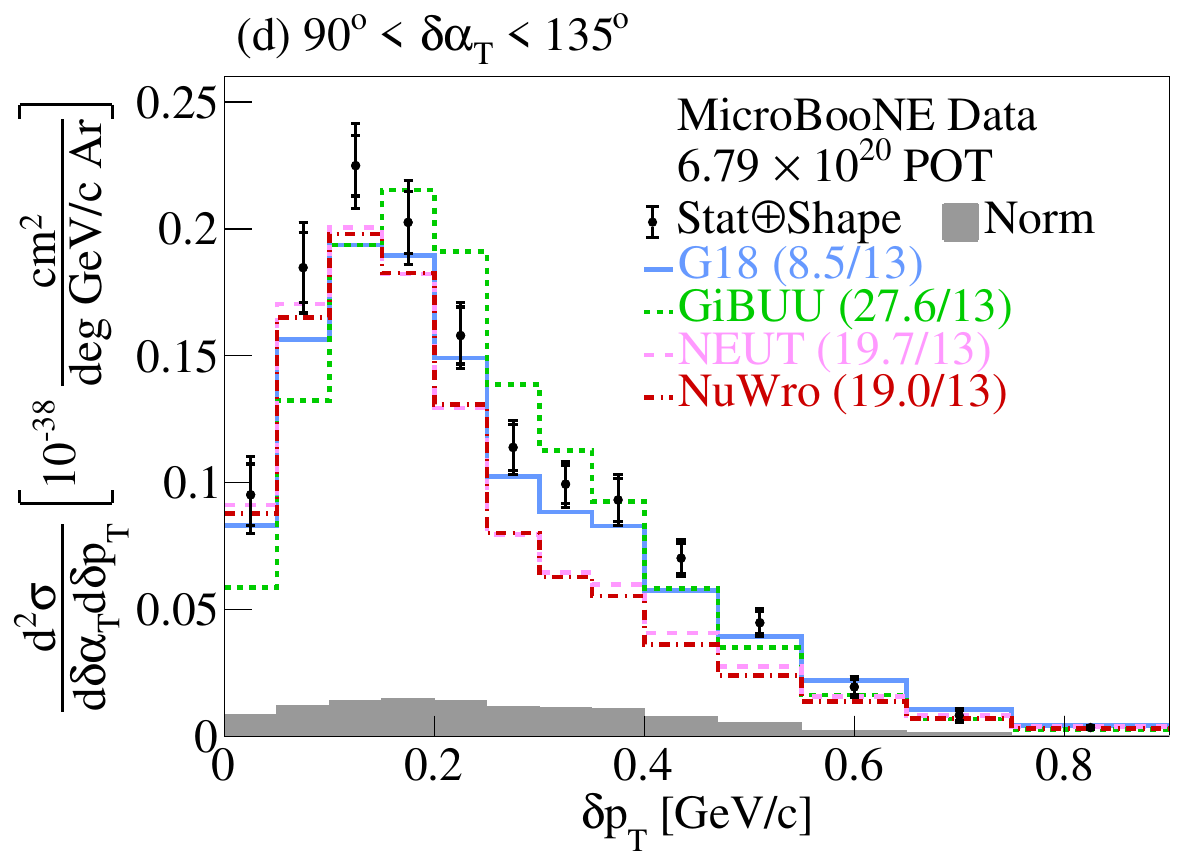}
\includegraphics[width=0.49\linewidth]{\figures 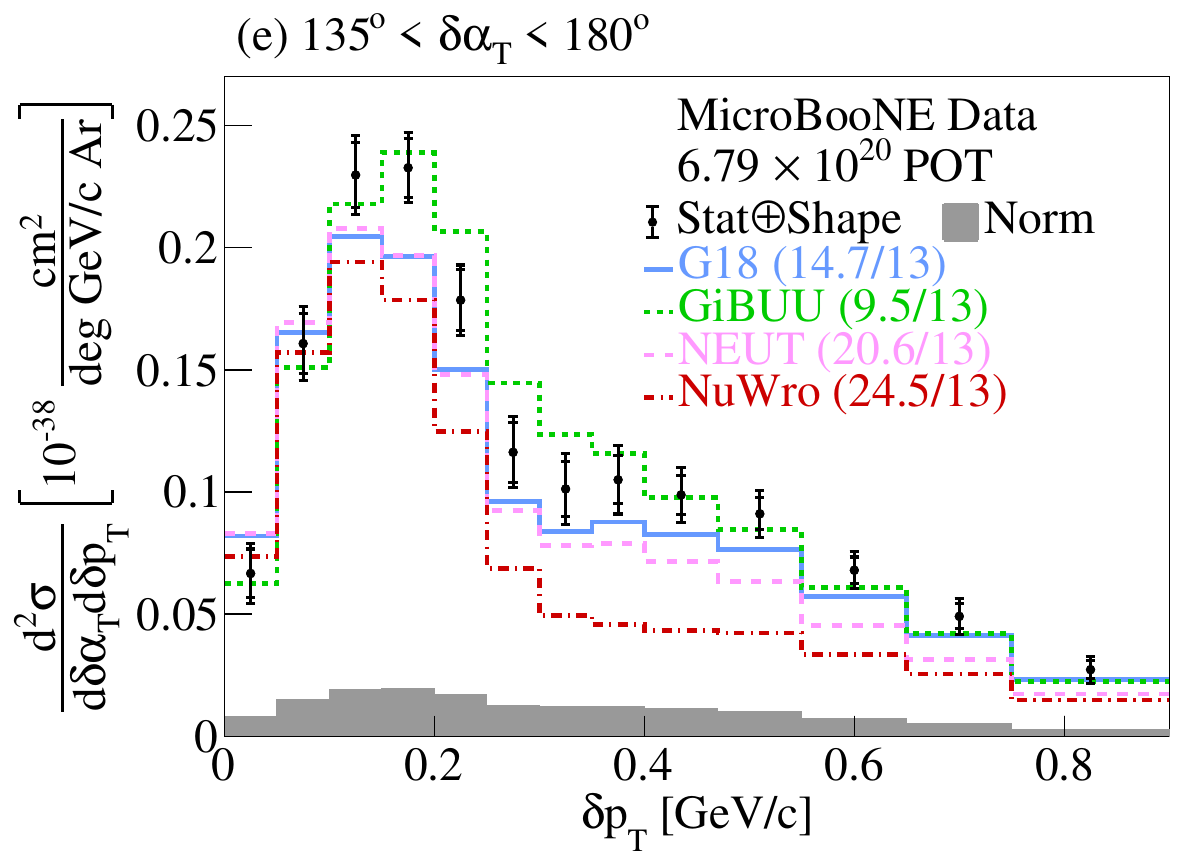}\\
\caption{
The flux-integrated (a) single- and (b-e) double- (in $\delta\alpha_{T}$ bins) differential cross sections as a function of $\delta p_{T}$. 
Inner and outer error bars show the statistical and total (statistical and shape systematic) uncertainty at the 1$\sigma$, or 68\%, confidence level. 
The gray band shows the normalization systematic uncertainty.
Colored lines show the results of theoretical cross section calculations using the $\texttt{G18 GENIE}$ (blue), $\texttt{GiBUU}$ (green), $\texttt{NEUT}$ (pink), and $\texttt{NuWro}$ (red) event generators.
The numbers in parentheses show the $\chi^{2}$/bins calculation for each one of the predictions.
}
\label{DeltaPTInDeltaAlphaTGen}
\end{figure*}

\begin{figure*}[htb!]
\centering
\includegraphics[width=0.49\linewidth]{\figures 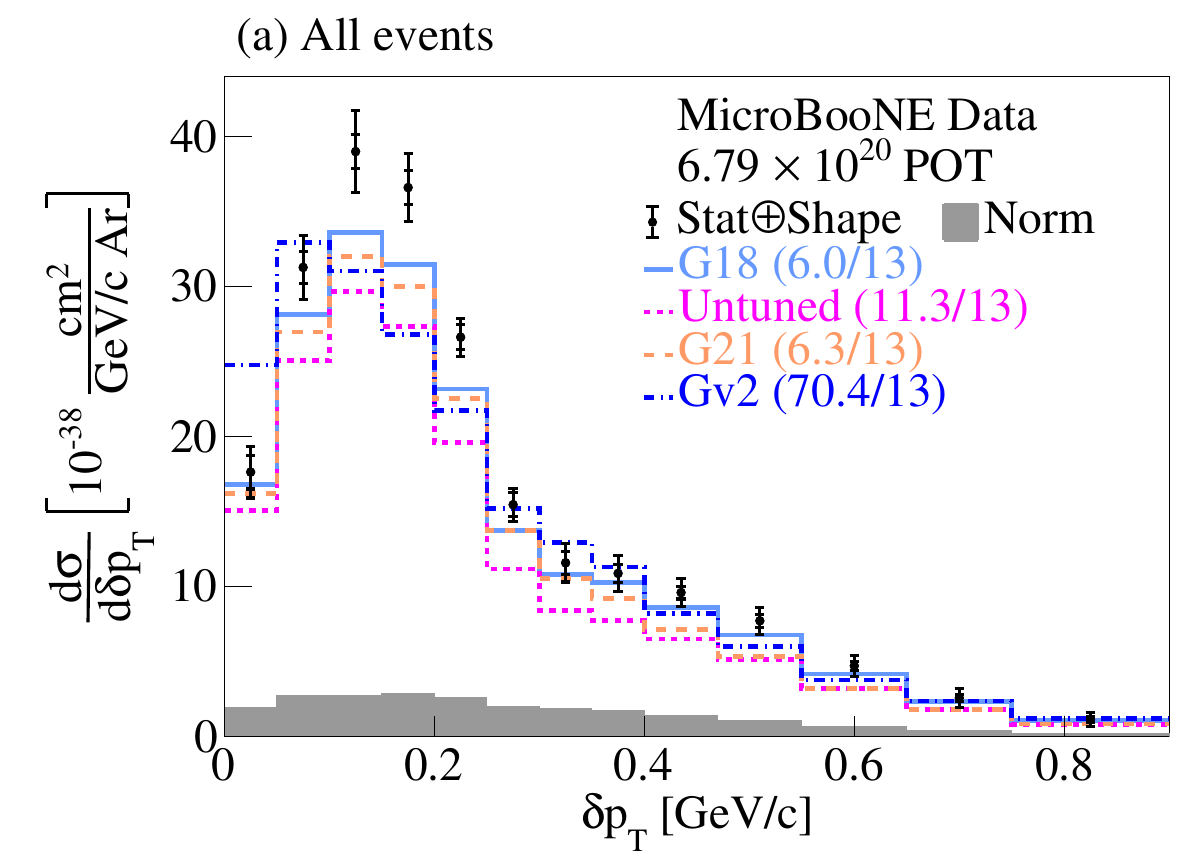}\\
\includegraphics[width=0.49\linewidth]{\figures 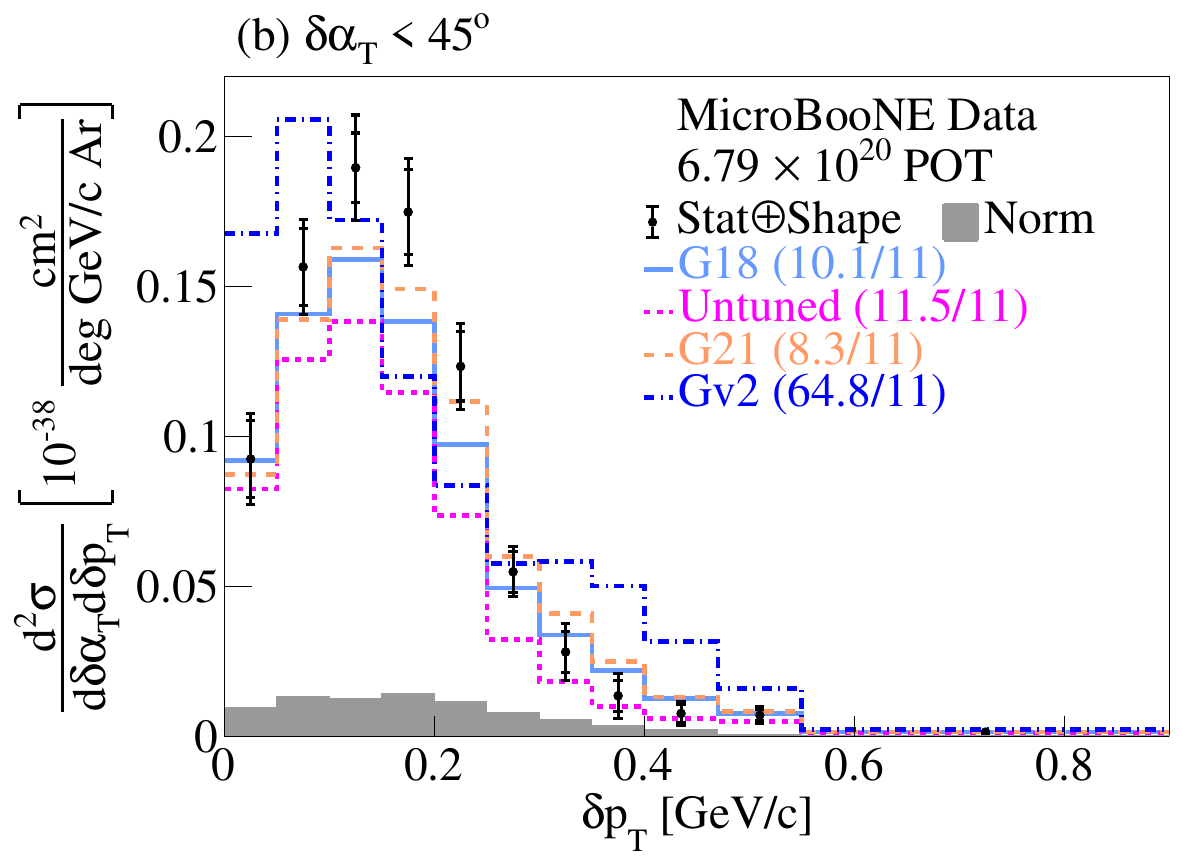}
\includegraphics[width=0.49\linewidth]{\figures 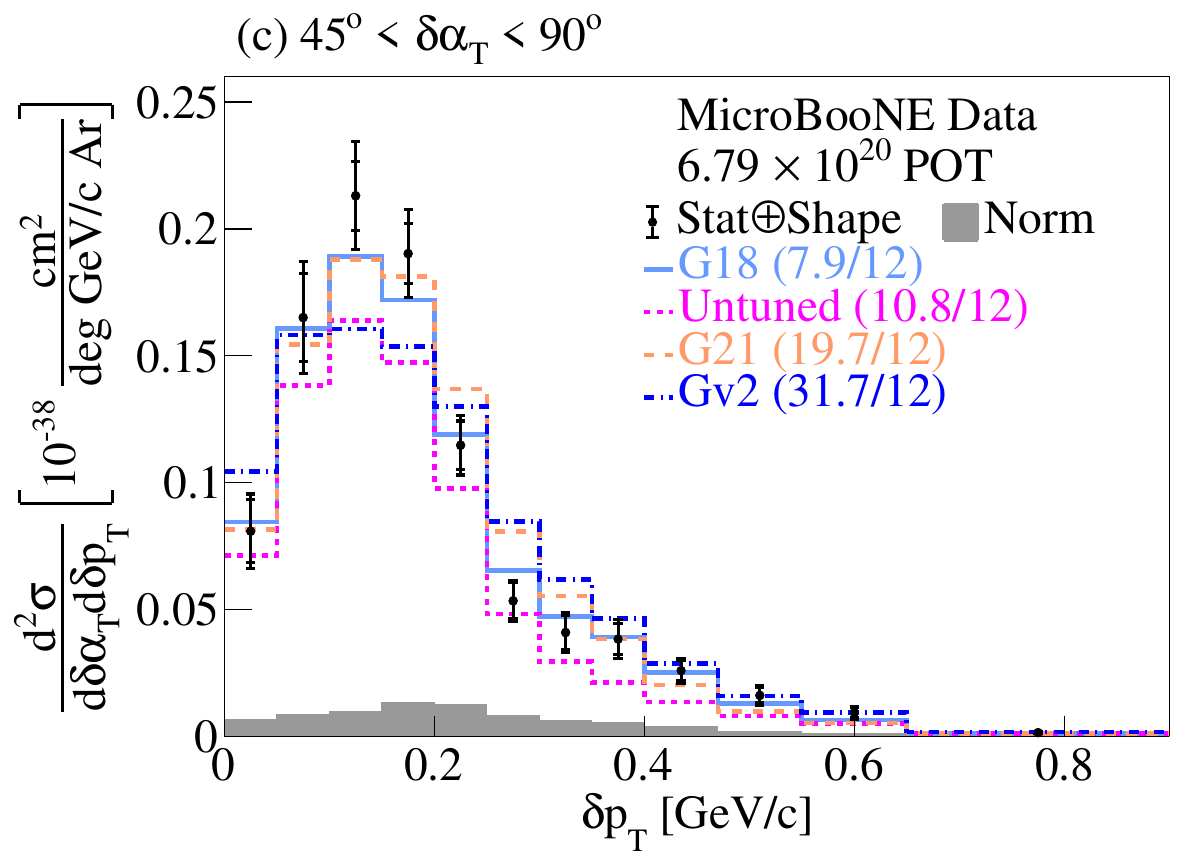}\\
\includegraphics[width=0.49\linewidth]{\figures 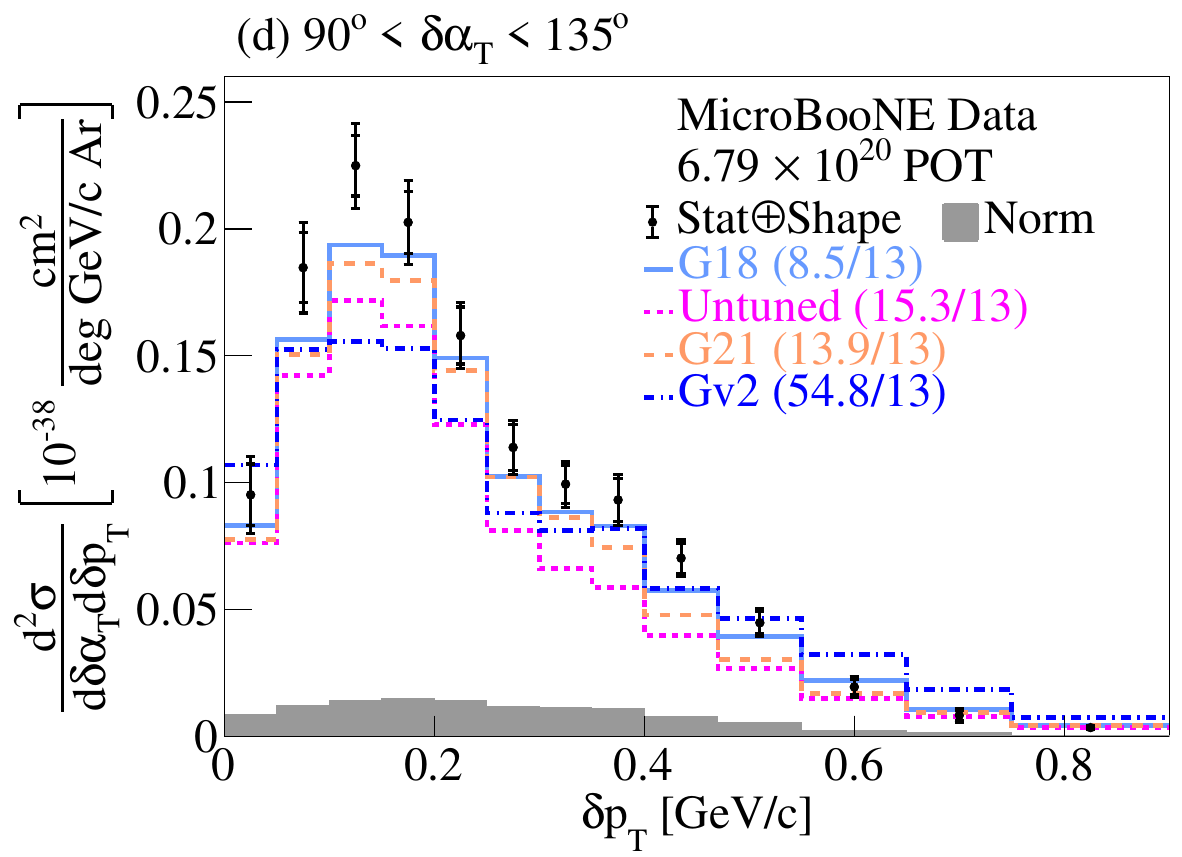}
\includegraphics[width=0.49\linewidth]{\figures 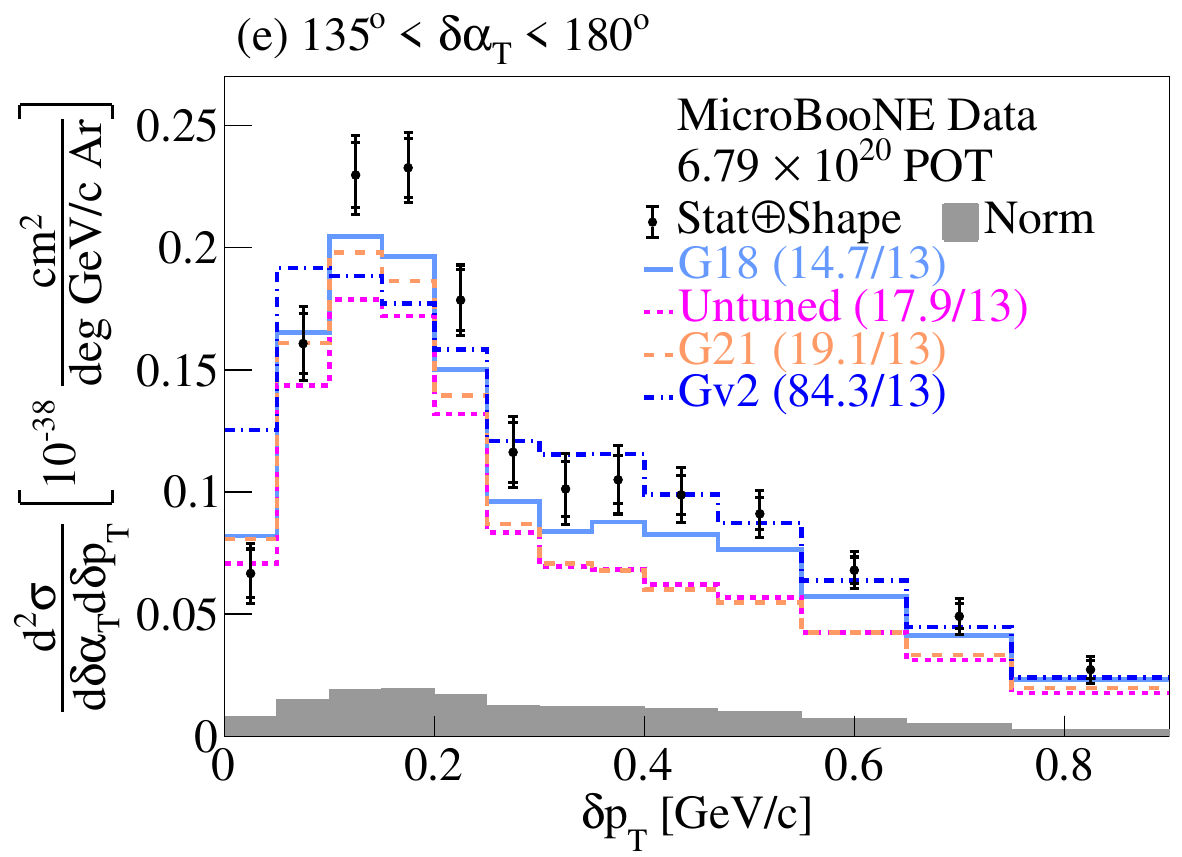}\\
\caption{
The flux-integrated (a) single- and (b-e) double- (in $\delta\alpha_{T}$ bins) differential cross sections as a function of $\delta p_{T}$. 
Inner and outer error bars show the statistical and total (statistical and shape systematic) uncertainty at the 1$\sigma$, or 68\%, confidence level. 
The gray band shows the normalization systematic uncertainty.
Colored lines show the results of theoretical cross section calculations using the $\texttt{G18}$ (light blue), $\texttt{Untuned}$ (magenta), $\texttt{G21}$ (orange), and $\texttt{Gv2}$ (dark blue) $\texttt{GENIE}$ configurations.
The numbers in parentheses show the $\chi^{2}$/bins calculation for each one of the predictions.
}
\label{DeltaPTInDeltaAlphaTGenie}
\end{figure*}

\begin{figure*}[htb!]
\centering 
\includegraphics[width=0.48\linewidth]{\figures 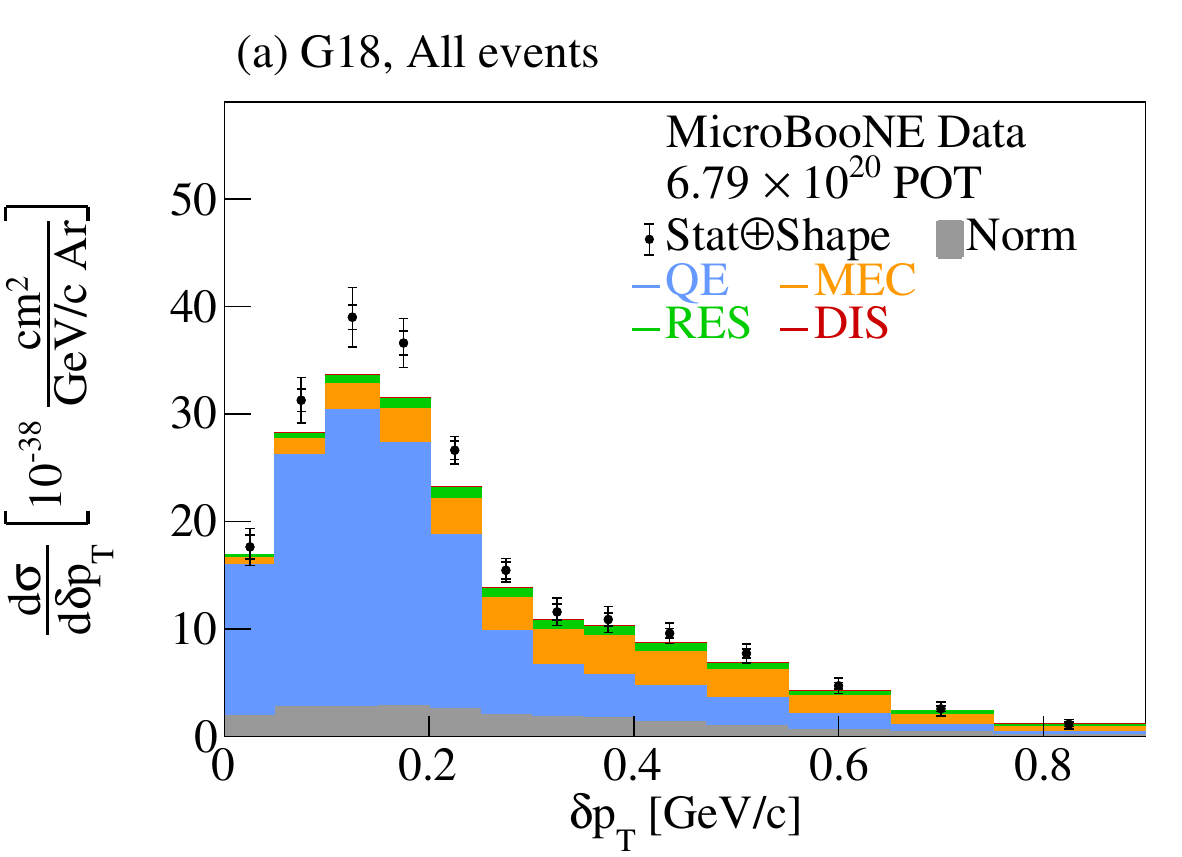}
\includegraphics[width=0.48\linewidth]{\figures 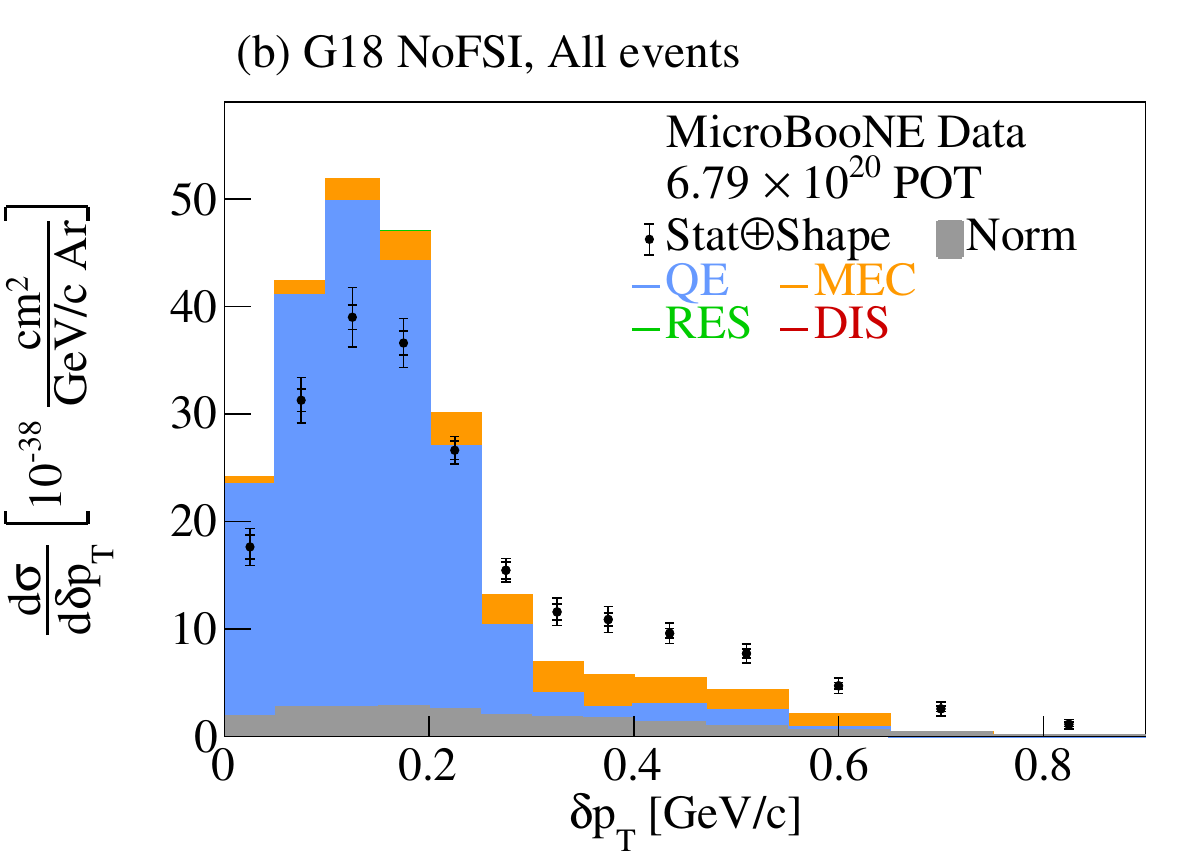}
\caption{Cross section interaction breakdown for the selected events for the G18 configuration (left) with FSI effects, and (right) without FSI effects as a function of $\delta p_{T}$.
}
\label{DeltaPTBreakdownPRD}
\end{figure*}

Figures~\ref{DeltaPTInDeltaAlphaTGen} and~\ref{DeltaPTInDeltaAlphaTGenie} show the measured single-differential cross sections as a function of $\delta p_{T}$ using all the events (panel a), as well as the double-differential results as a function of the same kinematic variable in $\delta\alpha_{T}$ bins (panels b-e).
In the presence of FSI, the proton can rescatter or be absorbed, yielding larger kinematic imbalances on the transverse plane and $\delta p_{T}$ values that extend beyond the Fermi momentum, as can be seen in Fig.~\ref{DeltaPTBreakdownPRD}.
Furthermore, the same extended tail can be obtained when pions produced due to multi-nucleon effects (MEC or RES) are either absorbed or below the detection threshold.
The single-differential result shows such a high-momentum tail that extends above 0.8\,GeV/$c$.
This picture is consistent with the results reported by the T2K and MINERvA collaborations~\cite{Avanzini:2021qlx,Abe:2018pwo,PhysRevLett.121.022504}.
Unlike the single-differential result, the double differential results with low $\delta\alpha_{T}$ extend only slightly above 0.4\,GeV/$c$.
That indicates that this region contains minimal FSI and multi-nucleon effects and the $\delta p_{T}$ distribution is driven by the nucleon Fermi motion.
On the other hand, the higher $\delta\alpha_{T}$ values correspond to $\delta p_{T}$ distributions that extend beyond 0.8\,GeV/$c$.
This behavior is indicative of the presence of FSI and multi-nucleon effects that smear the $\delta p_{T}$ distribution to higher values.
Future multi-differential results can help further disentangle the contributions from these effects. 
Figure~\ref{DeltaPTInDeltaAlphaTGen} shows the comparisons to a number of available neutrino event generators with $\texttt{NuWro}$ and $\texttt{G18}$ showing the best agreement over all events.
Figure~\ref{DeltaPTInDeltaAlphaTGenie} shows the same results compared to a number of $\texttt{GENIE}$ configurations illustrating that $\texttt{Gv2}$ is disfavored, an observation that is driven by the $\texttt{Gv2}$ low $\delta p_{T}$ behavior.
Furthermore, $\texttt{Untuned}$ shows a good $\chi^{2}$/bins performance across all slices but predicts lower values than data.
Additionally, Fig.~\ref{DeltaPTBreakdownPRD} shows the effect of final state interactions (FSI) on the CC1p0$\pi$ selection using the $\text{G18}$ configuration of $\text{GENIE}$.
The addition of FSI allows for more non-QE events to satisfy the CC1p0$\pi$ signal definition that smear the $\delta p_{T}$ distribution to higher values.



\begin{figure*}[htb!]
\centering 
\includegraphics[width=0.49\linewidth]{\figures 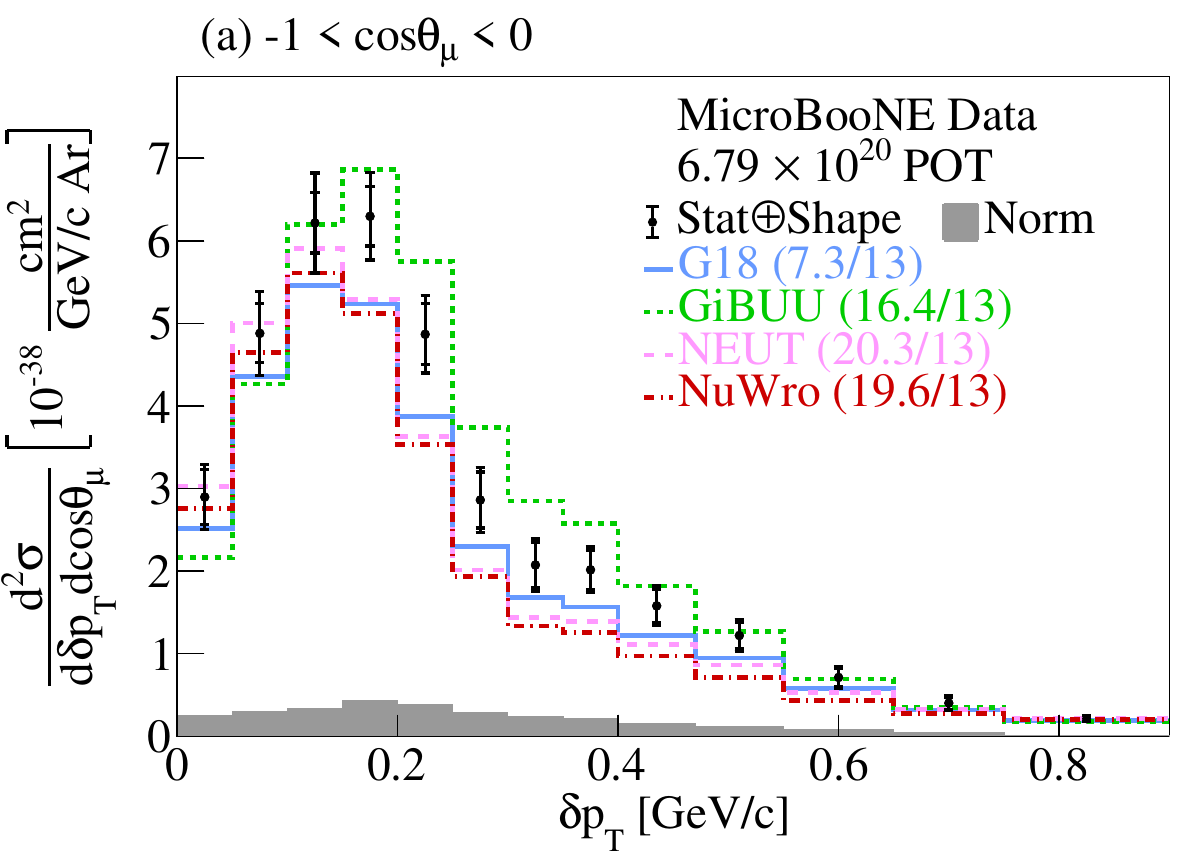}
\includegraphics[width=0.49\linewidth]{\figures 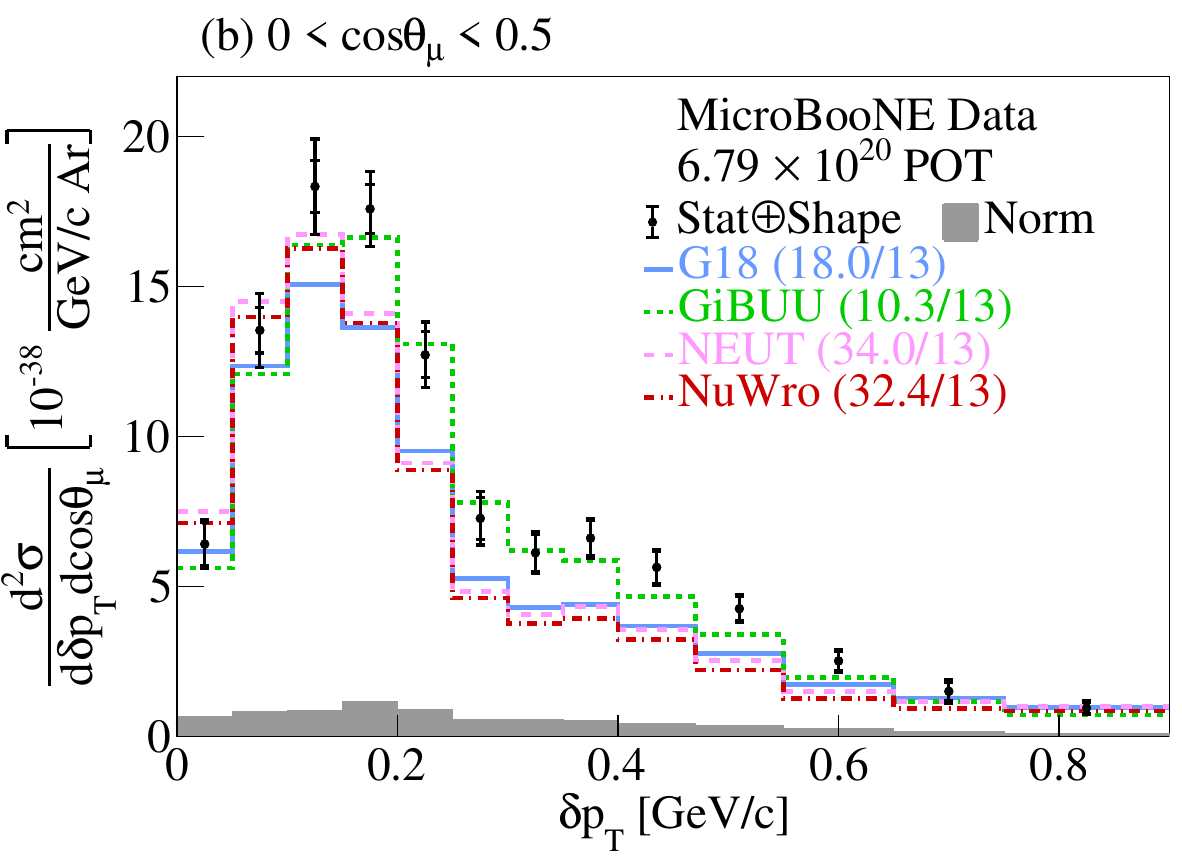}\\
\includegraphics[width=0.49\linewidth]{\figures 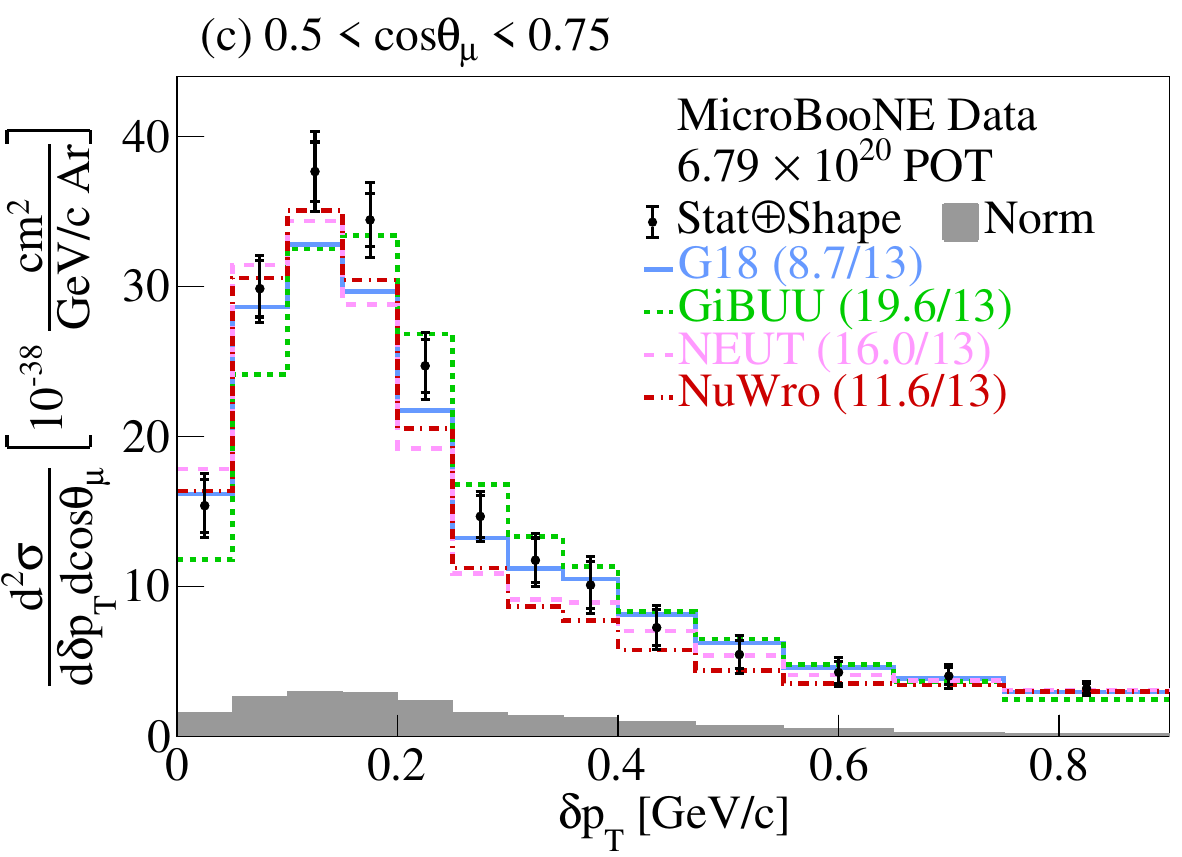}
\includegraphics[width=0.49\linewidth]{\figures 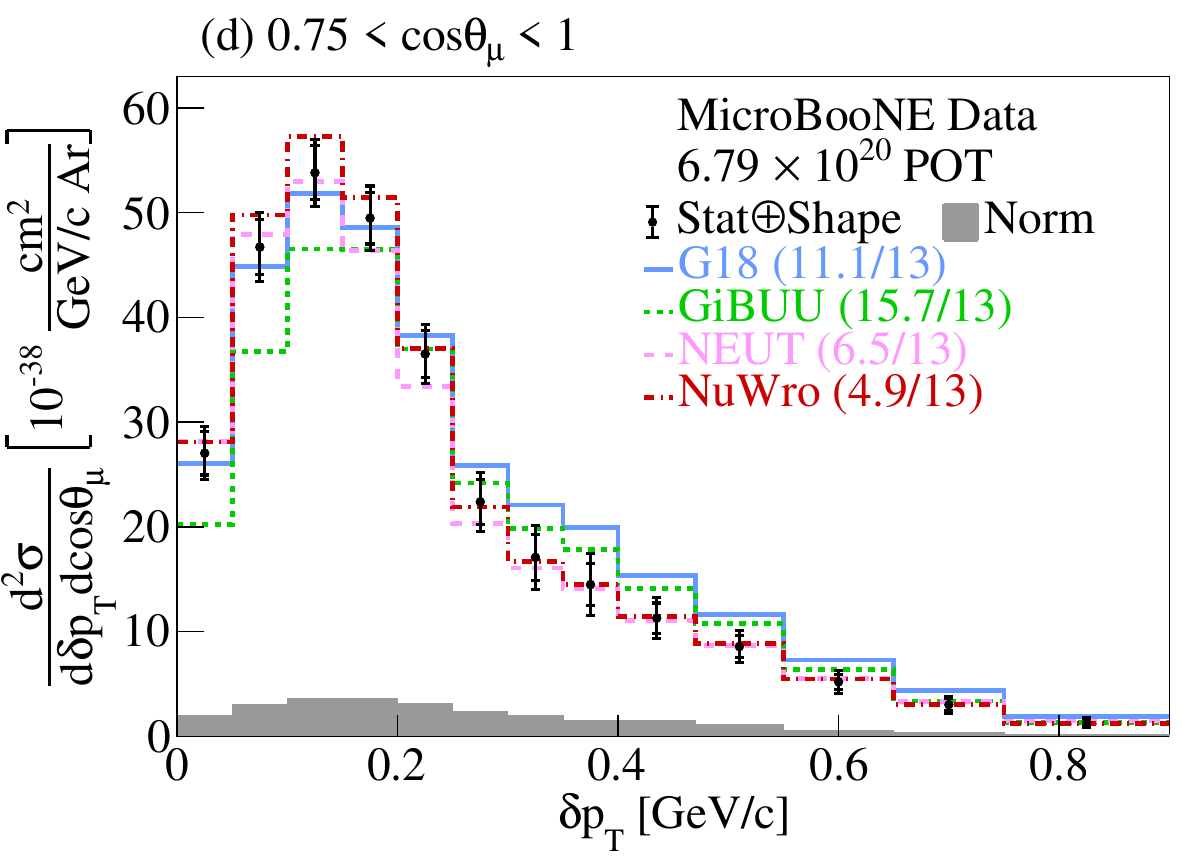}\\
\caption{
The flux-integrated double-differential cross sections as a function of $\delta p_{T}$ in cos$\theta_{\mu}$ bins. 
Inner and outer error bars show the statistical and total (statistical and shape systematic) uncertainty at the 1$\sigma$, or 68\%, confidence level. 
The gray band shows the normalization systematic uncertainty.
Colored lines show the results of theoretical cross section calculations using the $\texttt{G18 GENIE}$ (blue), $\texttt{GiBUU}$ (green), $\texttt{NEUT}$ (pink), and $\texttt{NuWro}$ (red) event generators.
The numbers in parentheses show the $\chi^{2}$/bins calculation for each one of the predictions. 
}
\label{DeltaPTInMuonCosThetaGen}
\end{figure*}

\begin{figure*}[htb!]
\centering 
\includegraphics[width=0.49\linewidth]{\figures 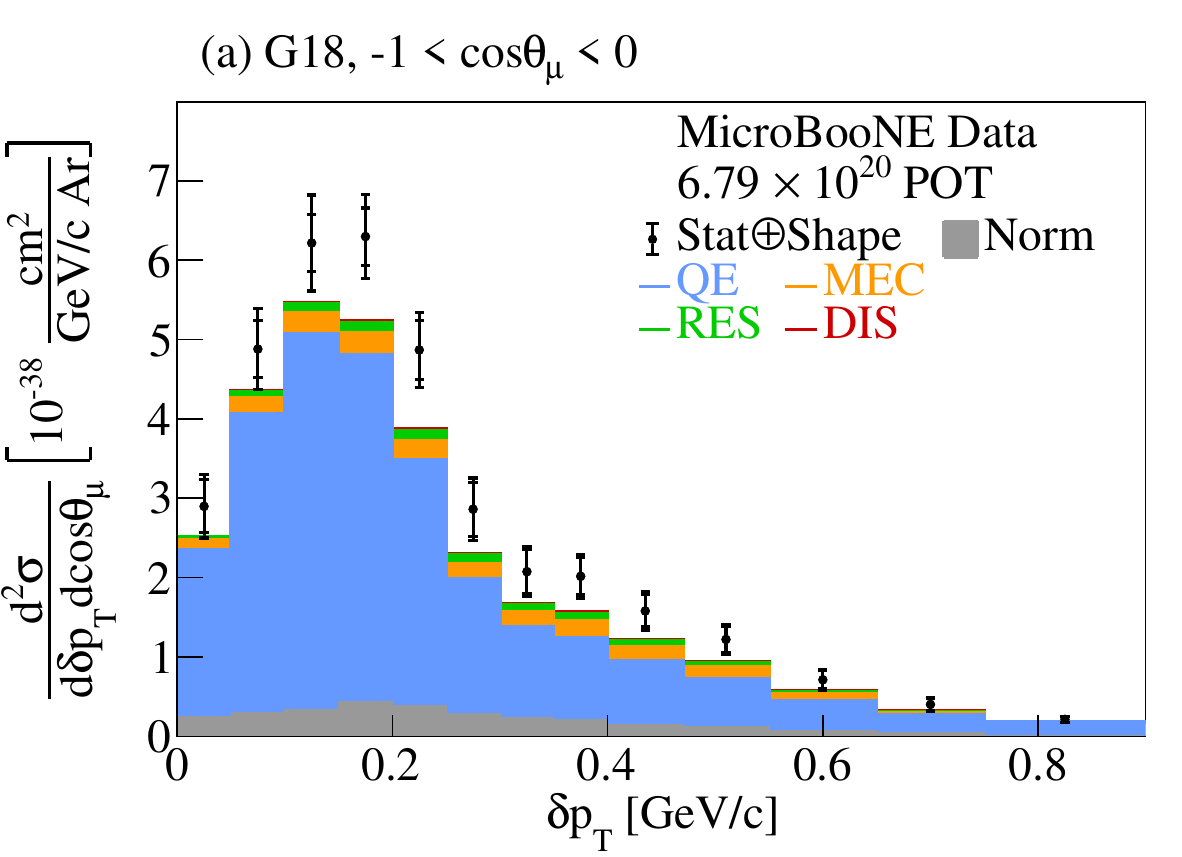}
\includegraphics[width=0.49\linewidth]{\figures 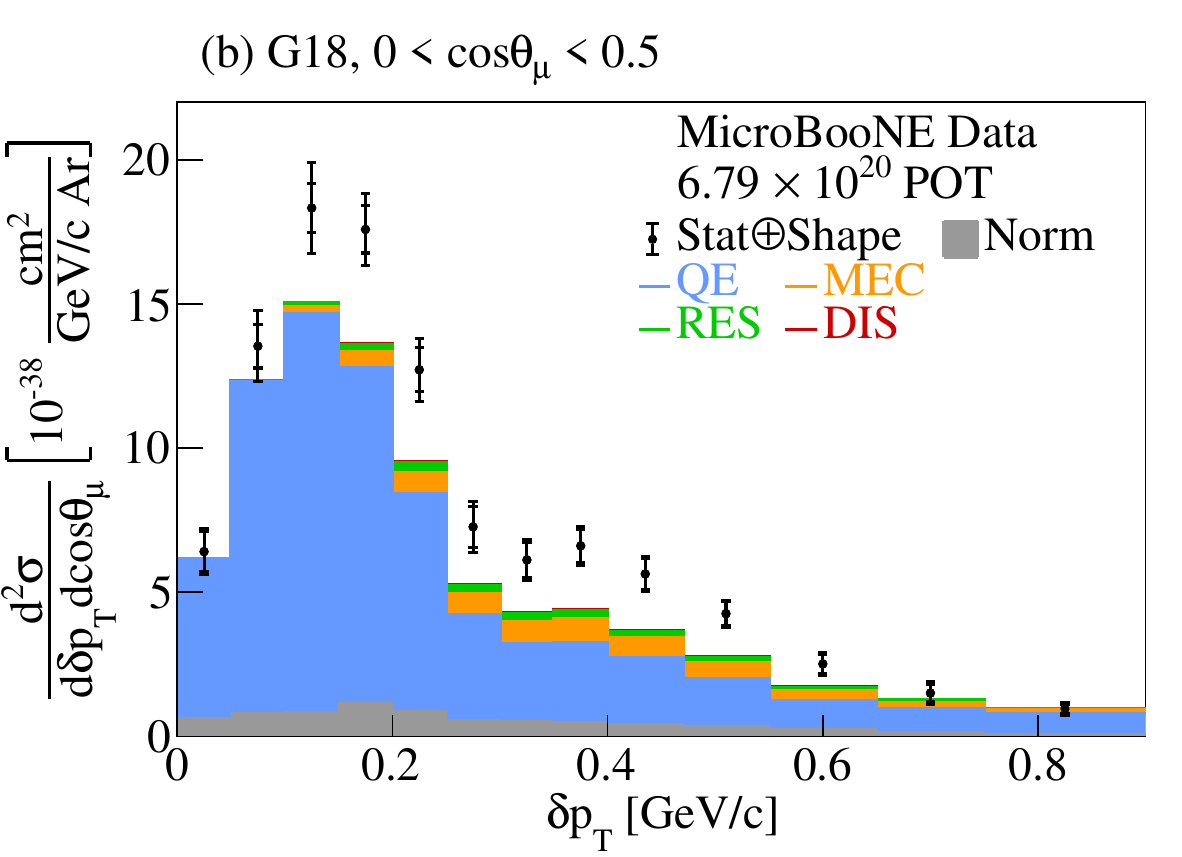}\\
\includegraphics[width=0.49\linewidth]{\figures 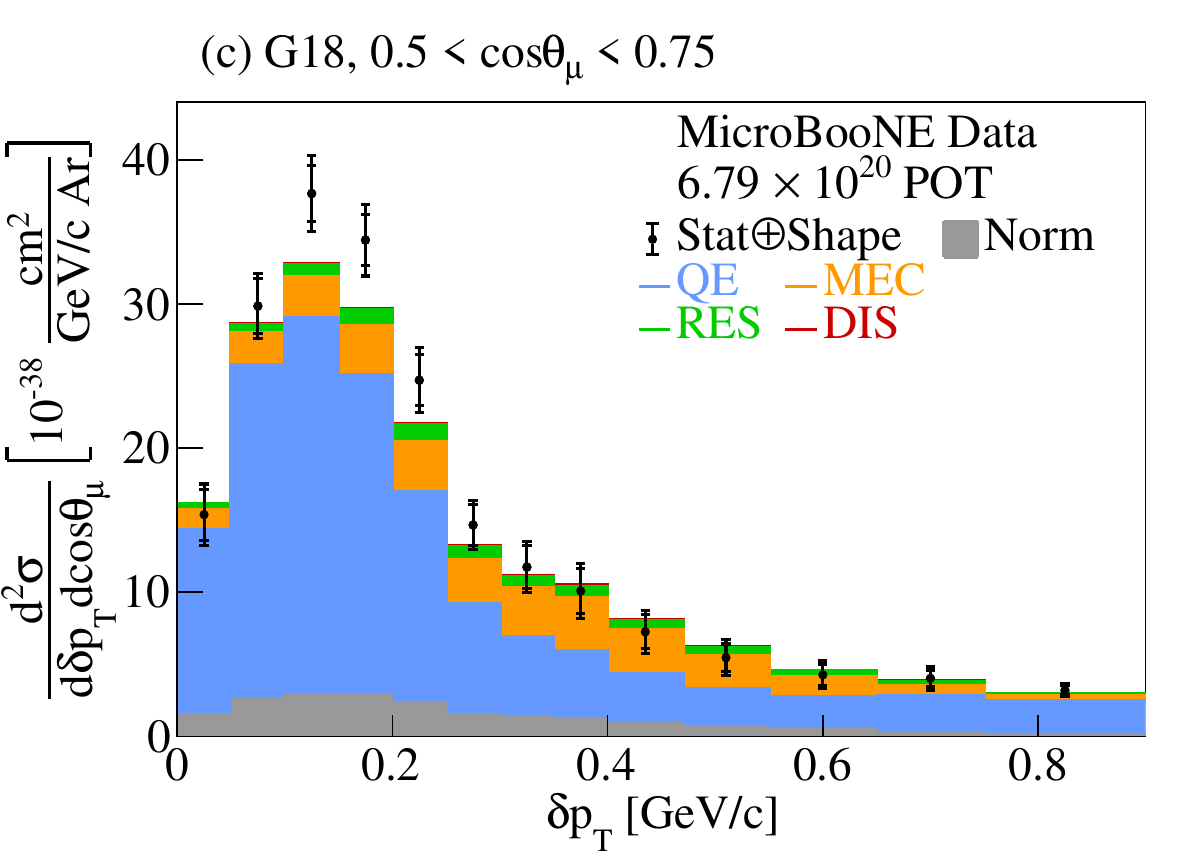}
\includegraphics[width=0.49\linewidth]{\figures 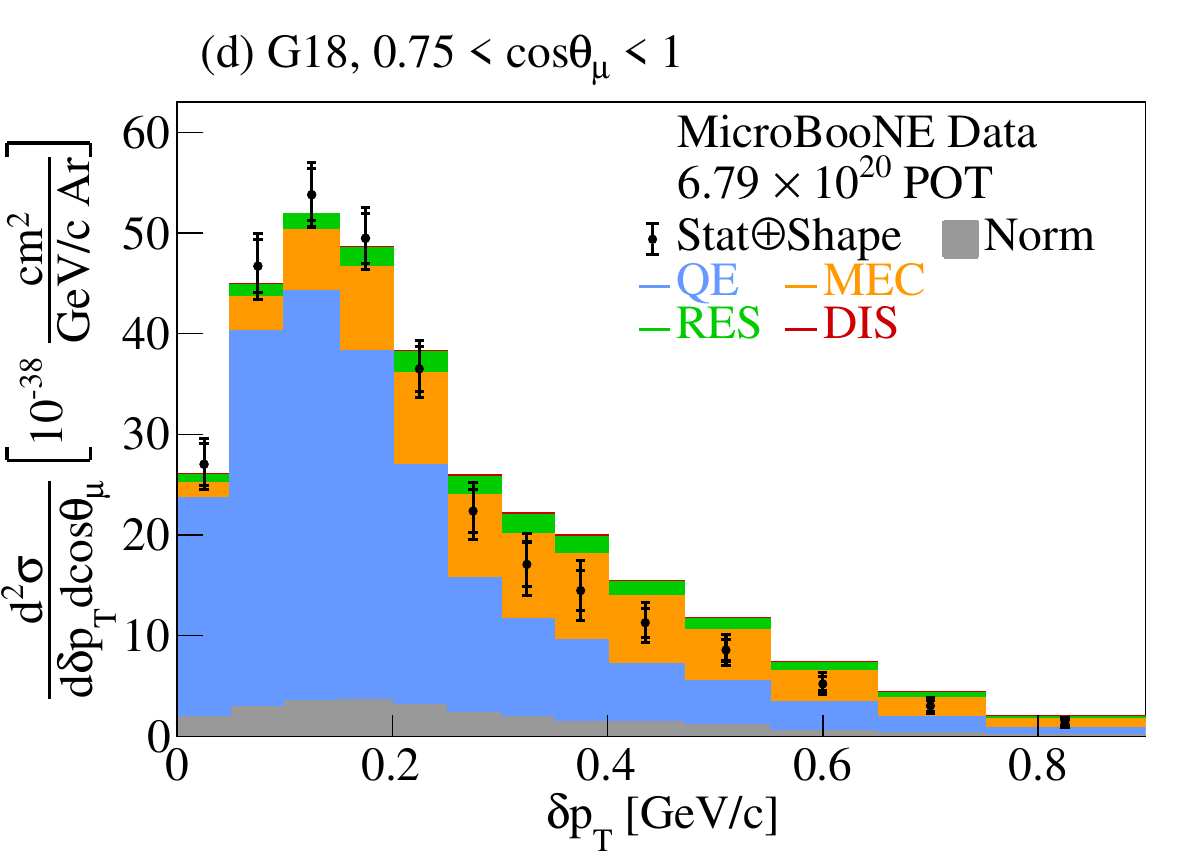}\\
\caption{
Comparison between the flux-integrated double-differential cross sections as a function of $\delta p_{T}$ for data and the $\texttt{G18 GENIE}$ prediction in cos$\theta_{\mu}$ bins. 
Inner and outer error bars show the statistical and total (statistical and shape systematic) uncertainty at the 1$\sigma$, or 68\%, confidence level. 
The gray band shows the normalization systematic uncertainty.
Colored stacked histograms show the results of theoretical cross section calculations using the $\texttt{G18 GENIE}$ prediction for QE (blue), MEC (orange), RES (green), and DIS (red) interactions.
}
\label{DeltaPTInMuonCosThetaInte}
\end{figure*}

\begin{figure*}[htb!]
\centering 
\includegraphics[width=0.49\linewidth]{\figures 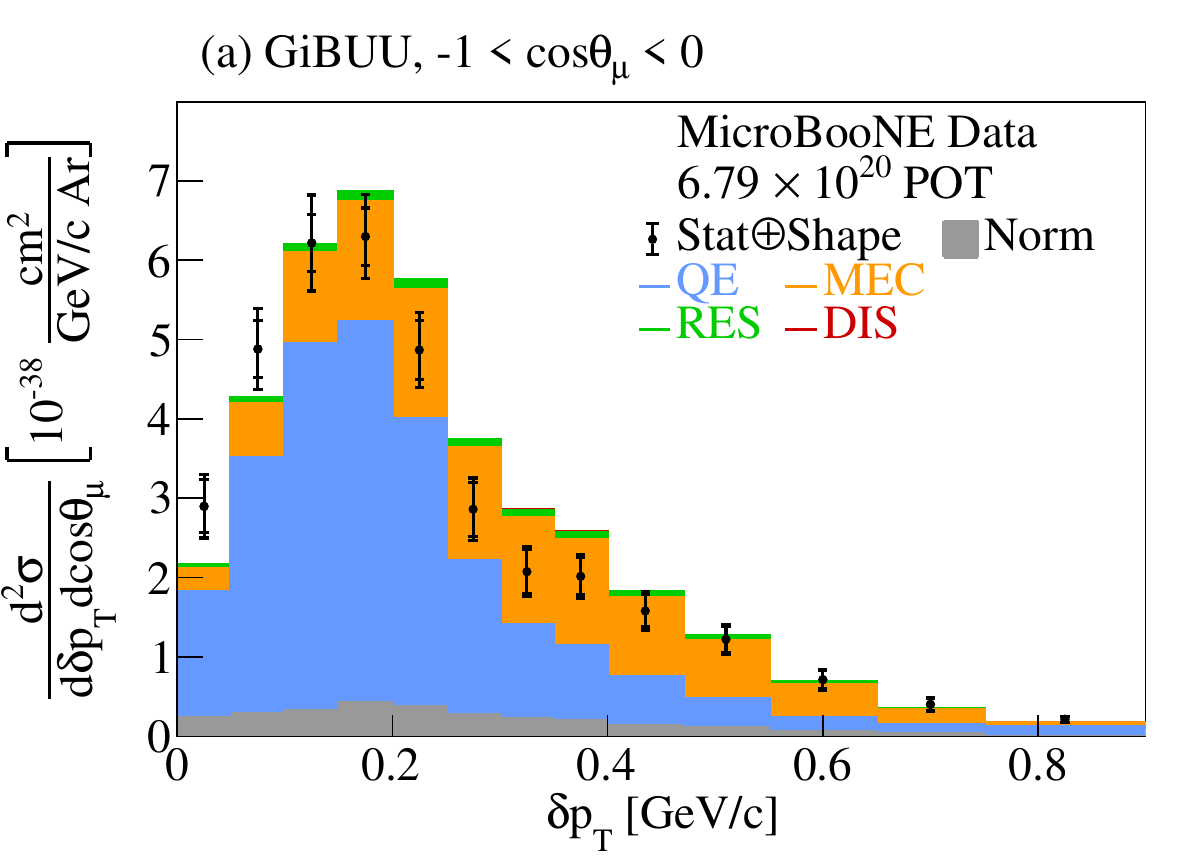}
\includegraphics[width=0.49\linewidth]{\figures 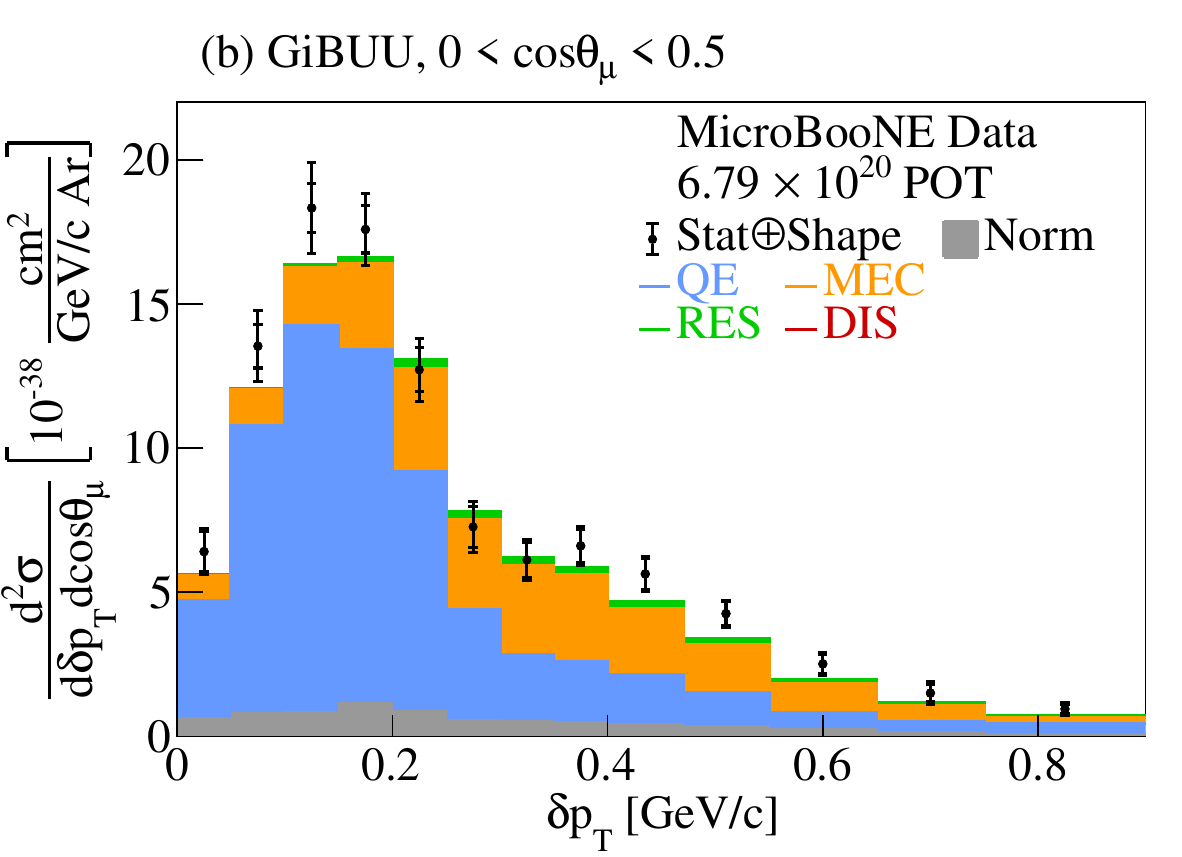}\\
\includegraphics[width=0.49\linewidth]{\figures 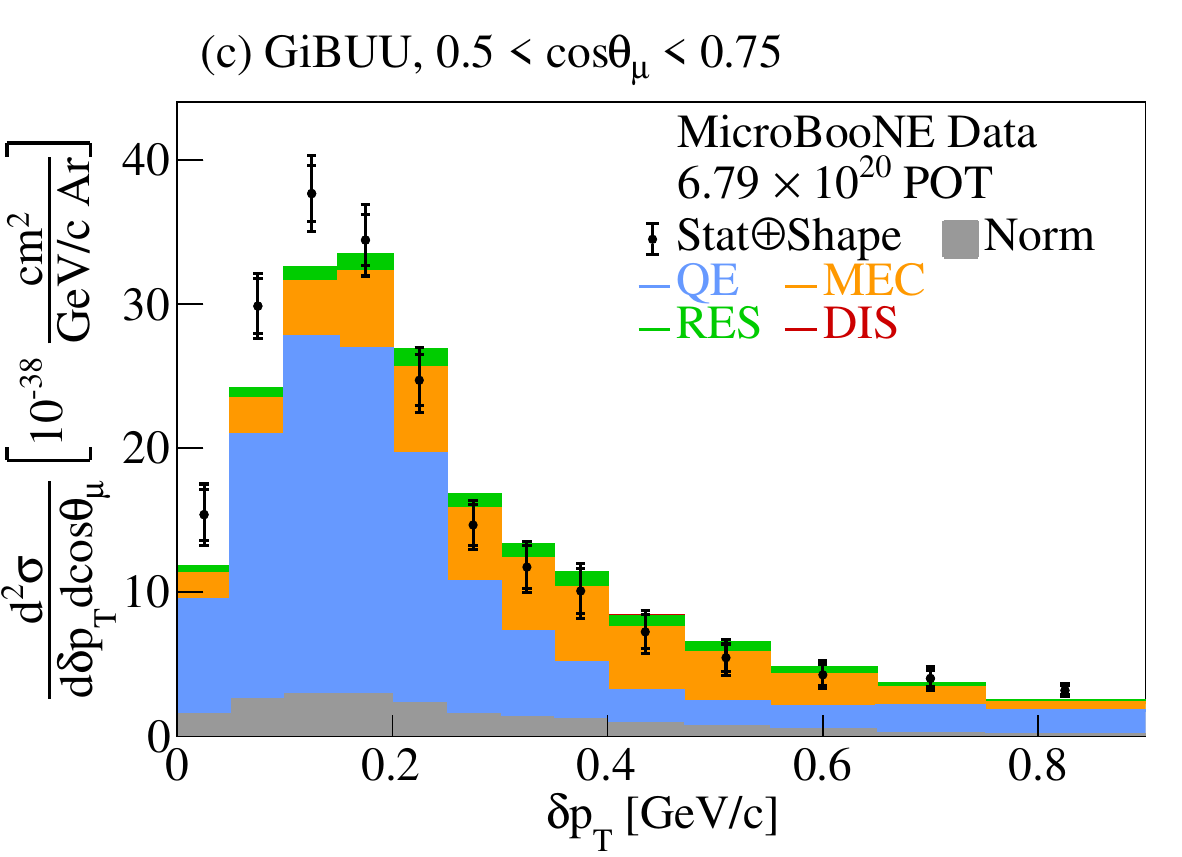}
\includegraphics[width=0.49\linewidth]{\figures 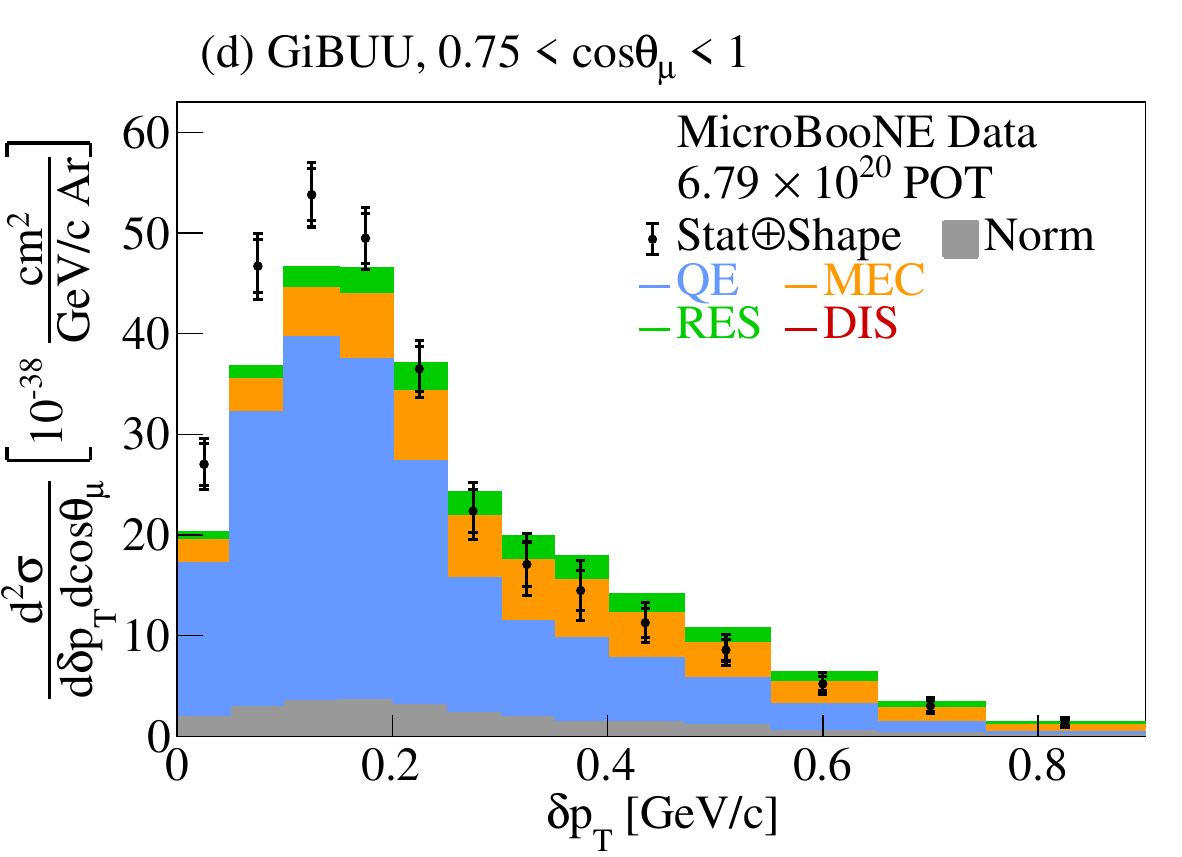}\\
\caption{
Comparison between the flux-integrated double-differential cross sections as a function of $\delta p_{T}$ for data and the GiBUU prediction in cos$\theta_{\mu}$ bins. 
Inner and outer error bars show the statistical and total (statistical and shape systematic) uncertainty at the 1$\sigma$, or 68\%, confidence level. 
The gray band shows the normalization systematic uncertainty.
Colored stacked histograms show the results of theoretical cross section calculations using the $\texttt{GiBUU}$ prediction for QE (blue), MEC (orange), RES (green), and DIS (red) interactions.
}
\label{DeltaPTInMuonCosThetaInteGiBUU}
\end{figure*}

\begin{figure*}[htb!]
\centering 
\includegraphics[width=0.49\linewidth]{\figures 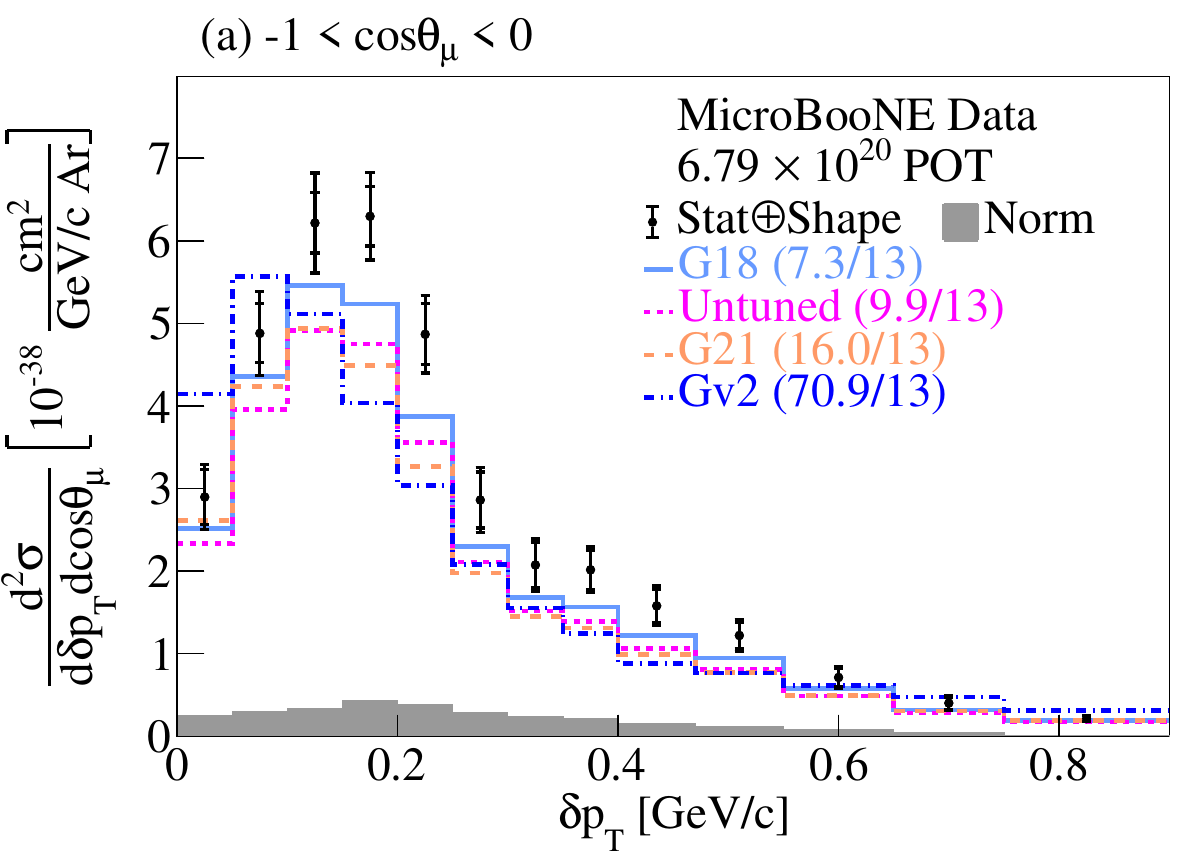}
\includegraphics[width=0.49\linewidth]{\figures 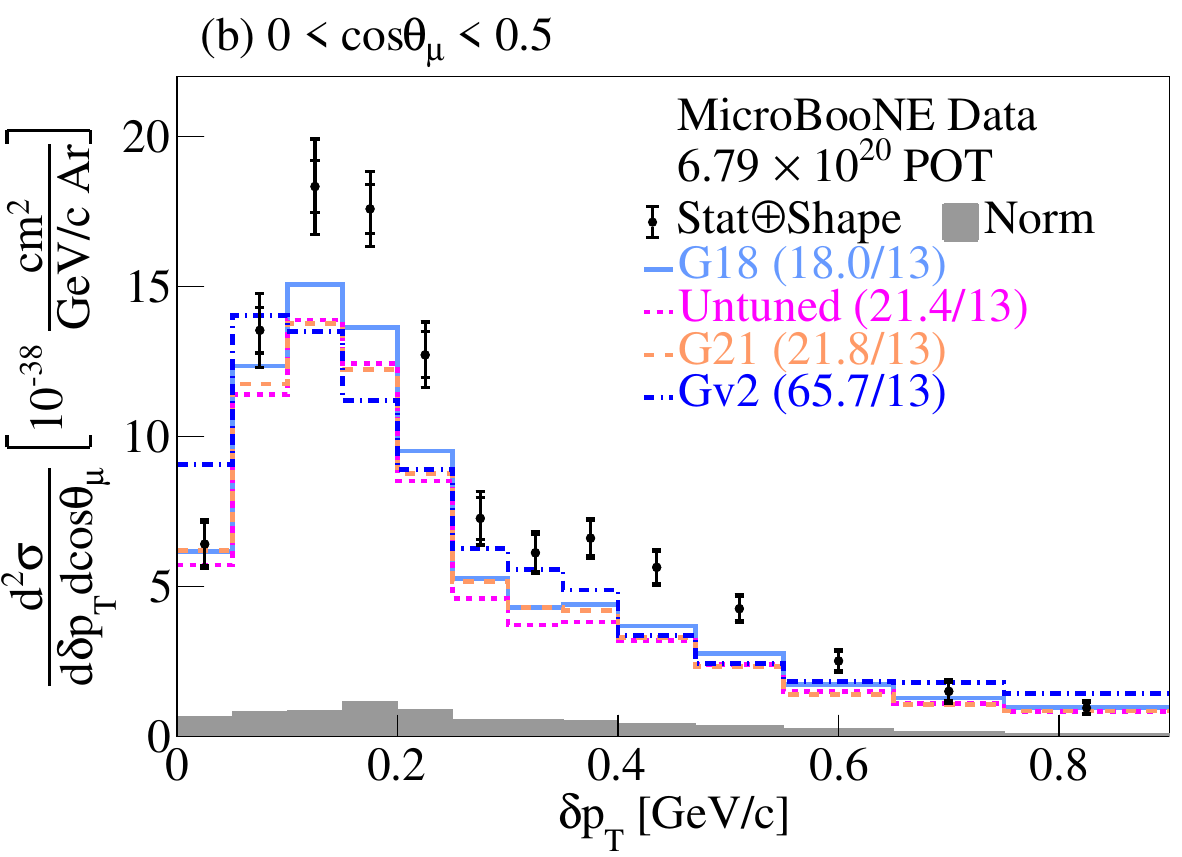}\\
\includegraphics[width=0.49\linewidth]{\figures 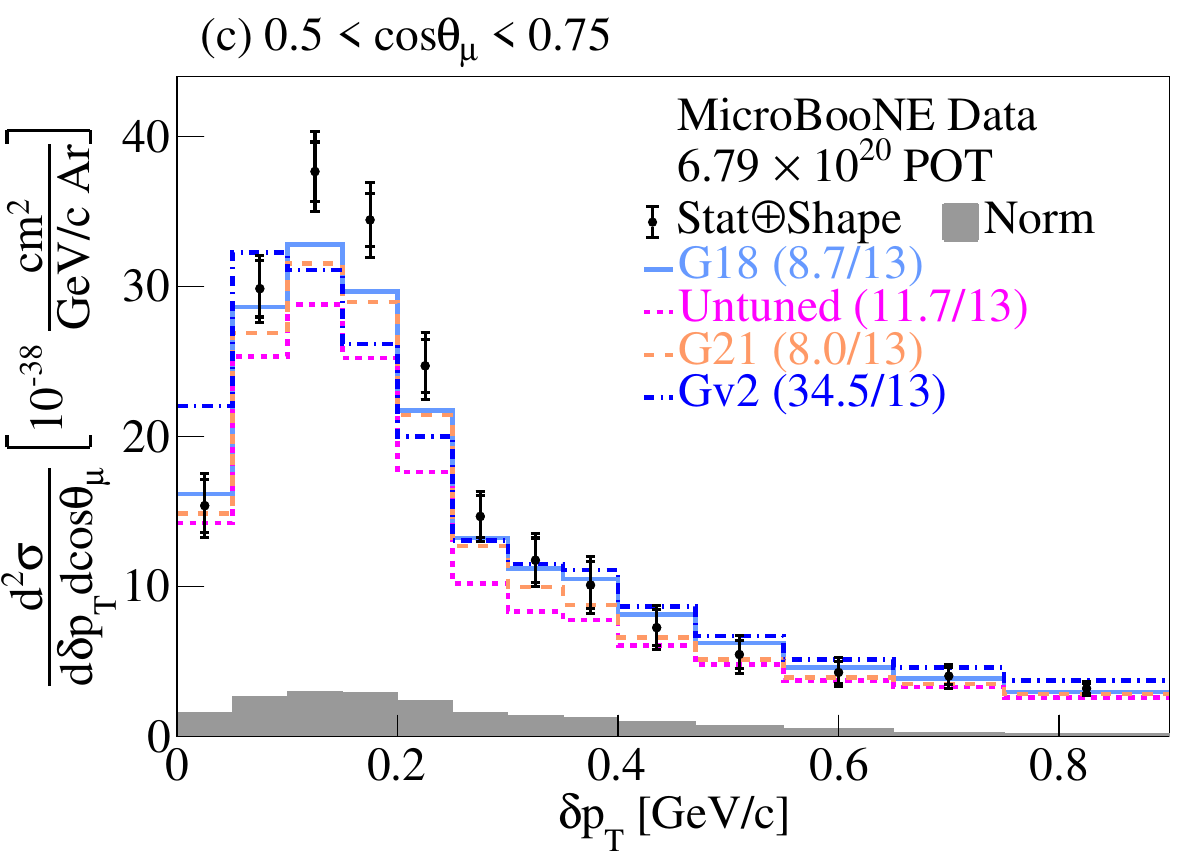}
\includegraphics[width=0.49\linewidth]{\figures 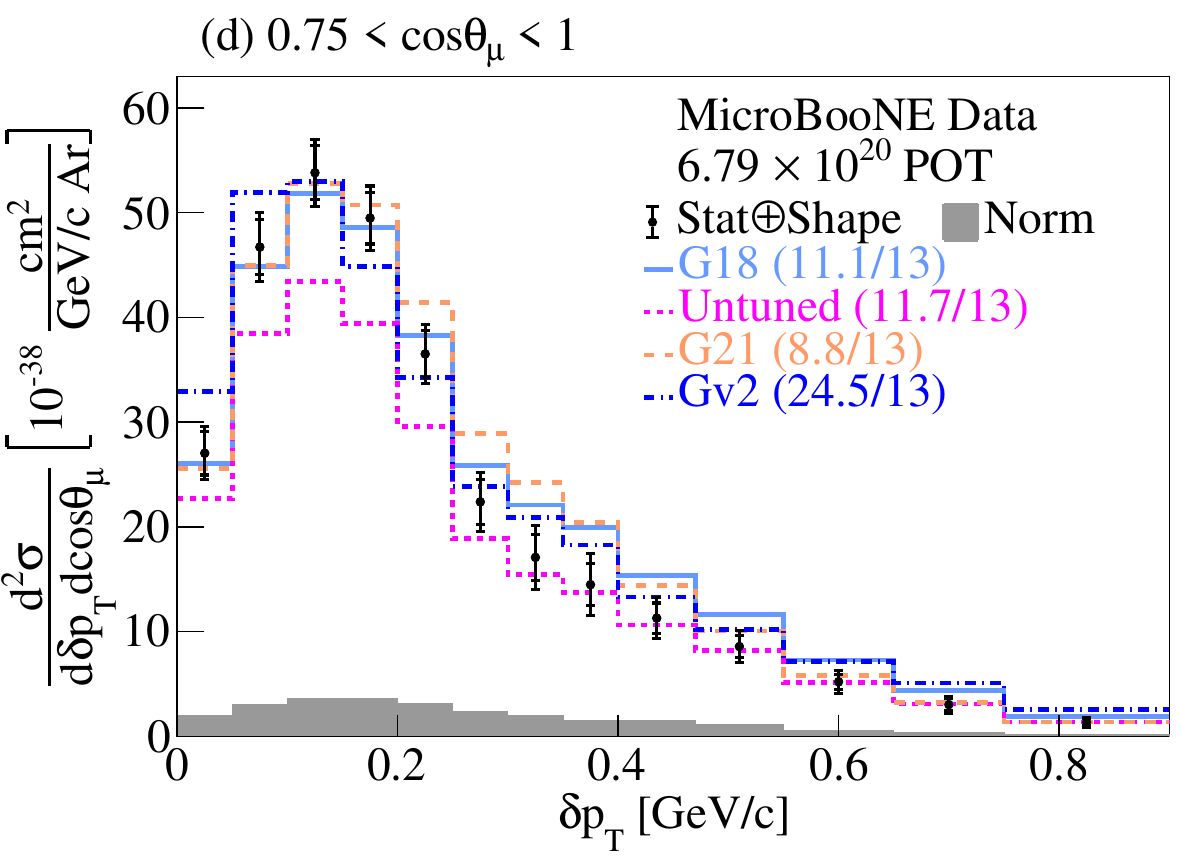}\\
\caption{
The flux-integrated double-differential cross sections as a function of $\delta p_{T}$ in cos$\theta_{\mu}$ bins. 
Inner and outer error bars show the statistical and total (statistical and shape systematic) uncertainty at the 1$\sigma$, or 68\%, confidence level. 
The gray band shows the normalization systematic uncertainty.
Colored lines show the results of theoretical cross section calculations using the $\texttt{G18}$ (light blue), $\texttt{Untuned}$ (magenta), $\texttt{G21}$ (orange), and $\texttt{Gv2}$ (dark blue) $\texttt{GENIE}$ configurations.
The numbers in parentheses show the $\chi^{2}$/bins calculation for each one of the predictions.
}
\label{DeltaPTInMuonCosThetaGenie}
\end{figure*}

Figure~\ref{DeltaPTInMuonCosThetaGen} shows the double-differential results as a function of $\delta p_{T}$ in cos$\theta_{\mu}$ bins.
In a factorized nuclear model such as the LFG, the Fermi motion part of $\delta p_{T}$ should stay constant in terms of the shape as a function of the outgoing lepton kinematics, since in such models the initial state nucleon momentum is a property of the nucleus that cannot be affected by the interaction momentum or energy transfer.
That is indeed the observed behavior in the reported results across all event generators and configurations, where no evidence of the inadequacy of the factorization approach is observed.
Figure~\ref{DeltaPTInMuonCosThetaGen} shows the comparisons to a number of available neutrino event generators, where the $\texttt{G18}$ prediction is favored based on the $\chi^{2}$/ndf results.
Apart from the factorization, a better separation between QE and non-QE can be gained depending on the cos$\theta_{\mu}$ region. 
As can be seen in Fig.~\ref{DeltaPTInMuonCosThetaInte} for $\texttt{G18}$, MEC events play a more pronounced role for forward muon scattering and in the high $\delta p_{T}$ tail, as opposed to backward scattering angles, which are much more strongly populated by QE events.
Furthermore, the $\texttt{G18}$ cross section prediction falls below the data in the -1 $<$ cos$\theta_{\mu}$ $<$ 0.5 region, as seen in Fig.~\ref{DeltaPTInMuonCosThetaInte}a and Fig.~\ref{DeltaPTInMuonCosThetaInte}b.
That could indicate that additional contribution from the QE part of the $\texttt{G18}$ prediction is needed beyond the MicroBooNE tune.
Figure~\ref{DeltaPTInMuonCosThetaInteGiBUU} shows the same interaction breakdown for $\texttt{GiBUU}$.
Unlike $\texttt{G18}$, $\texttt{GiBUU}$ illustrates a peak shift to the right, which becomes more pronounced in the backward direction.
This shift is driven by the enhanced MEC contribution in higher $\delta p_{T}$ values and the reduced QE contribution at smaller values.
In the backward direction, $\texttt{GiBUU}$ further shows a cross section excess driven by the MEC contribution.
Figure~\ref{DeltaPTInMuonCosThetaGenie} shows the same results compared to a number of $\texttt{GENIE}$ configurations illustrating that $\texttt{Gv2}$ is disfavored due to the low $\delta p_{T}$ bin behavior.



\begin{figure*}[htb!]
\centering 
\includegraphics[width=0.49\linewidth]{\figures 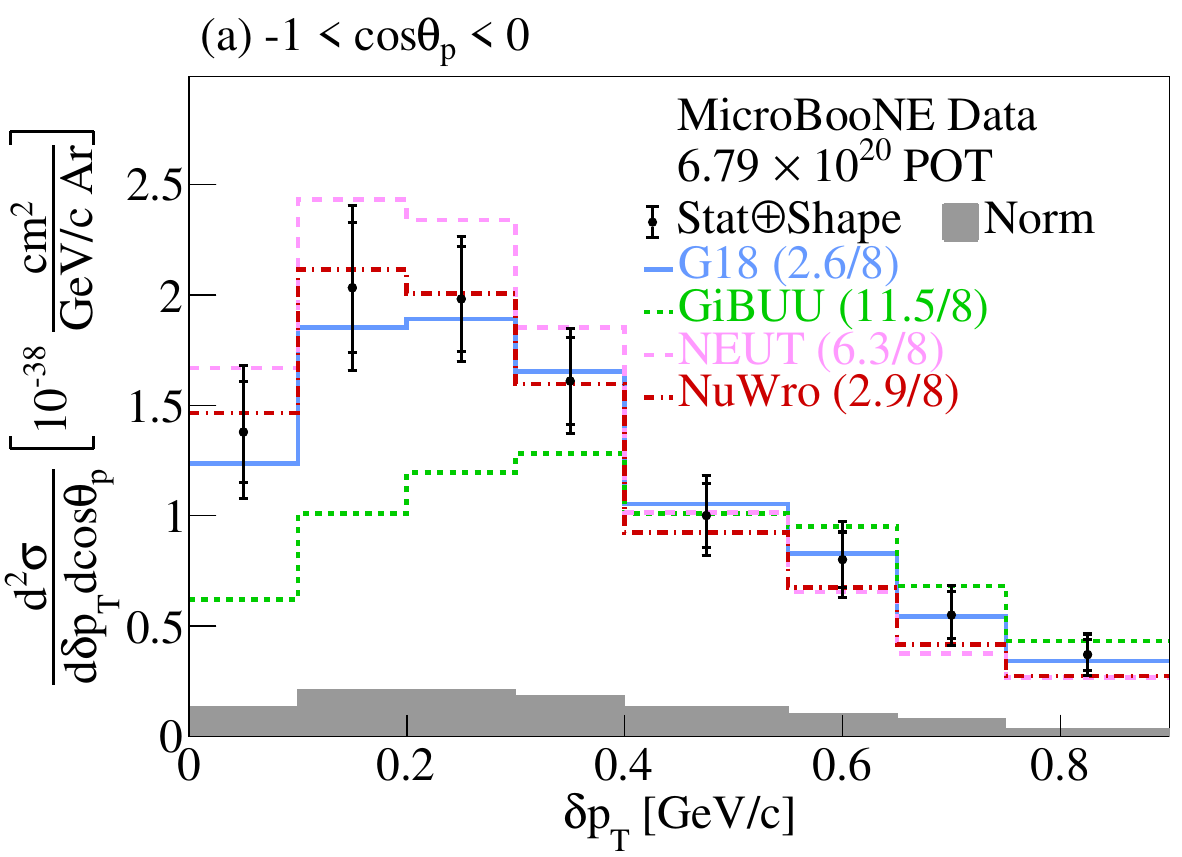}
\includegraphics[width=0.49\linewidth]{\figures 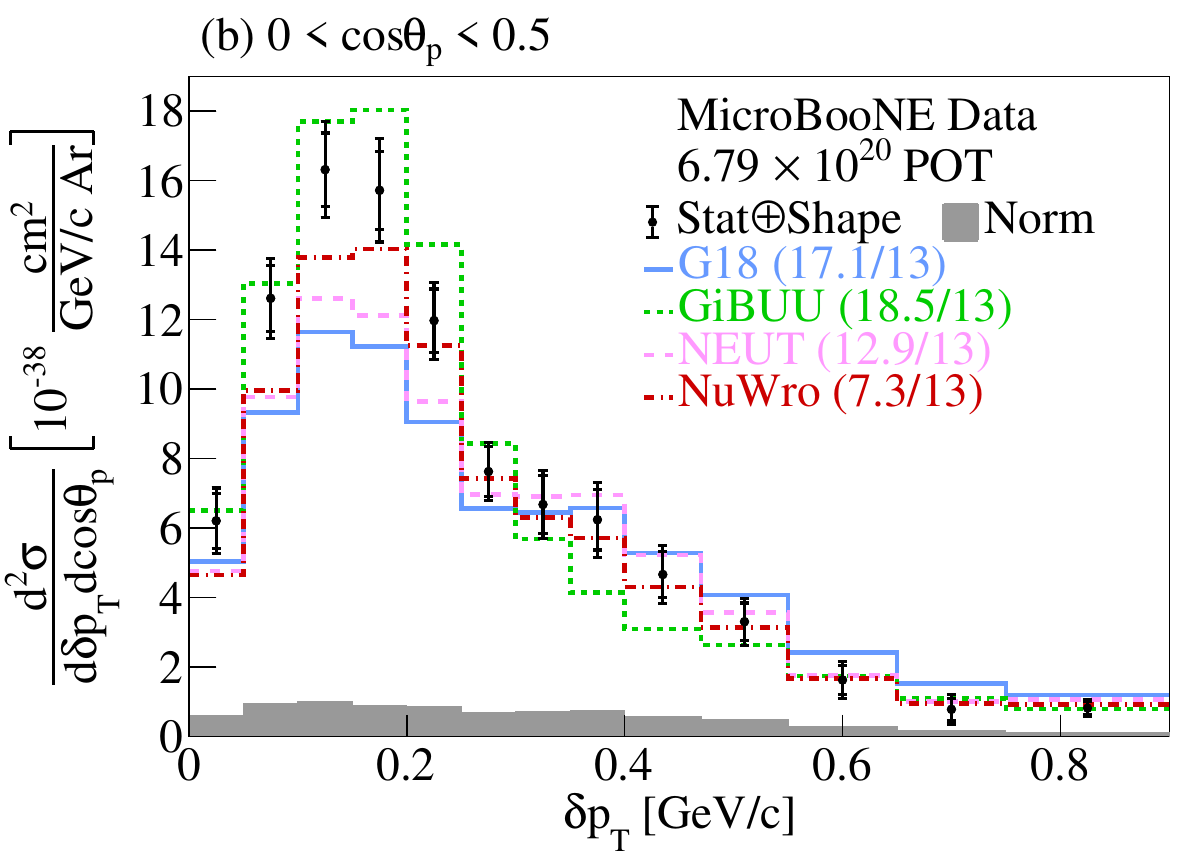}\\
\includegraphics[width=0.49\linewidth]{\figures 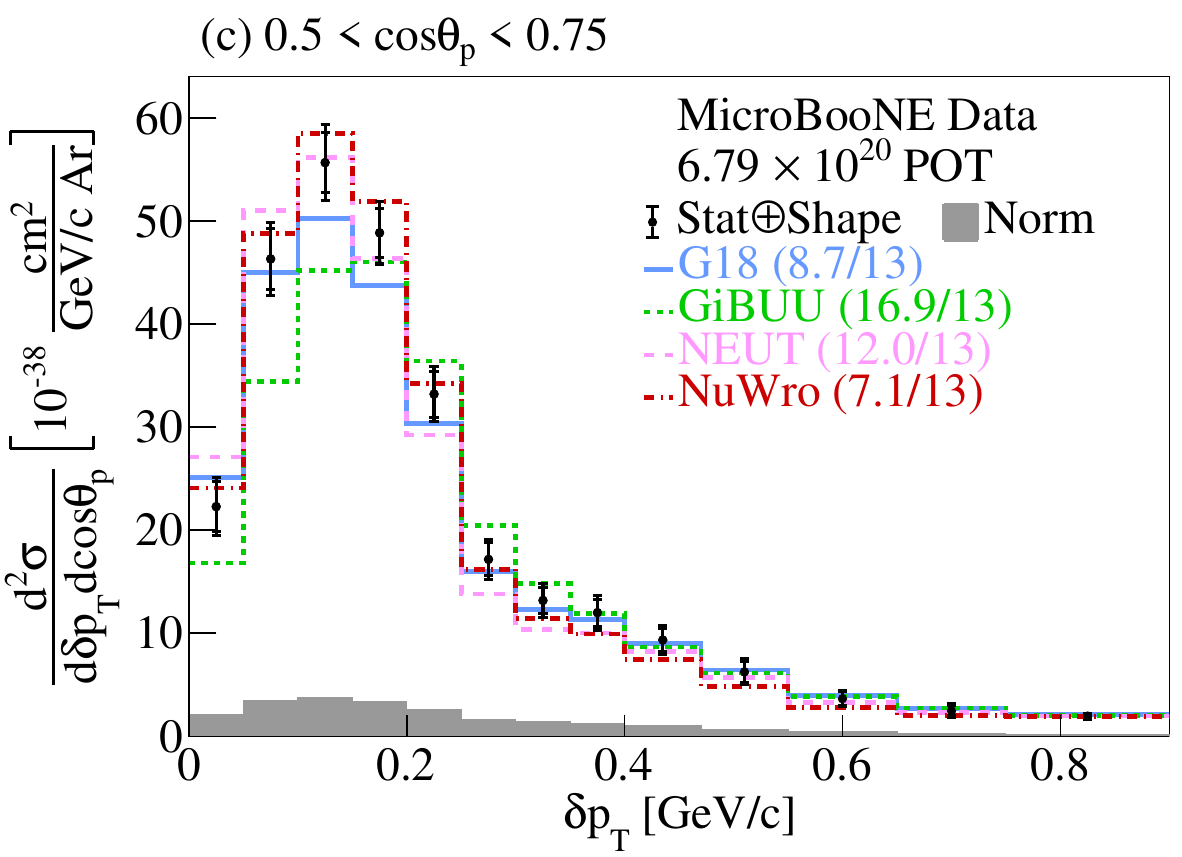}
\includegraphics[width=0.49\linewidth]{\figures 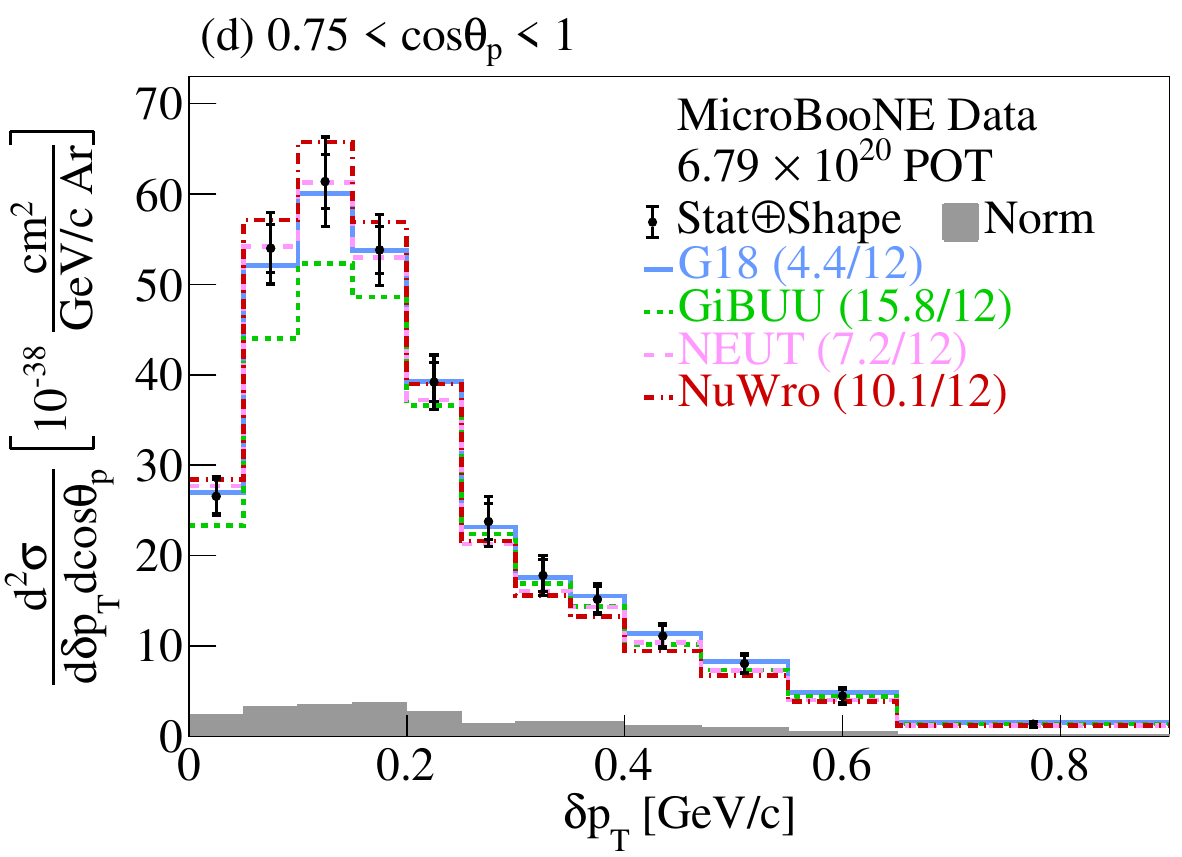}\\
\caption{
The flux-integrated double-differential cross sections as a function of $\delta p_{T}$ in cos$\theta_{p}$ bins. 
Inner and outer error bars show the statistical and total (statistical and shape systematic) uncertainty at the 1$\sigma$, or 68\%, confidence level. 
The gray band shows the normalization systematic uncertainty.
Colored lines show the results of theoretical cross section calculations using the $\texttt{G18 GENIE}$ (blue), $\texttt{GiBUU}$ (green), $\texttt{NEUT}$ (pink), and $\texttt{NuWro}$ (red) event generators.
The numbers in parentheses show the $\chi^{2}$/bins calculation for each one of the predictions.
}
\label{DeltaPTInProtonCosThetaGen}
\end{figure*}

\begin{figure*}[htb!]
\centering
\includegraphics[width=0.49\linewidth]{\figures 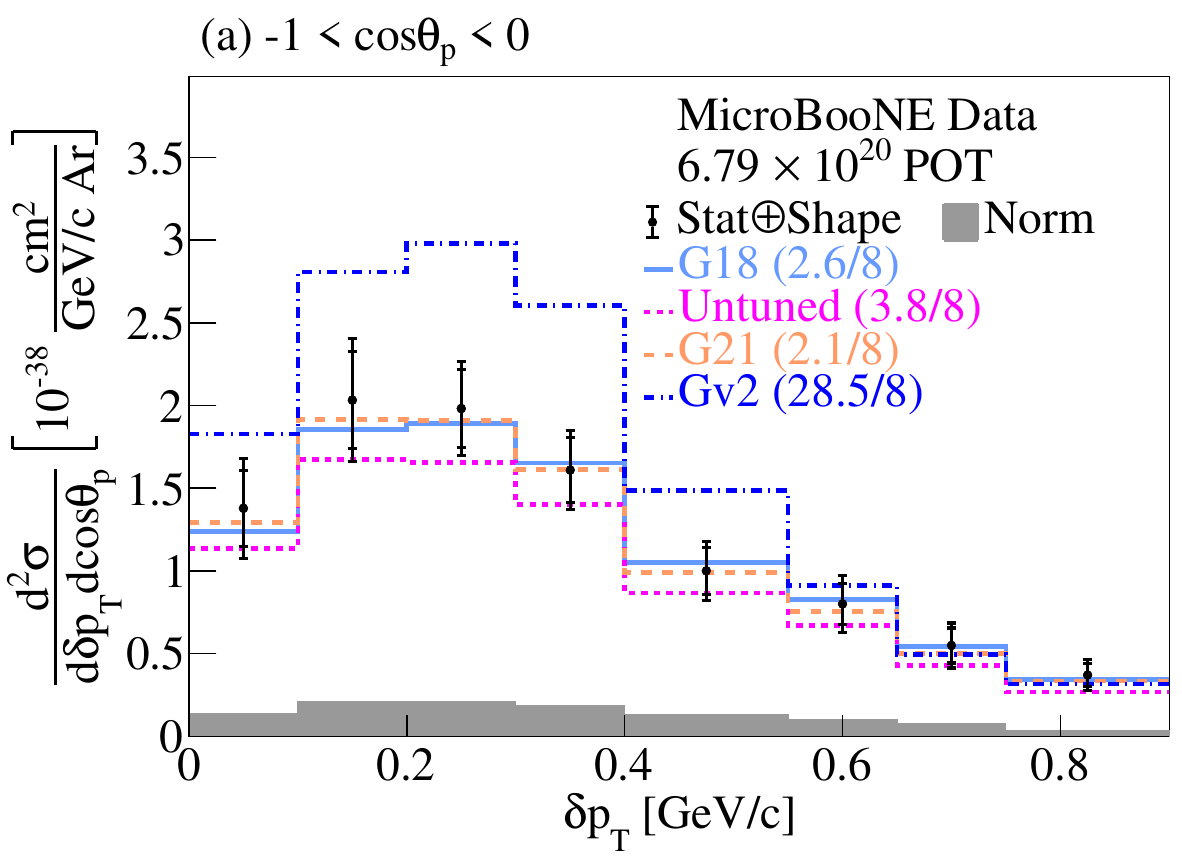}
\includegraphics[width=0.49\linewidth]{\figures 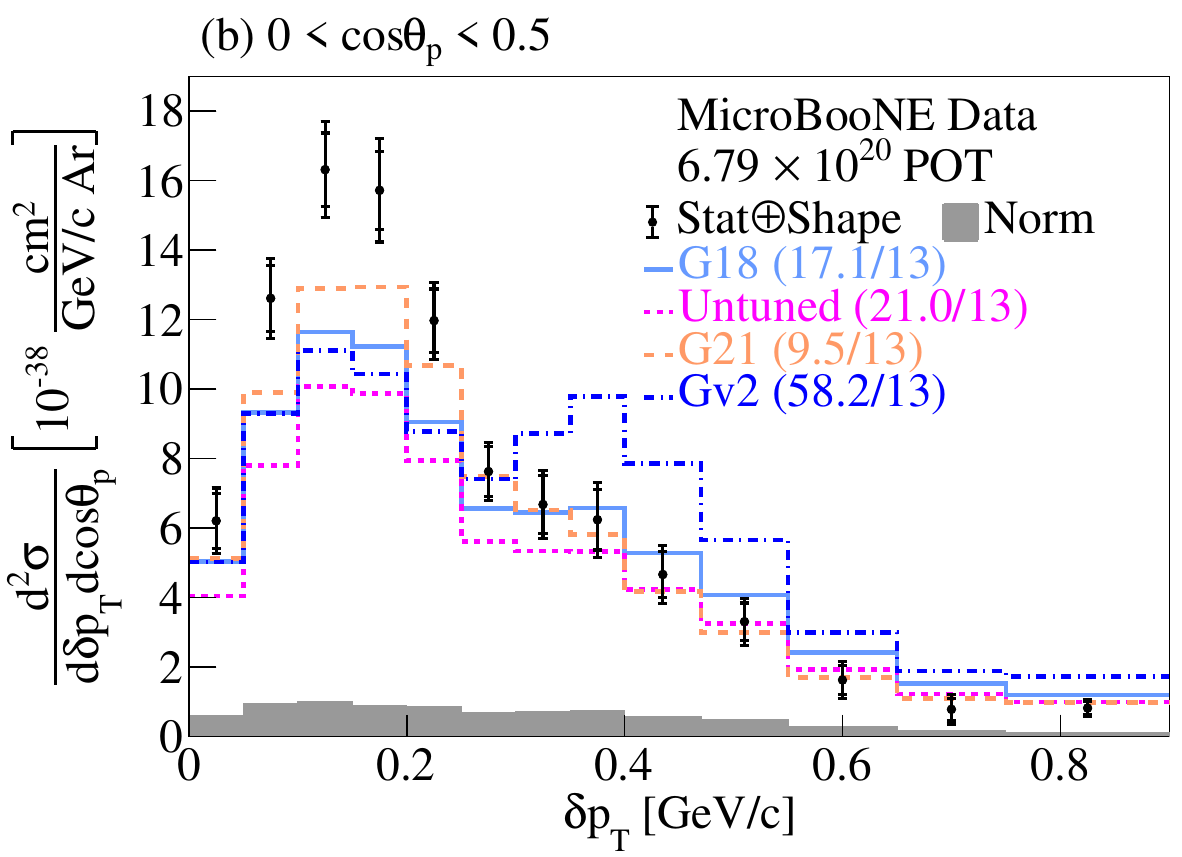}\\
\includegraphics[width=0.49\linewidth]{\figures 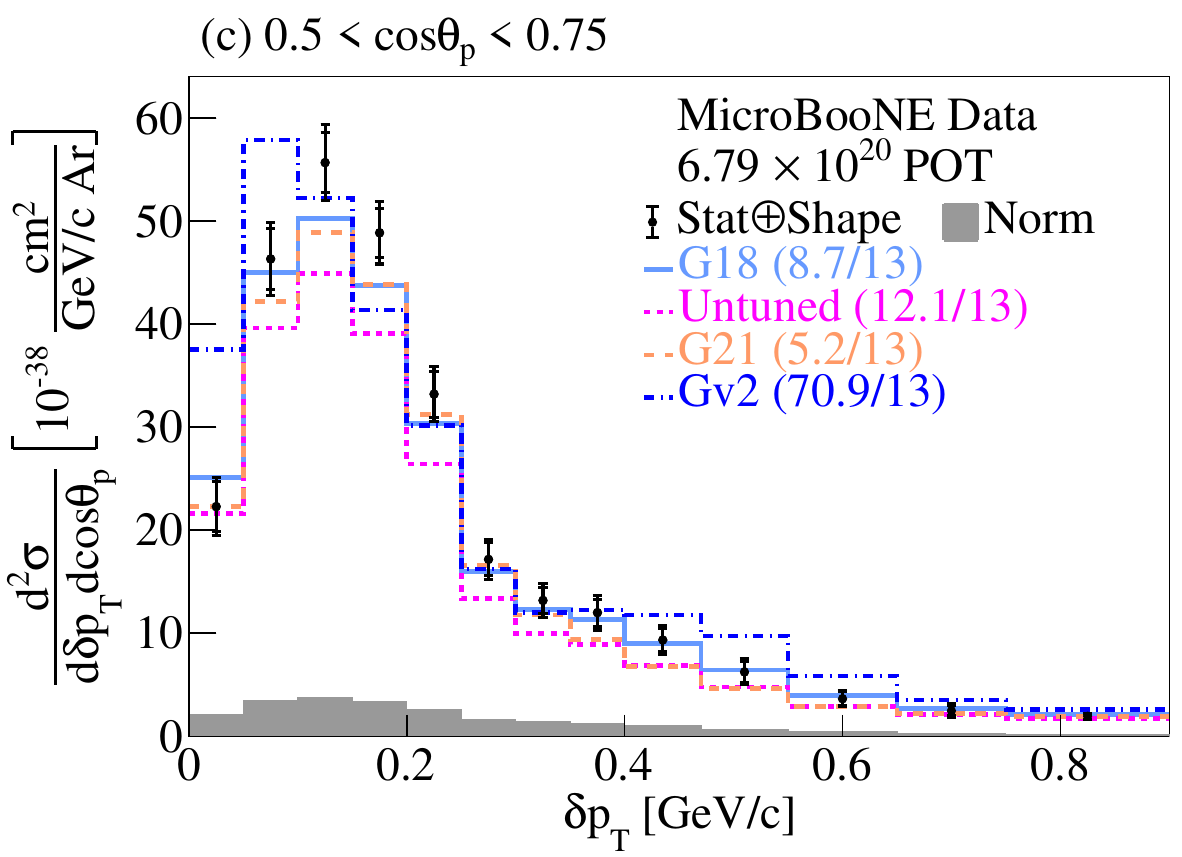}
\includegraphics[width=0.49\linewidth]{\figures 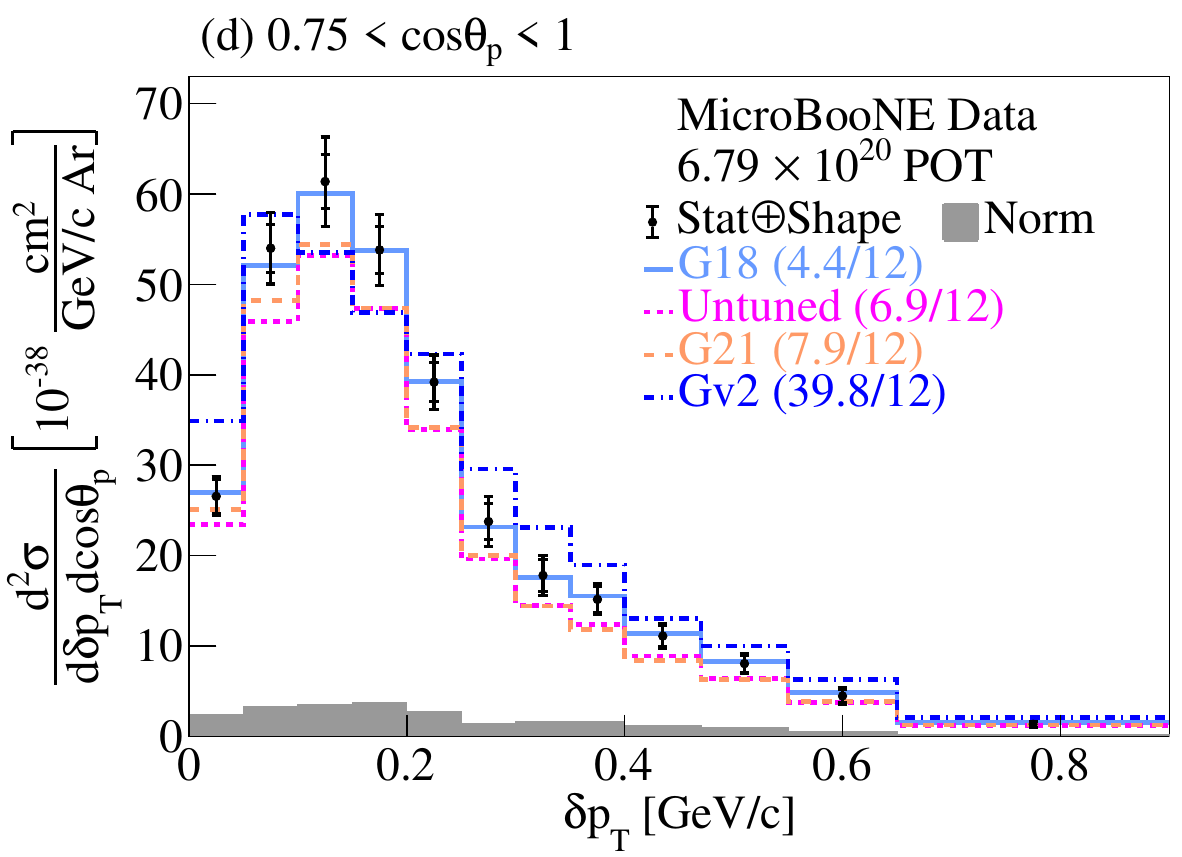}\\
\caption{
The flux-integrated double-differential cross sections as a function of $\delta p_{T}$ in cos$\theta_{p}$ bins. 
Inner and outer error bars show the statistical and total (statistical and shape systematic) uncertainty at the 1$\sigma$, or 68\%, confidence level. 
The gray band shows the normalization systematic uncertainty.
Colored lines show the results of theoretical cross section calculations using the $\texttt{G18}$ (light blue), $\texttt{Untuned}$ (magenta), $\texttt{G21}$ (orange), and $\texttt{Gv2}$ (dark blue) $\texttt{GENIE}$ configurations.
The numbers in parentheses show the $\chi^{2}$/bins calculation for each one of the predictions.
}
\label{DeltaPTInProtonCosThetaGenie}
\end{figure*}

\begin{figure*}[htb!]
\centering 
\includegraphics[width=0.49\linewidth]{\figures 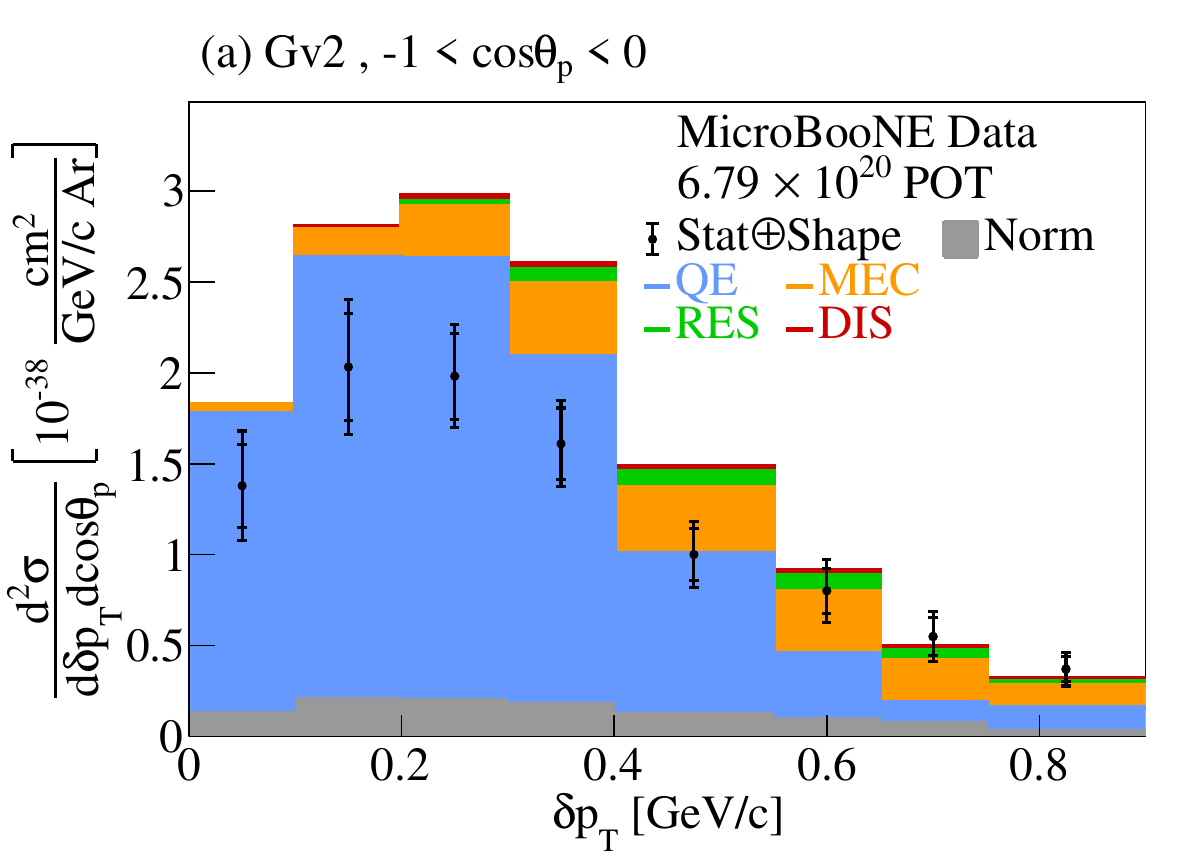}
\includegraphics[width=0.49\linewidth]{\figures 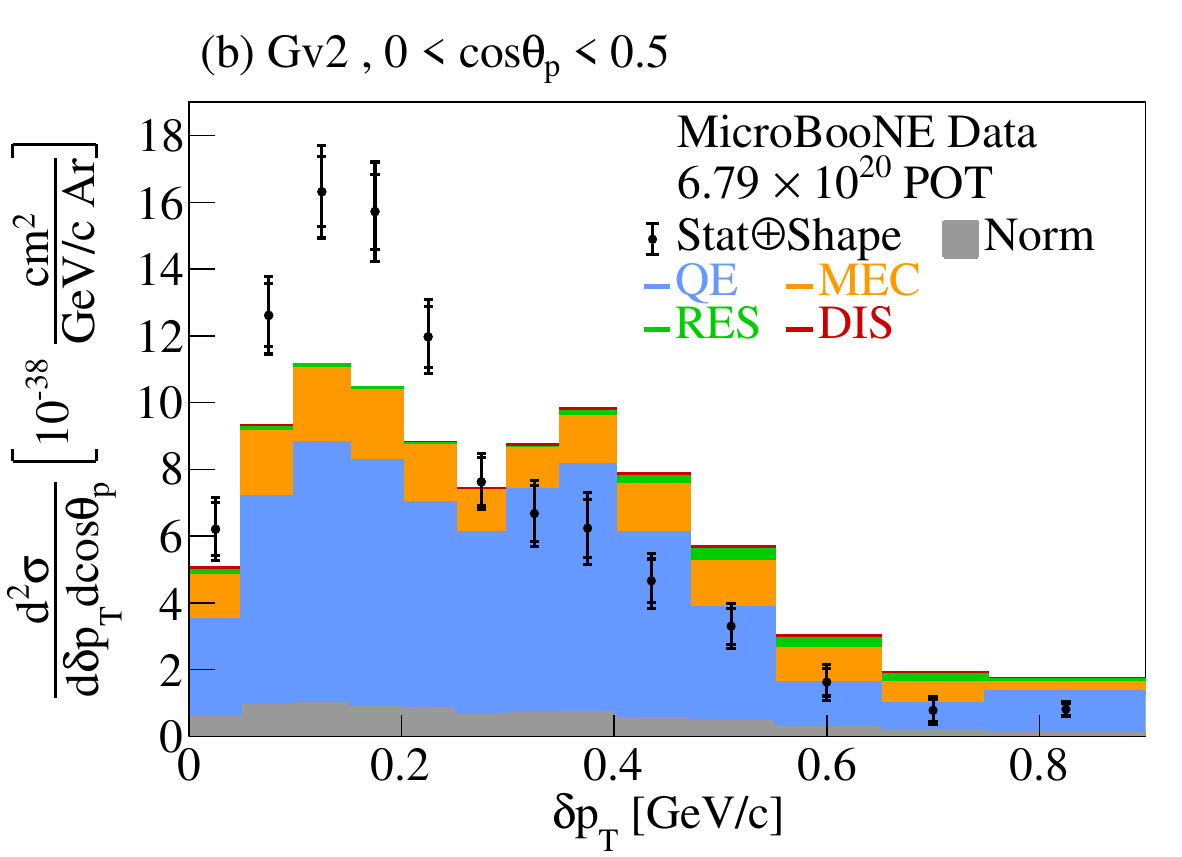}\\
\includegraphics[width=0.49\linewidth]{\figures 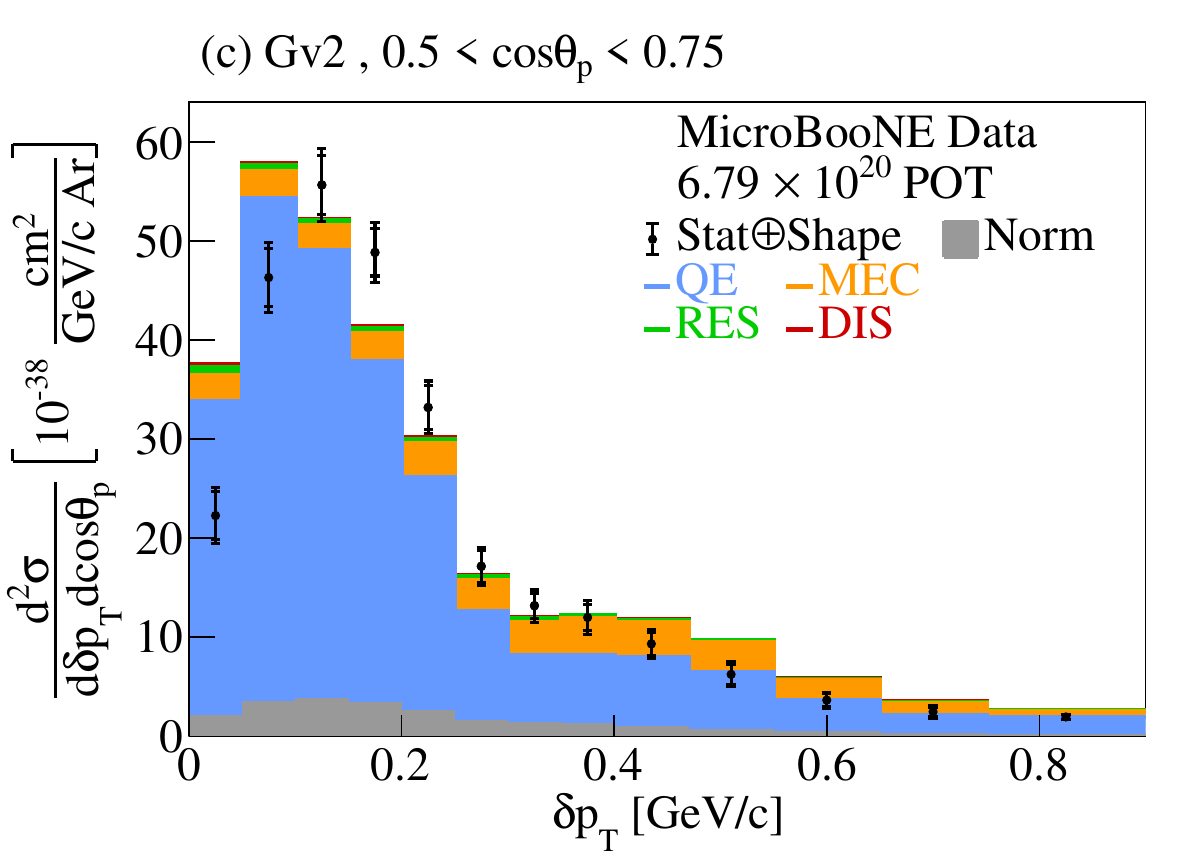}
\includegraphics[width=0.49\linewidth]{\figures 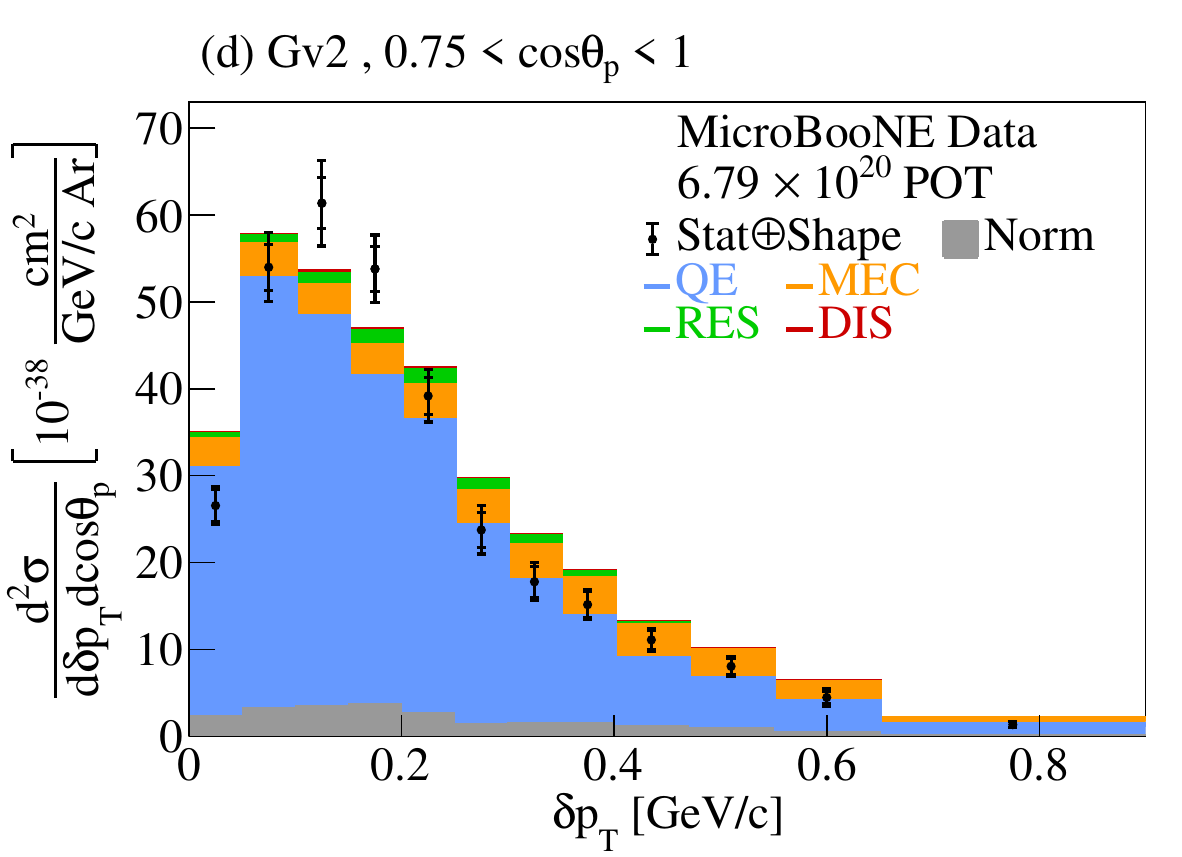}\\
\caption{
Comparison between the flux-integrated double-differential cross sections as a function of $\delta p_{T}$ for data and the Gv2 GENIE prediction in cos$\theta_{p}$ bins. 
Inner and outer error bars show the statistical and total (statistical and shape systematic) uncertainty at the 1$\sigma$, or 68\%, confidence level. 
The gray band shows the normalization systematic uncertainty.
Colored stacked histograms show the results of theoretical cross section calculations using the $\texttt{Gv2 GENIE}$ prediction for QE (blue), MEC (orange), RES (green), and DIS (red) interactions.
}
\label{DeltaPTInProtonCosThetaInte}
\end{figure*}

Figures~\ref{DeltaPTInProtonCosThetaGen} and~\ref{DeltaPTInProtonCosThetaGenie} show the double-differential cross section as a function of $\delta p_{T}$ in cos$\theta_{p}$ bins.
The factorization of the nuclear motion is mostly preserved in cos$\theta_{p}$ bins, analogously to the previous result in cos$\theta_{\mu}$.
Figure~\ref{DeltaPTInProtonCosThetaGen} shows the comparisons to a number of available neutrino event generators.
The $\texttt{GiBUU}$ prediction is significantly lower than the data in the backward proton angle for low $\delta p_{T}$ values, as shown in Fig.~\ref{DeltaPTInProtonCosThetaGen}a.
Figure~\ref{DeltaPTInProtonCosThetaGenie} shows the same results compared to a number of $\texttt{GENIE}$ configurations illustrating that $\texttt{Gv2}$ is disfavored across all cos$\theta_{p}$ bins.
As can be seen in Fig.~\ref{DeltaPTInProtonCosThetaInte}, this particularly poor performance is driven by the QE contribution.
For backward scattering events (panel a), the QE contribution predicted by Llewellyn Smith is significantly overestimated.
For intermediate angles (0 $<$ cos$\theta_{p}$ $<$ 0.5), the same QE model results in an unphysical double peak.
For forward scattering (0.5 $<$ cos$\theta_{p}$ $<$ 1), the $\texttt{Gv2}$ QE prediction yields a pronounced contribution at lower values of $\delta p_{T}$ compared to the data.



\begin{figure*}[htb!]
\centering
\includegraphics[width=0.49\linewidth]{\figures 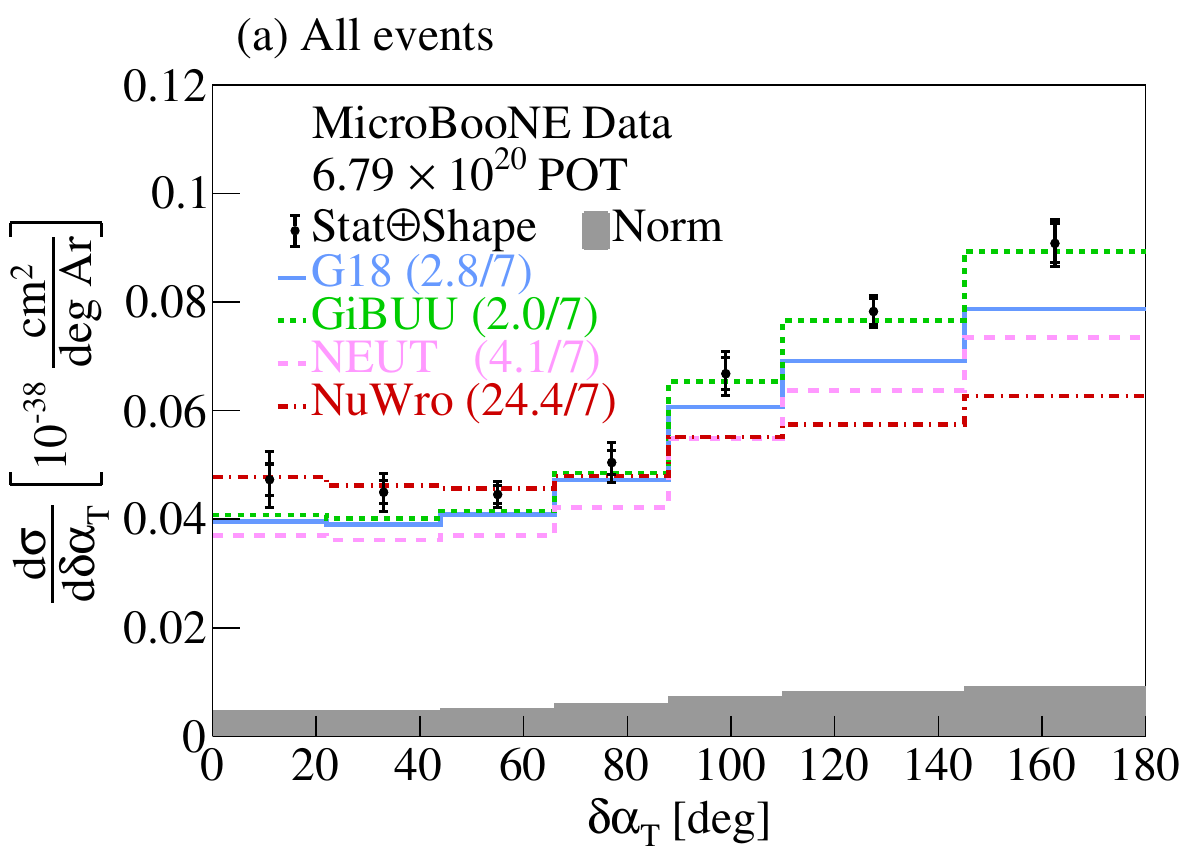}
\includegraphics[width=0.49\linewidth]{\figures 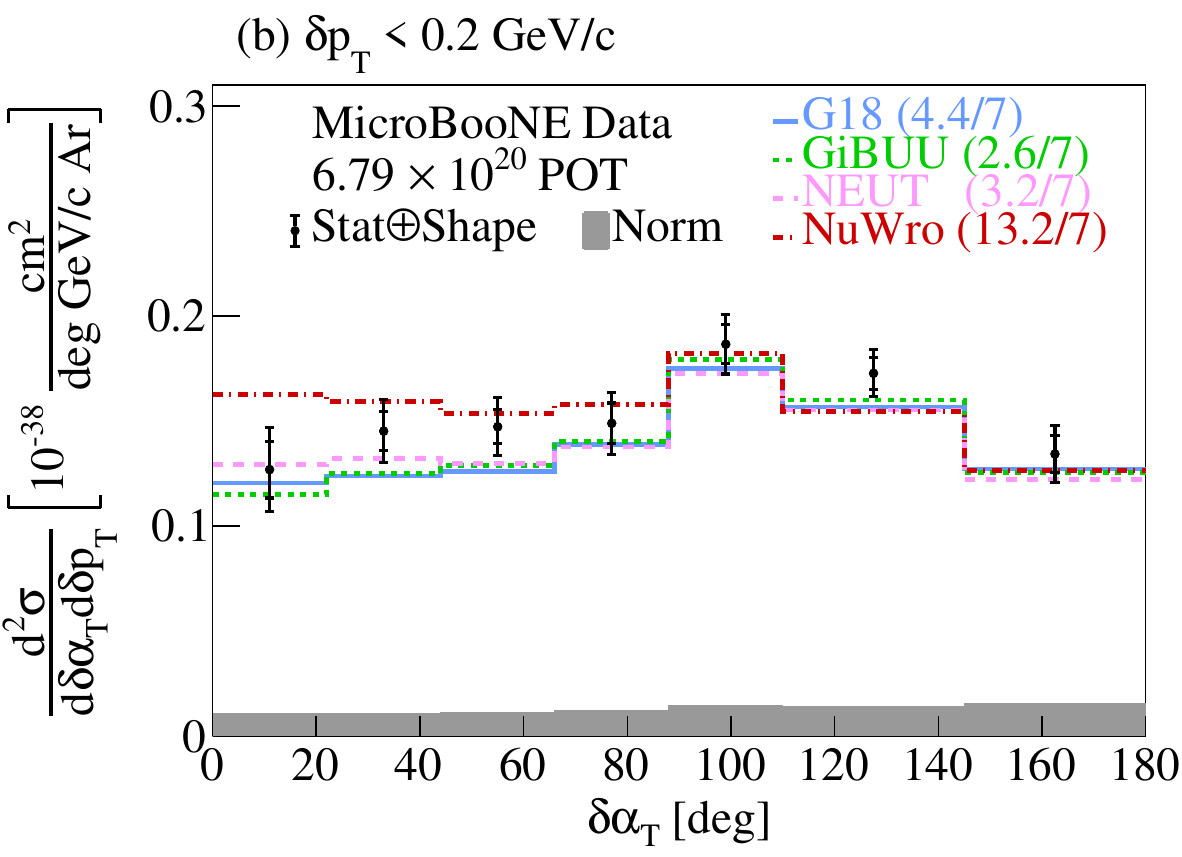}\\
\includegraphics[width=0.49\linewidth]{\figures 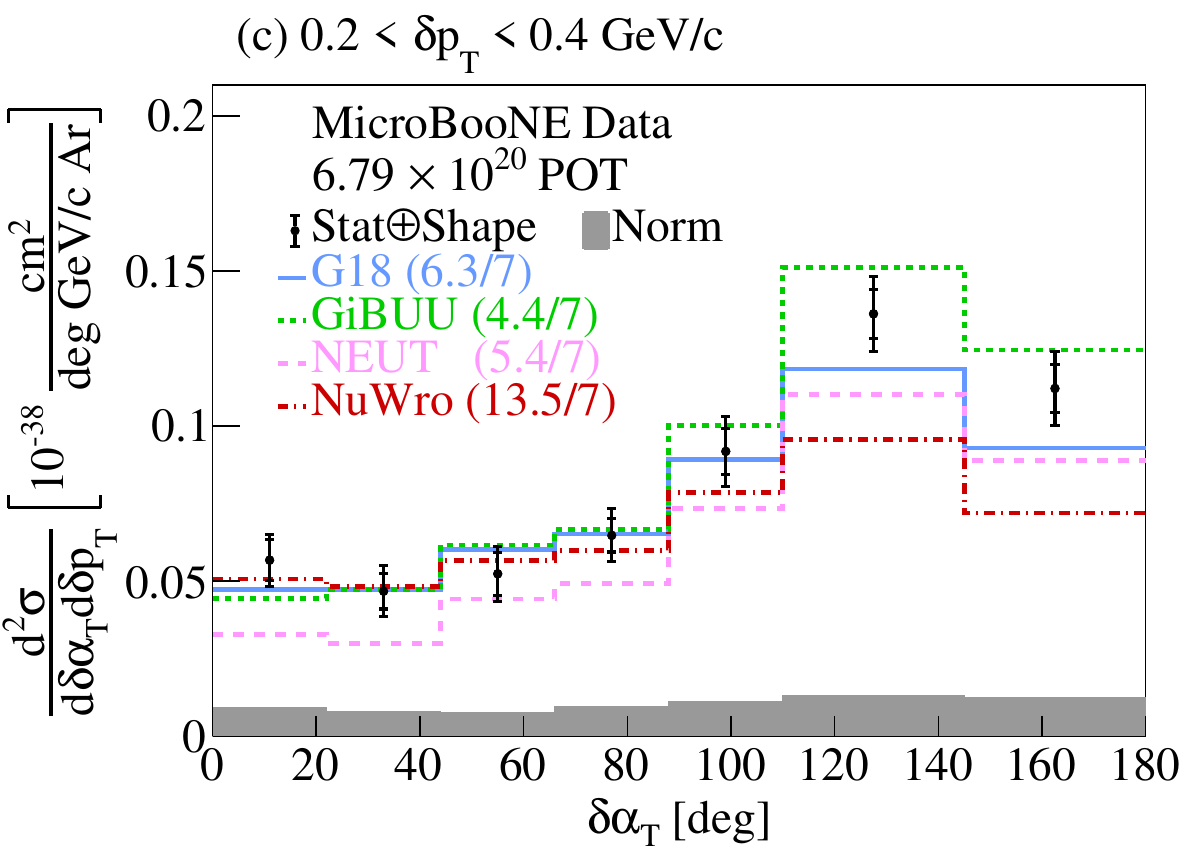}
\includegraphics[width=0.49\linewidth]{\figures 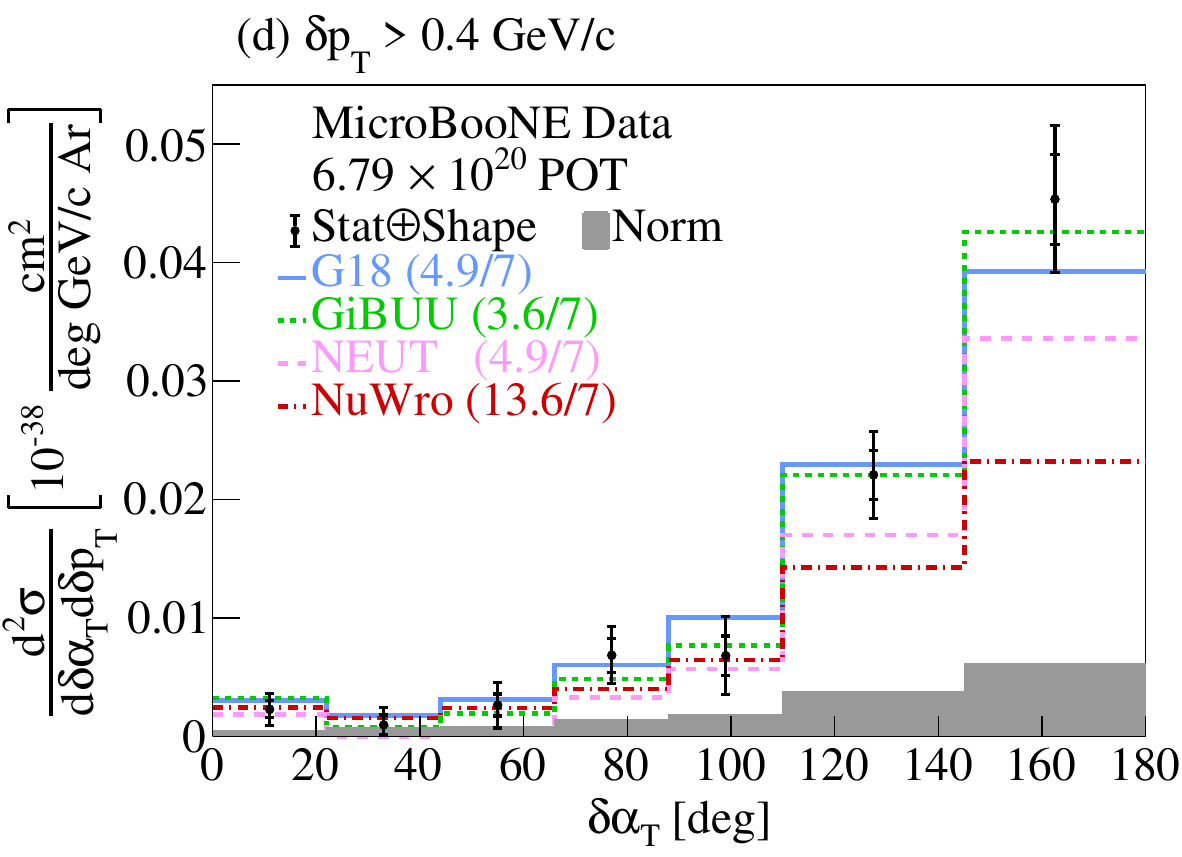}\\
\caption{
The flux-integrated (a) single- and (b-d) double- (in $\delta p_{T}$ bins) differential cross sections as a function of $\delta\alpha_{T}$. 
Inner and outer error bars show the statistical and total (statistical and shape systematic) uncertainty at the 1$\sigma$, or 68\%, confidence level. 
The gray band shows the normalization systematic uncertainty.
Colored lines show the results of theoretical cross section calculations using the $\texttt{G18 GENIE}$ (blue), $\texttt{GiBUU}$ (green), $\texttt{NEUT}$ (pink), and $\texttt{NuWro}$ (red) event generators.
The numbers in parentheses show the $\chi^{2}$/bins calculation for each one of the predictions.
}
\label{DeltaAlphaTInDeltaPTGen}
\end{figure*}

\begin{figure*}[htb!]
\centering
\includegraphics[width=0.49\linewidth]{\figures 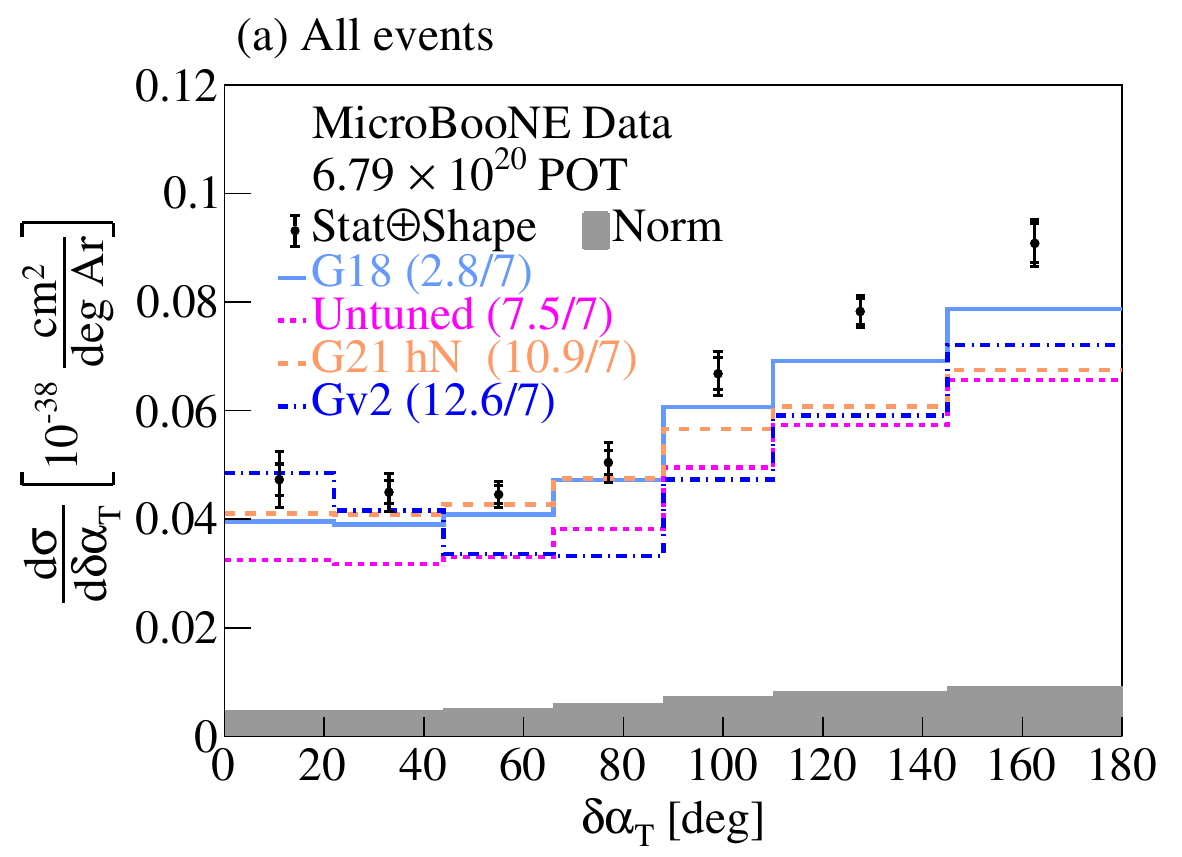}
\includegraphics[width=0.49\linewidth]{\figures 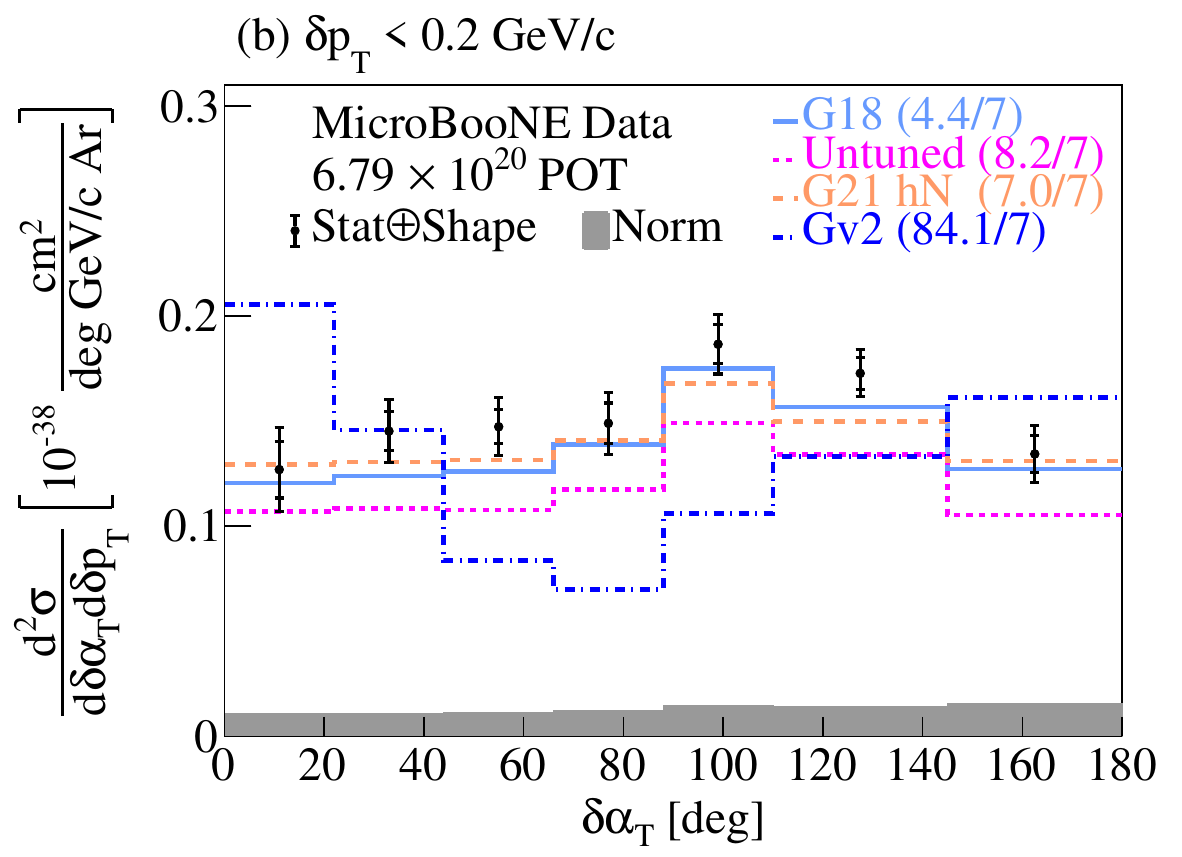}\\
\includegraphics[width=0.49\linewidth]{\figures 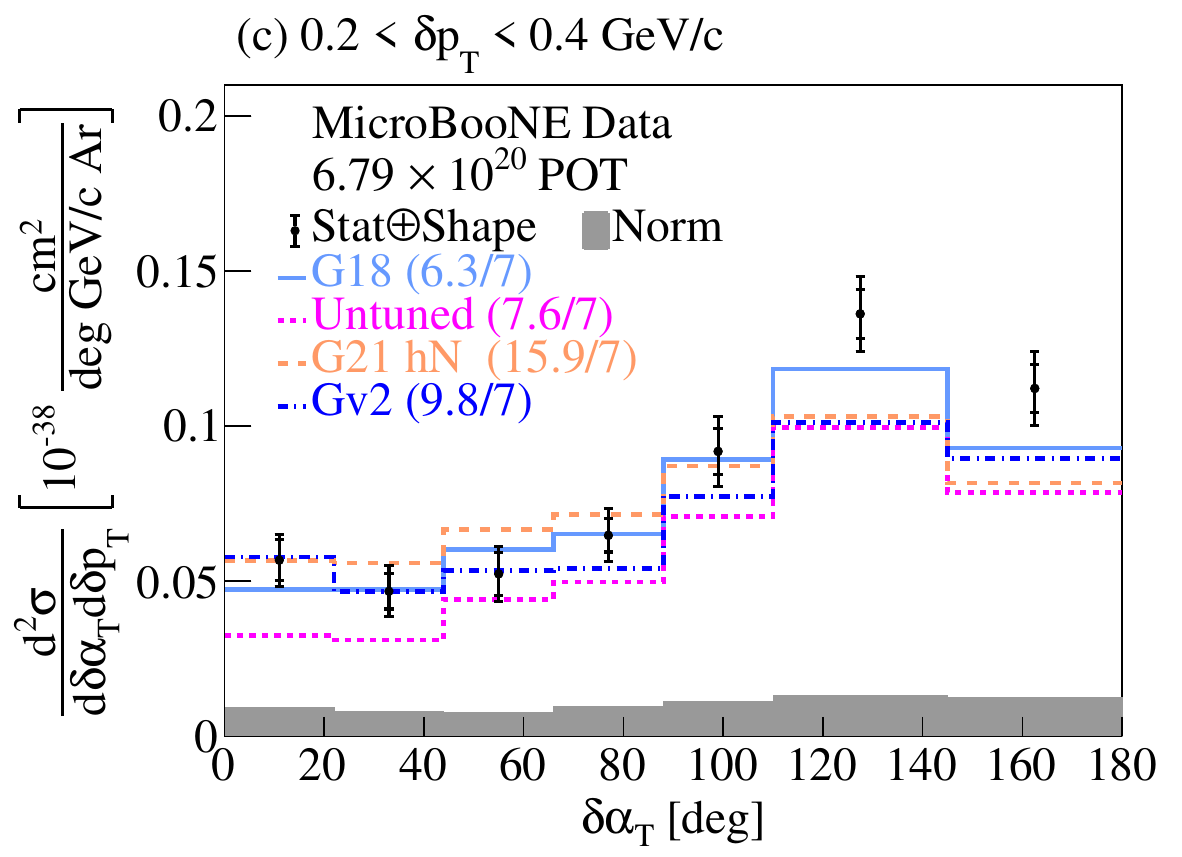}
\includegraphics[width=0.49\linewidth]{\figures 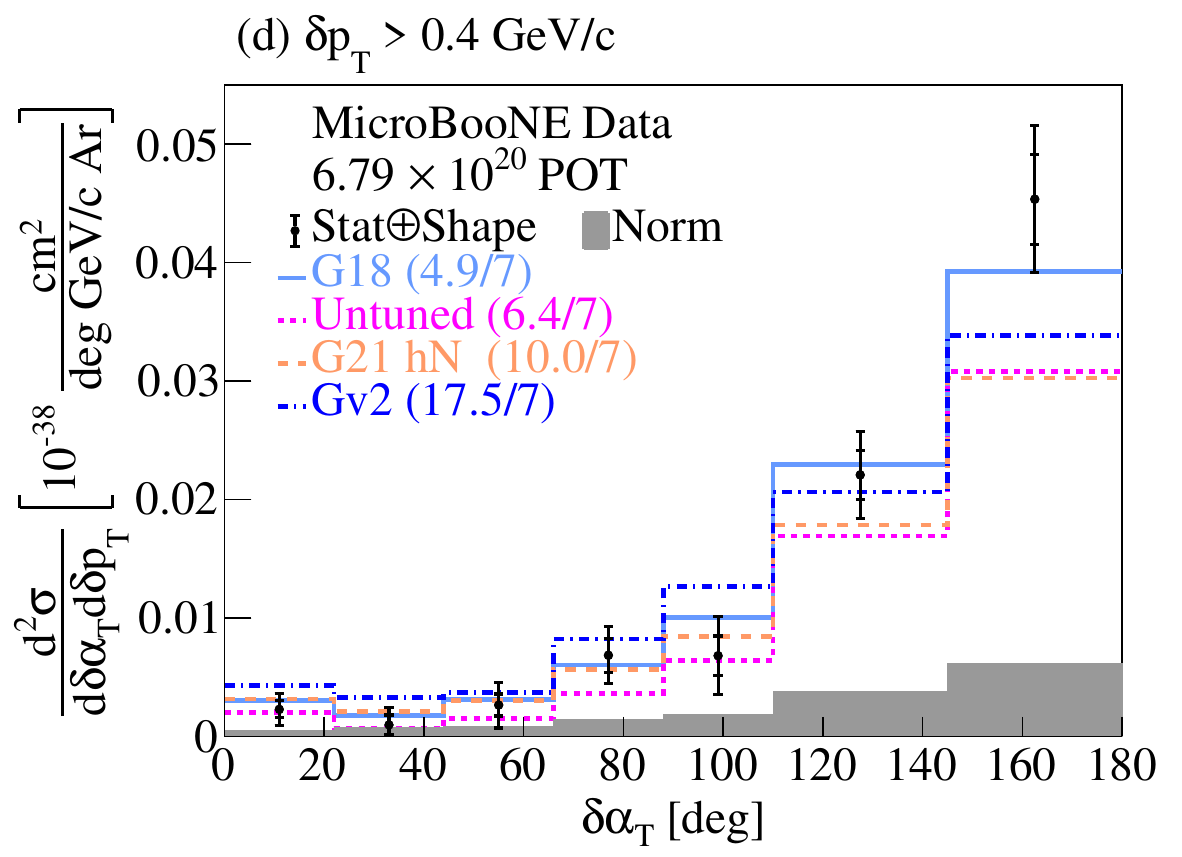}\\
\caption{
The flux-integrated (a) single- and (b-d) double- (in $\delta p_{T}$ bins) differential cross sections as a function of $\delta\alpha_{T}$. 
Inner and outer error bars show the statistical and total (statistical and shape systematic) uncertainty at the 1$\sigma$, or 68\%, confidence level. 
The gray band shows the normalization systematic uncertainty.
Colored lines show the results of theoretical cross section calculations using the $\texttt{G18}$ (light blue), $\texttt{Untuned}$ (magenta), $\texttt{G21}$ (orange), and $\texttt{Gv2}$ (dark blue) $\texttt{GENIE}$ configurations.
The numbers in parentheses show the $\chi^{2}$/bins calculation for each one of the predictions.
}
\label{DeltaAlphaTInDeltaPTGenie}
\end{figure*}

\begin{figure*}[htb!]
\centering 
\includegraphics[width=0.49\linewidth]{\figures 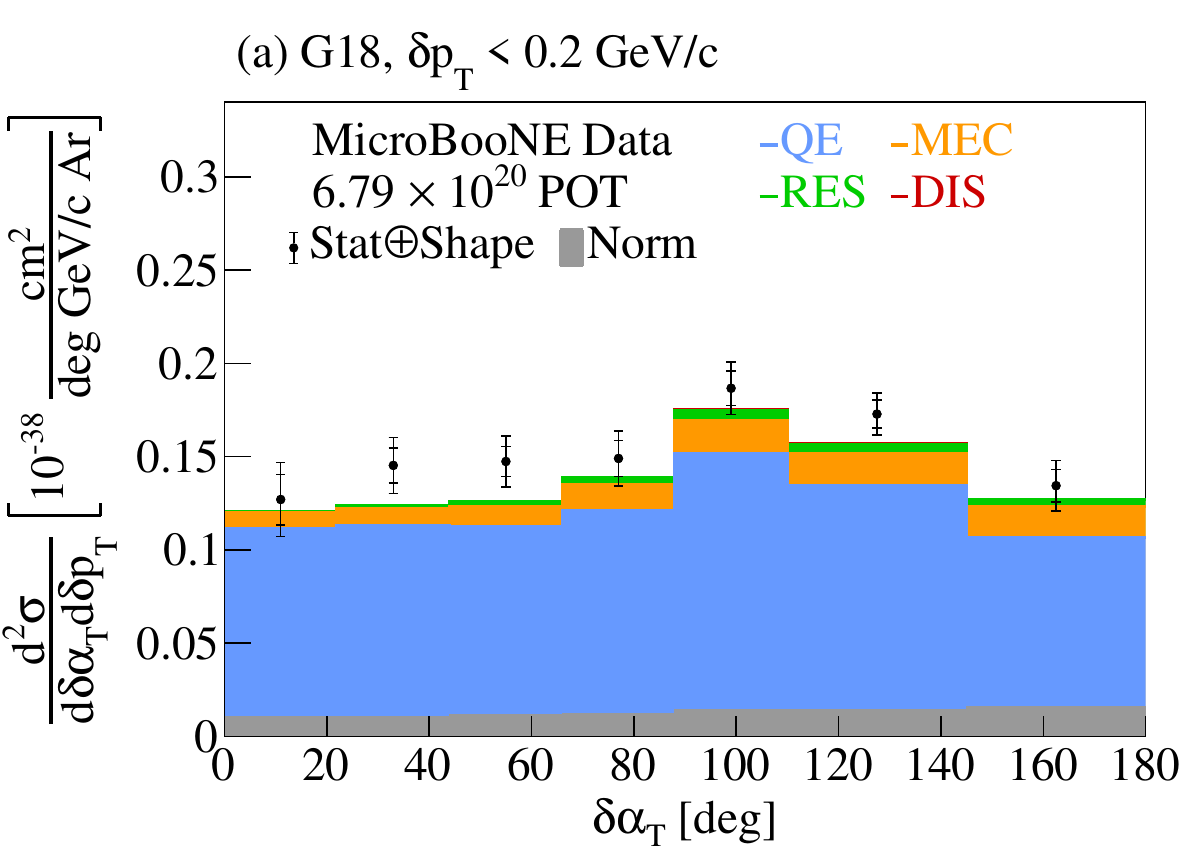}
\includegraphics[width=0.49\linewidth]{\figures 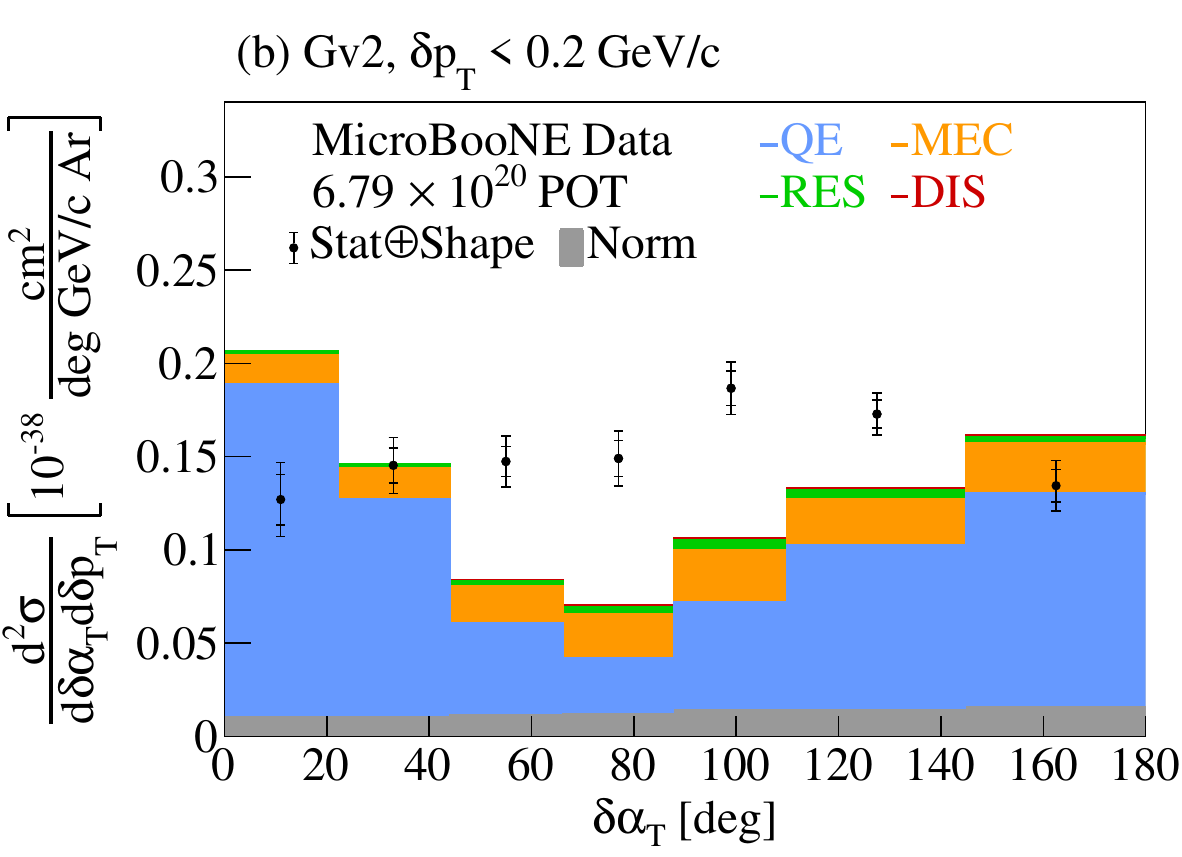}
\caption{
Comparison between the data flux-integrated double-differential cross section as a function of $\delta\alpha_{T}$ for events in the region $\delta p_{T}$ $<$ 0.2\,GeV/$c$ region against the $\texttt{G18}$ and $\texttt{Gv2 GENIE}$ predictions. 
Inner and outer error bars show the statistical and total (statistical and shape systematic) uncertainty at the 1$\sigma$, or 68\%, confidence level. 
The gray band shows the normalization systematic uncertainty.
Colored stacked histograms show the results of theoretical cross section calculations using the (a) $\texttt{G18}$ and (b) $\texttt{Gv2 GENIE}$ predictions for QE (blue), MEC (orange), RES (green), and DIS (red) interactions.
}
\label{DeltaAlphaTInDeltaPTInte}
\end{figure*}

\begin{figure*}[htb!]
\centering 
\includegraphics[width=0.49\linewidth]{\figures 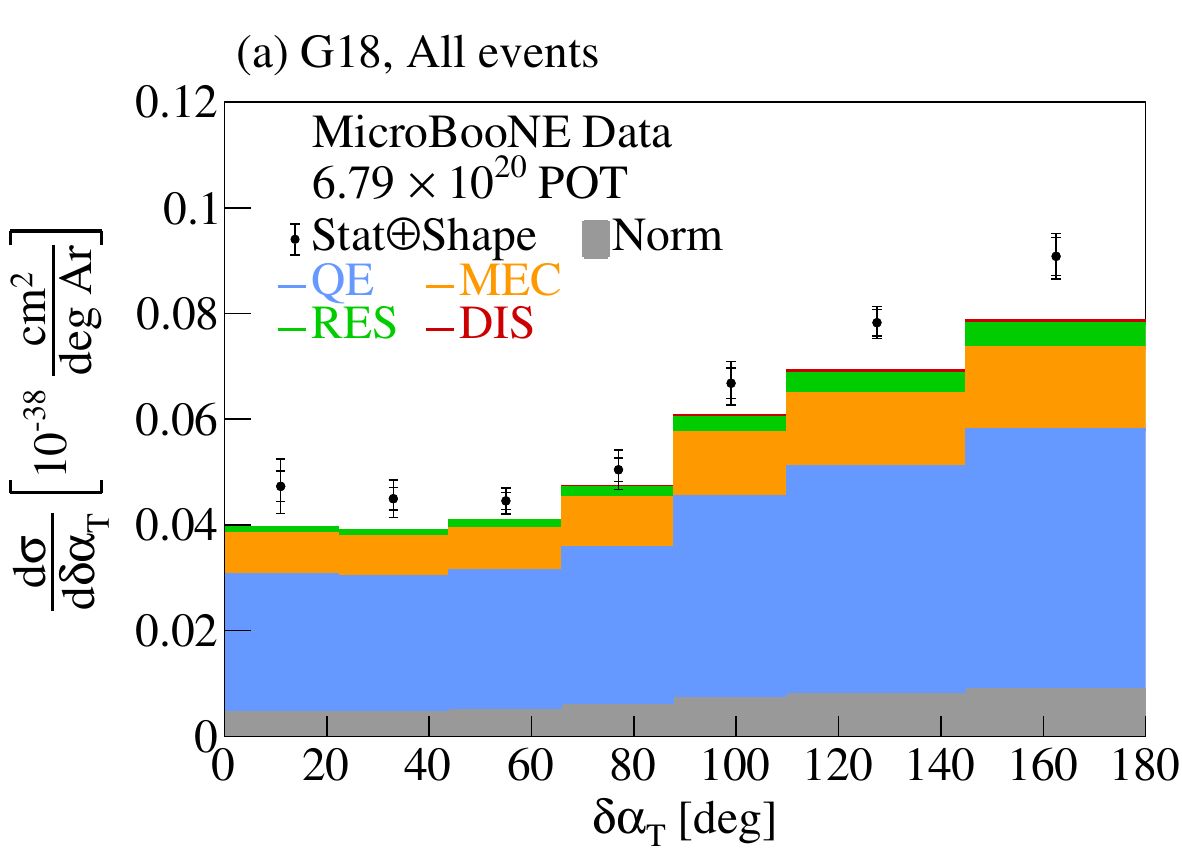}
\includegraphics[width=0.49\linewidth]{\figures 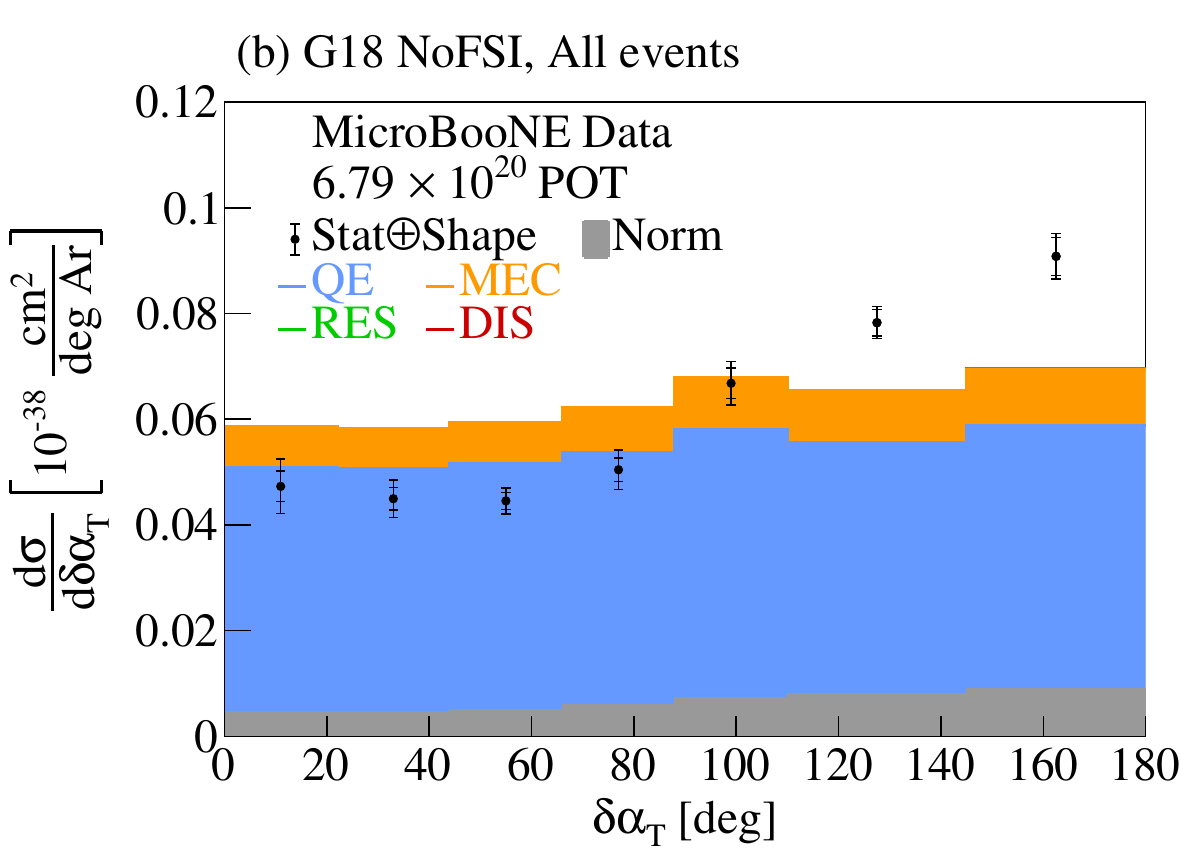}
\caption{Cross section interaction breakdown for the selected events for the G18 configuration (left) with FSI effects, and (right) without FSI effects as a function of $\delta\alpha_{T}$.
}
\label{DeltaAlphaTBreakdown}
\end{figure*}

Figures~\ref{DeltaAlphaTInDeltaPTGen} and~\ref{DeltaAlphaTInDeltaPTGenie} show the single-differential cross sections as a function of $\delta\alpha_{T}$ using all the events (panel a), as well as the double-differential results in the same kinematic variable in $\delta p_{T}$ bins (panels b-d).
The single-differential results shown in panel a yield some interesting observations when compared to the relevant T2K and MINERvA results~\cite{Avanzini:2021qlx,Abe:2018pwo,PhysRevLett.121.022504}.
Our distribution illustrates a slightly asymmetric behavior, similar to the one reported by the T2K collaboration at a comparable energy with MicroBooNE. 
Both the already-published T2K results on carbon and the ones presented in this work on argon mostly demonstrate data-MC agreement within the experimental uncertainties. 
Therefore, the mass-number dependence of the nuclear effects seems to be reasonably well-modeled.
Unlike our result, the measurement by MINERvA reports a more pronounced asymmetry on hydrocarbon.
The breakdown plots in Fig.~18 in Ref.~\cite{Avanzini:2021qlx} show that this behavior is driven by enhanced  pion-production rates due to the higher average beam energy.
Low $\delta p_{T}$ values result in a fairly uniform $\delta\alpha_{T}$ distribution indicative of the absence of FSI effects in that part of the phase-space.
On the other hand, higher $\delta p_{T}$ values result in a highly asymmetric $\delta\alpha_{T}$ distribution, which is driven by the strength of the FSI interactions.
Figure~\ref{DeltaAlphaTInDeltaPTGen} shows the comparisons to a number of available neutrino event generators, where $\texttt{NuWro}$ is the generator with the most conservative FSI strength.
Figure~\ref{DeltaAlphaTInDeltaPTGenie} shows the same results compared to a number of $\texttt{GENIE}$ configurations, where $\texttt{Gv2}$ yields the highest $\chi^{2}$/bins result, especially in the lowest $\delta p_{T}$ region.
As shown in Fig.~\ref{DeltaAlphaTInDeltaPTInte}, this is driven by the $\texttt{Gv2}$ QE performance, which results in peaks at the edges of the distribution, unlike the data result.
Additionally, Fig.~\ref{DeltaAlphaTBreakdown} shows the effect of FSI on the CC1p0$\pi$ selection using the $\text{G18}$ configuration of $\text{GENIE}$ that introduces an asymmetric behavior in $\delta\alpha_{T}$.



\begin{figure*}[htb!]
\centering 
\includegraphics[width=0.49\linewidth]{\figures 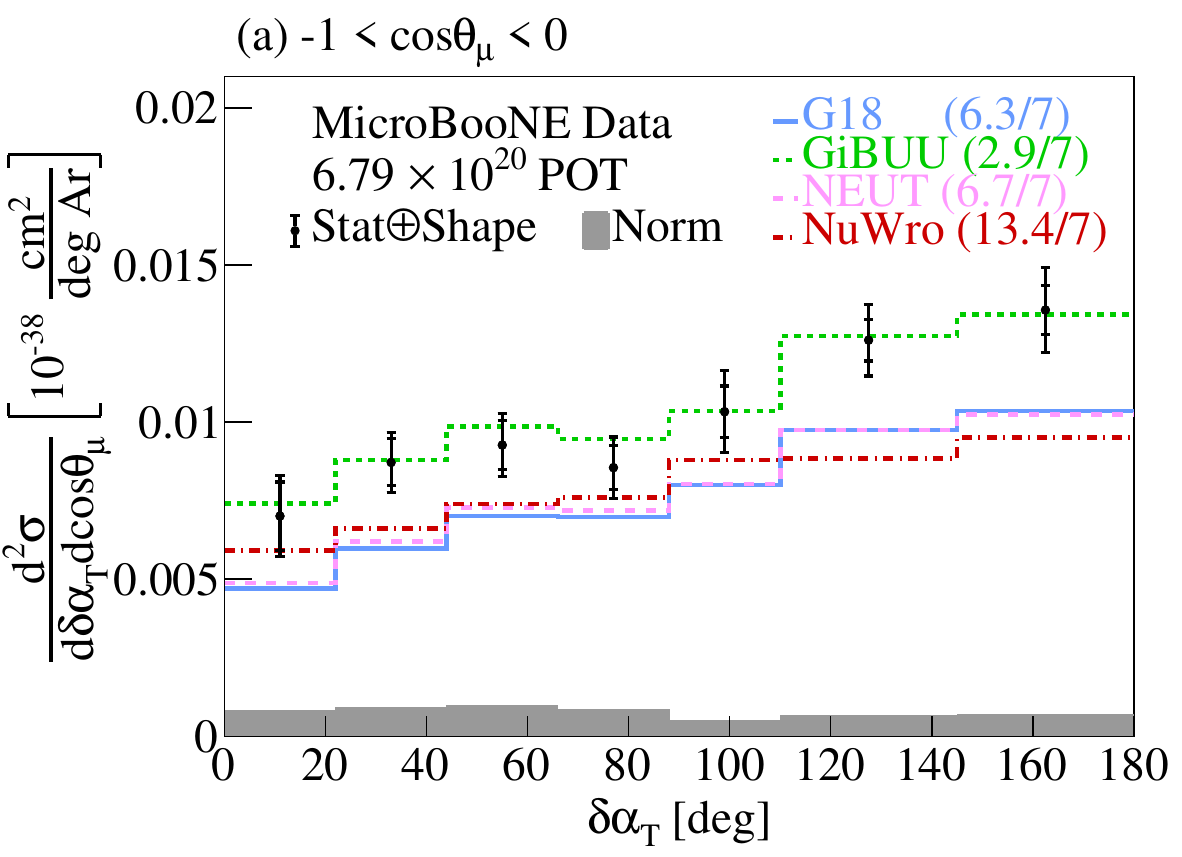}
\includegraphics[width=0.49\linewidth]{\figures 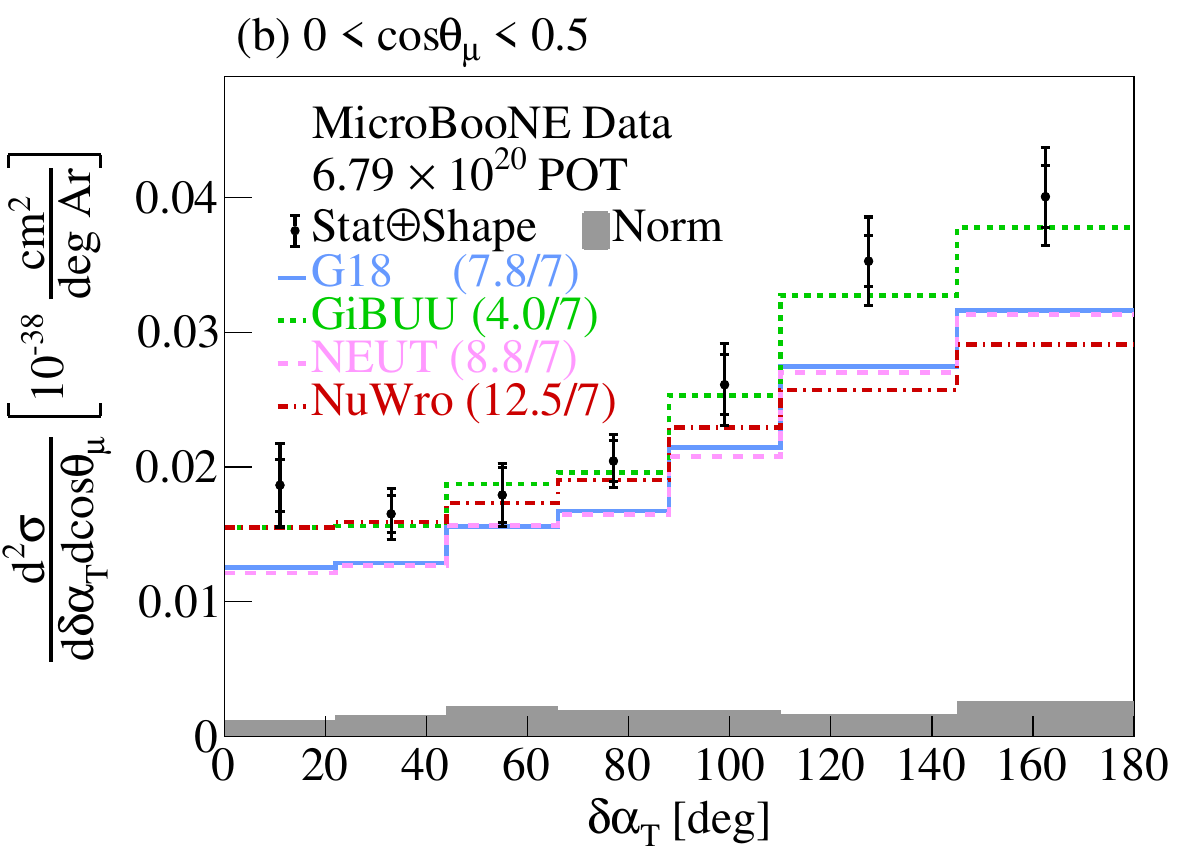}\\
\includegraphics[width=0.49\linewidth]{\figures 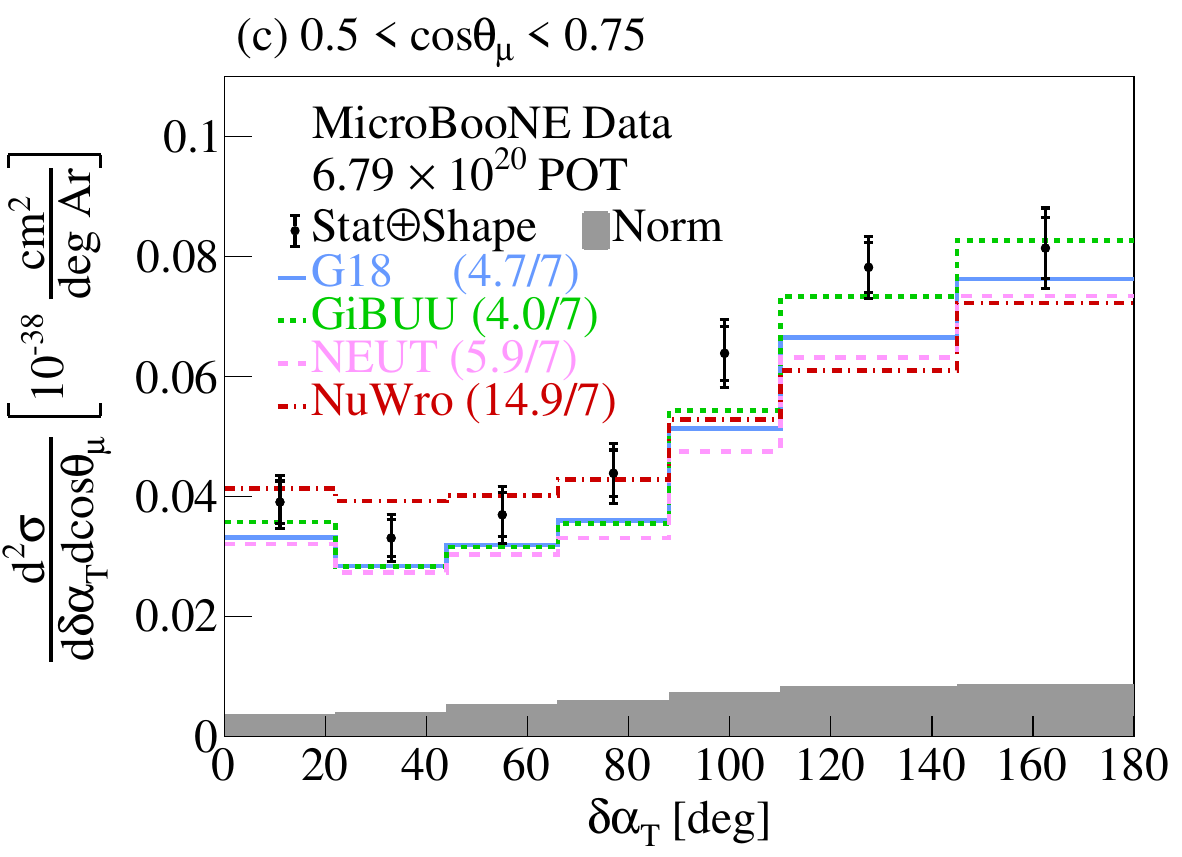}
\includegraphics[width=0.49\linewidth]{\figures 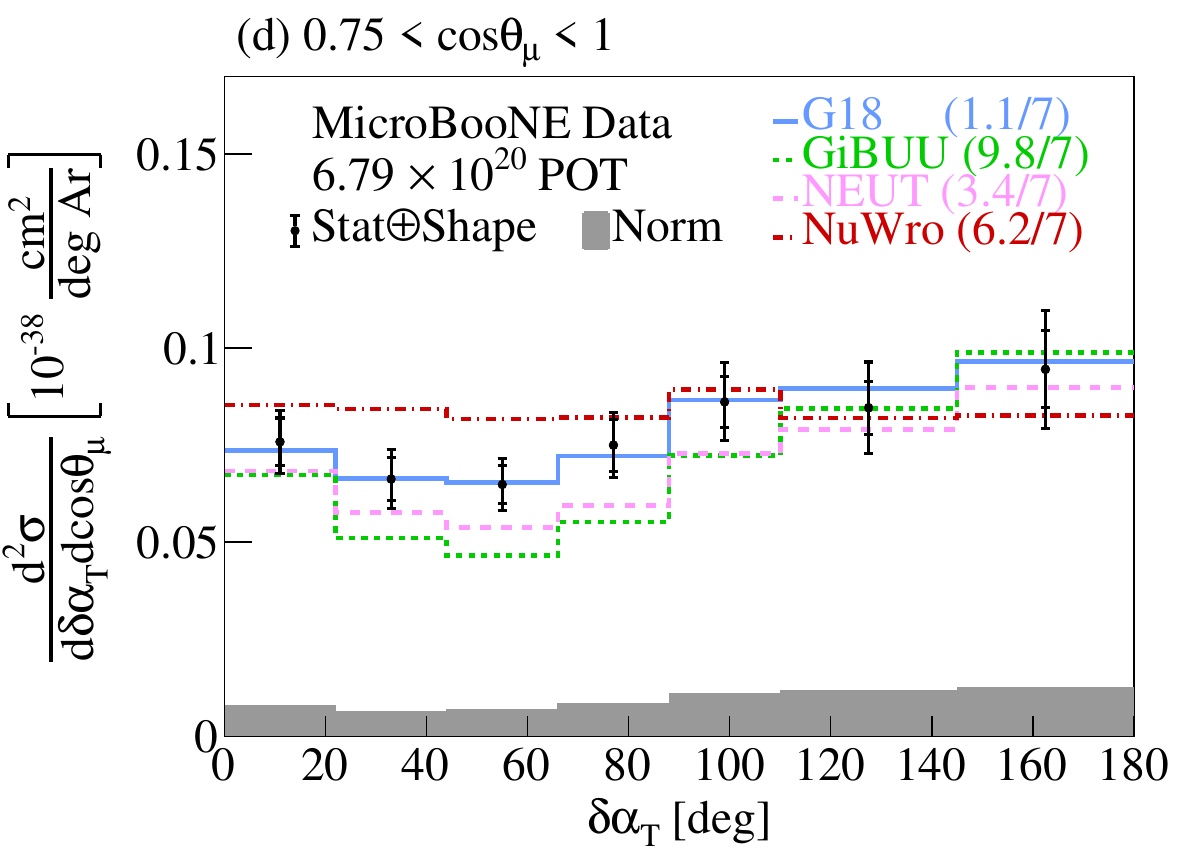}\\
\caption{
The flux-integrated double-differential cross sections as a function of $\delta\alpha_{T}$ in cos$\theta_{\mu}$ bins. 
Inner and outer error bars show the statistical and total (statistical and shape systematic) uncertainty at the 1$\sigma$, or 68\%, confidence level. 
The gray band shows the normalization systematic uncertainty.
Colored lines show the results of theoretical cross section calculations using the $\texttt{G18 GENIE}$ (blue), $\texttt{GiBUU}$ (green), $\texttt{NEUT}$ (pink), and $\texttt{NuWro}$ (red) event generators.
The numbers in parentheses show the $\chi^{2}$/bins calculation for each one of the predictions.
}
\label{DeltaAlphaTInMuonCosThetaGen}
\end{figure*}

\begin{figure*}[htb!]
\centering
\includegraphics[width=0.49\linewidth]{\figures 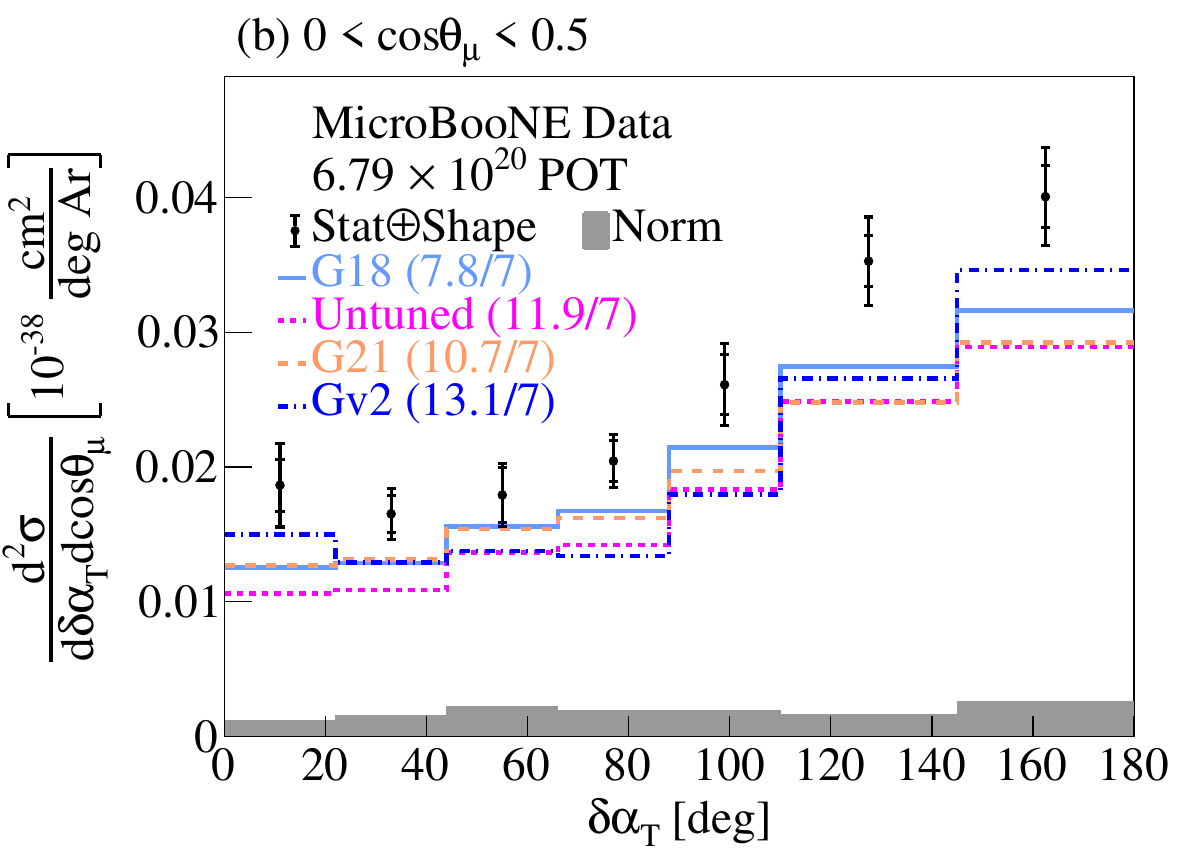}
\includegraphics[width=0.49\linewidth]{\figures 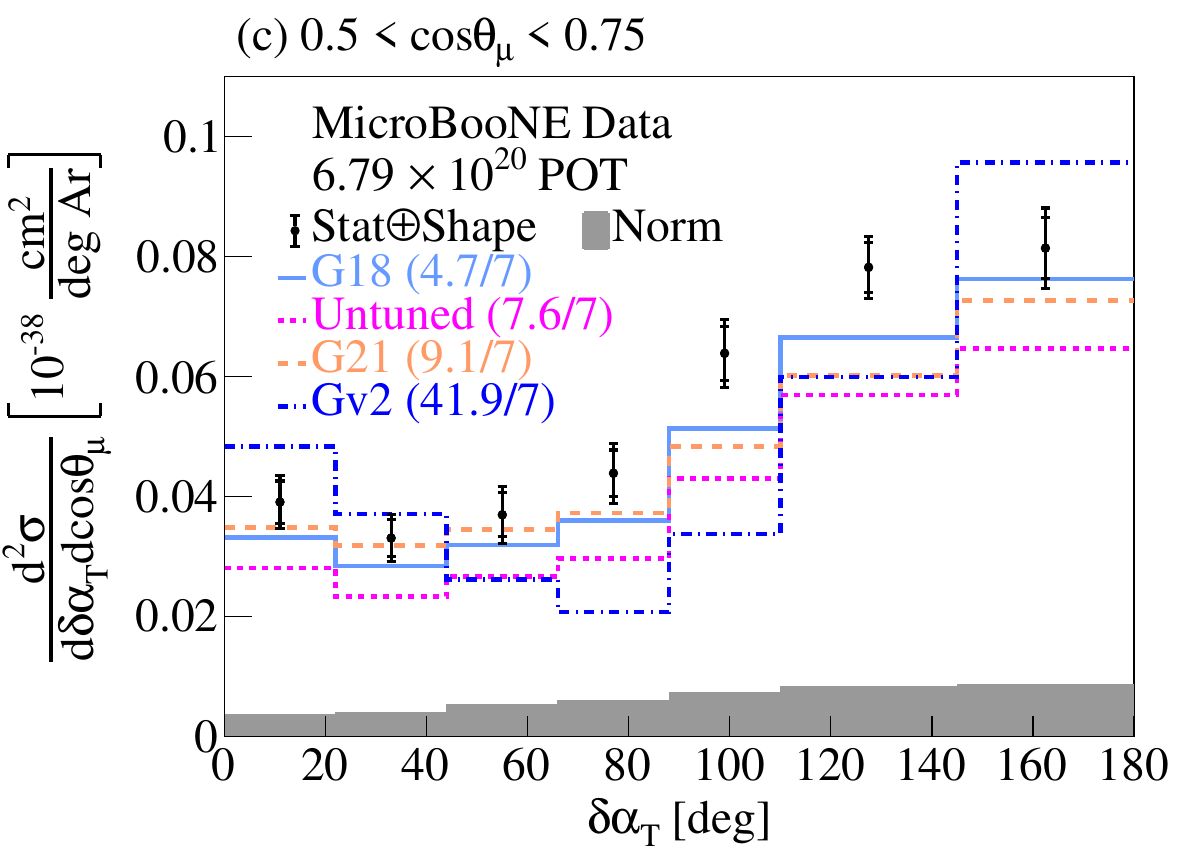}\\
\includegraphics[width=0.49\linewidth]{\figures 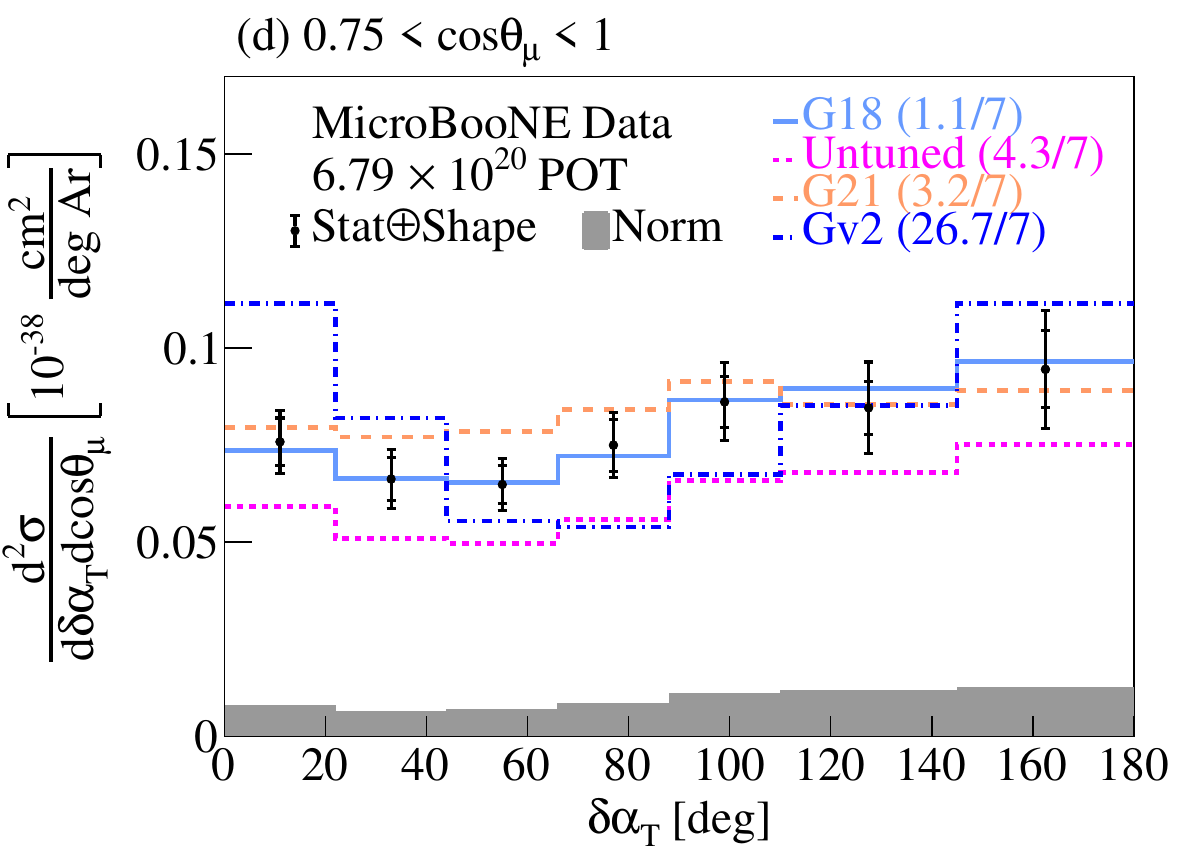}
\includegraphics[width=0.49\linewidth]{\figures 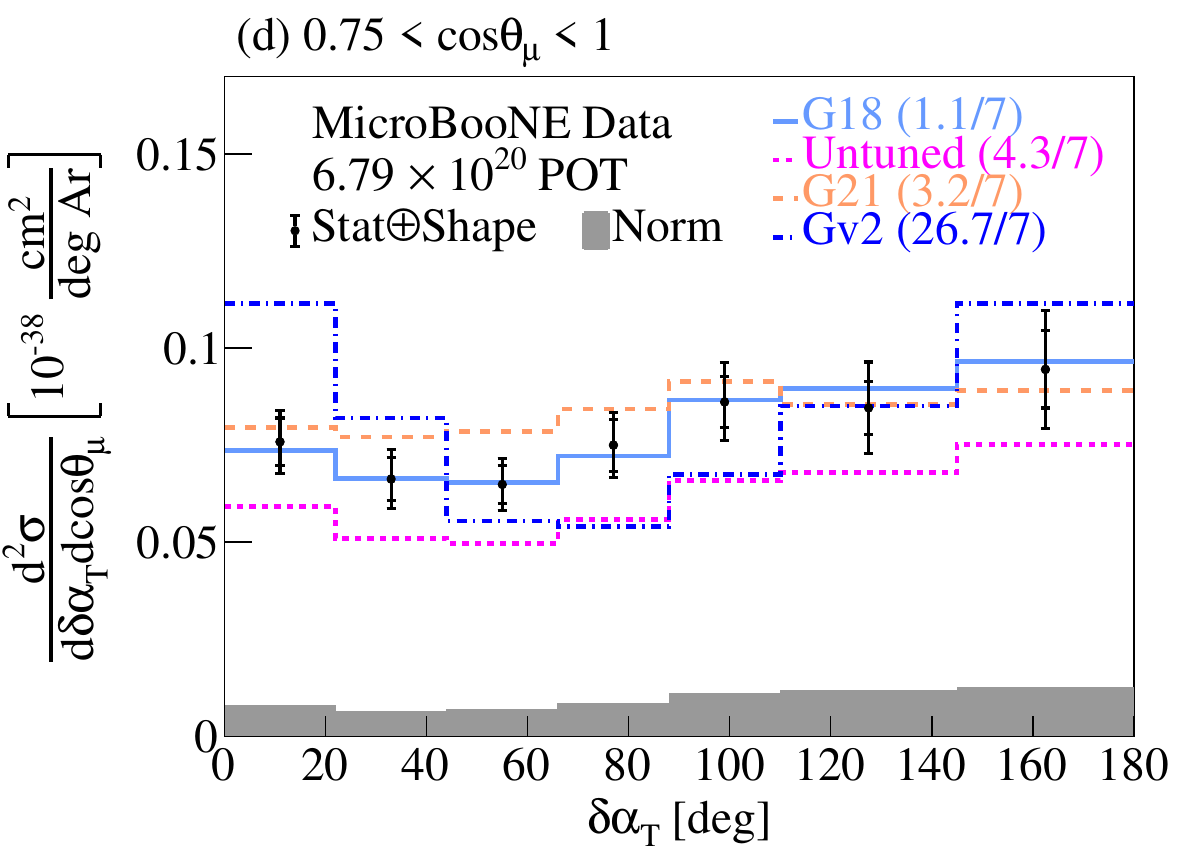}\\
\caption{
The flux-integrated double-differential cross sections as a function of $\delta\alpha_{T}$ in cos$\theta_{\mu}$ bins. 
Inner and outer error bars show the statistical and total (statistical and shape systematic) uncertainty at the 1$\sigma$, or 68\%, confidence level. 
The gray band shows the normalization systematic uncertainty.
Colored lines show the results of theoretical cross section calculations using the $\texttt{G18}$ (light blue), $\texttt{Untuned}$ (magenta), $\texttt{G21}$ (orange), and $\texttt{Gv2}$ (dark blue) $\texttt{GENIE}$ configurations.
The numbers in parentheses show the $\chi^{2}$/bins calculation for each one of the predictions.
}
\label{DeltaAlphaTInMuonCosThetaGenie}
\end{figure*}

Figures~\ref{DeltaAlphaTInMuonCosThetaGen} and~\ref{DeltaAlphaTInMuonCosThetaGenie} show the double-differential results as a function of $\delta\alpha_{T}$ in cos$\theta_{\mu}$ bins.
All the bins illustrate an asymmetric $\delta\alpha_{T}$ distribution, with the exception of the region where cos$\theta_{\mu}\approx$ 1, with the latter implying that this part of phase-space includes events with minimal FSI effects.
Figure~\ref{DeltaAlphaTInMuonCosThetaGen} shows the comparisons to a number of available neutrino event generators with $\texttt{GiBUU}$ giving the best performance.
Figure~\ref{DeltaAlphaTInMuonCosThetaGenie} shows the same results compared to a number of $\texttt{GENIE}$ configurations, illustrating that $\texttt{Gv2}$ is disfavored in the region where cos$\theta_{\mu} <$ 0.75.



\begin{figure*}[htb!]
\centering
\includegraphics[width=0.49\linewidth]{\figures 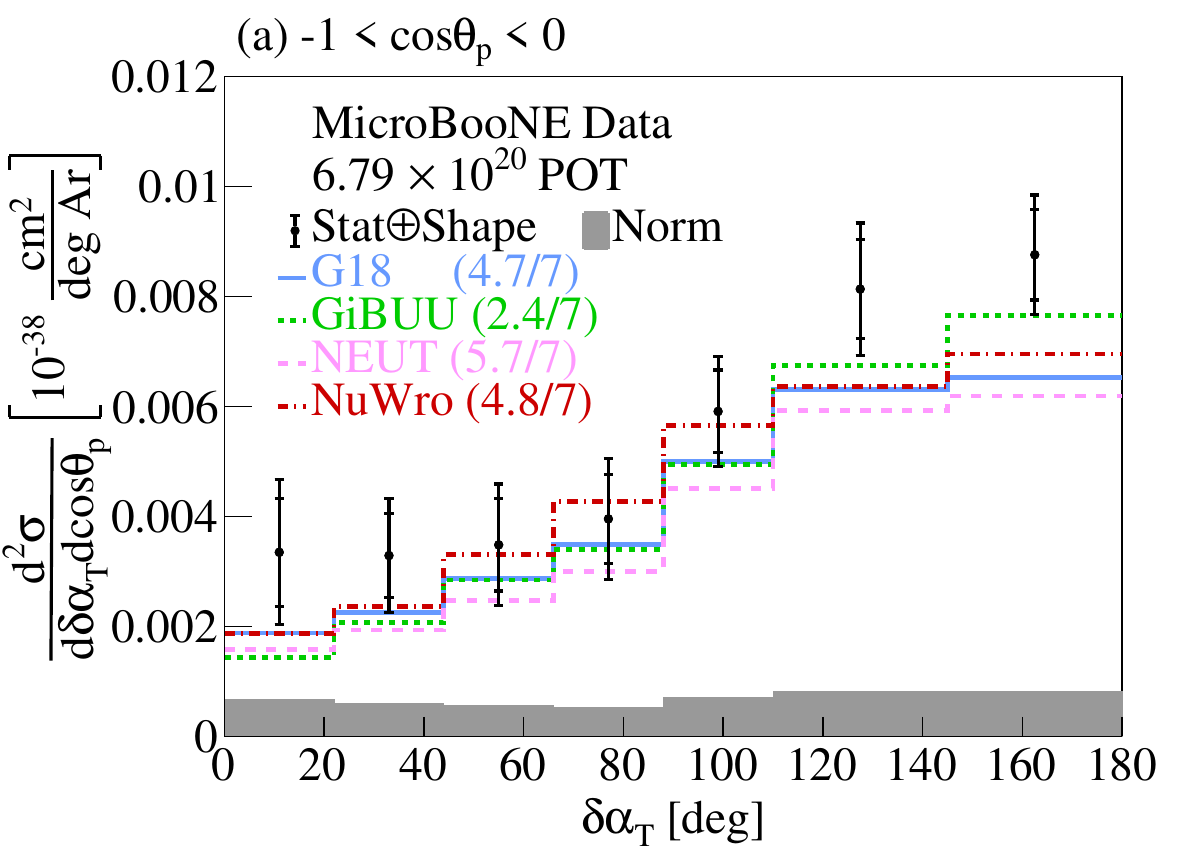}
\includegraphics[width=0.49\linewidth]{\figures 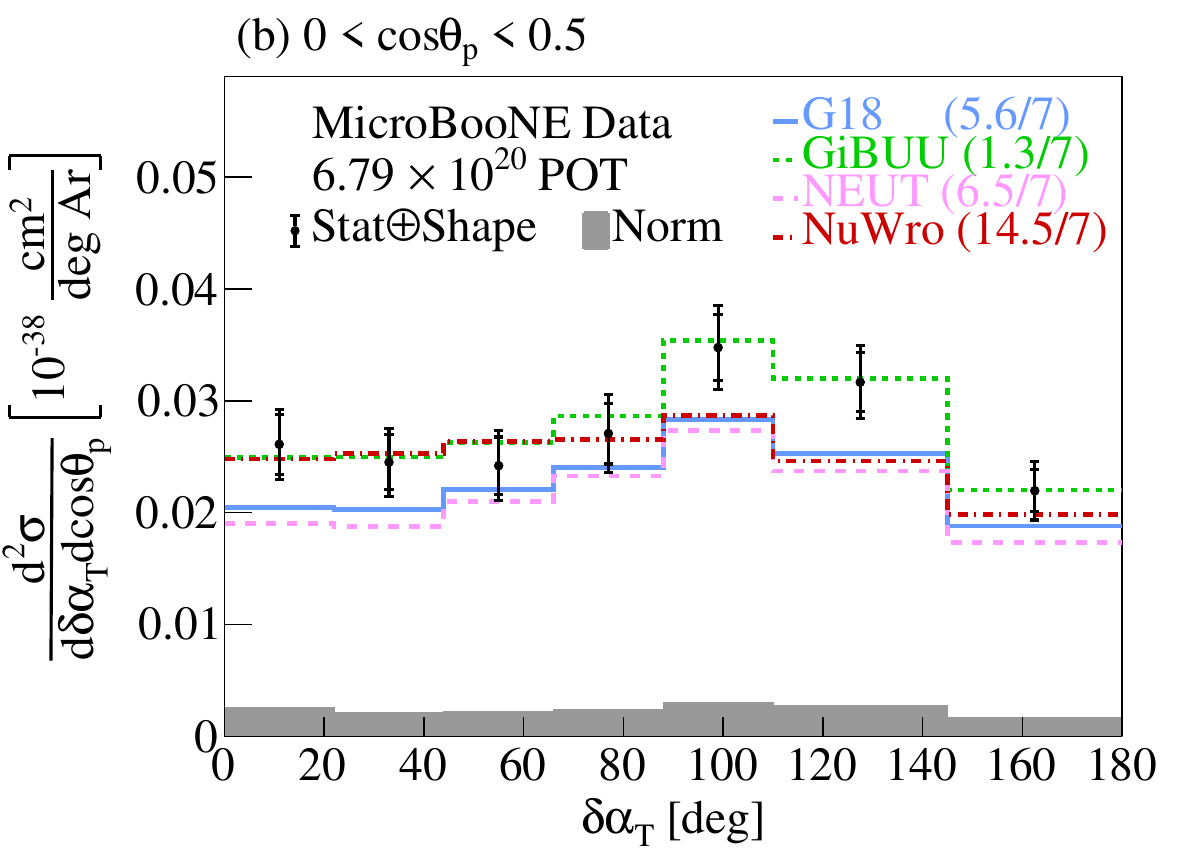}\\
\includegraphics[width=0.49\linewidth]{\figures 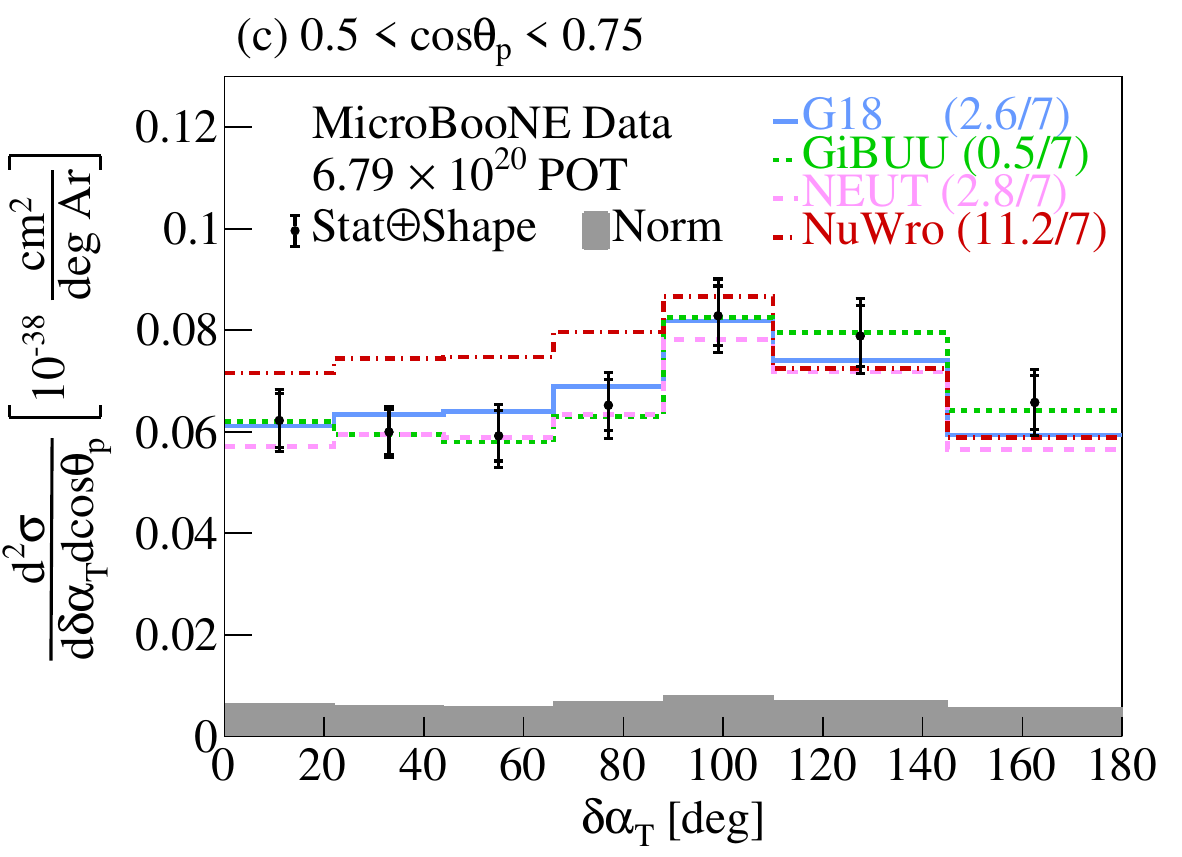}
\includegraphics[width=0.49\linewidth]{\figures 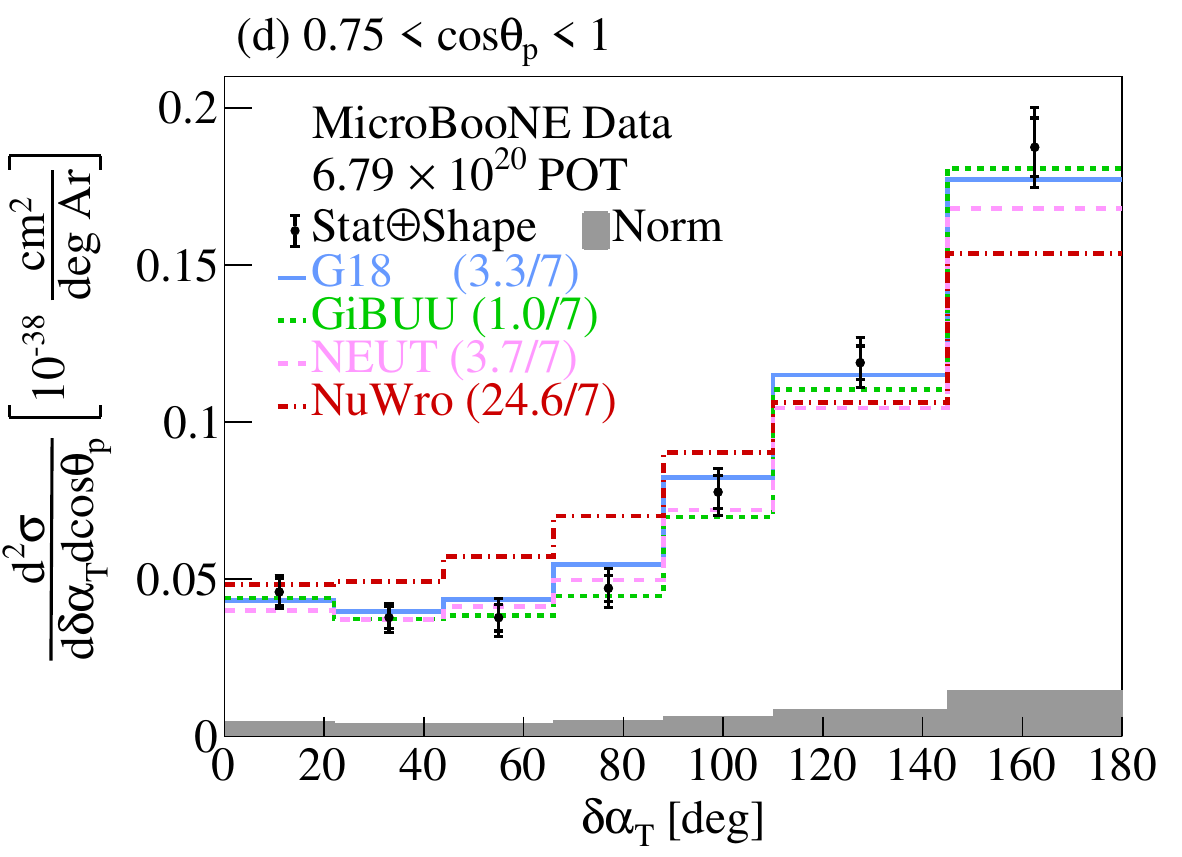}\\
\caption{
The flux-integrated double-differential cross sections as a function of $\delta\alpha_{T}$ in cos$\theta_{p}$ bins. 
Inner and outer error bars show the statistical and total (statistical and shape systematic) uncertainty at the 1$\sigma$, or 68\%, confidence level. 
The gray band shows the normalization systematic uncertainty.
Colored lines show the results of theoretical cross section calculations using the $\texttt{G18 GENIE}$ (blue), $\texttt{GiBUU}$ (green), $\texttt{NEUT}$ (pink), and $\texttt{NuWro}$ (red) event generators.
The numbers in parentheses show the $\chi^{2}$/bins calculation for each one of the predictions.
}
\label{DeltaAlphaTInProtonCosThetaGen}
\end{figure*}

\begin{figure*}[htb!]
\centering
\includegraphics[width=0.49\linewidth]{\figures 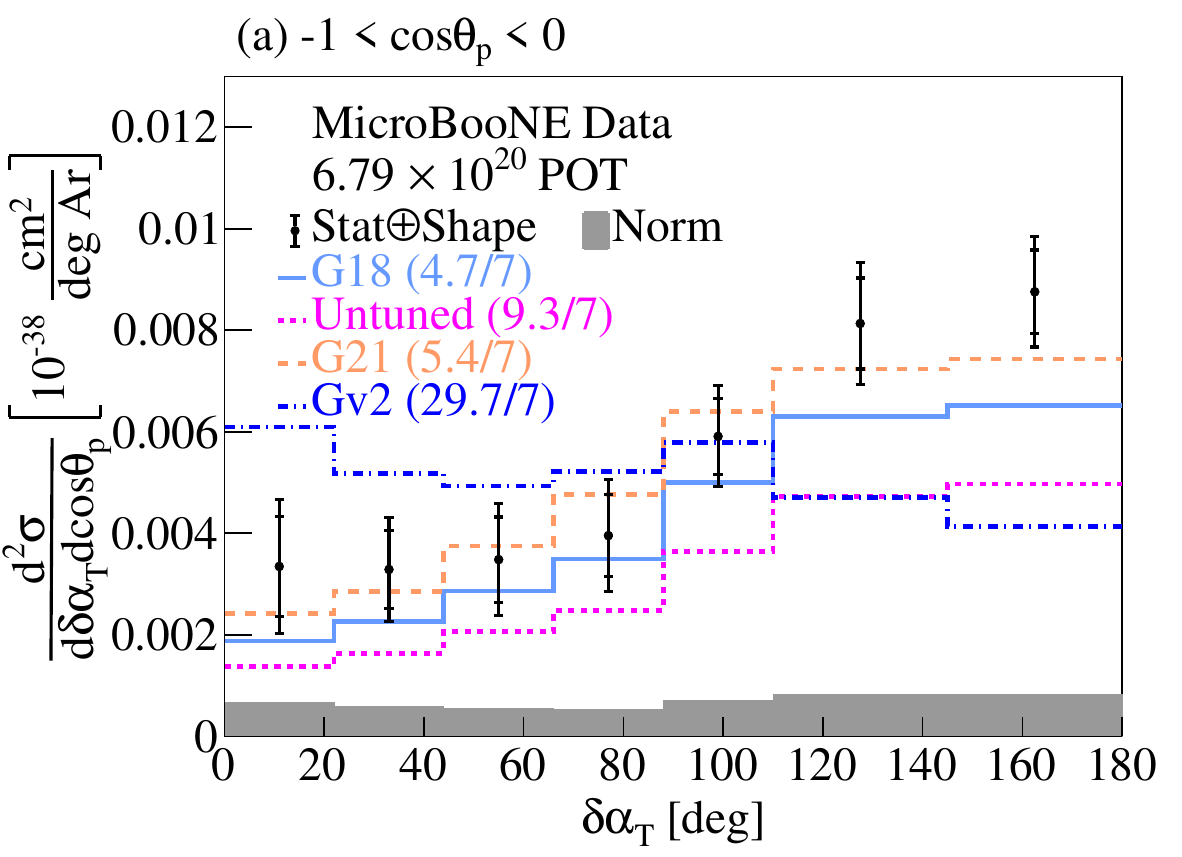}
\includegraphics[width=0.49\linewidth]{\figures 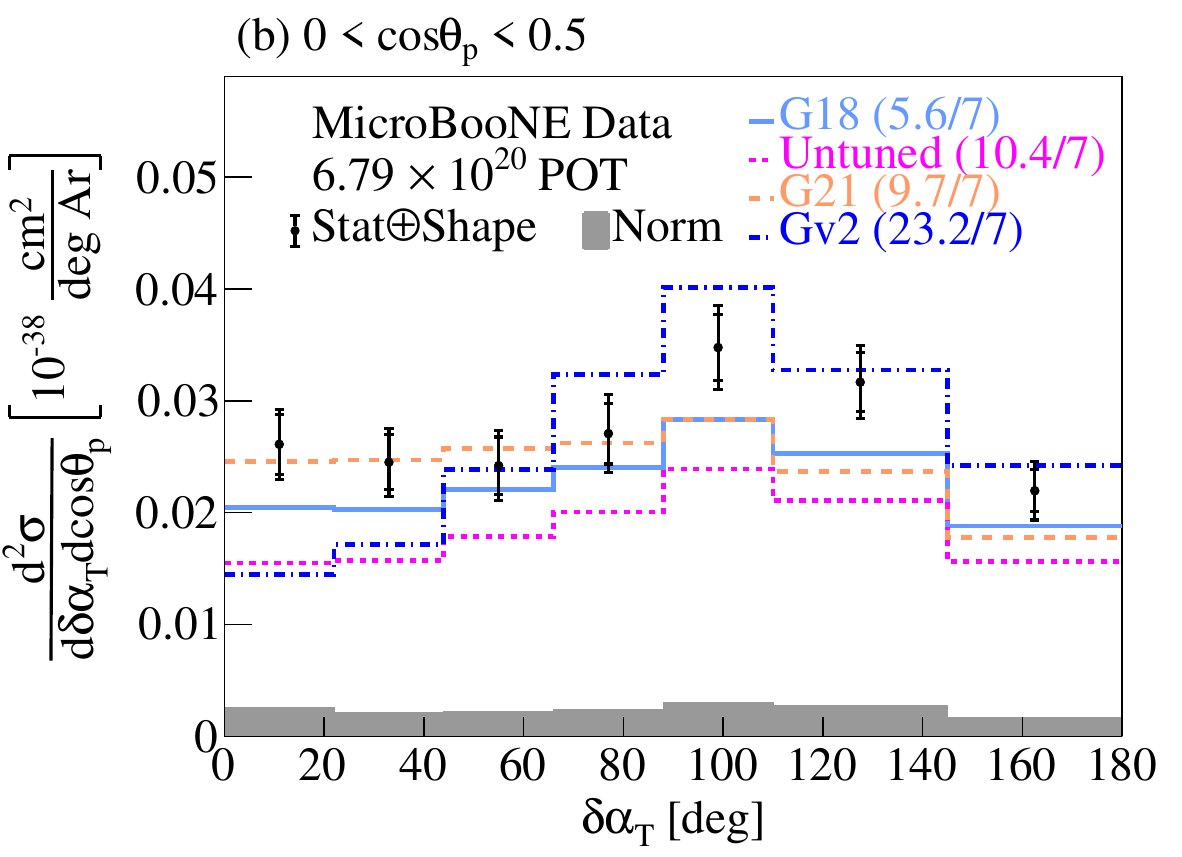}\\
\includegraphics[width=0.49\linewidth]{\figures 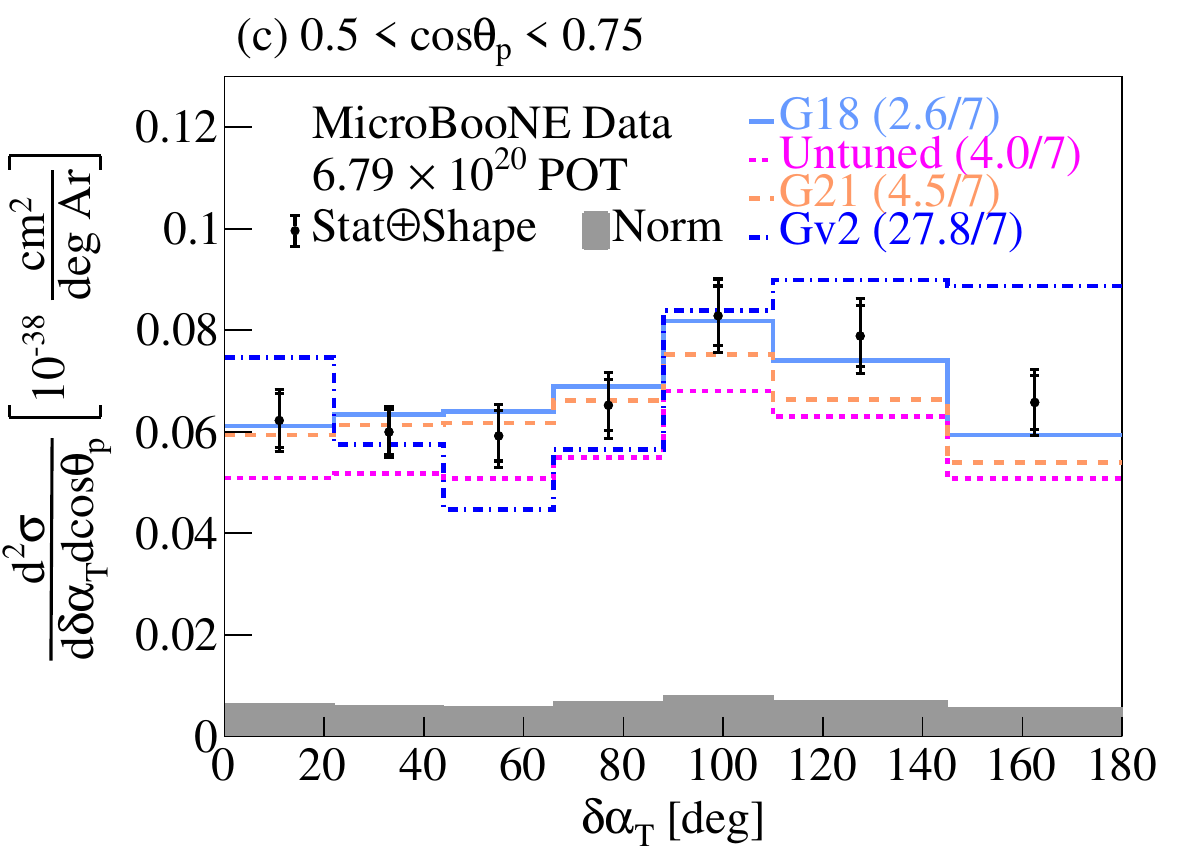}
\includegraphics[width=0.49\linewidth]{\figures 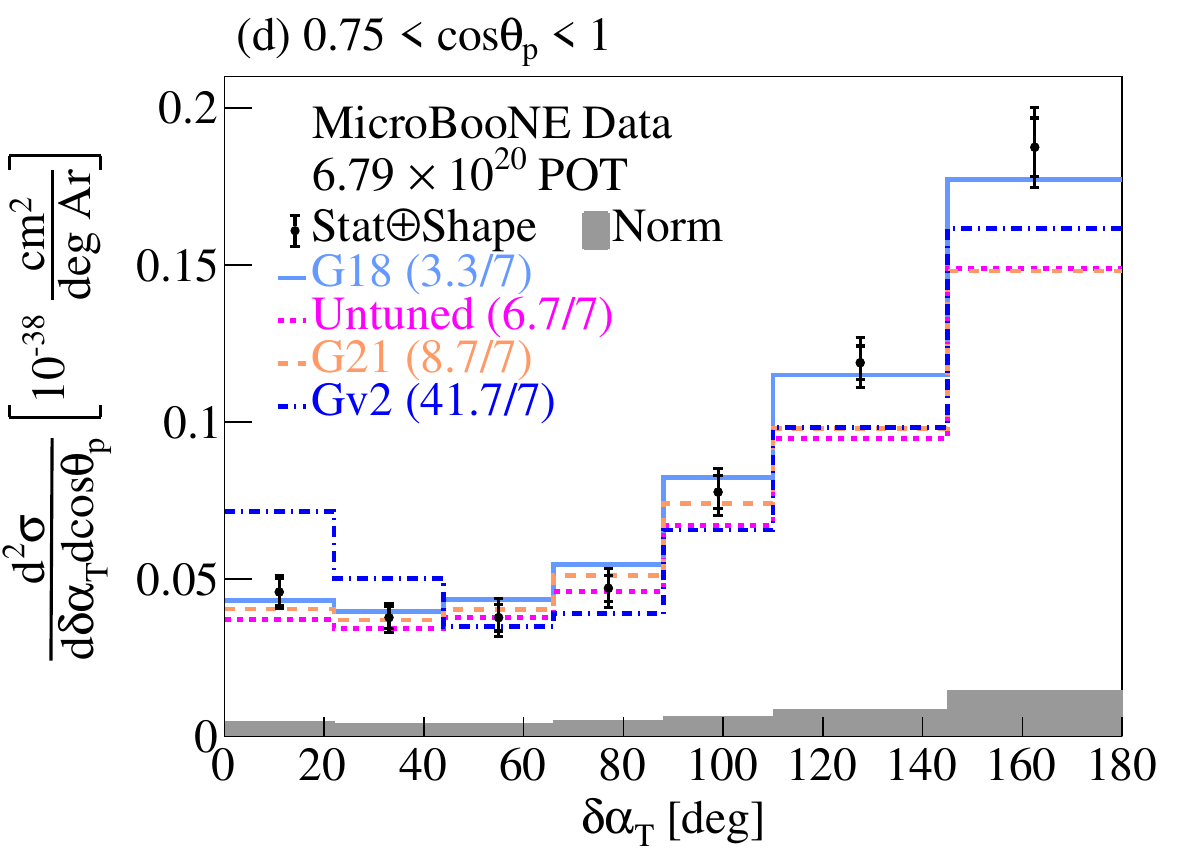}\\
\caption{
The flux-integrated double-differential cross sections as a function of $\delta\alpha_{T}$ in cos$\theta_{p}$ bins. 
Inner and outer error bars show the statistical and total (statistical and shape systematic) uncertainty at the 1$\sigma$, or 68\%, confidence level. 
The gray band shows the normalization systematic uncertainty.
Colored lines show the results of theoretical cross section calculations using the $\texttt{G18}$ (light blue), $\texttt{Untuned}$ (magenta), $\texttt{G21}$ (orange), and $\texttt{Gv2}$ (dark blue) $\texttt{GENIE}$ configurations.
The numbers in parentheses show the $\chi^{2}$/bins calculation for each one of the predictions.
}
\label{DeltaAlphaTInProtonCosThetaGenie}
\end{figure*}

\begin{figure*}[htb!]
\centering 
\includegraphics[width=0.49\linewidth]{\figures 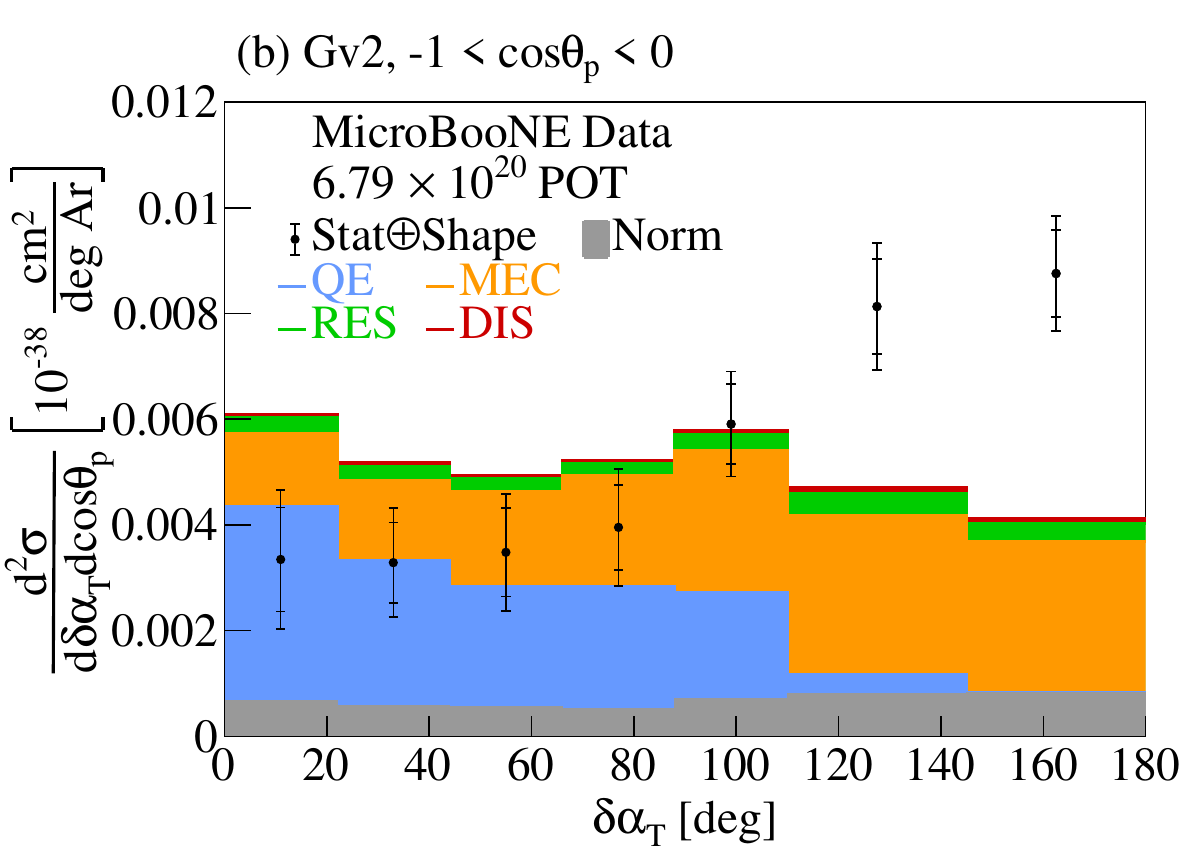}
\includegraphics[width=0.49\linewidth]{\figures 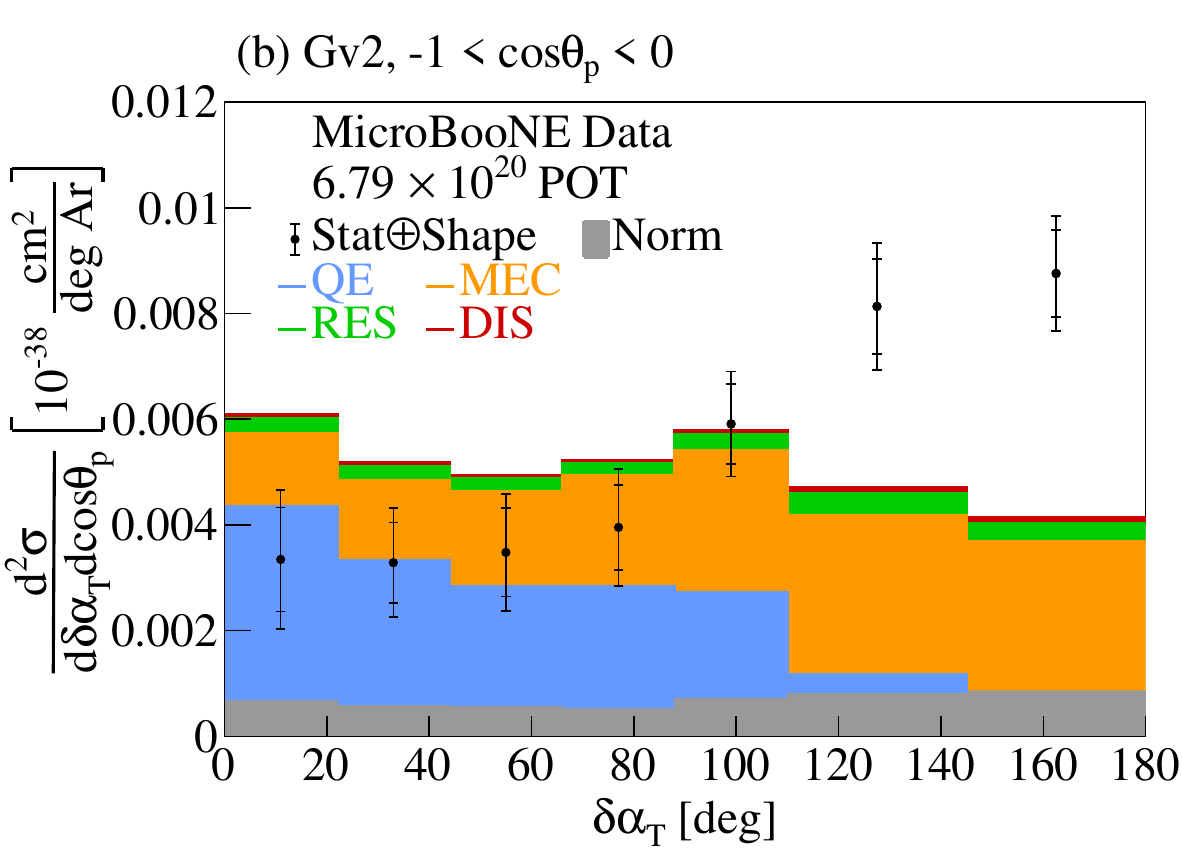}
\caption{
Comparison between the data flux-integrated double-differential cross section as a function of $\delta\alpha_{T}$ for events in the region -1 $<$ cos$\theta_{p}$ $<$ 0 region against the $\texttt{G18}$ and $\texttt{Gv2 GENIE}$ predictions. 
Inner and outer error bars show the statistical and total (statistical and shape systematic) uncertainty at the 1$\sigma$, or 68\%, confidence level. 
The gray band shows the normalization systematic uncertainty.
Colored stacked histograms show the results of theoretical cross section calculations using the (a) $\texttt{G18}$ and (b) $\texttt{Gv2 GENIE}$ predictions for QE (blue), MEC (orange), RES (green), and DIS (red) interactions.
}
\label{DeltaAlphaTInCosThetaPInte}
\end{figure*}

Figures~\ref{DeltaAlphaTInProtonCosThetaGen} and~\ref{DeltaAlphaTInProtonCosThetaGenie} show the double-differential cross sections as a function of $\delta\alpha_{T}$ in cos$\theta_{p}$ bins.
The results in the region with 0 $<$ cos$\theta_{p}$ $<$ 0.75 show a fairly flat distribution.
The cross section distributions corresponding to forward and backward proton scattering exhibit an FSI-driven asymmetric behavior. 
Figure~\ref{DeltaAlphaTInProtonCosThetaGen} shows the comparisons to a number of available neutrino event generators, where $\texttt{NuWro}$ yields a prediction that is disfavored for forward scattering.
Figure~\ref{DeltaAlphaTInProtonCosThetaGenie} shows the same results compared to a number of $\texttt{GENIE}$ configurations, illustrating that $\texttt{Gv2}$ is disfavored across all cos$\theta_{p}$ bins.
In the -1 $<$ cos$\theta_{p}$ $<$ 0 region shown in Fig.~\ref{DeltaAlphaTInProtonCosThetaGenie}a, all the predictions illustrate a peak close to 180$^{\circ}$ with the exception of $\texttt{Gv2}$.
The driving force for this difference is the $\texttt{Gv2}$ QE contribution, as can be seen in Fig.~\ref{DeltaAlphaTInCosThetaPInte}.
This is indicative of potential modeling issues in the Llewellyn Smith QE cross section and of the hA FSI performance used in the $\texttt{Gv2}$ prediction.
Unlike $\texttt{Gv2}$, the theory-driven $\texttt{GENIE v3}$ family of predictions ($\texttt{G18}$, $\texttt{Untuned}$, and $\texttt{G21}$) closely follow the data.



\begin{figure*}[htb!]
\centering 
\includegraphics[width=0.49\linewidth]{\figures 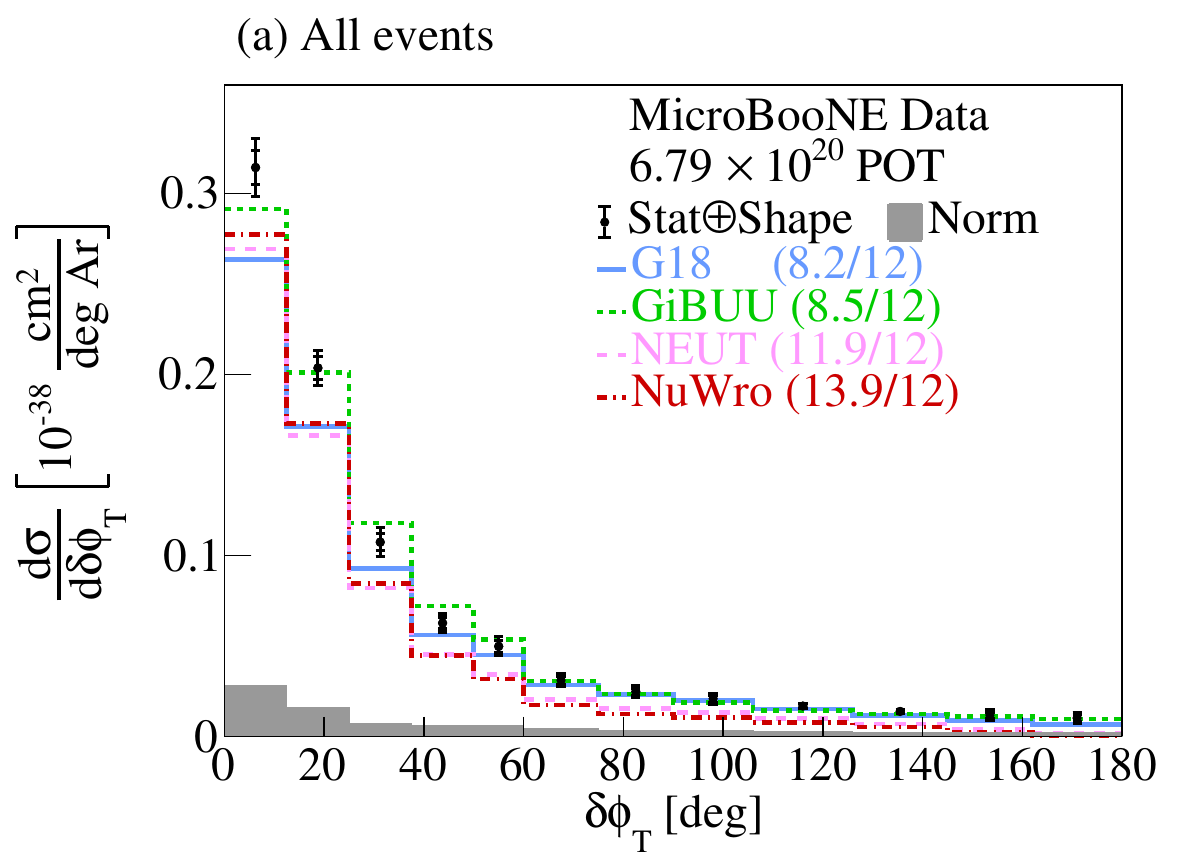}
\includegraphics[width=0.49\linewidth]{\figures 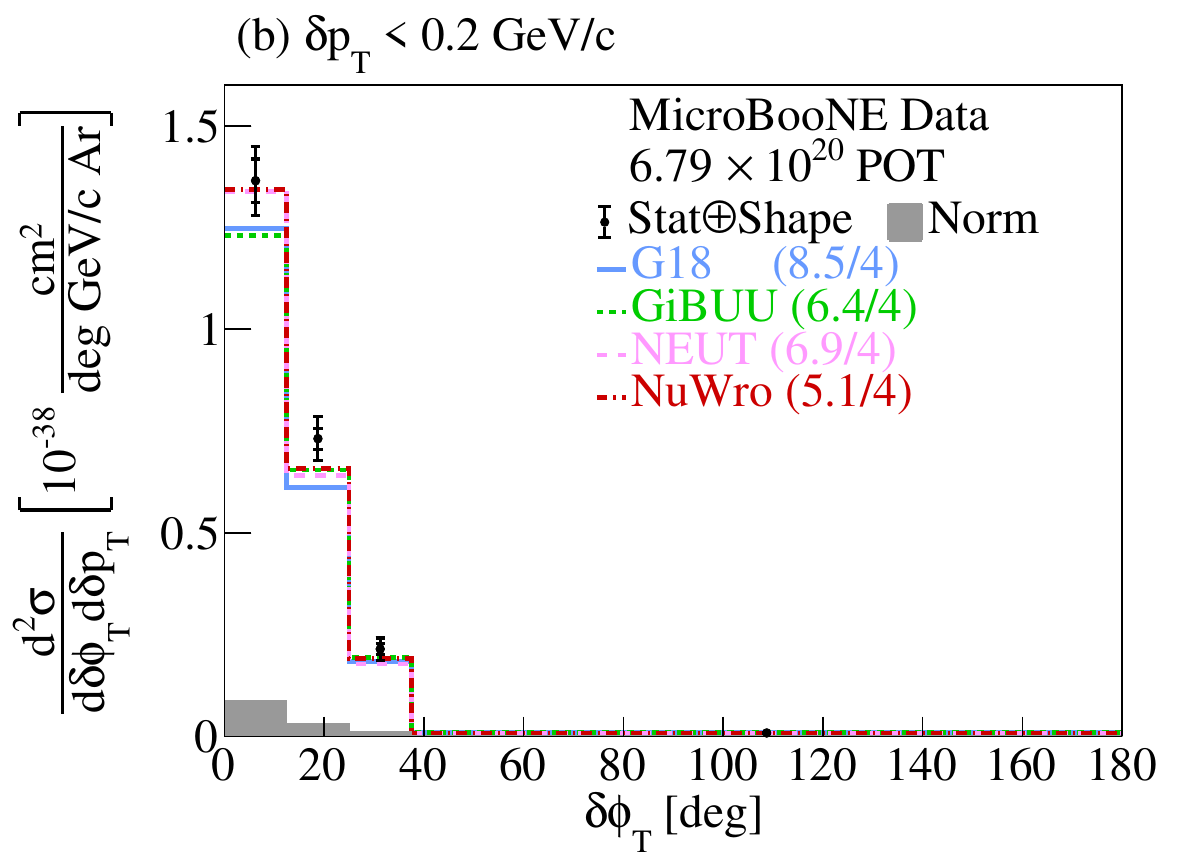}\\
\includegraphics[width=0.49\linewidth]{\figures 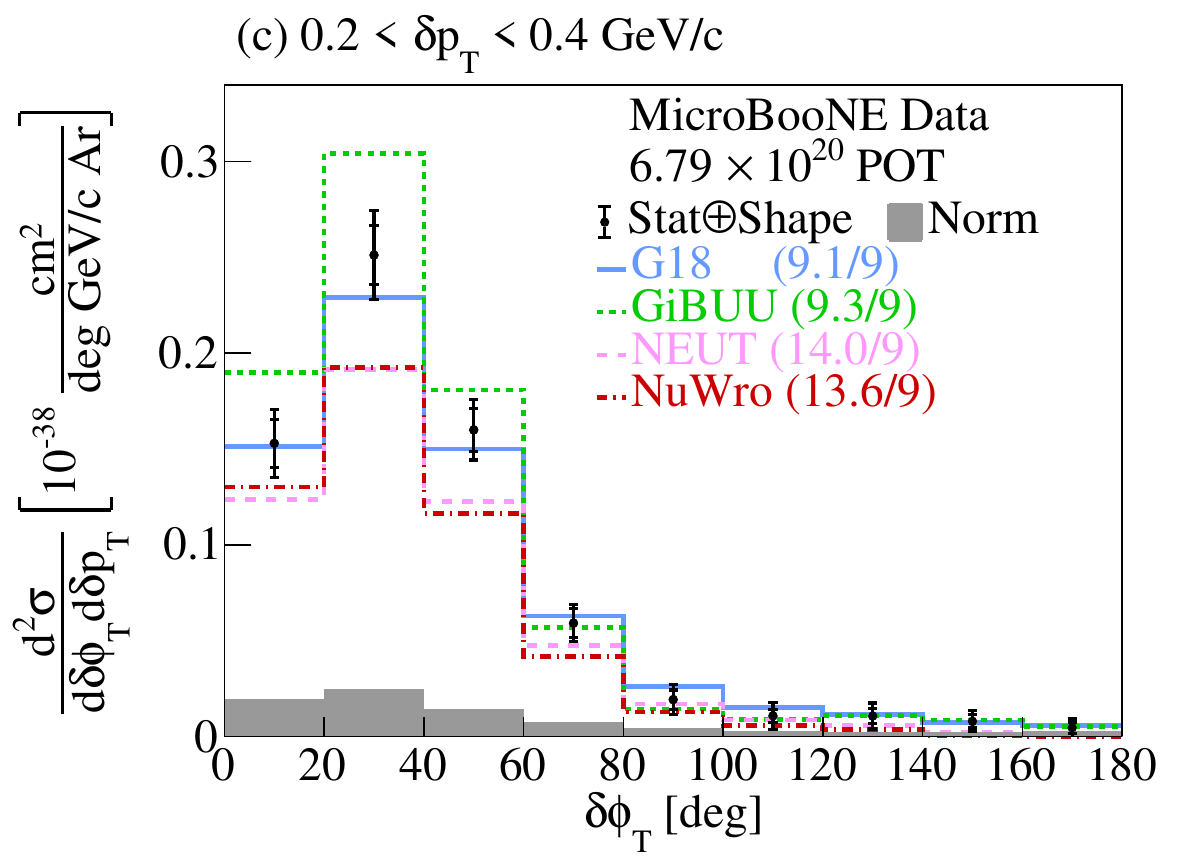}
\includegraphics[width=0.49\linewidth]{\figures 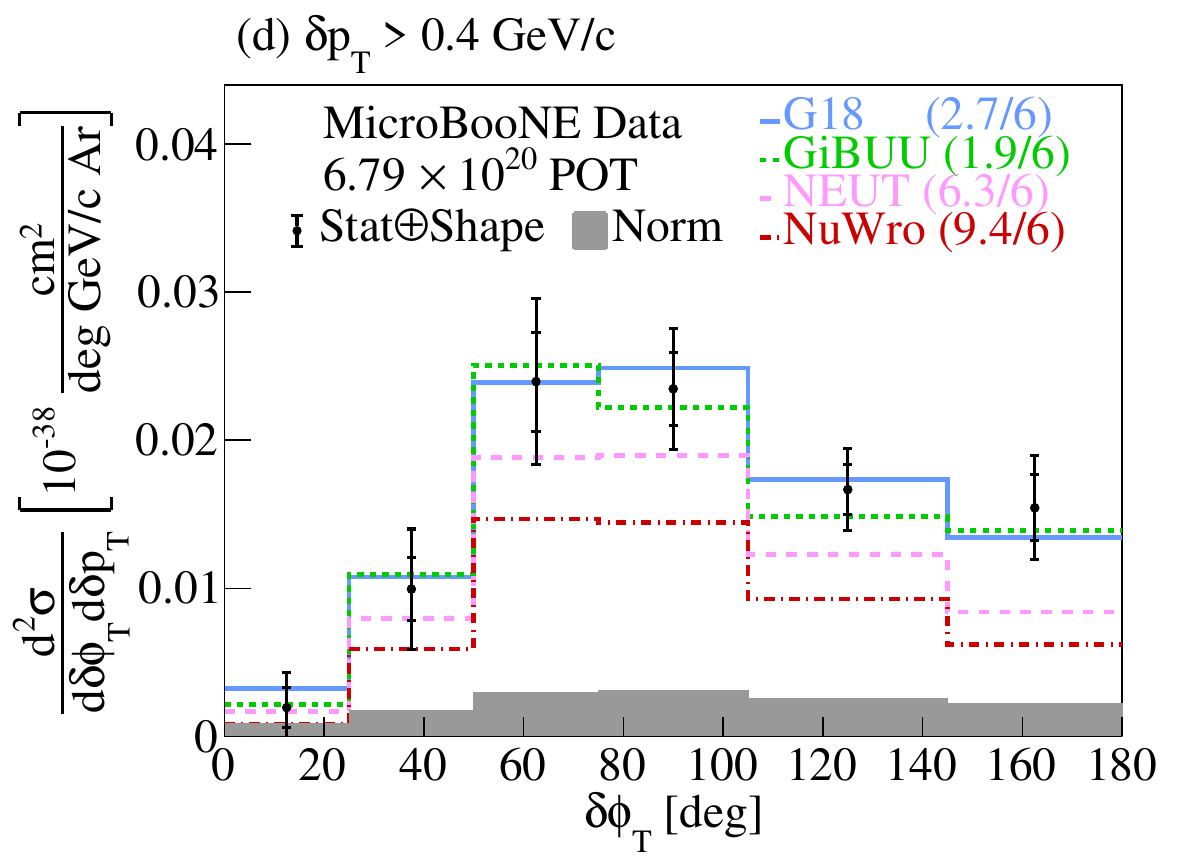}\\
\caption{
The flux-integrated (a) single- and (b-d) double- (in $\delta p_{T}$ bins) differential cross sections as a function of $\delta \phi_{T}$. 
Inner and outer error bars show the statistical and total (statistical and shape systematic) uncertainty at the 1$\sigma$, or 68\%, confidence level. 
The gray band shows the normalization systematic uncertainty.
Colored lines show the results of theoretical cross section calculations using the $\texttt{G18 GENIE}$ (blue), $\texttt{GiBUU}$ (green), $\texttt{NEUT}$ (pink), and $\texttt{NuWro}$ (red) event generators.
The numbers in parentheses show the $\chi^{2}$/bins calculation for each one of the predictions.
}
\label{DeltaPhiTInDeltaPTGen}
\end{figure*}

\begin{figure*}[htb!]
\centering
\includegraphics[width=0.49\linewidth]{\figures 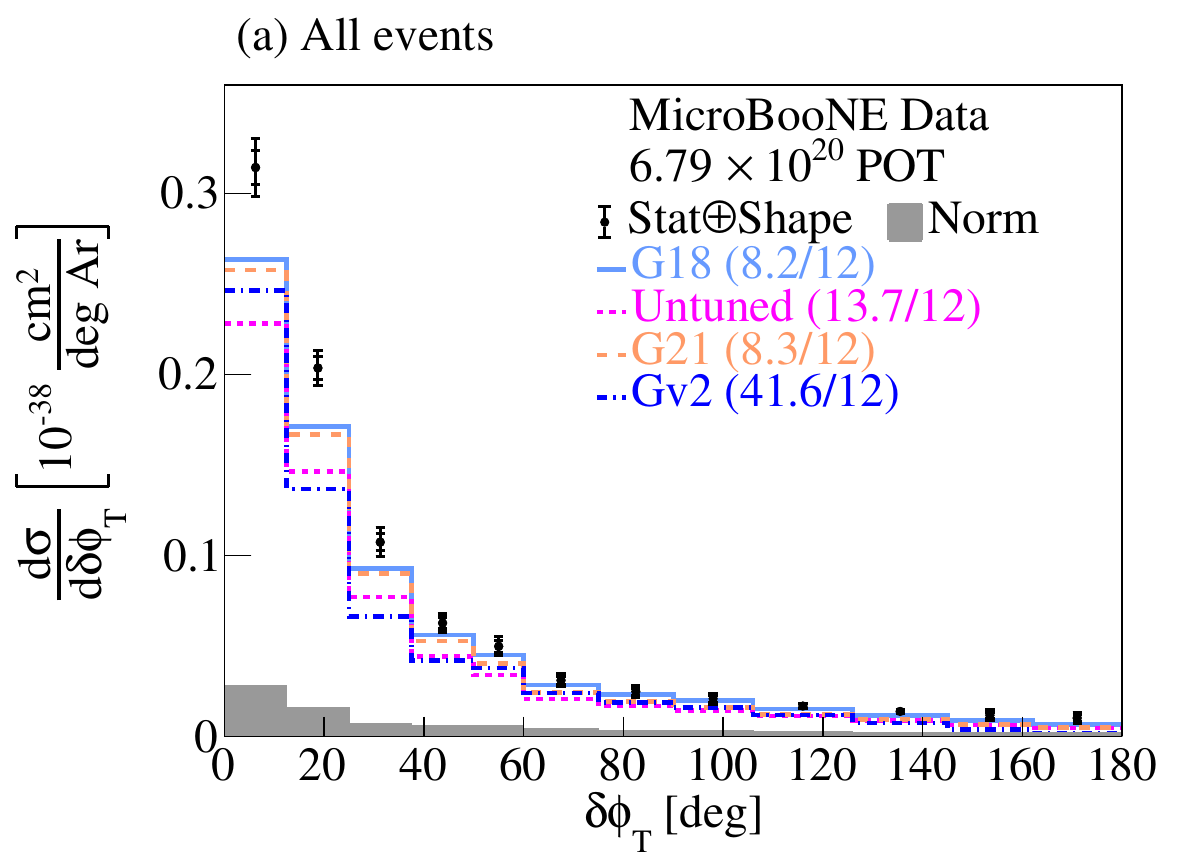}
\includegraphics[width=0.49\linewidth]{\figures 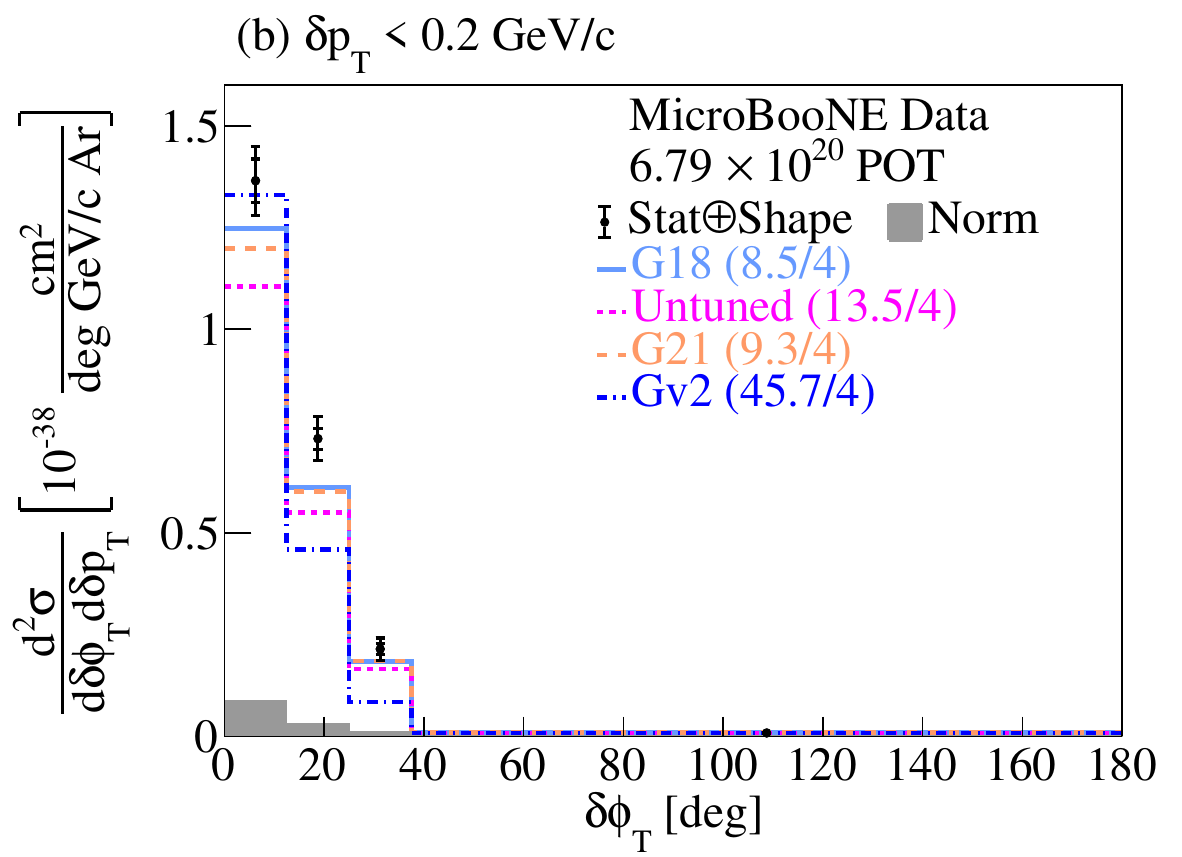}\\
\includegraphics[width=0.49\linewidth]{\figures 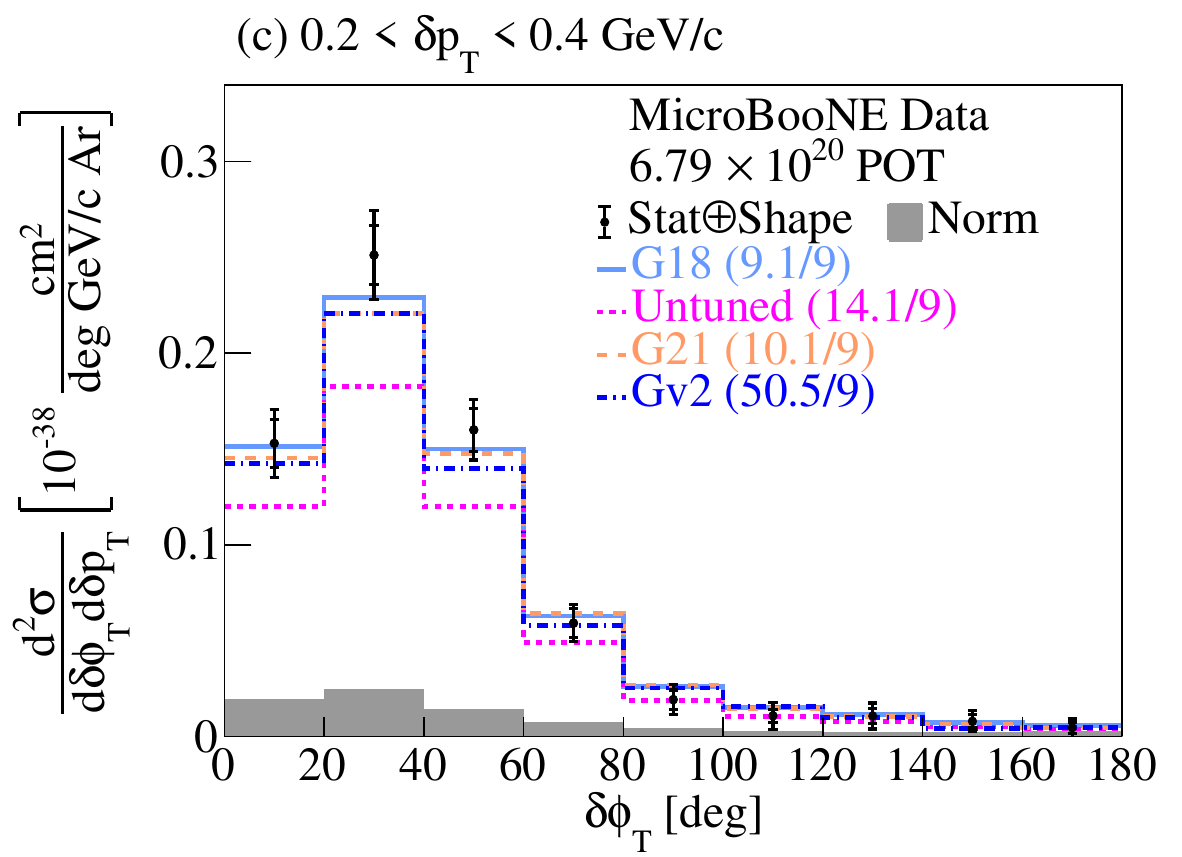}
\includegraphics[width=0.49\linewidth]{\figures 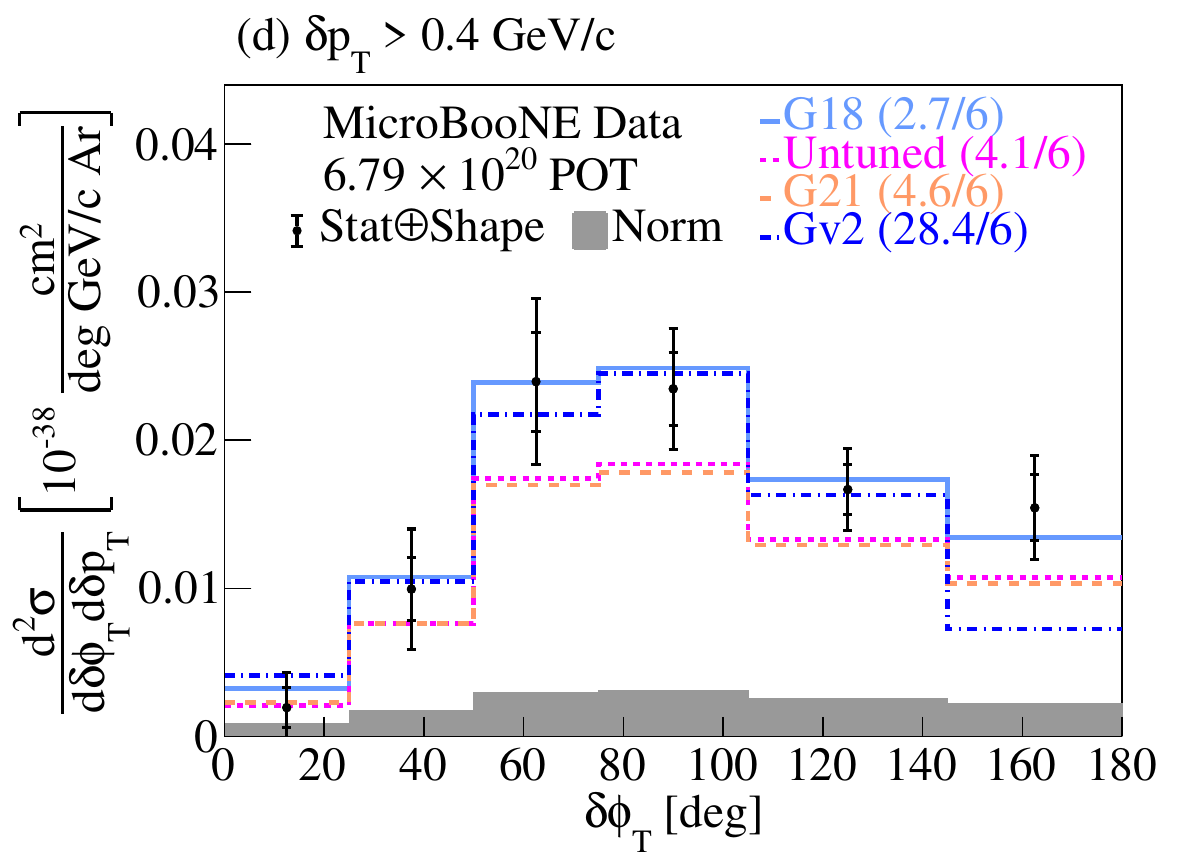}\\
\caption{
The flux-integrated (a) single- and (b-d) double- (in $\delta p_{T}$ bins) differential cross sections as a function of $\delta \phi_{T}$. 
Inner and outer error bars show the statistical and total (statistical and shape systematic) uncertainty at the 1$\sigma$, or 68\%, confidence level. 
The gray band shows the normalization systematic uncertainty.
Colored lines show the results of theoretical cross section calculations using the $\texttt{G18}$ (light blue), $\texttt{Untuned}$ (magenta), $\texttt{G21}$ (orange), and $\texttt{Gv2}$ (dark blue) $\texttt{GENIE}$ configurations.
The numbers in parentheses show the $\chi^{2}$/bins calculation for each one of the predictions.
}
\label{DeltaPhiTInDeltaPTGenie}
\end{figure*}

\begin{figure*}[htb!]
\centering 
\includegraphics[width=0.49\linewidth]{\figures 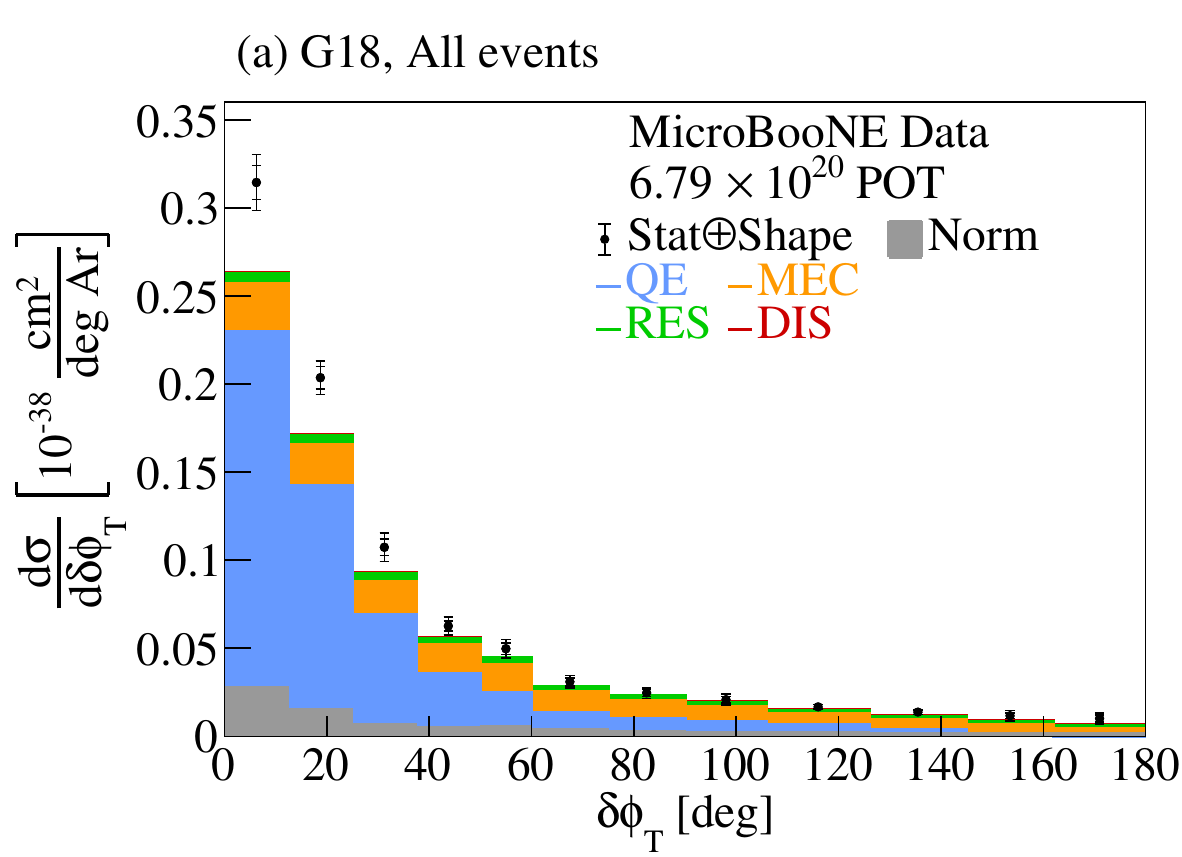}
\includegraphics[width=0.49\linewidth]{\figures 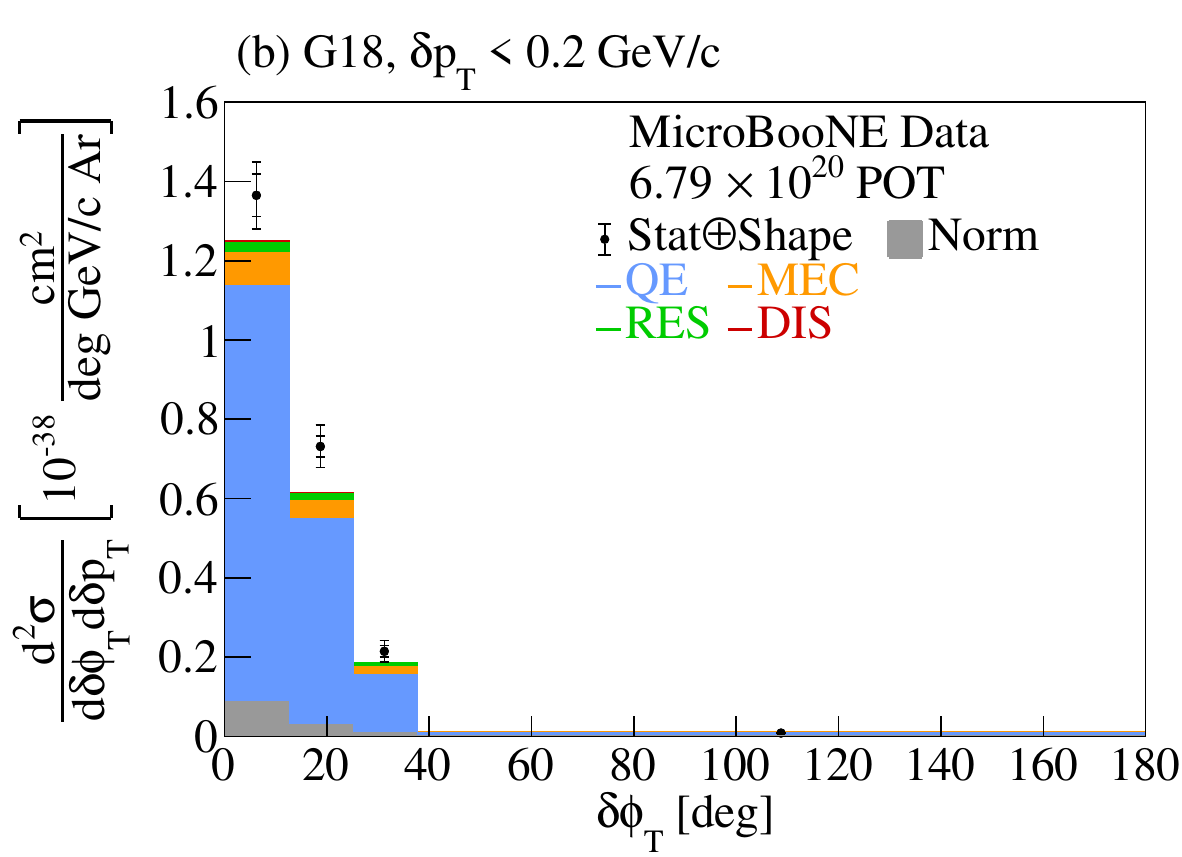}
\includegraphics[width=0.49\linewidth]{\figures 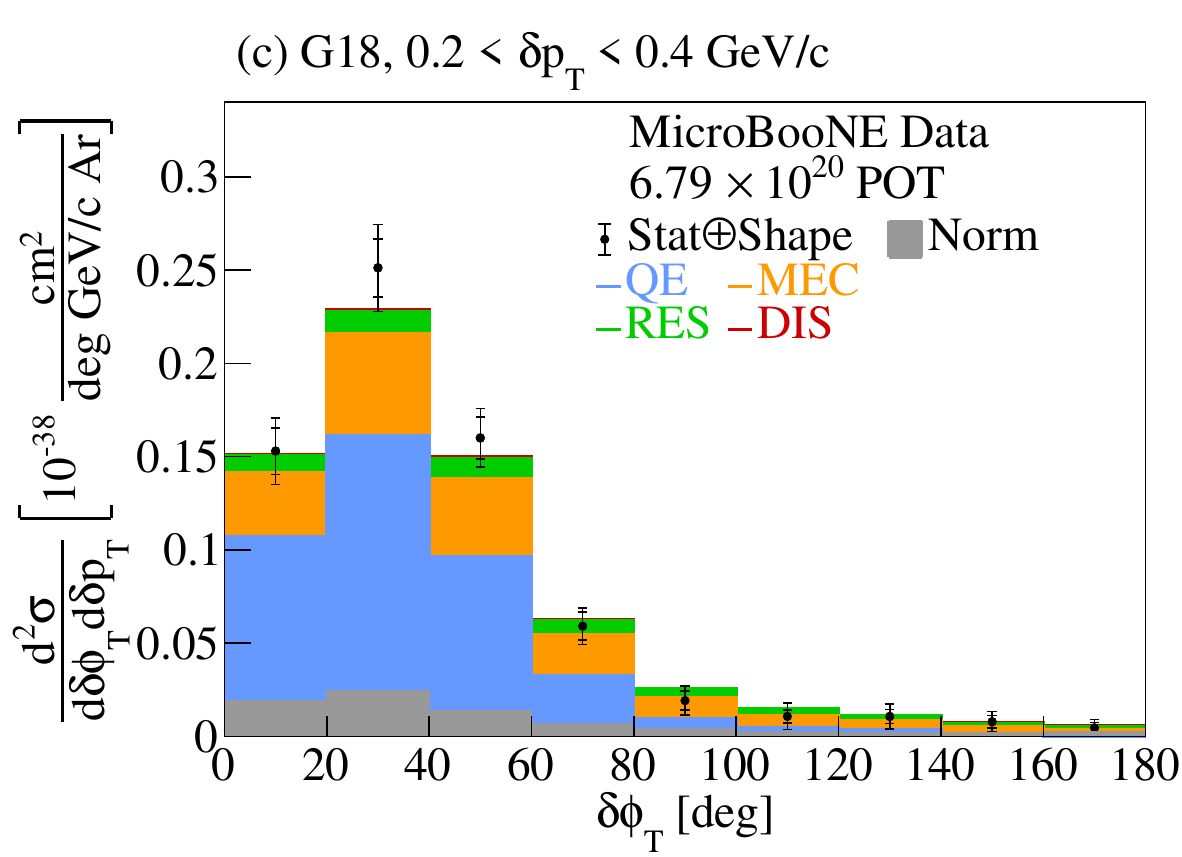}
\includegraphics[width=0.49\linewidth]{\figures 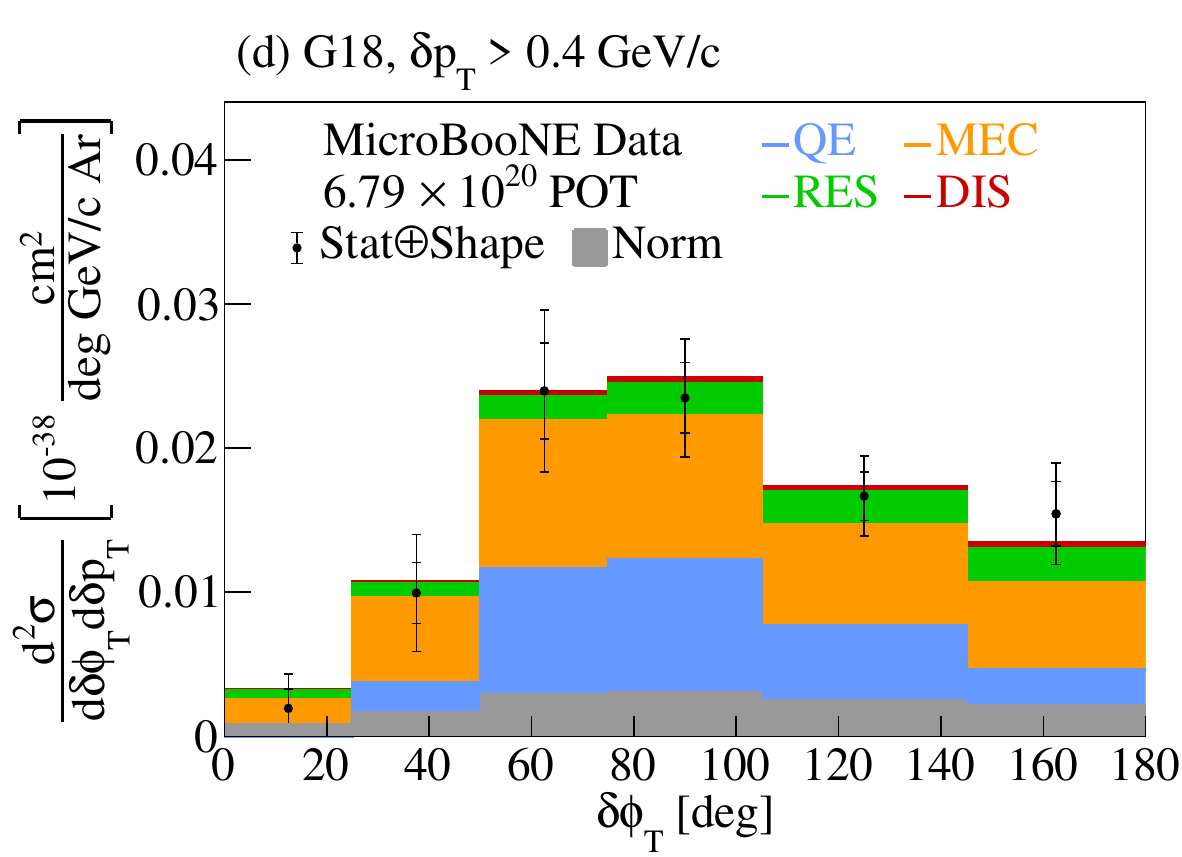}
\caption{
Comparison between the flux-integrated double- (in $\delta p_{T}$ bins) differential cross sections as a function of $\delta\phi_{T}$ for data and the $\texttt{G18 GENIE}$ prediction. 
Inner and outer error bars show the statistical and total (statistical and shape systematic) uncertainty at the 1$\sigma$, or 68\%, confidence level. 
The gray band shows the normalization systematic uncertainty.
Colored stacked histograms show the results of theoretical cross section calculations using the $\texttt{G18}$ prediction for QE (blue), MEC (orange), RES (green), and DIS (red) interactions.
}
\label{DeltaPhiTInDeltaPTInte}
\end{figure*}

Figures~\ref{DeltaPhiTInDeltaPTGen} and~\ref{DeltaPhiTInDeltaPTGenie} show the single-differential cross sections as a function of $\delta\phi_{T}$ using all the events (panel a), as well as the double-differential results as a function of the same kinematic variable in $\delta p_{T}$ bins (panels b-d).
Figure~\ref{DeltaPhiTInDeltaPTGen} shows the comparisons to a number of available neutrino event generators, with all the generators illustrating a fairly good performance.
This result is consistent with the one reported by the T2K collaboration~\cite{Avanzini:2021qlx,Abe:2018pwo}.
In the lowest $\delta p_{T}$ region shown in panel b, $\texttt{NuWro}$ is the generator with the best performance.
Figure~\ref{DeltaPhiTInDeltaPTGenie} shows the same results compared to a number of $\texttt{GENIE}$ configurations, where $\texttt{Gv2}$ is disfavored in all regions.
At small $\delta p_{T}$ values the cross section is decreasing and zero above $\approx$ 40$^{\circ}$ which indicates the absence of multi-nucleon and FSI effects. 
Higher $\delta p_{T}$ values lead to $\delta\phi_{T}$ cross sections that extend up to 180$^{\circ}$. 
This behavior is primarily driven by multi-body effects with hadrons below the detection threshold that introduce large kinematic imbalances, as can be seen in panels c-d of Fig.~\ref{DeltaPhiTInDeltaPTInte}.



\begin{figure*}[htb!]
\centering 
\includegraphics[width=0.49\linewidth]{\figures 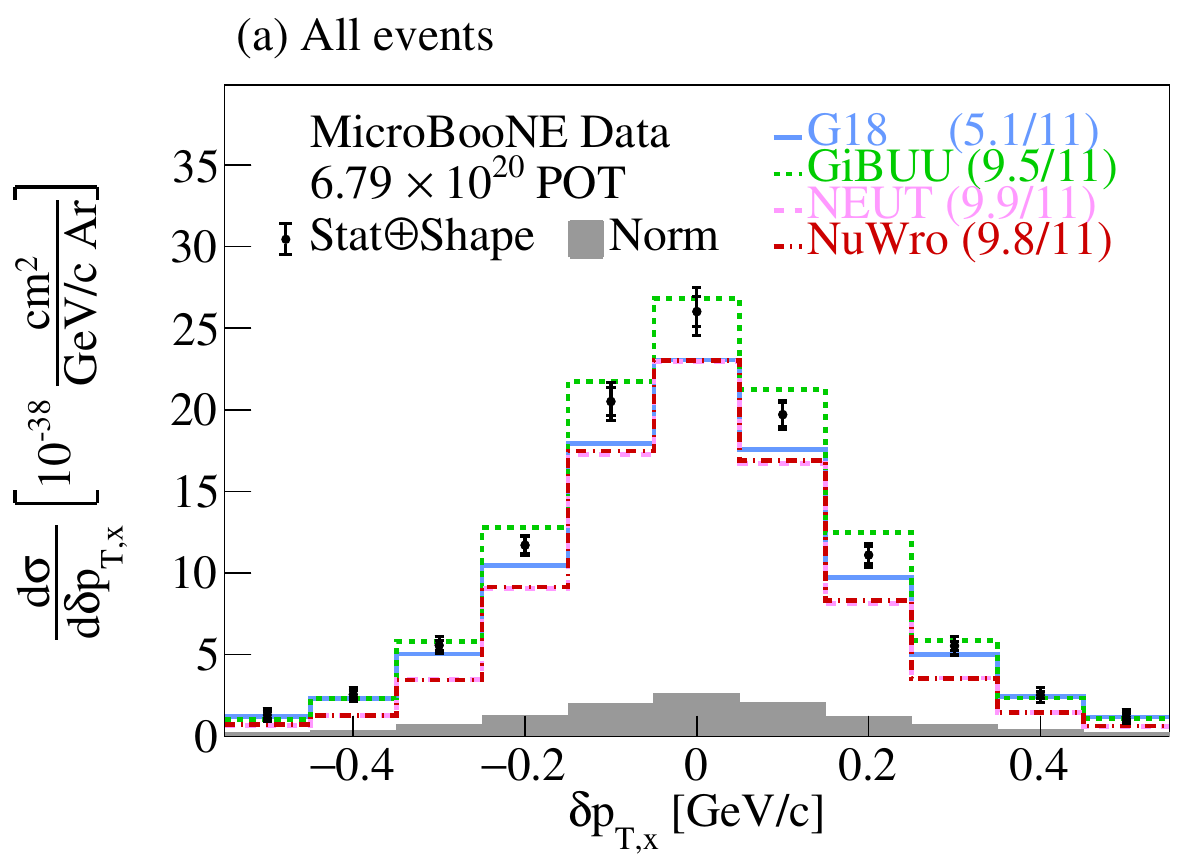}
\includegraphics[width=0.49\linewidth]{\figures 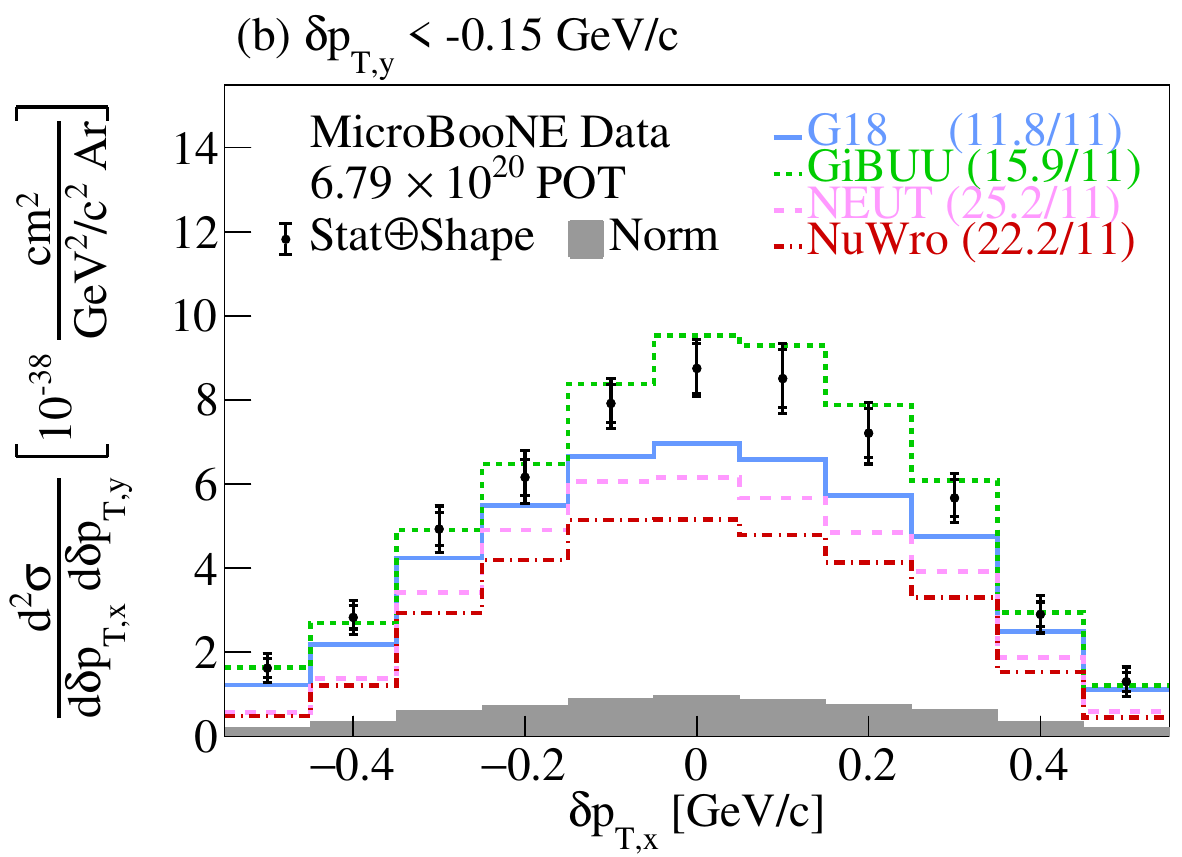}
\includegraphics[width=0.49\linewidth]{\figures 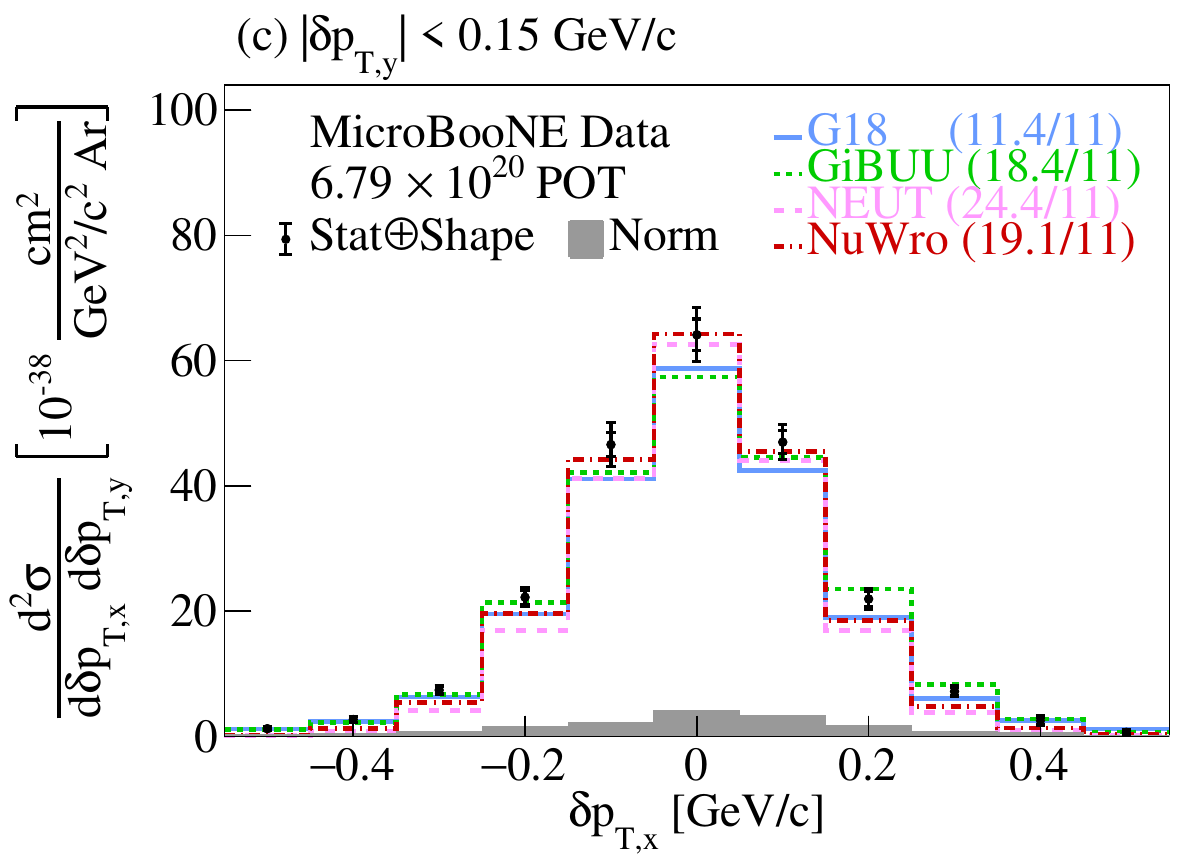}
\includegraphics[width=0.49\linewidth]{\figures 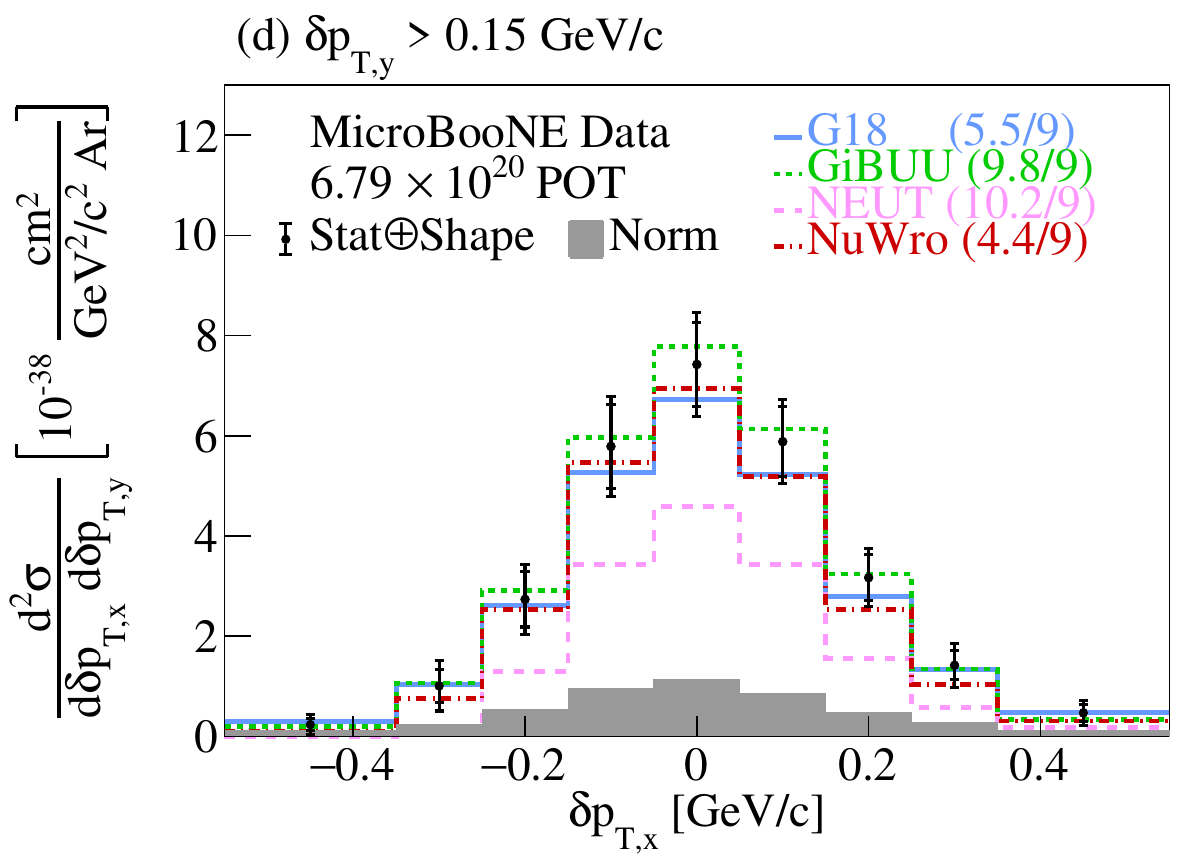}
\caption{
The flux-integrated (a) single- and (b-d) double- (in $\delta p_{T,y}$ bins) differential cross sections as a function of $\delta p_{T,x}$. 
Inner and outer error bars show the statistical and total (statistical and shape systematic) uncertainty at the 1$\sigma$, or 68\%, confidence level. 
The gray band shows the normalization systematic uncertainty.
Colored lines show the results of theoretical cross section calculations using the $\texttt{G18 GENIE}$ (blue), $\texttt{GiBUU}$ (green), $\texttt{NEUT}$ (pink), and $\texttt{NuWro}$ (red) event generators.
The numbers in parentheses show the $\chi^{2}$/bins calculation for each one of the predictions.
}
\label{DeltaPtxInDeltaPtyGen}
\end{figure*}

\begin{figure*}[htb!]
\centering 
\includegraphics[width=0.49\linewidth]{\figures 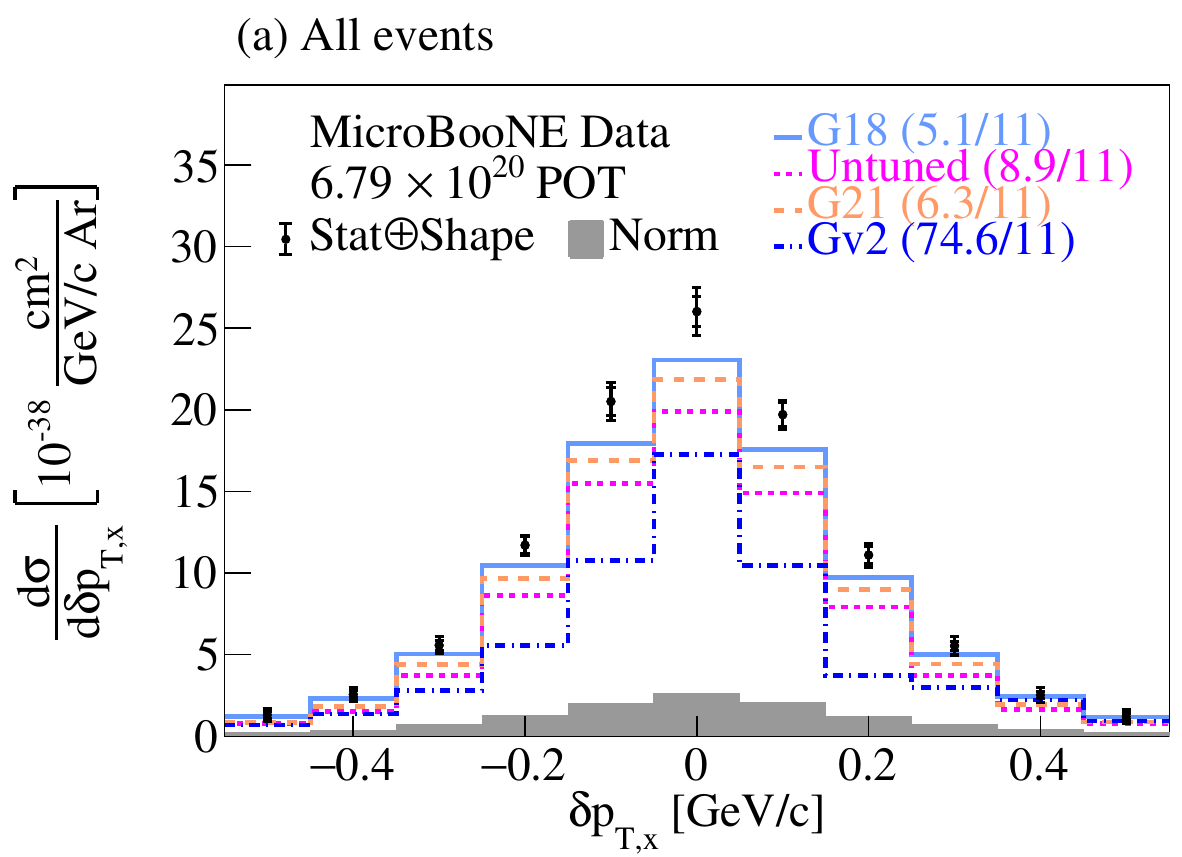}
\includegraphics[width=0.49\linewidth]{\figures 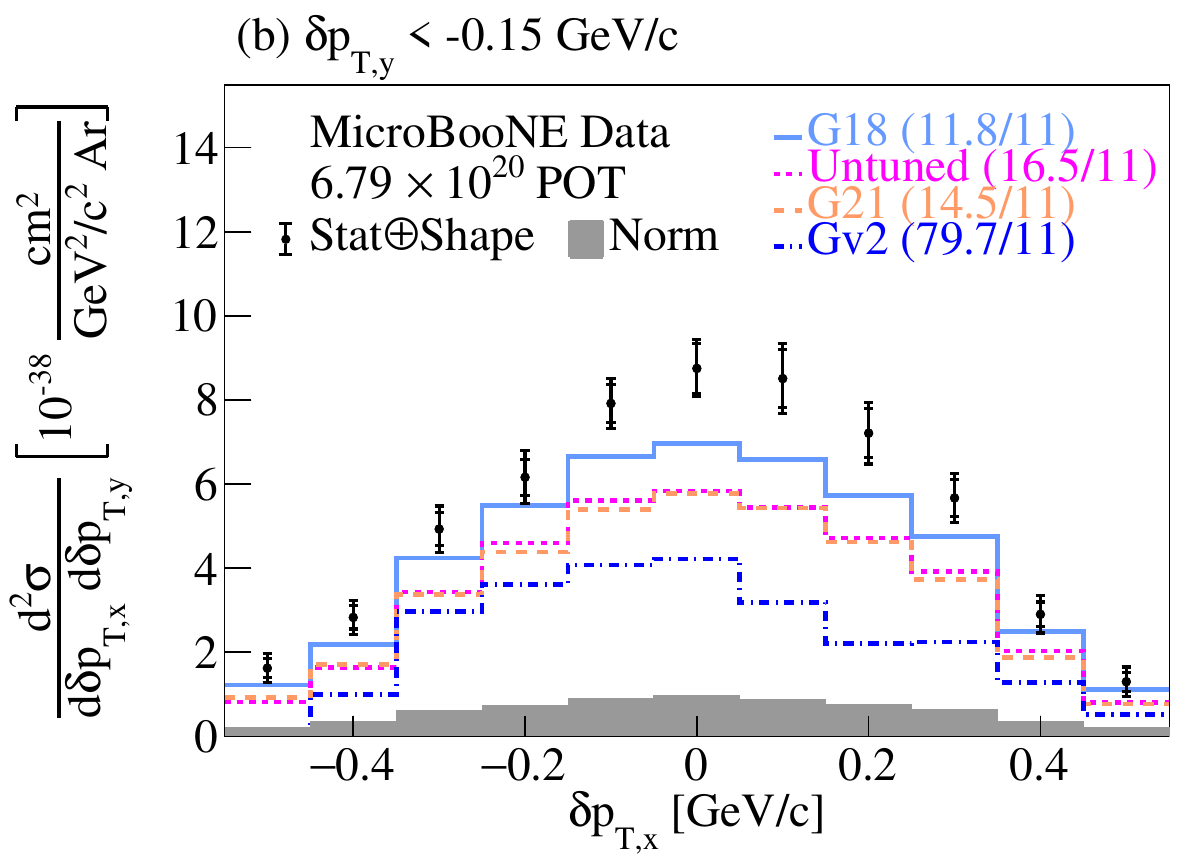}
\includegraphics[width=0.49\linewidth]{\figures 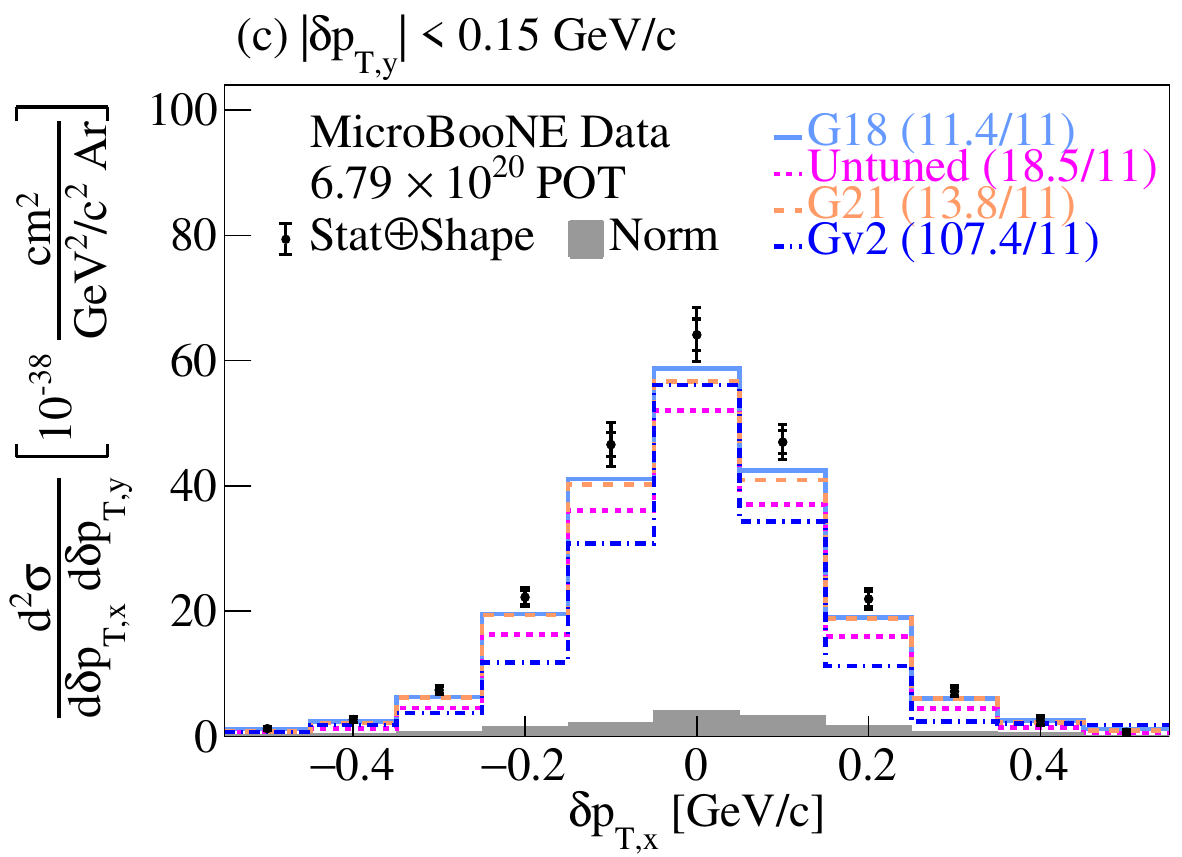}
\includegraphics[width=0.49\linewidth]{\figures 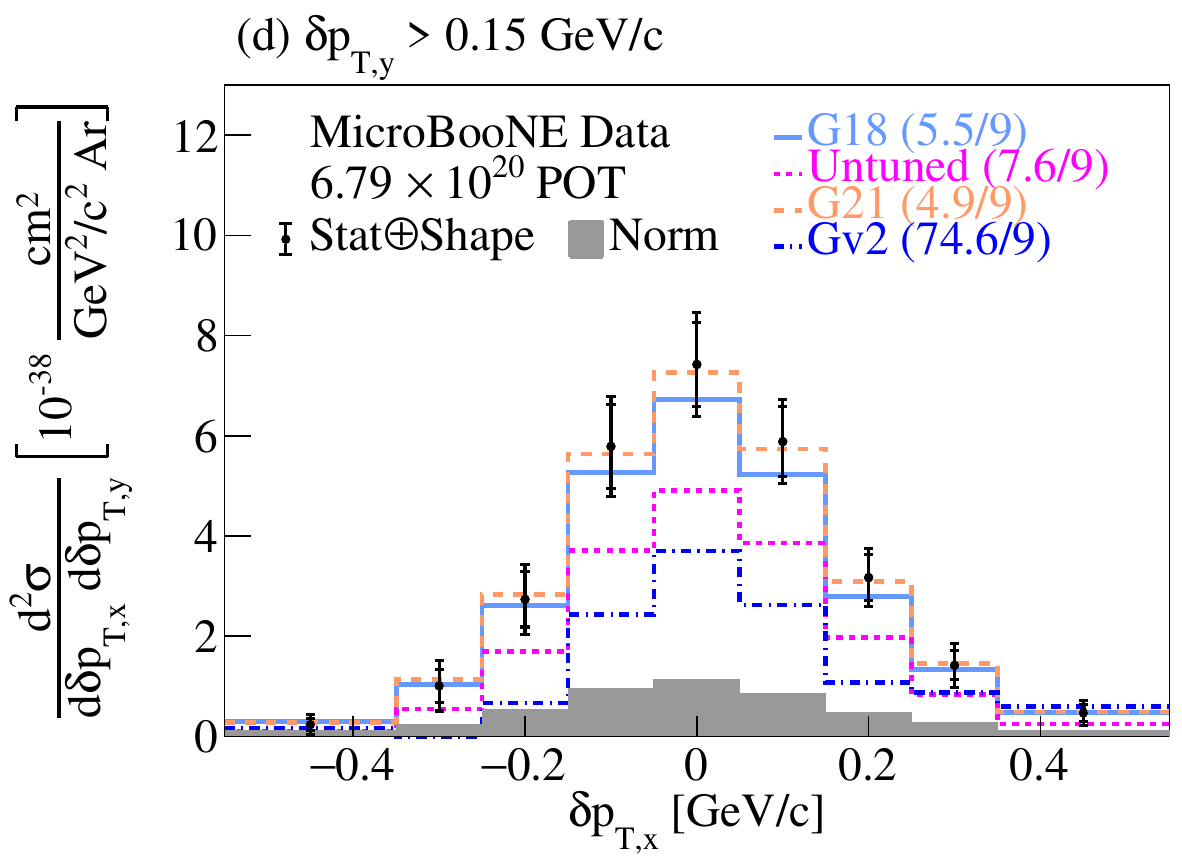}
\caption{
The flux-integrated (a) single- and (b-d) double- (in $\delta p_{T,y}$ bins) differential cross sections as a function of $\delta p_{T,x}$. 
Inner and outer error bars show the statistical and total (statistical and shape systematic) uncertainty at the 1$\sigma$, or 68\%, confidence level. 
The gray band shows the normalization systematic uncertainty.
Colored lines show the results of theoretical cross section calculations using the $\texttt{G18}$ (light blue), $\texttt{Untuned}$ (magenta), $\texttt{G21}$ (orange), and $\texttt{Gv2}$ (dark blue) $\texttt{GENIE}$ configurations.
The numbers in parentheses show the $\chi^{2}$/bins calculation for each one of the predictions.
}
\label{DeltaPtxInDeltaPtyGenie}
\end{figure*}

\begin{figure*}[htb!]
\centering 
\includegraphics[width=0.49\linewidth]{\figures 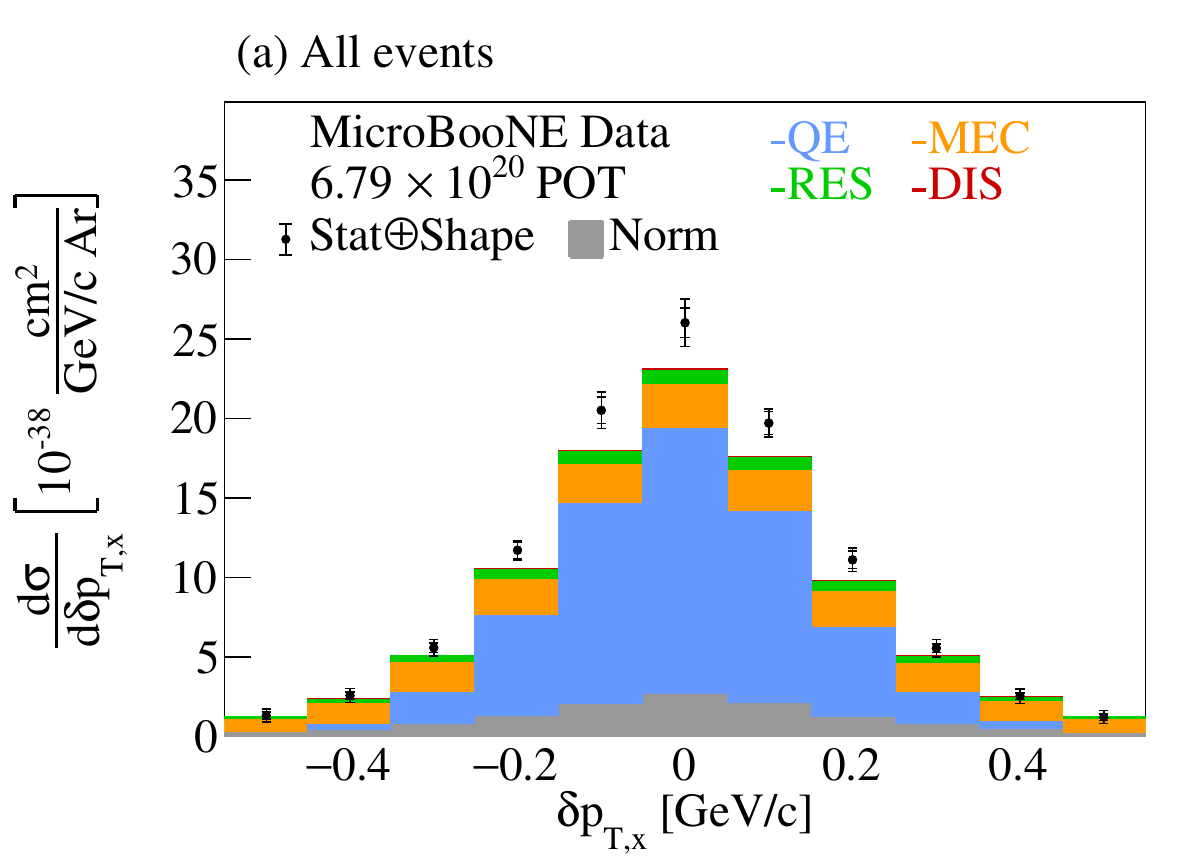}
\includegraphics[width=0.49\linewidth]{\figures 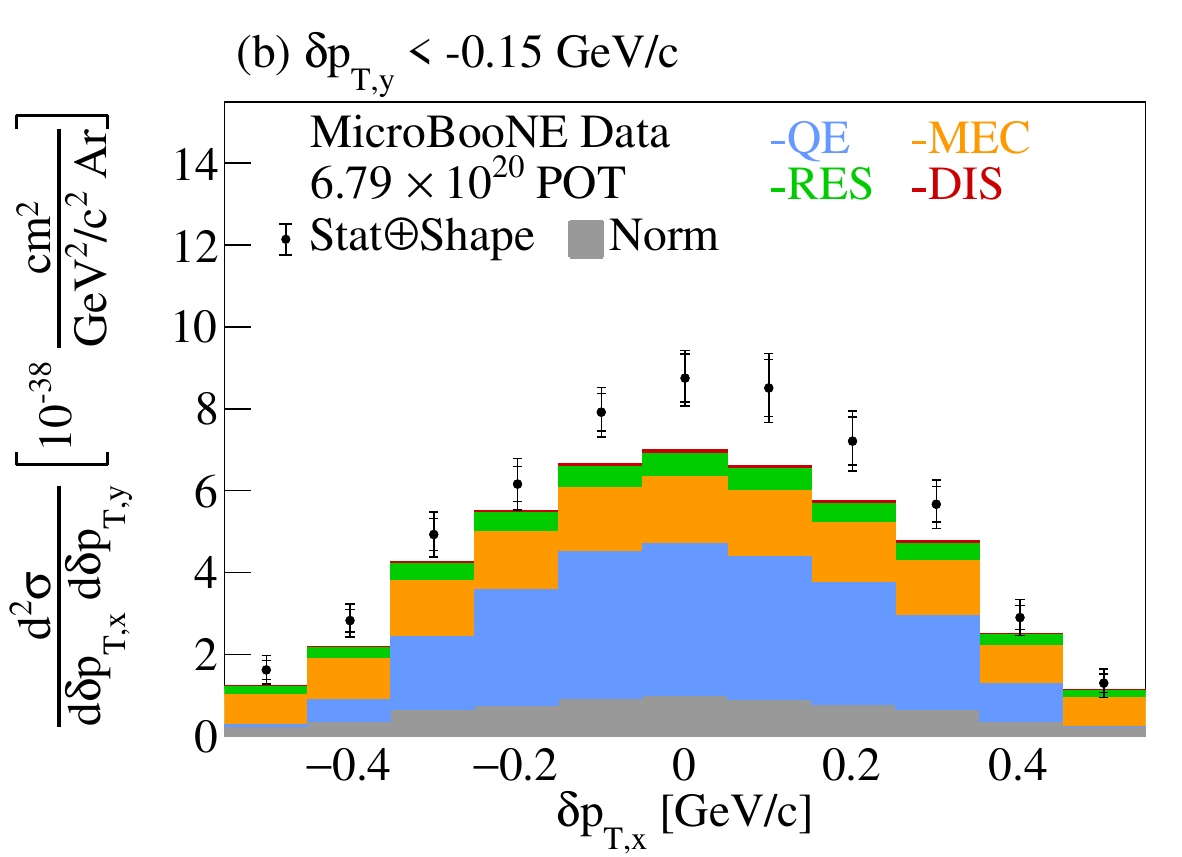}
\includegraphics[width=0.49\linewidth]{\figures 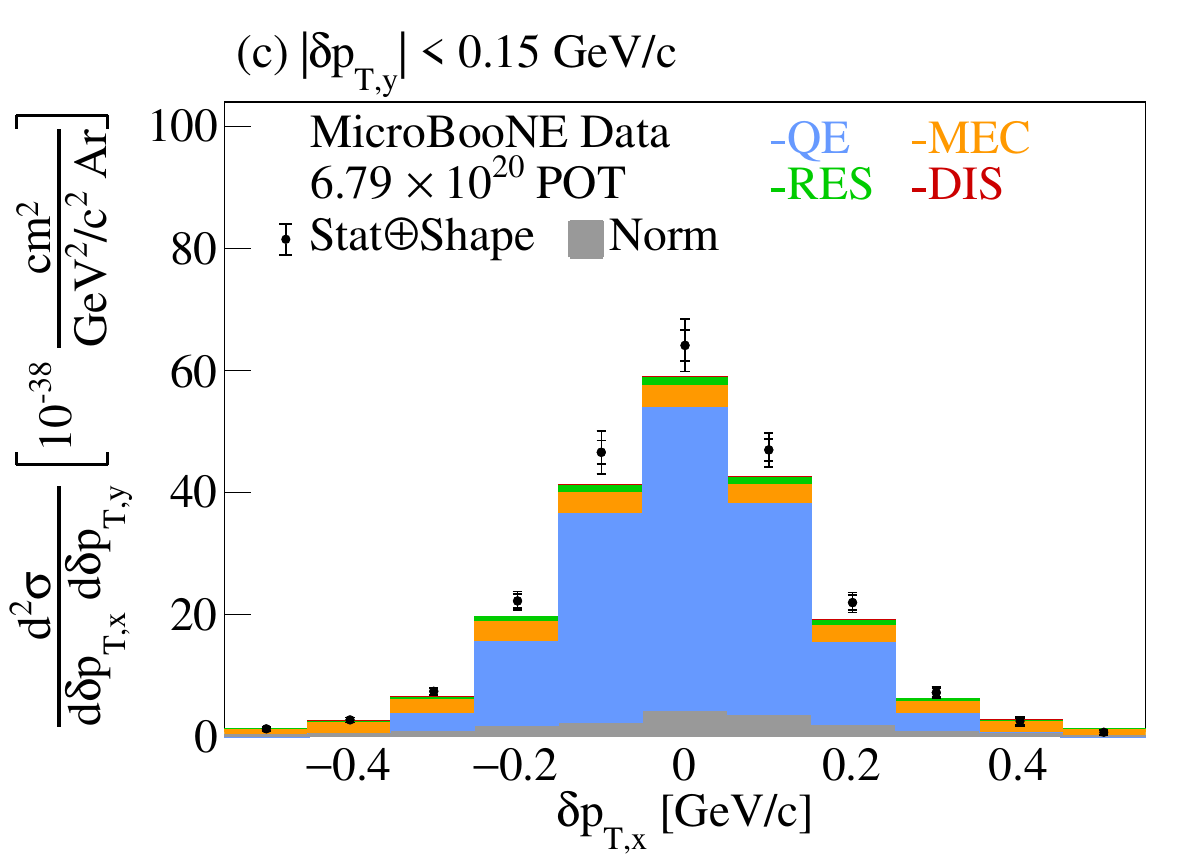}
\includegraphics[width=0.49\linewidth]{\figures 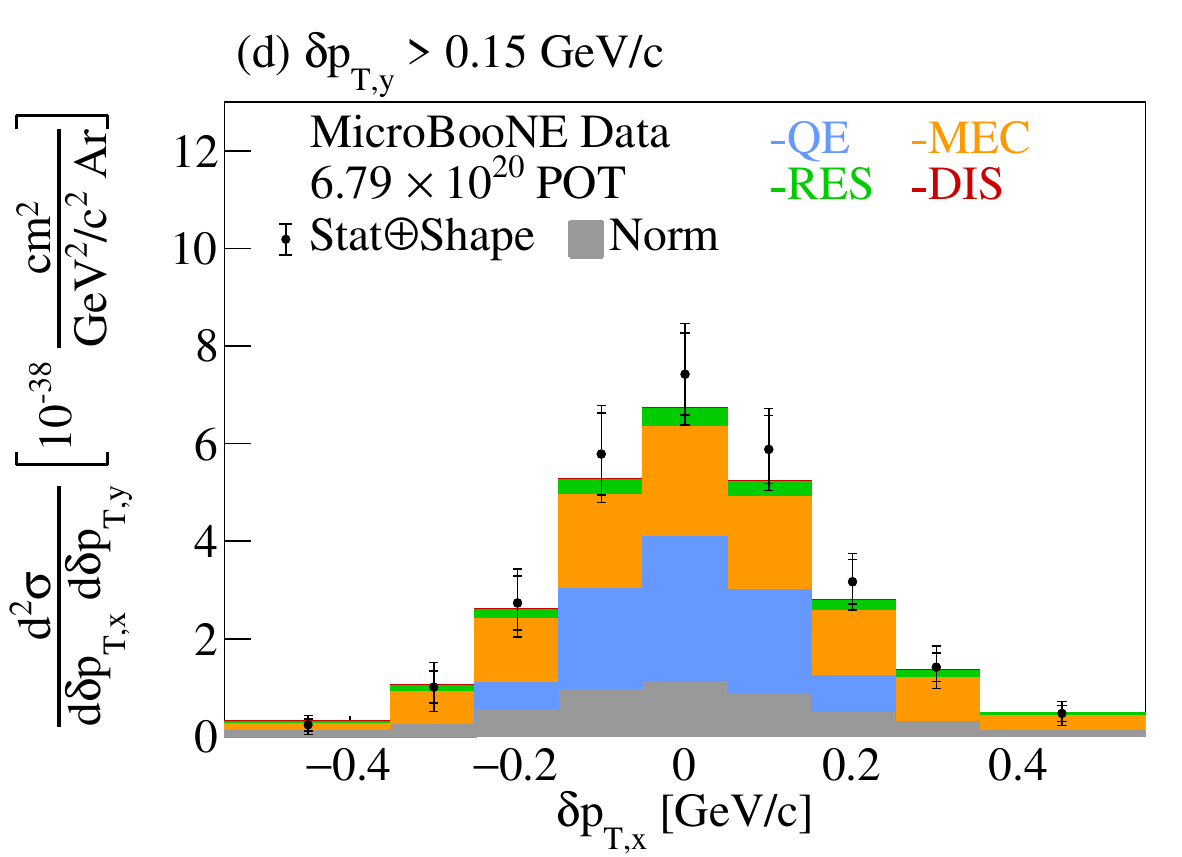}
\caption{
Comparison between the flux-integrated double- (in $\delta p_{T,y}$ bins) differential cross sections as a function of $\delta p_{T,x}$ for data and the $\texttt{G18 GENIE}$ prediction. 
Inner and outer error bars show the statistical and total (statistical and shape systematic) uncertainty at the 1$\sigma$, or 68\%, confidence level. 
The gray band shows the normalization systematic uncertainty.
Colored stacked histograms show the results of theoretical cross section calculations using the $\texttt{G18 GENIE}$ prediction for QE (blue), MEC (orange), RES (green), and DIS (red) interactions.
}
\label{DeltaPtxInDeltaPtyInte}
\end{figure*}

Figures~\ref{DeltaPtxInDeltaPtyGen} and~\ref{DeltaPtxInDeltaPtyGenie} show the single-differential cross sections as a function of $\delta p_{T,x}$ using all the events (panel a), as well as the double-differential results in the same kinematic variable in $\delta p_{T,y}$ slices (panels b-c).
Figure~\ref{DeltaPtxInDeltaPtyGen} shows the comparisons to a number of available neutrino event generators.
The central region with $|\delta p_{T,y}|<$ 0.15\,GeV/$c$ is dominated by QE interactions, while the broader distributions with $|\delta p_{T,y}|>$ 0.15\,GeV/$c$ are mainly driven by MEC events, as can be seen in Fig.~\ref{DeltaPtxInDeltaPtyInte}.
In the MEC dominated region of $\delta p_{T,y}<$ \mbox{-0.15}\,GeV/$c$, all the generators, apart from $\texttt{GiBUU}$, seem to be lacking in terms of the peak strength.
$\texttt{GiBUU}$ seems to be overestimating that MEC contribution in the $\delta p_{T,y}<$ -0.15\,GeV/$c$ bin.
With the exception of $\texttt{NEUT}$, all the event generators illustrate a good performance in the $|\delta p_{T,y}|<$ 0.15\,GeV/$c$ region.
Figure~\ref{DeltaPtxInDeltaPtyGenie} shows the same results compared to a number of $\texttt{GENIE}$ configurations, where $\texttt{Gv2}$ shows the worst performance.



The aforementioned results in kinematic imbalance variables illustrate significant differences across the event generators and configurations used for comparison, especially in the case of the double-differential studies.
Yet, the quantity that enters the oscillation probability is the true neutrino energy.
Neutrino energy estimators, such as the calorimetric energy $E^{Cal}$ defined in Eq.~\ref{ecaleq}, are used as a proxy for the true quantity.
The studies reported next present the results as a function of $E^{Cal}$ in bins of the kinematic imbalance variables.



\begin{figure*}[htb!]
\centering 
\includegraphics[width=0.49\linewidth]{\figures 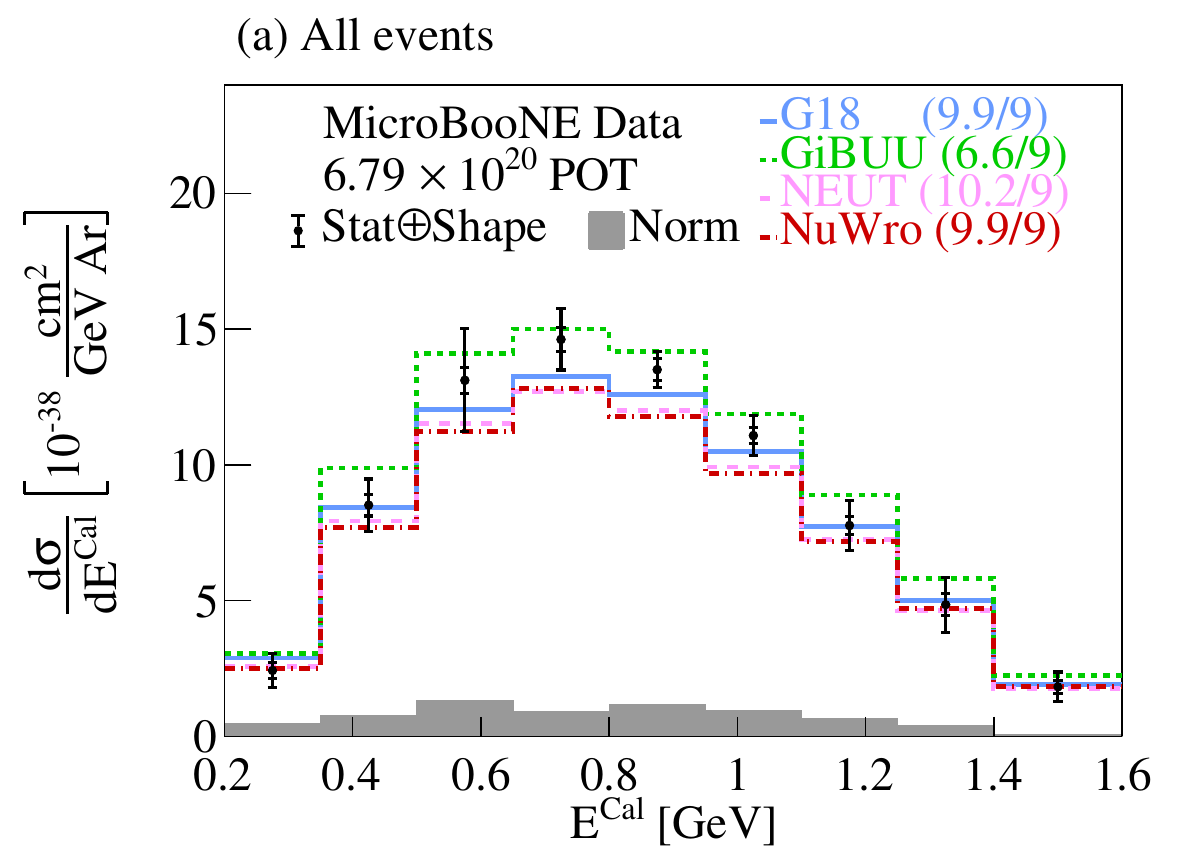}
\includegraphics[width=0.49\linewidth]{\figures 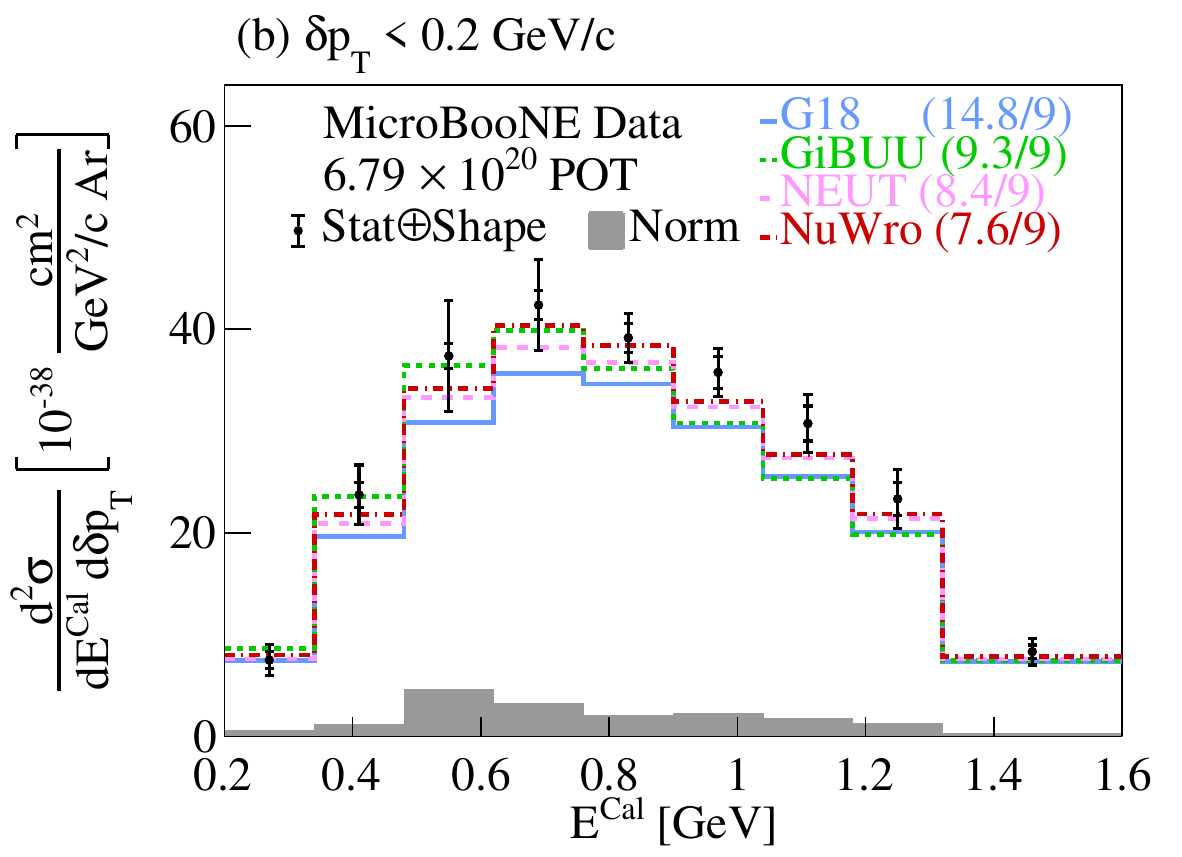}\\
\includegraphics[width=0.49\linewidth]{\figures 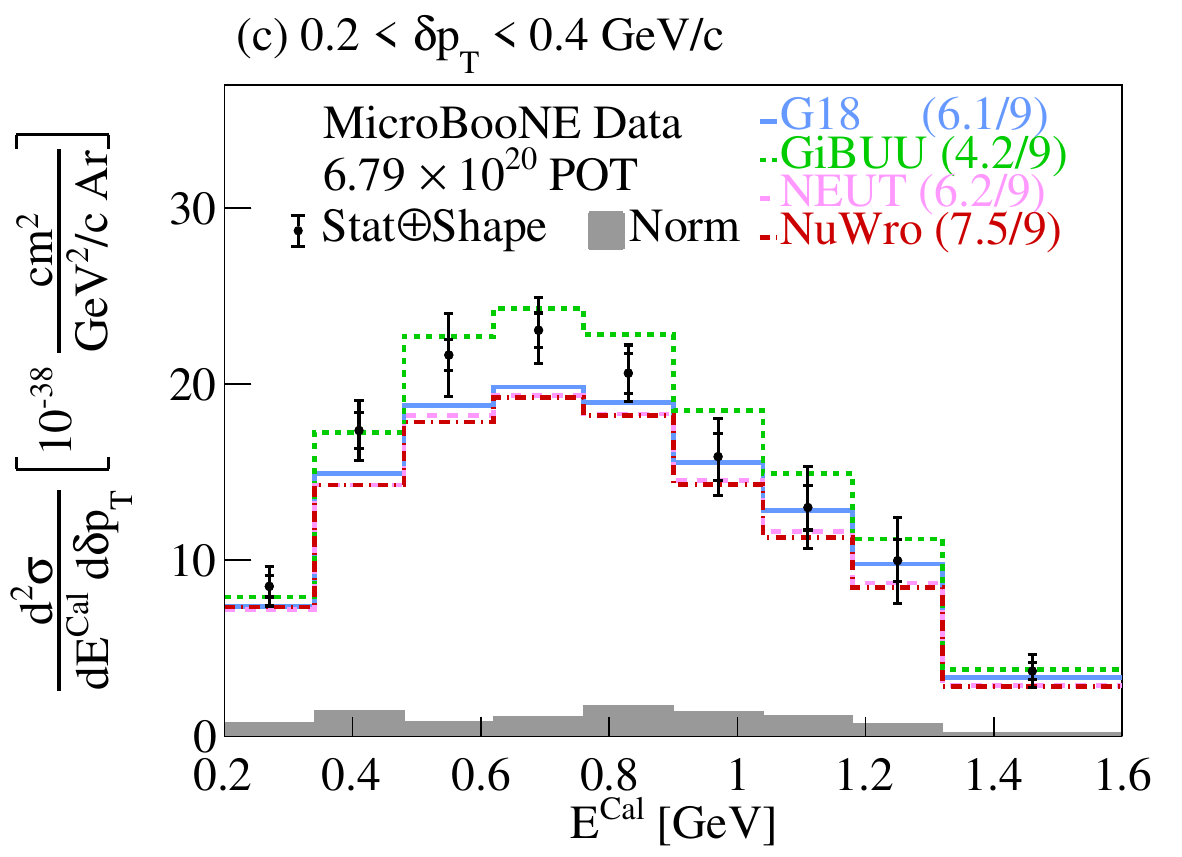}
\includegraphics[width=0.49\linewidth]{\figures 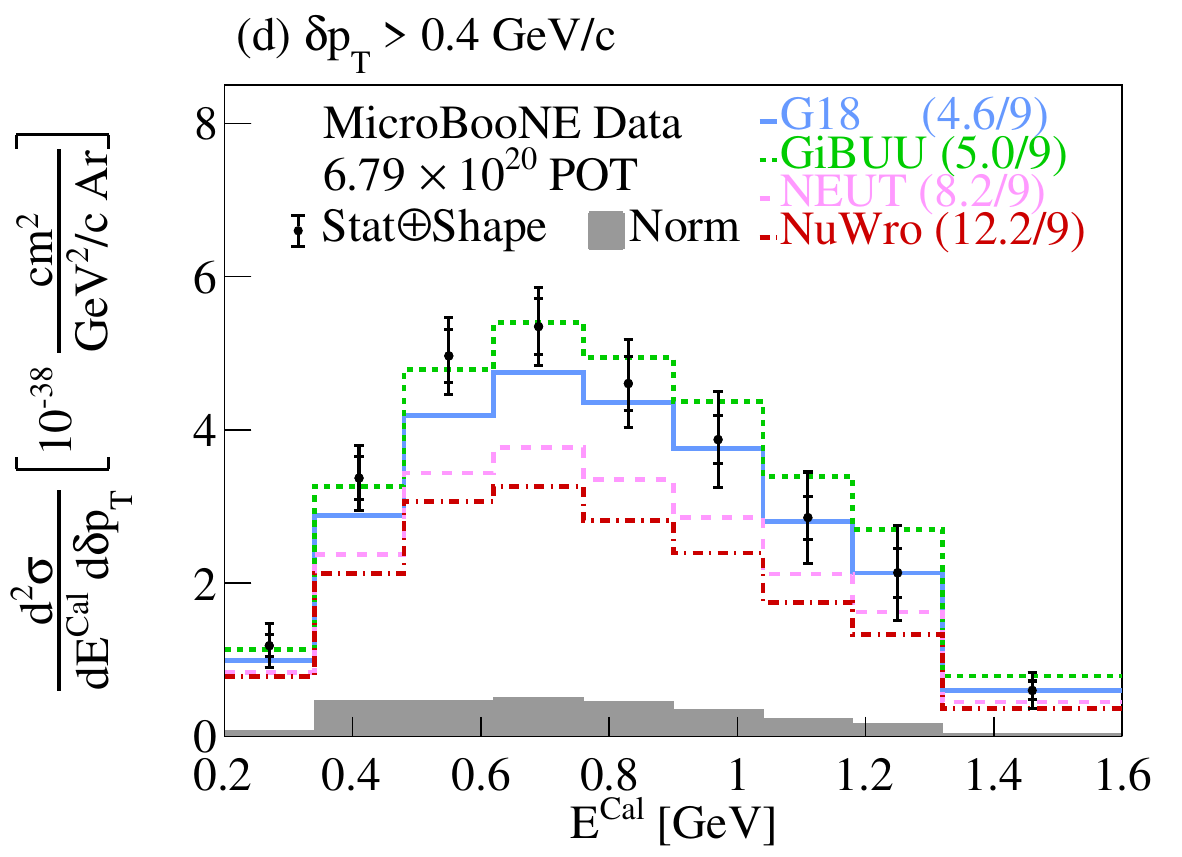}\\
\caption{
The flux-integrated (a) single- and (b-d) double- (in $\delta p_{T}$ bins) differential cross sections as a function of $E^{Cal}$. 
Inner and outer error bars show the statistical and total (statistical and shape systematic) uncertainty at the 1$\sigma$, or 68\%, confidence level. 
The gray band shows the normalization systematic uncertainty.
Colored lines show the results of theoretical cross section calculations using the $\texttt{G18 GENIE}$ (blue), $\texttt{GiBUU}$ (green), $\texttt{NEUT}$ (pink), and $\texttt{NuWro}$ (red) event generators.
The numbers in parentheses show the $\chi^{2}$/bins calculation for each one of the predictions.
}
\label{ECalInDeltaPTGen}
\end{figure*}

\begin{figure*}[htb!]
\centering
\includegraphics[width=0.49\linewidth]{\figures 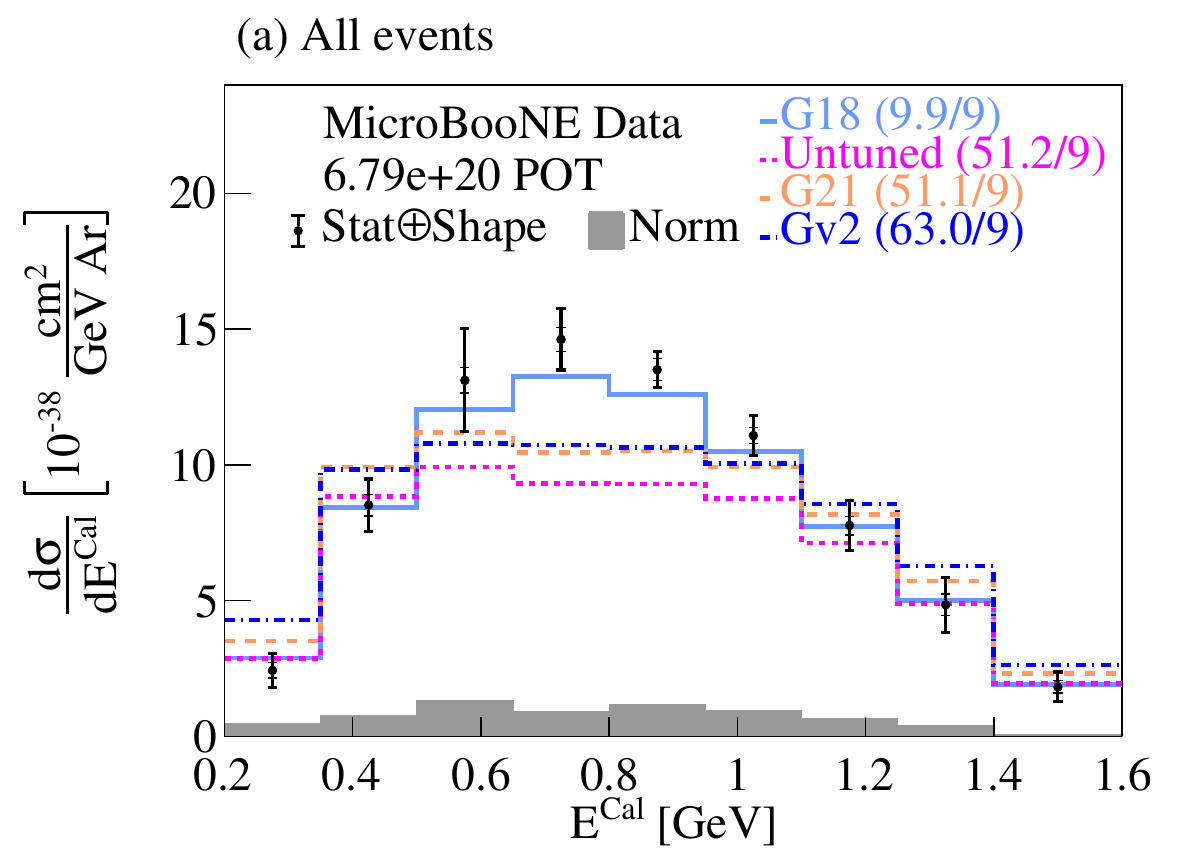}
\includegraphics[width=0.49\linewidth]{\figures 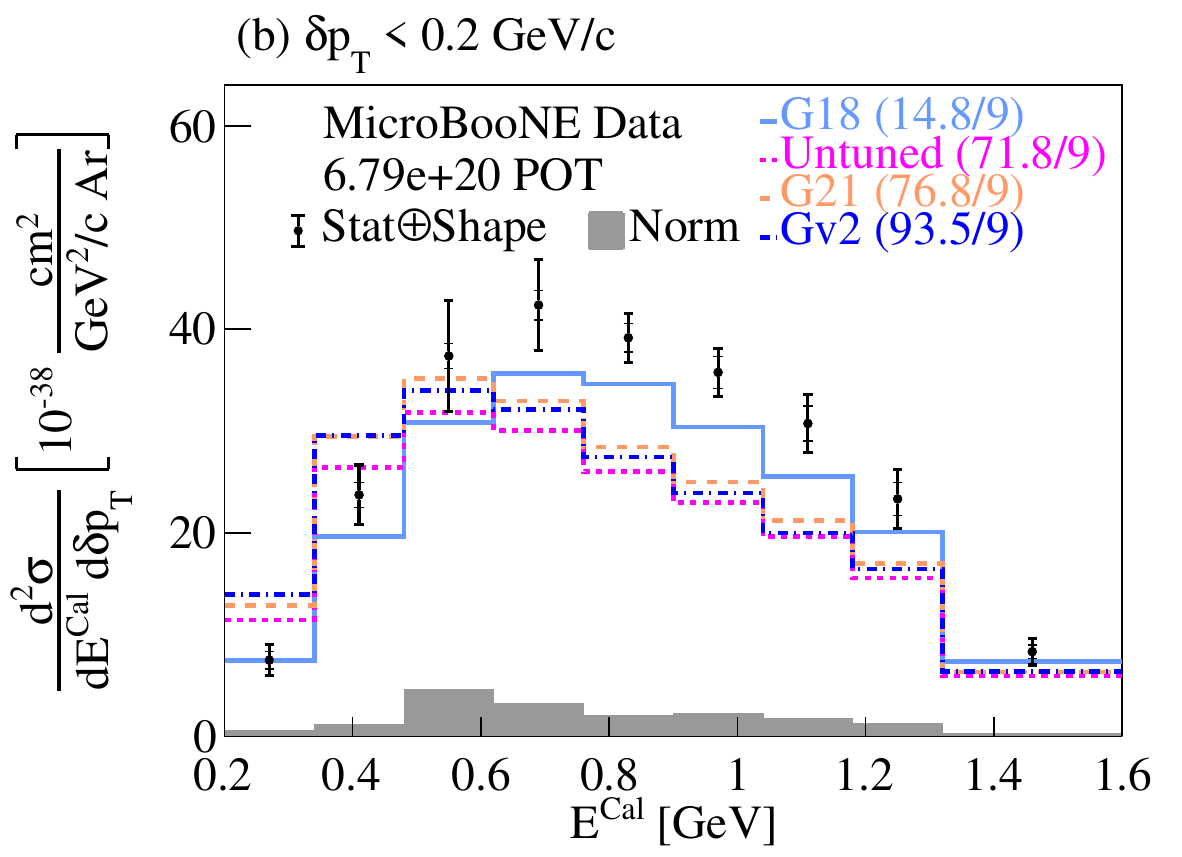}\\
\includegraphics[width=0.49\linewidth]{\figures 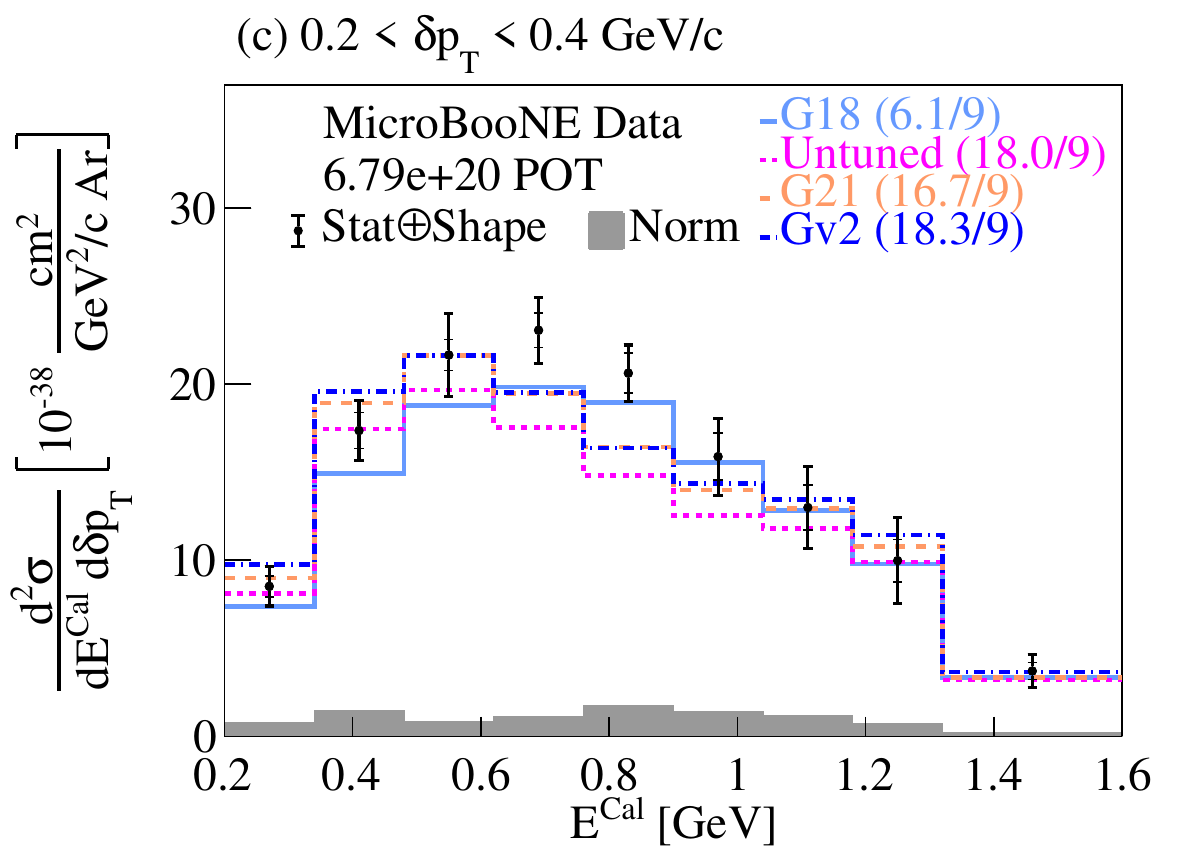}
\includegraphics[width=0.49\linewidth]{\figures 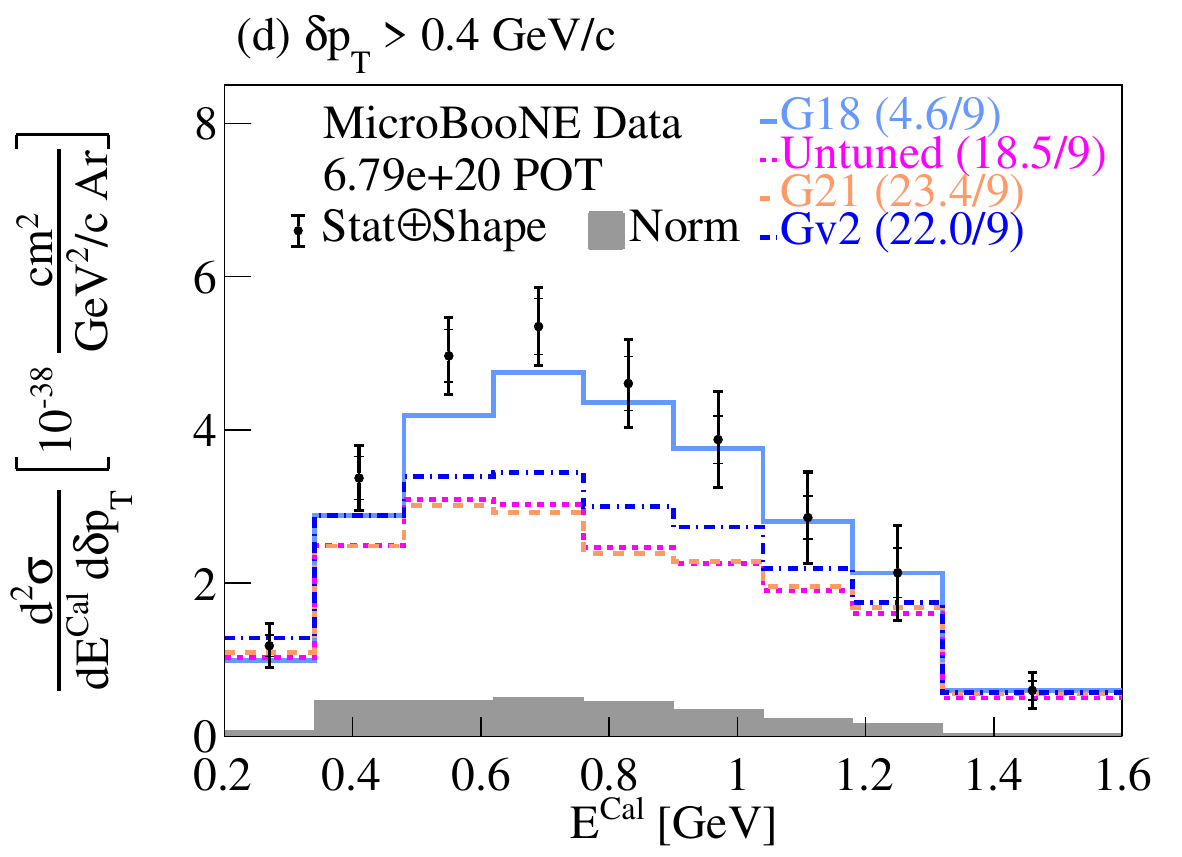}\\
\caption{
The flux-integrated (a) single- and (b-d) double- in $\delta p_{T}$ bins differential cross sections as a function of $E^{Cal}$. 
Inner and outer error bars show the statistical and total (statistical and shape systematic) uncertainty at the 1$\sigma$, or 68\%, confidence level. 
The gray band shows the normalization systematic uncertainty.
Colored lines show the results of theoretical cross section calculations using the $\texttt{G18}$ (light blue), $\texttt{Untuned}$ (magenta), $\texttt{G21}$ (orange), and $\texttt{Gv2}$ (dark blue) $\texttt{GENIE}$ configurations.
}
\label{ECalInDeltaPTGenie}
\end{figure*}

Figures~\ref{ECalInDeltaPTGen} and~\ref{ECalInDeltaPTGenie} show the single-differential cross sections as a function of $E^{Cal}$ using all the events (panel a), as well as the double-differential results in the same kinematic variable in $\delta p_{T}$ bins (panels b-d).
Figure~\ref{ECalInDeltaPTGen} shows the comparisons to a number of available neutrino event generators, where the $E^{Cal}$ distribution covers the same energy spectrum across all bins.
All the event generators illustrate an equally good performance in the lowest $\delta p_{T}$ bin.
$\texttt{NEUT}$ and $\texttt{NuWro}$ show a deficit relative to the data in the highest $\delta p_{T}$ bins.
Figure~\ref{ECalInDeltaPTGenie} shows the same results compared to a number of $\texttt{GENIE}$ configurations, where $\texttt{G18}$ illustrates the best performance.
In the lowest $\delta p_{T}$ bin, the different configurations illustrate a shift to the left compared to the data, unlike \texttt{G18}, which drives the significantly higher $\chi^{2}$ values.
Interestingly, all the alternative $\texttt{GENIE}$ configurations illustrate a plateau in the highest $\delta p_{T}$ bin that also yields high $\chi^{2}$/bins ratios.



\begin{figure*}[htb!]
\centering 
\includegraphics[width=0.49\linewidth]{\figures 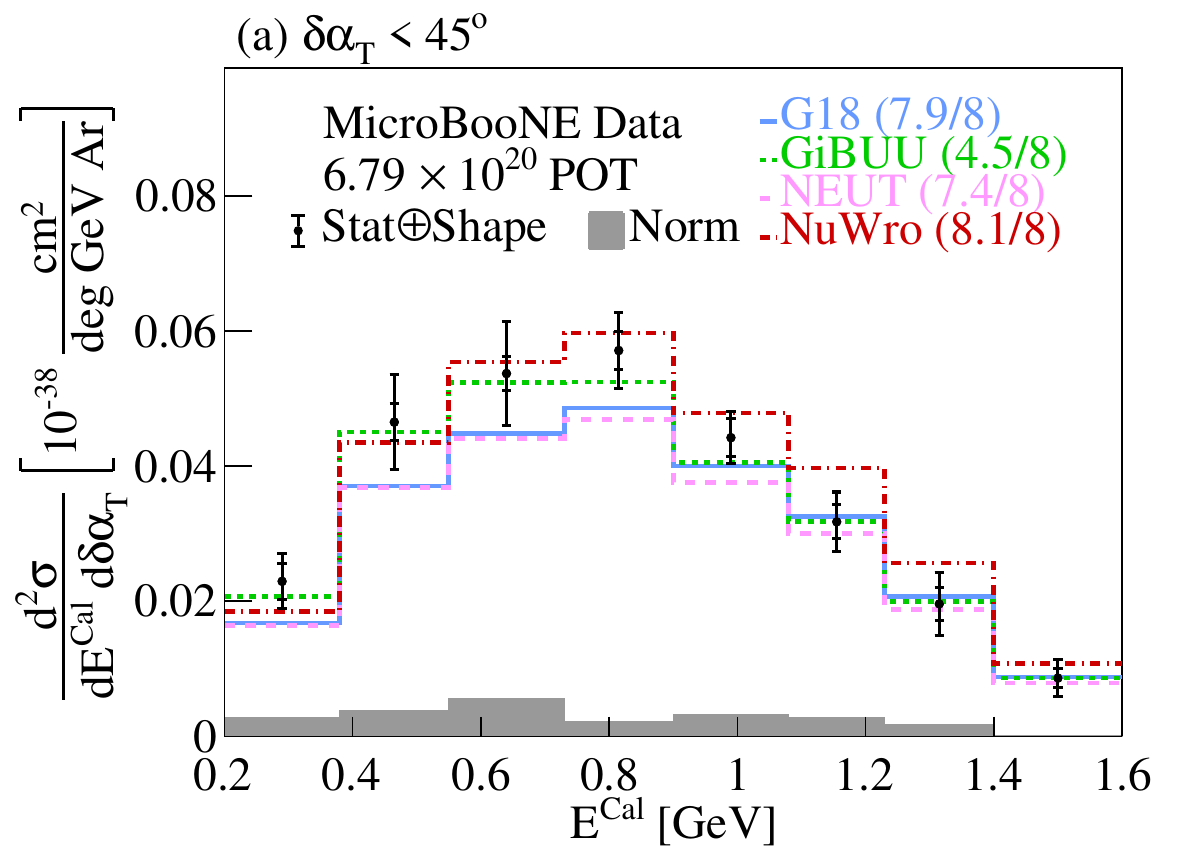}
\includegraphics[width=0.49\linewidth]{\figures 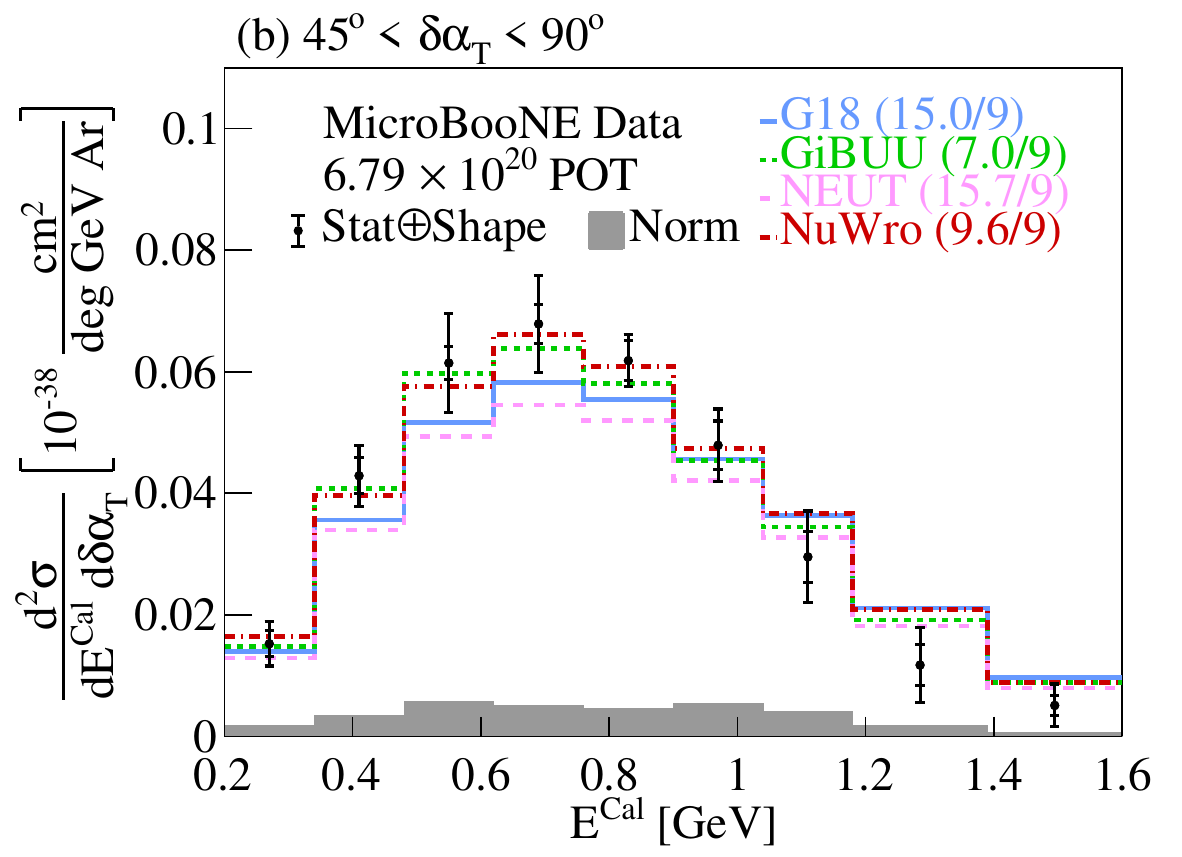}\\
\includegraphics[width=0.49\linewidth]{\figures 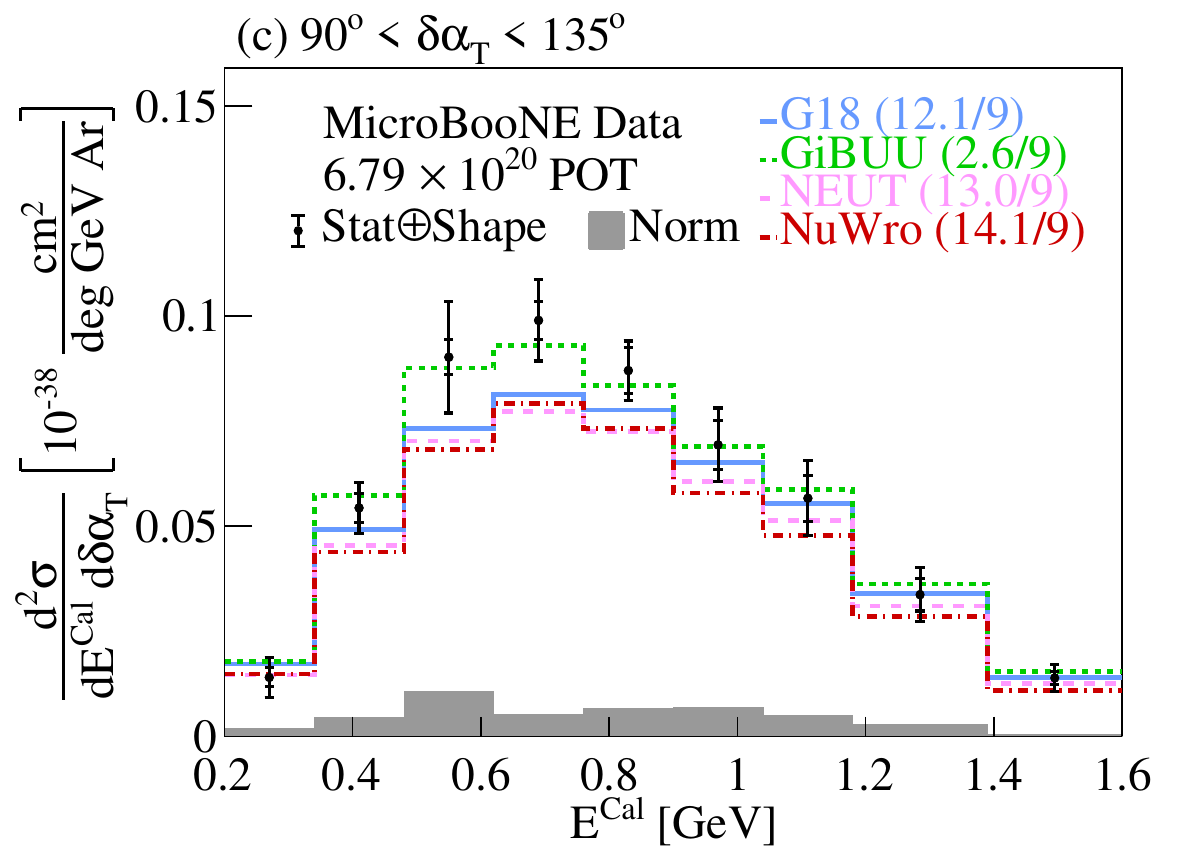}
\includegraphics[width=0.49\linewidth]{\figures 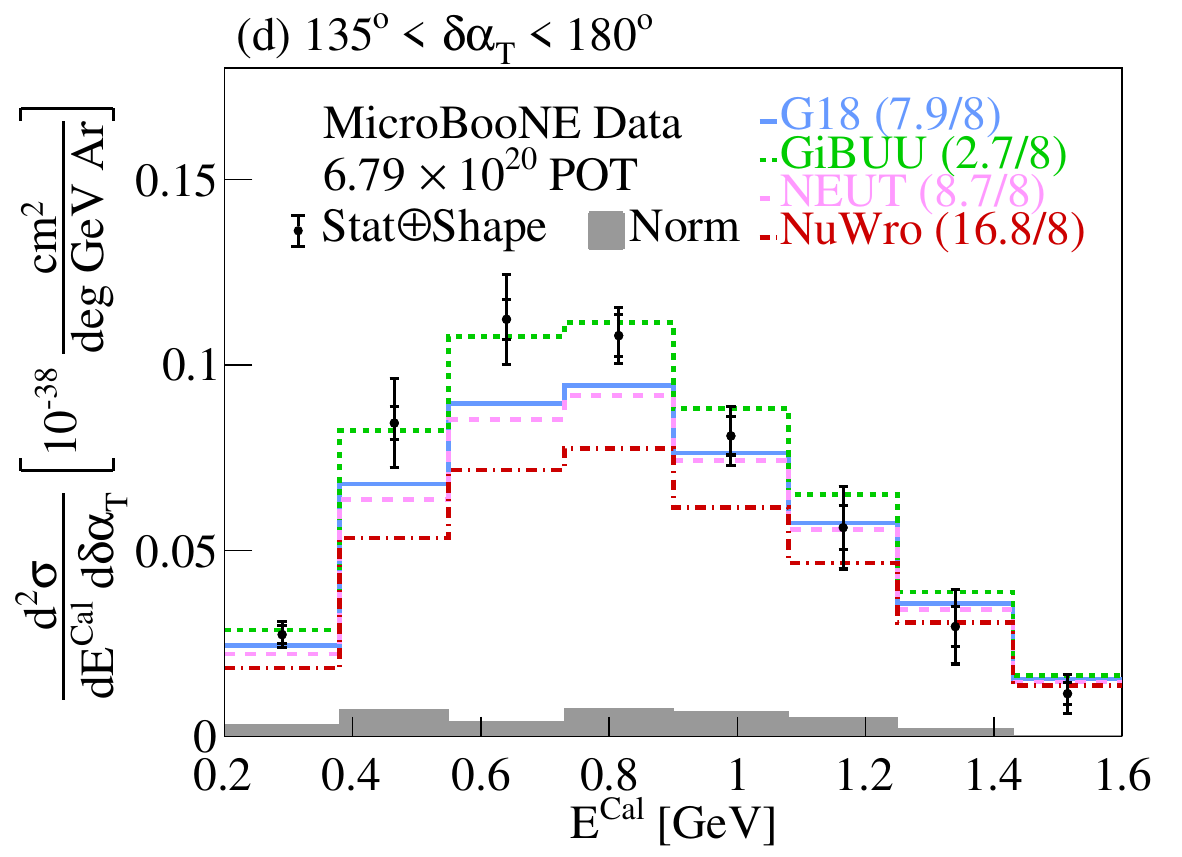}
\caption{
The flux-integrated double-differential cross sections as a function of $E^{Cal}$ in $\delta\alpha_{T}$ bins. 
Inner and outer error bars show the statistical and total (statistical and shape systematic) uncertainty at the 1$\sigma$, or 68\%, confidence level. 
The gray band shows the normalization systematic uncertainty.
Colored lines show the results of theoretical cross section calculations using the $\texttt{G18 GENIE}$ (blue), $\texttt{GiBUU}$ (green), $\texttt{NEUT}$ (pink), and $\texttt{NuWro}$ (red) event generators.
The numbers in parentheses show the $\chi^{2}$/bins calculation for each one of the predictions.
}
\label{ECalInDeltaAlphaTGen}
\end{figure*}

\begin{figure*}[htb!]
\centering
\includegraphics[width=0.49\linewidth]{\figures 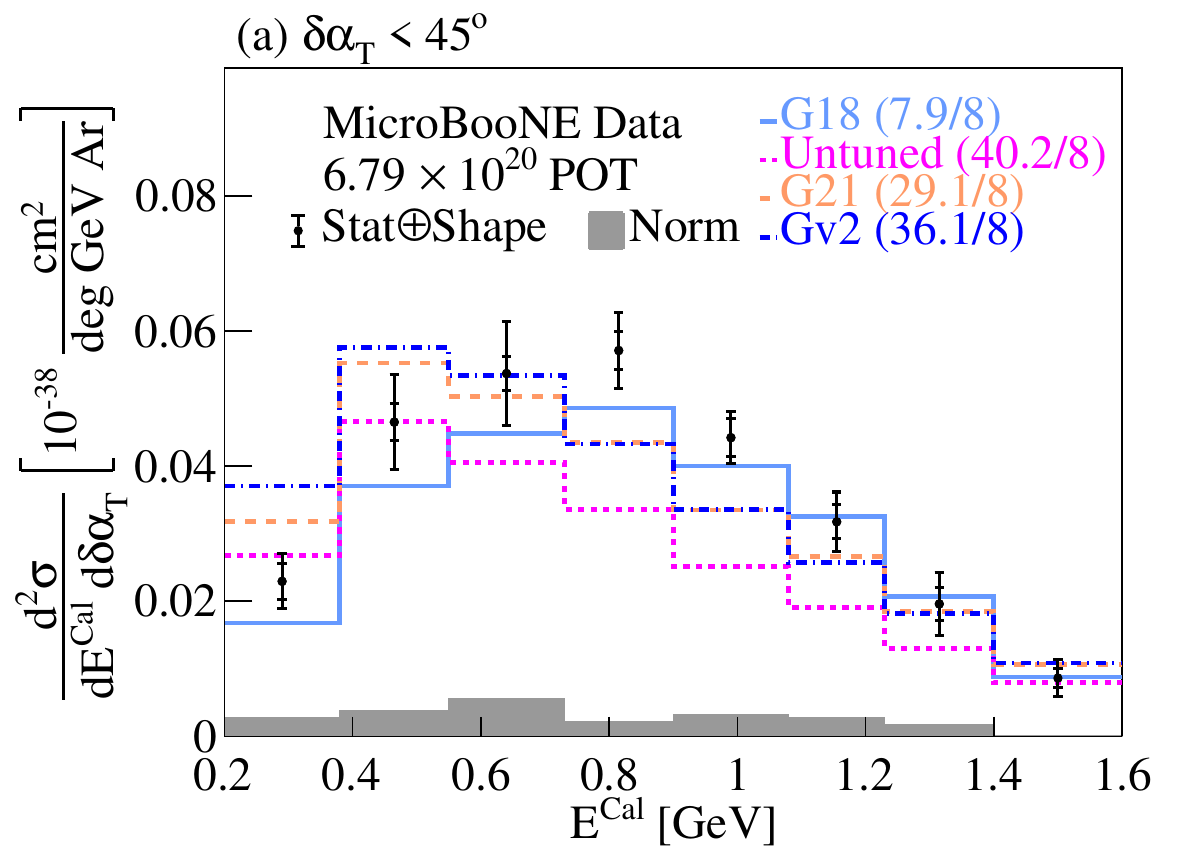}
\includegraphics[width=0.49\linewidth]{\figures 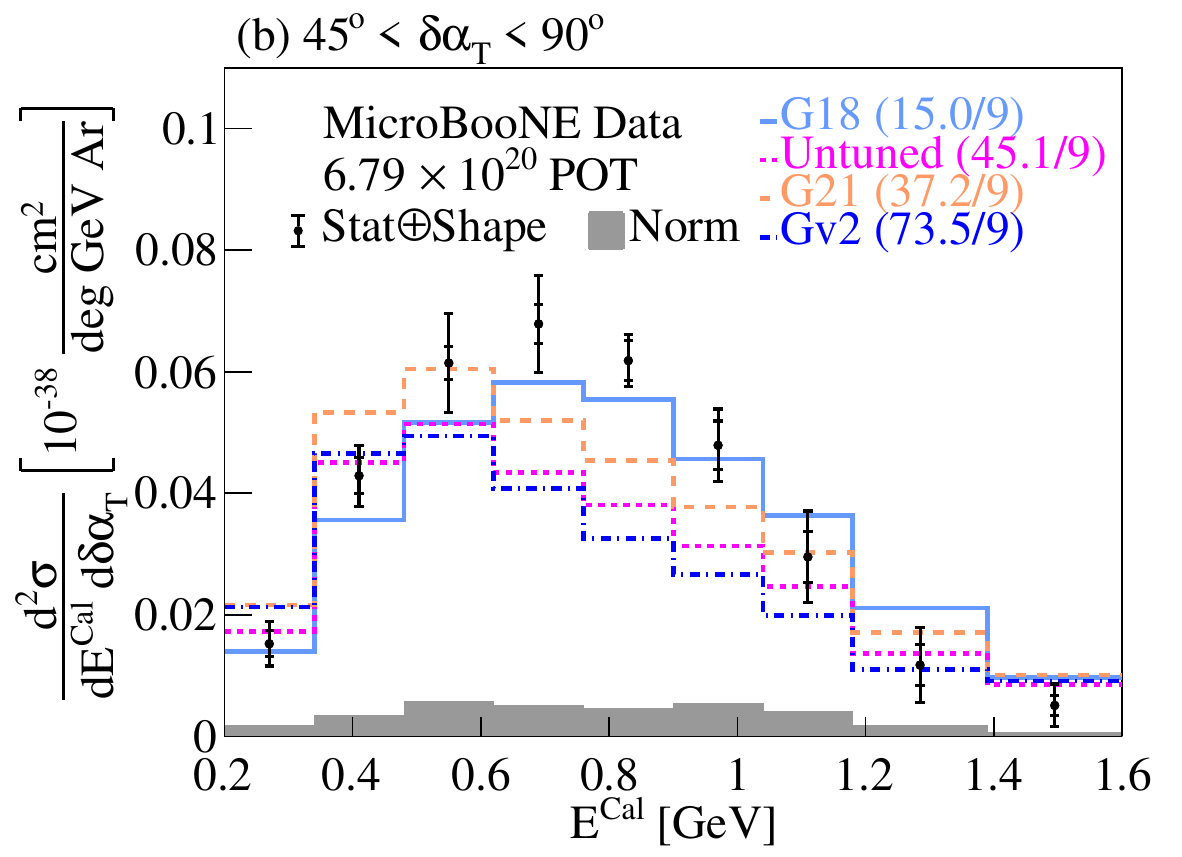}\\
\includegraphics[width=0.49\linewidth]{\figures 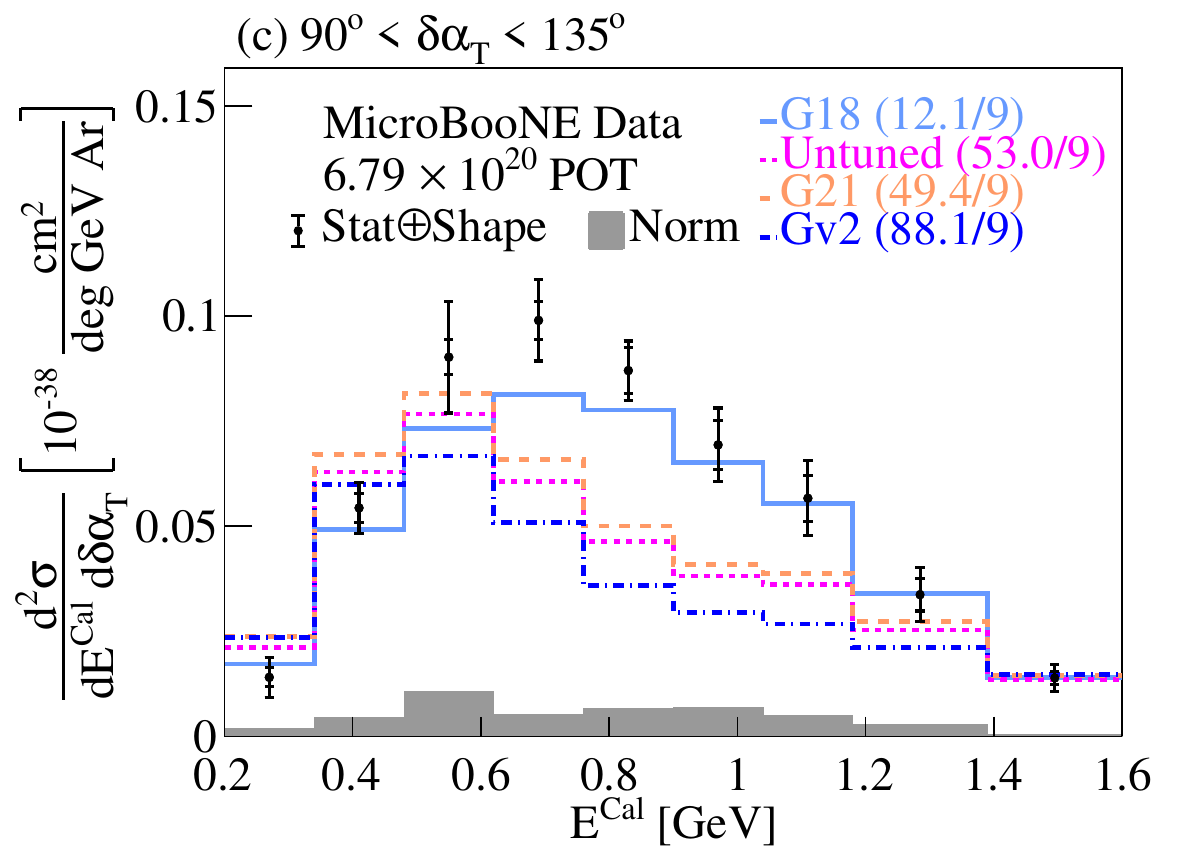}
\includegraphics[width=0.49\linewidth]{\figures 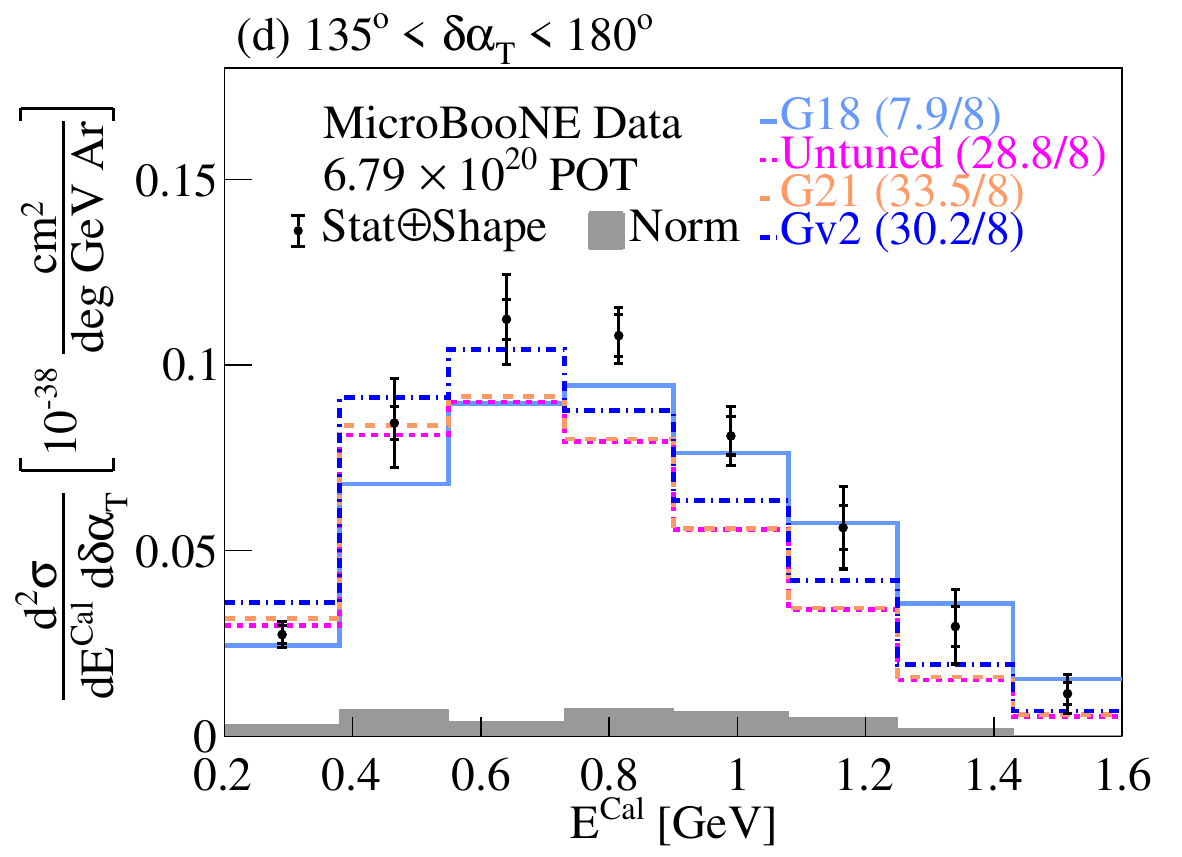}\\
\caption{
The flux-integrated double-differential cross sections as a function of $E^{Cal}$ in $\delta\alpha_{T}$ bins. 
Inner and outer error bars show the statistical and total (statistical and shape systematic) uncertainty at the 1$\sigma$, or 68\%, confidence level. 
The gray band shows the normalization systematic uncertainty.
Colored lines show the results of theoretical cross section calculations using the $\texttt{G18}$ (light blue), $\texttt{Untuned}$ (magenta), $\texttt{G21}$ (orange), and $\texttt{Gv2}$ (dark blue) $\texttt{GENIE}$ configurations.
The numbers in parentheses show the $\chi^{2}$/bins calculation for each one of the predictions.
}
\label{ECalInDeltaAlphaTGenie}
\end{figure*}

Figures~\ref{ECalInDeltaAlphaTGen} and~\ref{ECalInDeltaAlphaTGenie} show the double-differential results as a function of $E^{Cal}$ in $\delta\alpha_{T}$ bins.
Figure~\ref{ECalInDeltaAlphaTGen} shows the comparisons to a number of available neutrino event generators.
Once again, the $E^{Cal}$ distribution covers the same energy spectrum across all of our results and all the event generators show fairly good behavior.
\texttt{NuWro} illustrates a mild deficit in the 135$^{\circ}$ $< \delta\alpha_{T} <$ 180$^{\circ}$ bin, which is also reflected in the $\chi^{2}$/bins ratio.
Figure~\ref{ECalInDeltaAlphaTGenie} shows the same results compared to a number of $\texttt{GENIE}$ configurations, where all the $\texttt{GENIE}$ configurations except for $\texttt{G18}$ illustrate shape and strength differences.



\begin{figure*}[htb!]
\centering
\includegraphics[width=0.49\linewidth]{\figures 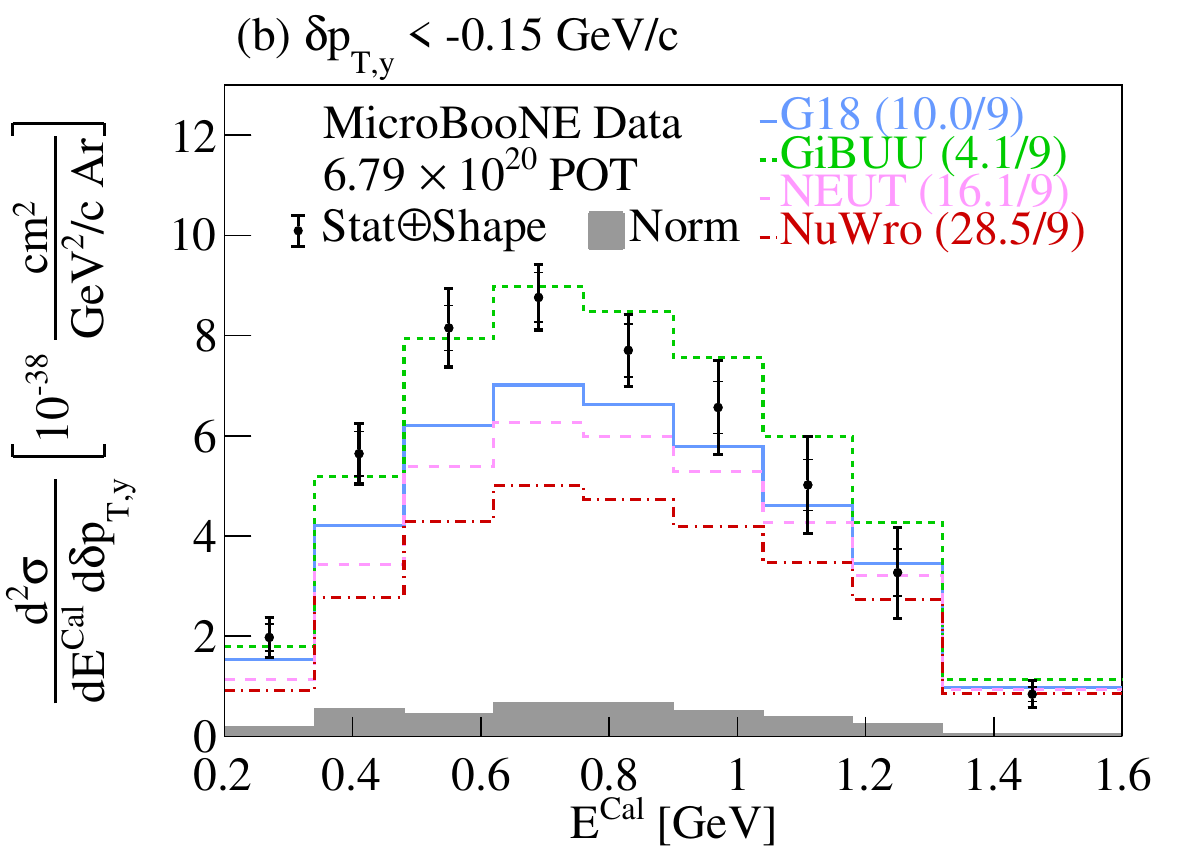}\\
\includegraphics[width=0.49\linewidth]{\figures 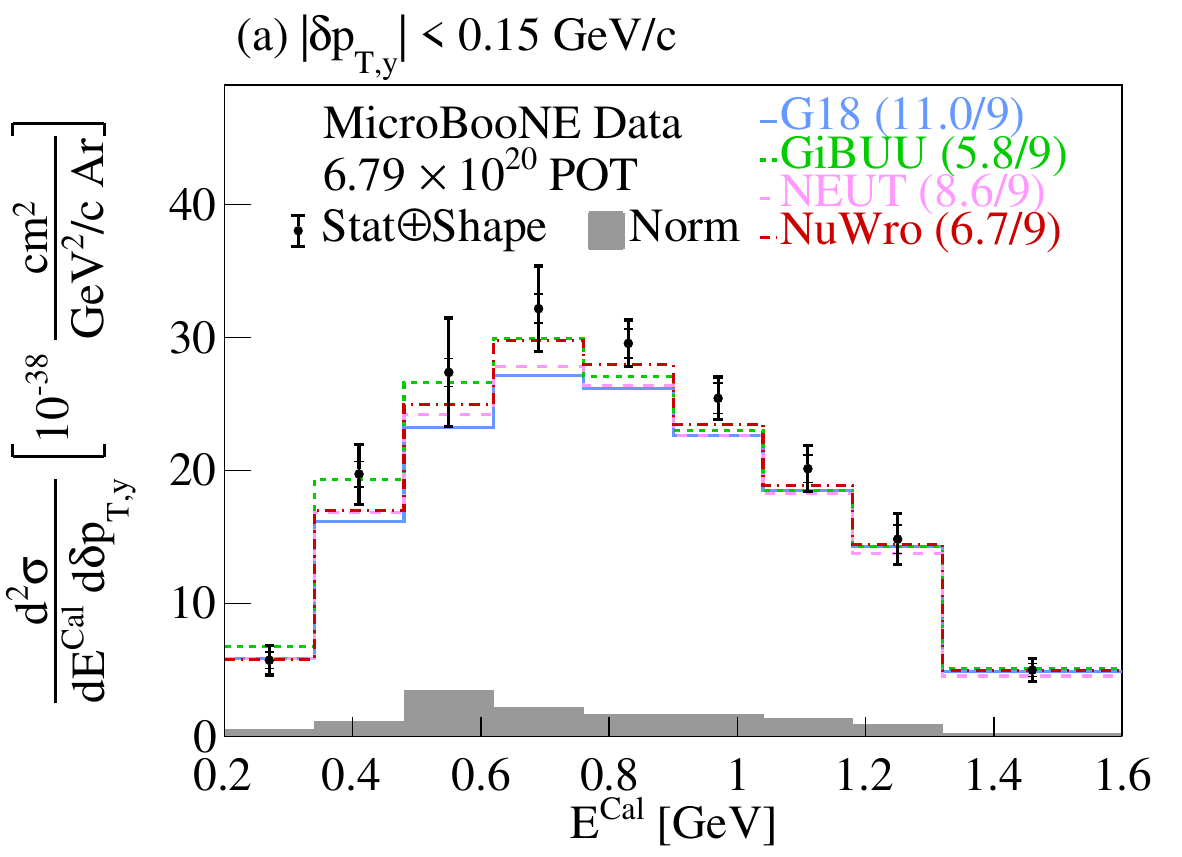}
\includegraphics[width=0.49\linewidth]{\figures 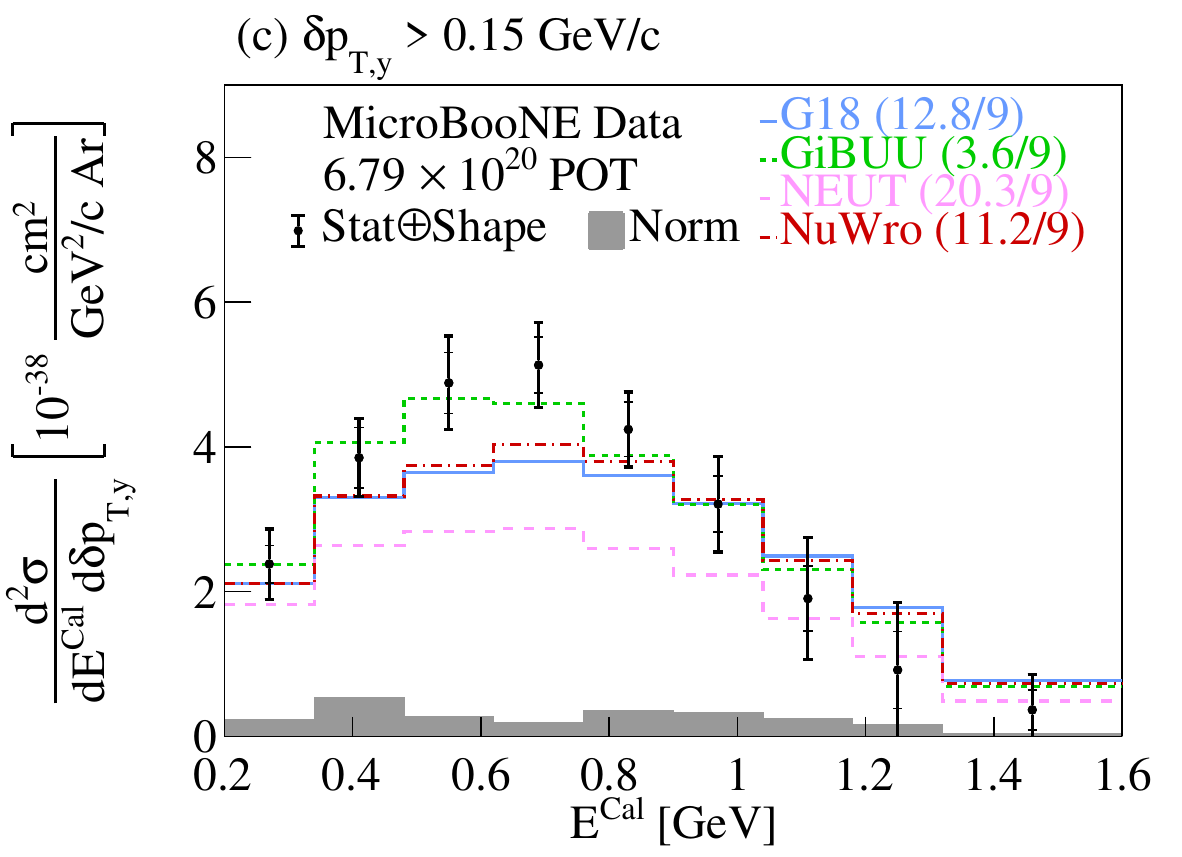}\\
\caption{
The flux-integrated double-differential cross sections as a function of $E^{Cal}$ in $\delta p_{T,y}$ bins. 
Inner and outer error bars show the statistical and total (statistical and shape systematic) uncertainty at the 1$\sigma$, or 68\%, confidence level. 
The gray band shows the normalization systematic uncertainty.
Colored lines show the results of theoretical cross section calculations using the $\texttt{G18 GENIE}$ (blue), $\texttt{GiBUU}$ (green), $\texttt{NEUT}$ (pink), and $\texttt{NuWro}$ (red) event generators.
The numbers in parentheses show the $\chi^{2}$/bins calculation for each one of the predictions.
}
\label{ECalInDeltaPtyGen}
\end{figure*}

\begin{figure*}[htb!]
\centering
\includegraphics[width=0.49\linewidth]{\figures 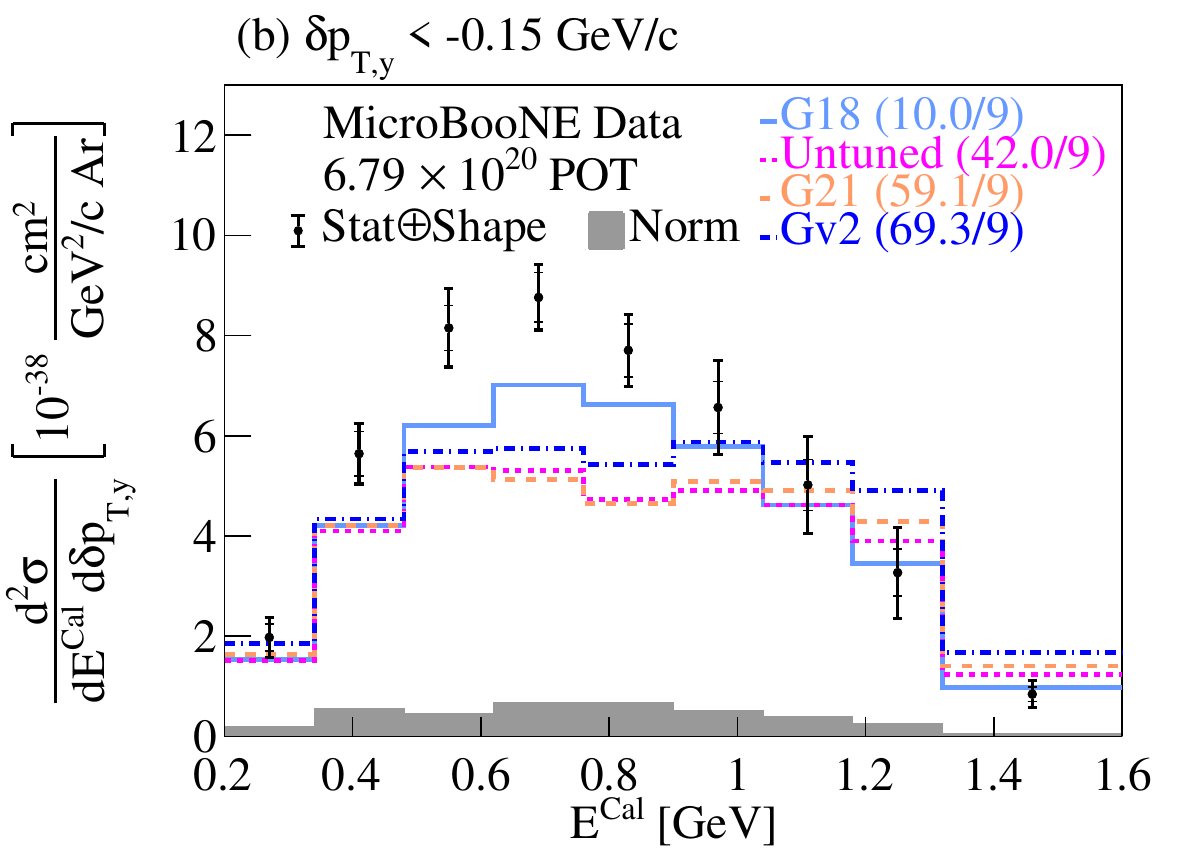}\\
\includegraphics[width=0.49\linewidth]{\figures 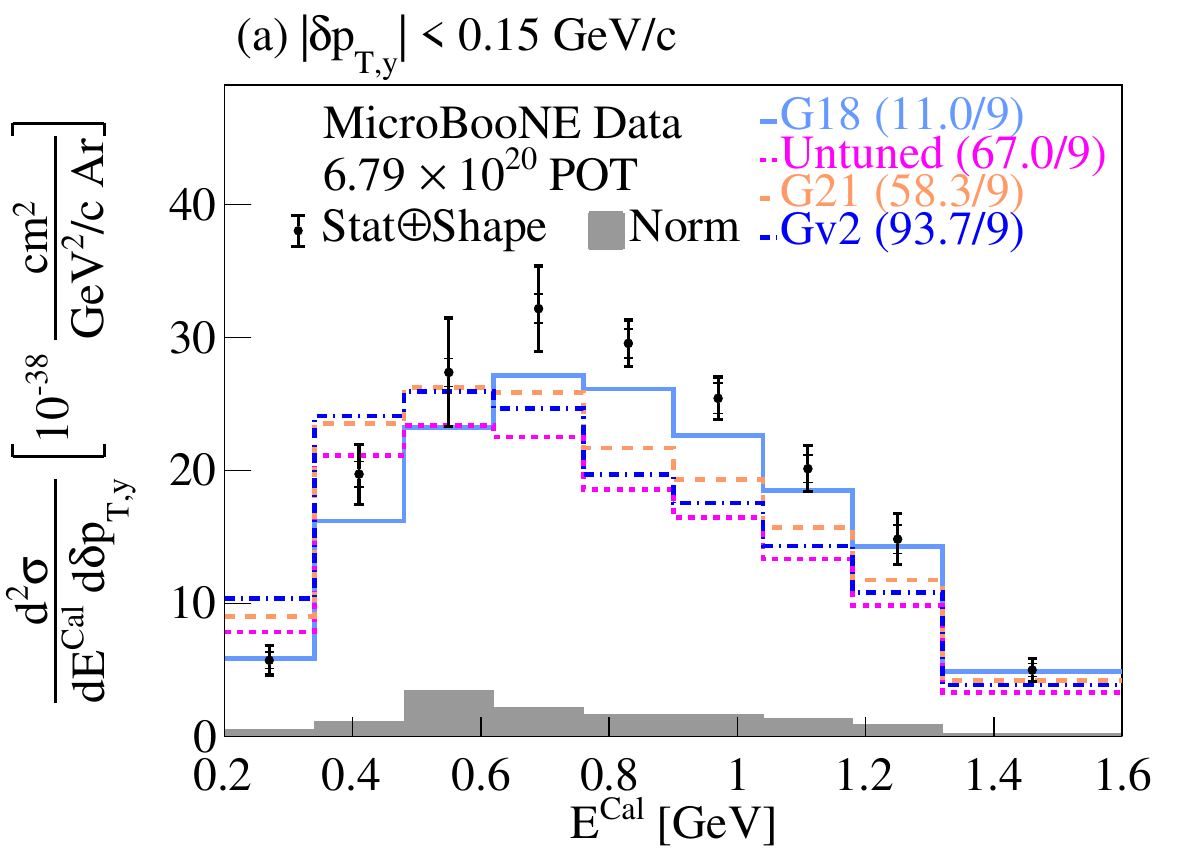}
\includegraphics[width=0.49\linewidth]{\figures 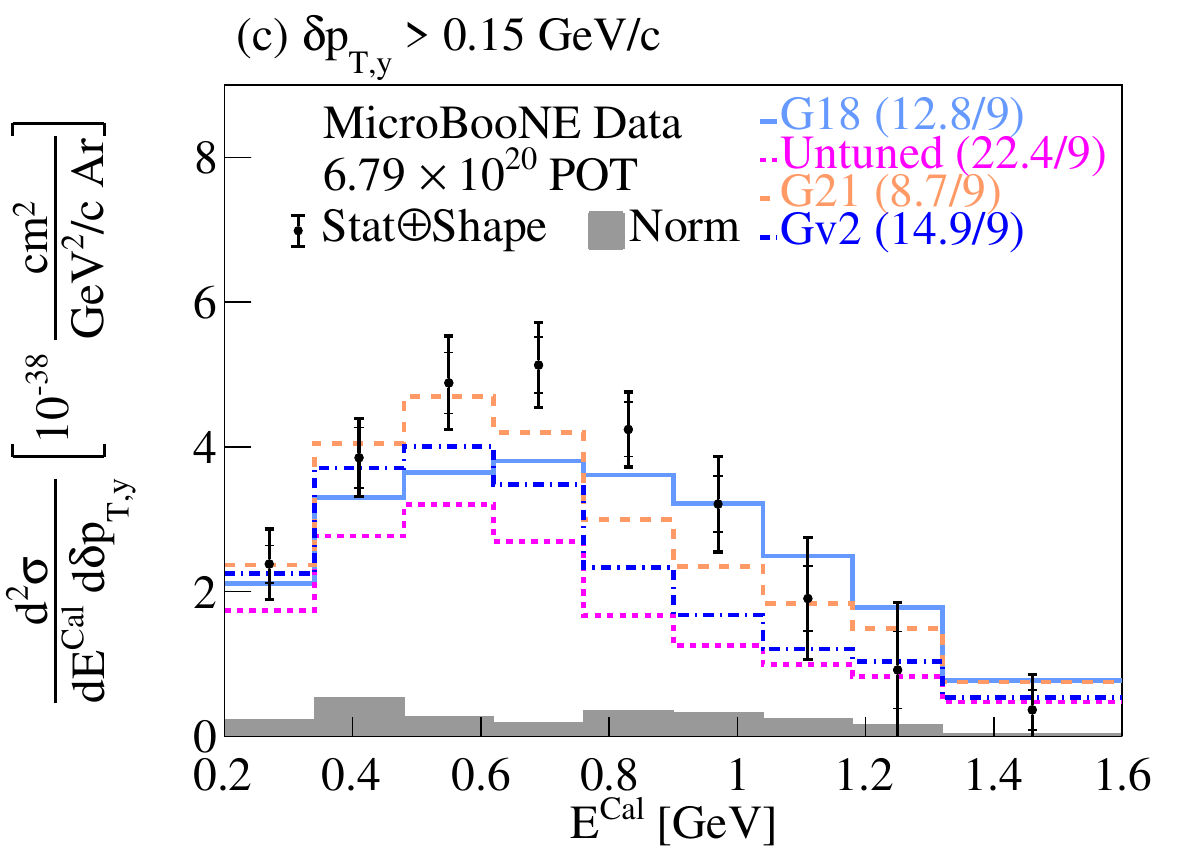}\\
\caption{
The flux-integrated double-differential cross sections as a function of $E^{Cal}$ in $\delta p_{T,y}$ bins. 
Inner and outer error bars show the statistical and total (statistical and shape systematic) uncertainty at the 1$\sigma$, or 68\%, confidence level. 
The gray band shows the normalization systematic uncertainty.
Colored lines show the results of theoretical cross section calculations using the $\texttt{G18}$ (light blue), $\texttt{Untuned}$ (magenta), $\texttt{G21}$ (orange), and $\texttt{Gv2}$ (dark blue) $\texttt{GENIE}$ configurations.
The numbers in parentheses show the $\chi^{2}$/bins calculation for each one of the predictions.
}
\label{ECalInDeltaPtyGenie}
\end{figure*}

Figures~\ref{ECalInDeltaPtyGen} and~\ref{ECalInDeltaPtyGenie} show the double-differential results as a function of $E^{Cal}$ in $\delta p_{T,y}$ bins.
Figure~\ref{ECalInDeltaPtyGen} shows the comparisons to a number of available neutrino event generators.
All event generators predict very similar cross sections for -0.15 $<$ $\delta p_{T,y}$ $<$ 0.15\,GeV/$c$ (panel a).
Unlike this central region, the $|\delta p_{T,y}|$ $>$ 0.15\,GeV/$c$ results yield a wide spread across the generator predictions (panels b-c).
Furthermore, apart from $\texttt{GiBUU}$, all the predictions lack strength in the $\delta p_{T,y}$ $<$ -0.15\,GeV/$c$ bin (panel b).
Additionally, $\texttt{NEUT}$ illustrates the same deficit in the $\delta p_{T,y}$ $>$ 0.15\,GeV/$c$ bin (panel c).
Figure~\ref{ECalInDeltaPtyGenie} shows the same results compared to a number of $\texttt{GENIE}$ configurations, where all the $\texttt{GENIE}$ configurations but $\texttt{G18}$ illustrate a poor performance due to shape and strength issues.


\section{Conclusions}\label{concl}

This work reports on measurements of flux-integrated differential cross sections for event topologies with a single muon and a single proton detected in the final state using the Booster Neutrino Beam at Fermi National Accelerator Laboratory and the \uB\ detector.
The data were studied for the first time in the form of single-differential cross sections in kinematic imbalance variables on argon.
Furthermore, the first double-differential cross sections in these variables were reported on the same nucleus.
Additionally, novel double-differential cross section measurements of a neutrino energy estimator in bins of these variables were presented.  
The results were compared to a number of event generators and model configurations.
The predictions as a function of the energy estimator across all generators and model configurations remain mostly unchanged regardless of the kinematic variable used for the double-differential measurements.
Based on the reported $\chi^{2}$/ bins, the good agreement observed across the calorimetric energy distributions suggests that the energy dependence is largely well-modeled across most predictions.
Unlike the energy estimator results, we found that the measured kinematic imbalance cross sections in different phase-space regions are sensitive to nuclear effects.
The performance of the event generators and configurations varies depending on the observable of interest.
Overall, the $\texttt{GENIE v3.0.6 G18\_10a\_02\_11a}$ cross section predictions with the MicroBooNE-specific tuning ($\texttt{G18}$) fit the data well.
On the other hand, the $\texttt{GENIE v2.12.10}$ ($\texttt{Gv2}$) cross section predictions are systematically a poor fit to data with significant shape differences across all variables of interest. 
The $\texttt{GENIE v3.0.6 G18\_10a\_02\_11a}$ configuration without additional tuning ($\texttt{Untuned}$) shows a systematic deficit of $\sim$ 20\% which necessitated the development of the aforementioned tune.
The $\texttt{GENIE v3.2.0 G21\_11b\_00\_000}$ configuration ($\texttt{G21}$) serves as an example of a theory-driven GENIE configuration that shows good agreement with data in most variables without the need for additional tuning.
$\texttt{GiBUU 2021}$ ($\texttt{GiBUU}$) shows good agreement with data in most kinematic variables, with the exception of $\delta p_{T}$, where a systematic shift to higher values of $\delta p_T$ has been identified.
A potential source of this shift is due to the $\texttt{GiBUU}$ MEC modeling.
The $\texttt{NuWro v19.02.2}$ ($\texttt{NuWro}$) prediction falls bellow the data due to poor FSI modeling and shows significant shape differences in FSI-dominated parts of the phase-space.
$\texttt{NEUT v5.4.0}$ ($\texttt{NEUT}$) also results in predictions mostly falling below the data points.
This mismodeling remains largely unnoticed when combined into the calorimetric energy estimator.
Yet, future neutrino oscillation measurements will rely on accurate cross section predictions and a precise mapping between measured and true neutrino energies. 
Therefore, such mismodeling effects might impact their experimental sensitivity.
The reported results both provide precision data to benchmark neutrino-nucleus interaction models and establish phase-space regions where precise reaction modeling is still needed.


\section{Acknowledgments}\label{ack}

This document was prepared by the MicroBooNE collaboration using the
resources of the Fermi National Accelerator Laboratory (Fermilab), a
U.S. Department of Energy, Office of Science, HEP User Facility.
Fermilab is managed by Fermi Research Alliance, LLC (FRA), acting
under Contract No. DE-AC02-07CH11359.  
This material is based upon work supported by Laboratory Directed Research and Development (LDRD) funding from Argonne National Laboratory, provided by the Director, Office of Science, of the U.S. Department of Energy under Contract No. DE-AC02-06CH11357.
MicroBooNE is supported by the
following: the U.S. Department of Energy, Office of Science, Offices
of High Energy Physics and Nuclear Physics; the U.S. National Science
Foundation; the Swiss National Science Foundation; the Science and
Technology Facilities Council (STFC), part of the United Kingdom Research 
and Innovation; the Royal Society (United Kingdom); the UK Research 
and Innovation (UKRI) Future Leaders Fellowship; and The European 
Union’s Horizon 2020 Marie Sklodowska-Curie Actions. Additional support 
for the laser calibration system and cosmic ray tagger was provided by 
the Albert Einstein Center for Fundamental Physics, Bern, Switzerland. 
We also acknowledge the contributions of technical and scientific staff 
to the design, construction, and operation of the MicroBooNE detector 
as well as the contributions of past collaborators to the development 
of MicroBooNE analyses, without whom this work would not have been 
possible. For the purpose of open access, the authors have applied a 
Creative Commons Attribution (CC BY) license to any Author Accepted 
Manuscript version arising from this submission.


\clearpage
\bibliography{main}


\end{document}


\centering
\large{\textbf{Multi-Differential Cross-Section Measurements in \texorpdfstring{$\nu_{\mu}$}{numu}-Argon\texorpdfstring{\\}{}Quasielastic-like Reactions with the MicroBooNE Detector}}

(Dated: \today)



\justify
\section{Data Release}

Overflow (underflow) values are included in the last (first) bin.
The additional smearing matrix $A_{C}$ should first be applied to the event distribution of an independent theoretical prediction when a comparison is performed to the data reported herein, and then divided by the bin width.
The $A_{C}$ matrices are dimensionless.
The double-differential cross sections include correlations between the phase-space slices.
The data release with the data results, the covariance matrices, and the additional smearing matrices is included in the DataRelease.root file.
Instructions on how to use the data release and the description of the binning scheme are included in the README file.

\raggedbottom

\begin{table}[H]
\raggedright
\begin{adjustbox}{width=\textwidth}
\small
\begin{tabular}{ |c|c|c|c|c| }
\hline
\multicolumn{5}{|c|}{Cross Section $\delta p_{T}$, $All\,events$} \\
\hline
\hline
Bin \# & Low edge [GeV/$\textit{c}$] & High edge [GeV/$\textit{c}$] & Cross Section [$10^{-38}\frac{cm^{2}}{(GeV/\textit{c})\,^{40}Ar}$] & Uncertainty [$10^{-38}\frac{cm^{2}}{(GeV/\textit{c})\,^{40}Ar}$] \\
\hline
\hline
1 & 0 & 0.05 & 17.620355 & 2.5914346\\
2 & 0.05 & 0.1 & 31.26762 & 3.4668697\\
3 & 0.1 & 0.15 & 38.993278 & 3.8931128\\
4 & 0.15 & 0.2 & 36.59134 & 3.6679221\\
5 & 0.2 & 0.25 & 26.614842 & 2.9078439\\
6 & 0.25 & 0.3 & 15.445038 & 2.285042\\
7 & 0.3 & 0.35 & 11.568891 & 2.2446977\\
8 & 0.35 & 0.4 & 10.864251 & 2.1028483\\
9 & 0.4 & 0.47 & 9.5845691 & 1.6583338\\
10 & 0.47 & 0.55 & 7.702555 & 1.3718607\\
11 & 0.55 & 0.65 & 4.6959675 & 0.97287196\\
12 & 0.65 & 0.75 & 2.5540892 & 0.77974472\\
13 & 0.75 & 0.9 & 1.1170042 & 0.50573303\\
\hline
\end{tabular}
\end{adjustbox}
\end{table}

\begin{table}[H]
\raggedright
\begin{adjustbox}{width=\textwidth}
\small
\begin{tabular}{ |c|c|c|c|c| }
\hline
\multicolumn{5}{|c|}{Cross Section $\delta p_{T}$, $\delta\alpha_{T}\,<\,45^{o}$} \\
\hline
\hline
Bin \# & Low edge [GeV/$\textit{c}$] & High edge [GeV/$\textit{c}$] & Cross Section [$10^{-38}\frac{cm^{2}}{deg\,(GeV/\textit{c})\,^{40}Ar}$] & Uncertainty [$10^{-38}\frac{cm^{2}}{deg\,(GeV/\textit{c})\,^{40}Ar}$] \\
\hline
\hline
1 & 0 & 0.05 & 0.092432219 & 0.017847411\\
2 & 0.05 & 0.1 & 0.15650347 & 0.020583601\\
3 & 0.1 & 0.15 & 0.18956302 & 0.021759872\\
4 & 0.15 & 0.2 & 0.17478665 & 0.022881942\\
5 & 0.2 & 0.25 & 0.12333558 & 0.018392123\\
6 & 0.25 & 0.3 & 0.054849811 & 0.011418205\\
7 & 0.3 & 0.35 & 0.028103987 & 0.010930512\\
8 & 0.35 & 0.4 & 0.013505763 & 0.0082180275\\
9 & 0.4 & 0.47 & 0.0076037373 & 0.00469917\\
10 & 0.47 & 0.55 & 0.0070471199 & 0.0029156727\\
11 & 0.55 & 0.9 & 0.0013794438 & 0.00037878852\\
\hline
\end{tabular}
\end{adjustbox}
\end{table}

\begin{table}[H]
\raggedright
\begin{adjustbox}{width=\textwidth}
\small
\begin{tabular}{ |c|c|c|c|c| }
\hline
\multicolumn{5}{|c|}{Cross Section $\delta p_{T}$, $45^{o}\,<\,\delta\alpha_{T}\,<\,90^{o}$} \\
\hline
\hline
Bin \# & Low edge [GeV/$\textit{c}$] & High edge [GeV/$\textit{c}$] & Cross Section [$10^{-38}\frac{cm^{2}}{deg\,(GeV/\textit{c})\,^{40}Ar}$] & Uncertainty [$10^{-38}\frac{cm^{2}}{deg\,(GeV/\textit{c})\,^{40}Ar}$] \\
\hline
\hline
1 & 0 & 0.05 & 0.080839958 & 0.016091714\\
2 & 0.05 & 0.1 & 0.16500821 & 0.023786632\\
3 & 0.1 & 0.15 & 0.21299492 & 0.0233775\\
4 & 0.15 & 0.2 & 0.19020215 & 0.021715469\\
5 & 0.2 & 0.25 & 0.11468038 & 0.017230934\\
6 & 0.25 & 0.3 & 0.053372907 & 0.011484295\\
7 & 0.3 & 0.35 & 0.040883617 & 0.010114144\\
8 & 0.35 & 0.4 & 0.03832816 & 0.0094356404\\
9 & 0.4 & 0.47 & 0.025847002 & 0.0062119236\\
10 & 0.47 & 0.55 & 0.016095979 & 0.0044185631\\
11 & 0.55 & 0.65 & 0.0093466585 & 0.0027899017\\
12 & 0.65 & 0.9 & 0.0014243277 & 0.00037605569\\
\hline
\end{tabular}
\end{adjustbox}
\end{table}

\begin{table}[H]
\raggedright
\begin{adjustbox}{width=\textwidth}
\small
\begin{tabular}{ |c|c|c|c|c| }
\hline
\multicolumn{5}{|c|}{Cross Section $\delta p_{T}$, $90^{o}\,<\,\delta\alpha_{T}\,<\,135^{o}$} \\
\hline
\hline
Bin \# & Low edge [GeV/$\textit{c}$] & High edge [GeV/$\textit{c}$] & Cross Section [$10^{-38}\frac{cm^{2}}{deg\,(GeV/\textit{c})\,^{40}Ar}$] & Uncertainty [$10^{-38}\frac{cm^{2}}{deg\,(GeV/\textit{c})\,^{40}Ar}$] \\
\hline
\hline
1 & 0 & 0.05 & 0.095157009 & 0.017451739\\
2 & 0.05 & 0.1 & 0.18468234 & 0.021502271\\
3 & 0.1 & 0.15 & 0.22487235 & 0.021799237\\
4 & 0.15 & 0.2 & 0.20251233 & 0.02233952\\
5 & 0.2 & 0.25 & 0.15791306 & 0.019451585\\
6 & 0.25 & 0.3 & 0.11383164 & 0.015791141\\
7 & 0.3 & 0.35 & 0.099356533 & 0.014450469\\
8 & 0.35 & 0.4 & 0.093123487 & 0.014717311\\
9 & 0.4 & 0.47 & 0.0701963 & 0.010694715\\
10 & 0.47 & 0.55 & 0.044745679 & 0.0077857583\\
11 & 0.55 & 0.65 & 0.01944616 & 0.0048998765\\
12 & 0.65 & 0.75 & 0.0082805748 & 0.0031612527\\
13 & 0.75 & 0.9 & 0.0034429473 & 0.0013177862\\
\hline
\end{tabular}
\end{adjustbox}
\end{table}

\begin{table}[H]
\raggedright
\begin{adjustbox}{width=\textwidth}
\small
\begin{tabular}{ |c|c|c|c|c| }
\hline
\multicolumn{5}{|c|}{Cross Section $\delta p_{T}$, $135^{o}\,<\,\delta\alpha_{T}\,<\,180^{o}$} \\
\hline
\hline
Bin \# & Low edge [GeV/$\textit{c}$] & High edge [GeV/$\textit{c}$] & Cross Section [$10^{-38}\frac{cm^{2}}{deg\,(GeV/\textit{c})\,^{40}Ar}$] & Uncertainty [$10^{-38}\frac{cm^{2}}{deg\,(GeV/\textit{c})\,^{40}Ar}$] \\
\hline
\hline
1 & 0 & 0.05 & 0.066626447 & 0.014818726\\
2 & 0.05 & 0.1 & 0.16070794 & 0.021216529\\
3 & 0.1 & 0.15 & 0.22967355 & 0.025077241\\
4 & 0.15 & 0.2 & 0.2326529 & 0.024400816\\
5 & 0.2 & 0.25 & 0.17848331 & 0.022261407\\
6 & 0.25 & 0.3 & 0.11622475 & 0.019361849\\
7 & 0.3 & 0.35 & 0.10120987 & 0.019008044\\
8 & 0.35 & 0.4 & 0.104967 & 0.018704587\\
9 & 0.4 & 0.47 & 0.098770409 & 0.016175196\\
10 & 0.47 & 0.55 & 0.091046302 & 0.013985857\\
11 & 0.55 & 0.65 & 0.067951756 & 0.010618235\\
12 & 0.65 & 0.75 & 0.049075129 & 0.0090352476\\
13 & 0.75 & 0.9 & 0.027264037 & 0.0062171874\\
\hline
\end{tabular}
\end{adjustbox}
\end{table}

\begin{table}[H]
\raggedright
\begin{adjustbox}{width=\textwidth}
\small
\begin{tabular}{ |c|c|c|c|c| }
\hline
\multicolumn{5}{|c|}{Cross Section $\delta p_{T}$, $-1\,<\,cos\theta_{\mu}\,<\,0$} \\
\hline
\hline
Bin \# & Low edge [GeV/$\textit{c}$] & High edge [GeV/$\textit{c}$] & Cross Section [$10^{-38}\frac{cm^{2}}{(GeV/\textit{c})\,^{40}Ar}$] & Uncertainty [$10^{-38}\frac{cm^{2}}{(GeV/\textit{c})\,^{40}Ar}$] \\
\hline
\hline
1 & 0 & 0.05 & 2.8972789 & 0.47028924\\
2 & 0.05 & 0.1 & 4.8795088 & 0.59172953\\
3 & 0.1 & 0.15 & 6.2158841 & 0.69451593\\
4 & 0.15 & 0.2 & 6.2967622 & 0.68640789\\
5 & 0.2 & 0.25 & 4.8680159 & 0.60573656\\
6 & 0.25 & 0.3 & 2.8610903 & 0.48818178\\
7 & 0.3 & 0.35 & 2.0742146 & 0.39651533\\
8 & 0.35 & 0.4 & 2.0157949 & 0.35078711\\
9 & 0.4 & 0.47 & 1.5790889 & 0.2847846\\
10 & 0.47 & 0.55 & 1.2206528 & 0.22957788\\
11 & 0.55 & 0.65 & 0.71243255 & 0.15542827\\
12 & 0.65 & 0.75 & 0.40168493 & 0.10229892\\
13 & 0.75 & 0.9 & 0.21324573 & 0.049325249\\
\hline
\end{tabular}
\end{adjustbox}
\end{table}

\begin{table}[H]
\raggedright
\begin{adjustbox}{width=\textwidth}
\small
\begin{tabular}{ |c|c|c|c|c| }
\hline
\multicolumn{5}{|c|}{Cross Section $\delta p_{T}$, $0\,<\,cos\theta_{\mu}\,<\,0.5$} \\
\hline
\hline
Bin \# & Low edge [GeV/$\textit{c}$] & High edge [GeV/$\textit{c}$] & Cross Section [$10^{-38}\frac{cm^{2}}{(GeV/\textit{c})\,^{40}Ar}$] & Uncertainty [$10^{-38}\frac{cm^{2}}{(GeV/\textit{c})\,^{40}Ar}$] \\
\hline
\hline
1 & 0 & 0.05 & 6.4076331 & 1.0364626\\
2 & 0.05 & 0.1 & 13.536861 & 1.4879789\\
3 & 0.1 & 0.15 & 18.330102 & 1.8051206\\
4 & 0.15 & 0.2 & 17.58502 & 1.7108479\\
5 & 0.2 & 0.25 & 12.716533 & 1.4127618\\
6 & 0.25 & 0.3 & 7.2659058 & 1.0593244\\
7 & 0.3 & 0.35 & 6.1179115 & 0.89297058\\
8 & 0.35 & 0.4 & 6.6062262 & 0.85030913\\
9 & 0.4 & 0.47 & 5.6331547 & 0.7446876\\
10 & 0.47 & 0.55 & 4.2548941 & 0.5942413\\
11 & 0.55 & 0.65 & 2.5110257 & 0.47093354\\
12 & 0.65 & 0.75 & 1.498103 & 0.42196058\\
13 & 0.75 & 0.9 & 0.94575386 & 0.2502498\\
\hline
\end{tabular}
\end{adjustbox}
\end{table}

\begin{table}[H]
\raggedright
\begin{adjustbox}{width=\textwidth}
\small
\begin{tabular}{ |c|c|c|c|c| }
\hline
\multicolumn{5}{|c|}{Cross Section $\delta p_{T}$, $0.5\,<\,cos\theta_{\mu}\,<\,0.75$} \\
\hline
\hline
Bin \# & Low edge [GeV/$\textit{c}$] & High edge [GeV/$\textit{c}$] & Cross Section [$10^{-38}\frac{cm^{2}}{(GeV/\textit{c})\,^{40}Ar}$] & Uncertainty [$10^{-38}\frac{cm^{2}}{(GeV/\textit{c})\,^{40}Ar}$] \\
\hline
\hline
1 & 0 & 0.05 & 15.380561 & 2.6701845\\
2 & 0.05 & 0.1 & 29.852446 & 3.4608007\\
3 & 0.1 & 0.15 & 37.660908 & 3.9966594\\
4 & 0.15 & 0.2 & 34.433818 & 3.8608125\\
5 & 0.2 & 0.25 & 24.70568 & 3.2696812\\
6 & 0.25 & 0.3 & 14.660729 & 2.3292109\\
7 & 0.3 & 0.35 & 11.740573 & 2.2385472\\
8 & 0.35 & 0.4 & 10.084049 & 2.2760598\\
9 & 0.4 & 0.47 & 7.2463407 & 1.779621\\
10 & 0.47 & 0.55 & 5.4514811 & 1.4472018\\
11 & 0.55 & 0.65 & 4.2622024 & 1.1129072\\
12 & 0.65 & 0.75 & 4.0114928 & 0.84580123\\
13 & 0.75 & 0.9 & 3.1828755 & 0.49113742\\
\hline
\end{tabular}
\end{adjustbox}
\end{table}

\begin{table}[H]
\raggedright
\begin{adjustbox}{width=\textwidth}
\small
\begin{tabular}{ |c|c|c|c|c| }
\hline
\multicolumn{5}{|c|}{Cross Section $\delta p_{T}$, $0.75\,<\,cos\theta_{\mu}\,<\,1$} \\
\hline
\hline
Bin \# & Low edge [GeV/$\textit{c}$] & High edge [GeV/$\textit{c}$] & Cross Section [$10^{-38}\frac{cm^{2}}{(GeV/\textit{c})\,^{40}Ar}$] & Uncertainty [$10^{-38}\frac{cm^{2}}{(GeV/\textit{c})\,^{40}Ar}$] \\
\hline
\hline
1 & 0 & 0.05 & 27.028925 & 3.2175885\\
2 & 0.05 & 0.1 & 46.712216 & 4.4403746\\
3 & 0.1 & 0.15 & 53.809965 & 4.8124714\\
4 & 0.15 & 0.2 & 49.48201 & 4.7392997\\
5 & 0.2 & 0.25 & 36.508467 & 4.2067223\\
6 & 0.25 & 0.3 & 22.380246 & 3.6766652\\
7 & 0.3 & 0.35 & 17.079853 & 3.6890718\\
8 & 0.35 & 0.4 & 14.479513 & 3.3118675\\
9 & 0.4 & 0.47 & 11.284078 & 2.4680123\\
10 & 0.47 & 0.55 & 8.5691441 & 1.8983666\\
11 & 0.55 & 0.65 & 5.2053077 & 1.2526561\\
12 & 0.65 & 0.75 & 3.0276476 & 0.90112721\\
13 & 0.75 & 0.9 & 1.3485971 & 0.50599433\\
\hline
\end{tabular}
\end{adjustbox}
\end{table}

\begin{table}[H]
\raggedright
\begin{adjustbox}{width=\textwidth}
\small
\begin{tabular}{ |c|c|c|c|c| }
\hline
\multicolumn{5}{|c|}{Cross Section $\delta p_{T}$, $-1\,<\,cos\theta_{p}\,<\,0$} \\
\hline
\hline
Bin \# & Low edge [GeV/$\textit{c}$] & High edge [GeV/$\textit{c}$] & Cross Section [$10^{-38}\frac{cm^{2}}{(GeV/\textit{c})\,^{40}Ar}$] & Uncertainty [$10^{-38}\frac{cm^{2}}{(GeV/\textit{c})\,^{40}Ar}$] \\
\hline
\hline
1 & 0 & 0.1 & 1.378803 & 0.33041234\\
2 & 0.1 & 0.2 & 2.0326442 & 0.42979728\\
3 & 0.2 & 0.3 & 1.9819565 & 0.35209247\\
4 & 0.3 & 0.4 & 1.6099007 & 0.30137645\\
5 & 0.4 & 0.55 & 1.0003658 & 0.22409022\\
6 & 0.55 & 0.65 & 0.80078906 & 0.2020649\\
7 & 0.65 & 0.75 & 0.54911148 & 0.15740353\\
8 & 0.75 & 0.9 & 0.37003264 & 0.1000518\\
\hline
\end{tabular}
\end{adjustbox}
\end{table}

\begin{table}[H]
\raggedright
\begin{adjustbox}{width=\textwidth}
\small
\begin{tabular}{ |c|c|c|c|c| }
\hline
\multicolumn{5}{|c|}{Cross Section $\delta p_{T}$, $0\,<\,cos\theta_{p}\,<\,0.5$} \\
\hline
\hline
Bin \# & Low edge [GeV/$\textit{c}$] & High edge [GeV/$\textit{c}$] & Cross Section [$10^{-38}\frac{cm^{2}}{(GeV/\textit{c})\,^{40}Ar}$] & Uncertainty [$10^{-38}\frac{cm^{2}}{(GeV/\textit{c})\,^{40}Ar}$] \\
\hline
\hline
1 & 0 & 0.05 & 6.2047395 & 1.1218596\\
2 & 0.05 & 0.1 & 12.611858 & 1.4924083\\
3 & 0.1 & 0.15 & 16.317472 & 1.7120459\\
4 & 0.15 & 0.2 & 15.721539 & 1.7369845\\
5 & 0.2 & 0.25 & 11.970312 & 1.4026407\\
6 & 0.25 & 0.3 & 7.6230327 & 1.0931229\\
7 & 0.3 & 0.35 & 6.6756946 & 1.2203851\\
8 & 0.35 & 0.4 & 6.2360016 & 1.304357\\
9 & 0.4 & 0.47 & 4.6592316 & 1.0078739\\
10 & 0.47 & 0.55 & 3.2994268 & 0.83046501\\
11 & 0.55 & 0.65 & 1.6218589 & 0.6115073\\
12 & 0.65 & 0.75 & 0.77615555 & 0.4547838\\
13 & 0.75 & 0.9 & 0.81279288 & 0.25752135\\
\hline
\end{tabular}
\end{adjustbox}
\end{table}

\begin{table}[H]
\raggedright
\begin{adjustbox}{width=\textwidth}
\small
\begin{tabular}{ |c|c|c|c|c| }
\hline
\multicolumn{5}{|c|}{Cross Section $\delta p_{T}$, $0.5\,<\,cos\theta_{p}\,<\,0.75$} \\
\hline
\hline
Bin \# & Low edge [GeV/$\textit{c}$] & High edge [GeV/$\textit{c}$] & Cross Section [$10^{-38}\frac{cm^{2}}{(GeV/\textit{c})\,^{40}Ar}$] & Uncertainty [$10^{-38}\frac{cm^{2}}{(GeV/\textit{c})\,^{40}Ar}$] \\
\hline
\hline
1 & 0 & 0.05 & 22.277226 & 3.5211665\\
2 & 0.05 & 0.1 & 46.301076 & 4.9714461\\
3 & 0.1 & 0.15 & 55.646232 & 5.3160007\\
4 & 0.15 & 0.2 & 48.837166 & 4.5634458\\
5 & 0.2 & 0.25 & 33.188962 & 3.7314142\\
6 & 0.25 & 0.3 & 17.158152 & 2.5264334\\
7 & 0.3 & 0.35 & 13.178919 & 2.1835041\\
8 & 0.35 & 0.4 & 11.992574 & 2.120296\\
9 & 0.4 & 0.47 & 9.3306926 & 1.7638469\\
10 & 0.47 & 0.55 & 6.2473566 & 1.4376167\\
11 & 0.55 & 0.65 & 3.6497802 & 0.93500365\\
12 & 0.65 & 0.75 & 2.4738763 & 0.72885201\\
13 & 0.75 & 0.9 & 1.9513767 & 0.43560868\\
\hline
\end{tabular}
\end{adjustbox}
\end{table}

\begin{table}[H]
\raggedright
\begin{adjustbox}{width=\textwidth}
\small
\begin{tabular}{ |c|c|c|c|c| }
\hline
\multicolumn{5}{|c|}{Cross Section $\delta p_{T}$, $0.75\,<\,cos\theta_{p}\,<\,1$} \\
\hline
\hline
Bin \# & Low edge [GeV/$\textit{c}$] & High edge [GeV/$\textit{c}$] & Cross Section [$10^{-38}\frac{cm^{2}}{(GeV/\textit{c})\,^{40}Ar}$] & Uncertainty [$10^{-38}\frac{cm^{2}}{(GeV/\textit{c})\,^{40}Ar}$] \\
\hline
\hline
1 & 0 & 0.05 & 26.556627 & 3.2406928\\
2 & 0.05 & 0.1 & 53.997273 & 5.1475536\\
3 & 0.1 & 0.15 & 61.374239 & 6.0588105\\
4 & 0.15 & 0.2 & 53.806285 & 5.4451484\\
5 & 0.2 & 0.25 & 39.182829 & 4.0469823\\
6 & 0.25 & 0.3 & 23.752503 & 3.1179207\\
7 & 0.3 & 0.35 & 17.786513 & 2.7239434\\
8 & 0.35 & 0.4 & 15.15475 & 2.369565\\
9 & 0.4 & 0.47 & 11.097626 & 1.8273797\\
10 & 0.47 & 0.55 & 8.0519323 & 1.502272\\
11 & 0.55 & 0.65 & 4.465837 & 1.1345191\\
12 & 0.65 & 0.9 & 1.341167 & 0.47864332\\
\hline
\end{tabular}
\end{adjustbox}
\end{table}

\begin{table}[H]
\raggedright
\begin{adjustbox}{width=\textwidth}
\small
\begin{tabular}{ |c|c|c|c|c| }
\hline
\multicolumn{5}{|c|}{Cross Section $\delta\alpha_{T}$, $All\,events$} \\
\hline
\hline
Bin \# & Low edge [deg] & High edge [deg] & Cross Section [$10^{-38}\frac{cm^{2}}{deg\,^{40}Ar}$] & Uncertainty [$10^{-38}\frac{cm^{2}}{deg\,^{40}Ar}$] \\
\hline
\hline
1 & 0 & 22 & 0.047295775 & 0.0069867236\\
2 & 22 & 44 & 0.044962975 & 0.0058600051\\
3 & 44 & 66 & 0.044549171 & 0.0056673462\\
4 & 66 & 88 & 0.050441444 & 0.0070400151\\
5 & 88 & 110 & 0.066811717 & 0.008404137\\
6 & 110 & 145 & 0.078273462 & 0.0087227652\\
7 & 145 & 180 & 0.090827721 & 0.01011356\\
\hline
\end{tabular}
\end{adjustbox}
\end{table}

\begin{table}[H]
\raggedright
\begin{adjustbox}{width=\textwidth}
\small
\begin{tabular}{ |c|c|c|c|c| }
\hline
\multicolumn{5}{|c|}{Cross Section $\delta\alpha_{T}$, $\delta p_{T}\,<\,0.2\,GeV/\textit{c}$} \\
\hline
\hline
Bin \# & Low edge [deg] & High edge [deg] & Cross Section [$10^{-38}\frac{cm^{2}}{deg\,(GeV/\textit{c})\,^{40}Ar}$] & Uncertainty [$10^{-38}\frac{cm^{2}}{deg\,(GeV/\textit{c})\,^{40}Ar}$] \\
\hline
\hline
1 & 0 & 22 & 0.12695878 & 0.022679165\\
2 & 22 & 44 & 0.14524611 & 0.018476406\\
3 & 44 & 66 & 0.14734652 & 0.017919708\\
4 & 66 & 88 & 0.14897757 & 0.019169822\\
5 & 88 & 110 & 0.1866162 & 0.020386908\\
6 & 110 & 145 & 0.17277607 & 0.018093391\\
7 & 145 & 180 & 0.13438066 & 0.02078805\\
\hline
\end{tabular}
\end{adjustbox}
\end{table}

\begin{table}[H]
\raggedright
\begin{adjustbox}{width=\textwidth}
\small
\begin{tabular}{ |c|c|c|c|c| }
\hline
\multicolumn{5}{|c|}{Cross Section $\delta\alpha_{T}$, $0.2\,<\,\delta p_{T}\,<\,0.4\,GeV/\textit{c}$} \\
\hline
\hline
Bin \# & Low edge [deg] & High edge [deg] & Cross Section [$10^{-38}\frac{cm^{2}}{deg\,(GeV/\textit{c})\,^{40}Ar}$] & Uncertainty [$10^{-38}\frac{cm^{2}}{deg\,(GeV/\textit{c})\,^{40}Ar}$] \\
\hline
\hline
1 & 0 & 22 & 0.056799224 & 0.012487662\\
2 & 22 & 44 & 0.04678748 & 0.011520484\\
3 & 44 & 66 & 0.052397281 & 0.011723828\\
4 & 66 & 88 & 0.064796542 & 0.012797715\\
5 & 88 & 110 & 0.091862658 & 0.015913016\\
6 & 110 & 145 & 0.13622086 & 0.017895931\\
7 & 145 & 180 & 0.11216705 & 0.017442804\\
\hline
\end{tabular}
\end{adjustbox}
\end{table}

\begin{table}[H]
\raggedright
\begin{adjustbox}{width=\textwidth}
\small
\begin{tabular}{ |c|c|c|c|c| }
\hline
\multicolumn{5}{|c|}{Cross Section $\delta\alpha_{T}$, $\delta p_{T}\,>\,0.4\,GeV/\textit{c}$} \\
\hline
\hline
Bin \# & Low edge [deg] & High edge [deg] & Cross Section [$10^{-38}\frac{cm^{2}}{deg\,(GeV/\textit{c})\,^{40}Ar}$] & Uncertainty [$10^{-38}\frac{cm^{2}}{deg\,(GeV/\textit{c})\,^{40}Ar}$] \\
\hline
\hline
1 & 0 & 22 & 0.00227052 & 0.001440055\\
2 & 22 & 44 & 0.00095479865 & 0.0016242054\\
3 & 44 & 66 & 0.0026413791 & 0.0021136415\\
4 & 66 & 88 & 0.0068414187 & 0.0027824891\\
5 & 88 & 110 & 0.0068027421 & 0.0037867296\\
6 & 110 & 145 & 0.022075943 & 0.0052628707\\
7 & 145 & 180 & 0.045358218 & 0.0087483334\\
\hline
\end{tabular}
\end{adjustbox}
\end{table}

\begin{table}[H]
\raggedright
\begin{adjustbox}{width=\textwidth}
\small
\begin{tabular}{ |c|c|c|c|c| }
\hline
\multicolumn{5}{|c|}{Cross Section $\delta\alpha_{T}$, $-1\,<\,cos\theta_{\mu}\,<\,0$} \\
\hline
\hline
Bin \# & Low edge [deg] & High edge [deg] & Cross Section [$10^{-38}\frac{cm^{2}}{deg\,^{40}Ar}$] & Uncertainty [$10^{-38}\frac{cm^{2}}{deg\,^{40}Ar}$] \\
\hline
\hline
1 & 0 & 22 & 0.0070094933 & 0.0015308661\\
2 & 22 & 44 & 0.0087167783 & 0.0013189905\\
3 & 44 & 66 & 0.00926925 & 0.0014036153\\
4 & 66 & 88 & 0.008550577 & 0.001298917\\
5 & 88 & 110 & 0.010329485 & 0.0014025224\\
6 & 110 & 145 & 0.012608986 & 0.0013252542\\
7 & 145 & 180 & 0.013565859 & 0.0015269307\\
\hline
\end{tabular}
\end{adjustbox}
\end{table}

\begin{table}[H]
\raggedright
\begin{adjustbox}{width=\textwidth}
\small
\begin{tabular}{ |c|c|c|c|c| }
\hline
\multicolumn{5}{|c|}{Cross Section $\delta\alpha_{T}$, $0\,<\,cos\theta_{\mu}\,<\,0.5$} \\
\hline
\hline
Bin \# & Low edge [deg] & High edge [deg] & Cross Section [$10^{-38}\frac{cm^{2}}{deg\,^{40}Ar}$] & Uncertainty [$10^{-38}\frac{cm^{2}}{deg\,^{40}Ar}$] \\
\hline
\hline
1 & 0 & 22 & 0.018647988 & 0.0028682921\\
2 & 22 & 44 & 0.016518253 & 0.0024557512\\
3 & 44 & 66 & 0.017919296 & 0.0032168139\\
4 & 66 & 88 & 0.020441798 & 0.0027424304\\
5 & 88 & 110 & 0.026109824 & 0.0036028398\\
6 & 110 & 145 & 0.035282063 & 0.0036682863\\
7 & 145 & 180 & 0.040071864 & 0.0044843425\\
\hline
\end{tabular}
\end{adjustbox}
\end{table}

\begin{table}[H]
\raggedright
\begin{adjustbox}{width=\textwidth}
\small
\begin{tabular}{ |c|c|c|c|c| }
\hline
\multicolumn{5}{|c|}{Cross Section $\delta\alpha_{T}$, $0.5\,<\,cos\theta_{\mu}\,<\,0.75$} \\
\hline
\hline
Bin \# & Low edge [deg] & High edge [deg] & Cross Section [$10^{-38}\frac{cm^{2}}{deg\,^{40}Ar}$] & Uncertainty [$10^{-38}\frac{cm^{2}}{deg\,^{40}Ar}$] \\
\hline
\hline
1 & 0 & 22 & 0.039046898 & 0.005749103\\
2 & 22 & 44 & 0.033052251 & 0.0056115837\\
3 & 44 & 66 & 0.036931702 & 0.0070911942\\
4 & 66 & 88 & 0.043853227 & 0.0078299826\\
5 & 88 & 110 & 0.063848291 & 0.0092902497\\
6 & 110 & 145 & 0.078181153 & 0.0098760067\\
7 & 145 & 180 & 0.081386797 & 0.010939914\\
\hline
\end{tabular}
\end{adjustbox}
\end{table}

\begin{table}[H]
\raggedright
\begin{adjustbox}{width=\textwidth}
\small
\begin{tabular}{ |c|c|c|c|c| }
\hline
\multicolumn{5}{|c|}{Cross Section $\delta\alpha_{T}$, $0.75\,<\,cos\theta_{\mu}\,<\,1$} \\
\hline
\hline
Bin \# & Low edge [deg] & High edge [deg] & Cross Section [$10^{-38}\frac{cm^{2}}{deg\,^{40}Ar}$] & Uncertainty [$10^{-38}\frac{cm^{2}}{deg\,^{40}Ar}$] \\
\hline
\hline
1 & 0 & 22 & 0.075867825 & 0.011265493\\
2 & 22 & 44 & 0.066287719 & 0.0098232164\\
3 & 44 & 66 & 0.064898625 & 0.0095195444\\
4 & 66 & 88 & 0.075013592 & 0.011945994\\
5 & 88 & 110 & 0.086169464 & 0.014835516\\
6 & 110 & 145 & 0.084650691 & 0.016739163\\
7 & 145 & 180 & 0.094544205 & 0.019711329\\
\hline
\end{tabular}
\end{adjustbox}
\end{table}

\begin{table}[H]
\raggedright
\begin{adjustbox}{width=\textwidth}
\small
\begin{tabular}{ |c|c|c|c|c| }
\hline
\multicolumn{5}{|c|}{Cross Section $\delta\alpha_{T}$, $-1\,<\,cos\theta_{p}\,<\,0$} \\
\hline
\hline
Bin \# & Low edge [deg] & High edge [deg] & Cross Section [$10^{-38}\frac{cm^{2}}{deg\,^{40}Ar}$] & Uncertainty [$10^{-38} \frac{cm^{2}}{deg\,^{40}Ar}$] \\
\hline
\hline
1 & 0 & 22 & 0.0033462261 & 0.0014791079\\
2 & 22 & 44 & 0.0032866125 & 0.0011874245\\
3 & 44 & 66 & 0.0034815573 & 0.0012390043\\
4 & 66 & 88 & 0.0039540905 & 0.0012217276\\
5 & 88 & 110 & 0.0059084898 & 0.0012229052\\
6 & 110 & 145 & 0.0081331169 & 0.0014535821\\
7 & 145 & 180 & 0.0087578079 & 0.0013635834\\
\hline
\end{tabular}
\end{adjustbox}
\end{table}

\begin{table}[H]
\raggedright
\begin{adjustbox}{width=\textwidth}
\small
\begin{tabular}{ |c|c|c|c|c| }
\hline
\multicolumn{5}{|c|}{Cross Section $\delta\alpha_{T}$, $0\,<\,cos\theta_{p}\,<\,0.5$} \\
\hline
\hline
Bin \# & Low edge [deg] & High edge [deg] & Cross Section [$10^{-38}\frac{cm^{2}}{deg\,^{40}Ar}$] & Uncertainty [$10^{-38} \frac{cm^{2}}{deg\,^{40}Ar}$] \\
\hline
\hline
1 & 0 & 22 & 0.02610793 & 0.0040358753\\
2 & 22 & 44 & 0.024509617 & 0.00373604\\
3 & 44 & 66 & 0.024198822 & 0.0038533857\\
4 & 66 & 88 & 0.02706767 & 0.004270476\\
5 & 88 & 110 & 0.034761104 & 0.0048532738\\
6 & 110 & 145 & 0.031658748 & 0.0042707914\\
7 & 145 & 180 & 0.021958891 & 0.0030897305\\
\hline
\end{tabular}
\end{adjustbox}
\end{table}

\begin{table}[H]
\raggedright
\begin{adjustbox}{width=\textwidth}
\small
\begin{tabular}{ |c|c|c|c|c| }
\hline
\multicolumn{5}{|c|}{Cross Section $\delta\alpha_{T}$, $0.5\,<\,cos\theta_{p}\,<\,0.75$} \\
\hline
\hline
Bin \# & Low edge [deg] & High edge [deg] & Cross Section [$10^{-38}\frac{cm^{2}}{deg\,^{40}Ar}$] & Uncertainty [$10^{-38} \frac{cm^{2}}{deg\,^{40}Ar}$] \\
\hline
\hline
1 & 0 & 22 & 0.0622161 & 0.0089635735\\
2 & 22 & 44 & 0.059950002 & 0.0078953421\\
3 & 44 & 66 & 0.059187243 & 0.0085398018\\
4 & 66 & 88 & 0.065220419 & 0.0093878885\\
5 & 88 & 110 & 0.08283717 & 0.010863812\\
6 & 110 & 145 & 0.078870599 & 0.0101726\\
7 & 145 & 180 & 0.065774373 & 0.0086854381\\
\hline
\end{tabular}
\end{adjustbox}
\end{table}

\begin{table}[H]
\raggedright
\begin{adjustbox}{width=\textwidth}
\small
\begin{tabular}{ |c|c|c|c|c| }
\hline
\multicolumn{5}{|c|}{Cross Section $\delta\alpha_{T}$, $0.75\,<\,cos\theta_{p}\,<\,1$} \\
\hline
\hline
Bin \# & Low edge [deg] & High edge [deg] & Cross Section [$10^{-38}\frac{cm^{2}}{deg\,^{40}Ar}$] & Uncertainty [$10^{-38} \frac{cm^{2}}{deg\,^{40}Ar}$] \\
\hline
\hline
1 & 0 & 22 & 0.045922492 & 0.0069121365\\
2 & 22 & 44 & 0.037776029 & 0.0062443018\\
3 & 44 & 66 & 0.037782878 & 0.0074232644\\
4 & 66 & 88 & 0.047116116 & 0.0079431241\\
5 & 88 & 110 & 0.077736146 & 0.0098339076\\
6 & 110 & 145 & 0.1189007 & 0.01172061\\
7 & 145 & 180 & 0.18747798 & 0.019376388\\
\hline
\end{tabular}
\end{adjustbox}
\end{table}

\begin{table}[H]
\raggedright
\begin{adjustbox}{width=\textwidth}
\small
\begin{tabular}{ |c|c|c|c|c| }
\hline
\multicolumn{5}{|c|}{Cross Section $\delta\phi_{T}$, $All\,events$} \\
\hline
\hline
Bin \# & Low edge [deg] & High edge [deg] & Cross Section [$10^{-38}\frac{cm^{2}}{deg\,^{40}Ar}$] & Uncertainty [$10^{-38}\frac{cm^{2}}{deg\,^{40}Ar}$] \\
\hline
\hline
1 & 0 & 12.5 & 0.31441812 & 0.032282057\\
2 & 12.5 & 25 & 0.20358925 & 0.018703326\\
3 & 25 & 37.5 & 0.10730283 & 0.01081705\\
4 & 37.5 & 50 & 0.062615831 & 0.0076582523\\
5 & 50 & 60 & 0.0497139 & 0.0079765306\\
6 & 60 & 75 & 0.030944447 & 0.0054822397\\
7 & 75 & 90 & 0.024592357 & 0.0046787268\\
8 & 90 & 106 & 0.020754713 & 0.0043201244\\
9 & 106 & 126 & 0.01667195 & 0.0034824536\\
10 & 126 & 145 & 0.013700149 & 0.0032522416\\
11 & 145 & 162 & 0.011648388 & 0.0034856785\\
12 & 162 & 180 & 0.010095183 & 0.0036015185\\
\hline
\end{tabular}
\end{adjustbox}
\end{table}

\begin{table}[H]
\raggedright
\begin{adjustbox}{width=\textwidth}
\small
\begin{tabular}{ |c|c|c|c|c| }
\hline
\multicolumn{5}{|c|}{Cross Section $\delta\phi_{T}$, $\delta p_{T}\,<\,0.2\,GeV/\textit{c}$} \\
\hline
\hline
Bin \# & Low edge [deg] & High edge [deg] & Cross Section [$10^{-38}\frac{cm^{2}}{deg\,(GeV/\textit{c})\,^{40}Ar}$] & Uncertainty [$10^{-38}\frac{cm^{2}}{deg\,(GeV/\textit{c})\,^{40}Ar}$] \\
\hline
\hline
1 & 0 & 12.5 & 1.3649882 & 0.12155958\\
2 & 12.5 & 25 & 0.73128946 & 0.061854637\\
3 & 25 & 37.5 & 0.21436706 & 0.029316416\\
4 & 37.5 & 180 & 0.0081390931 & 0.0018999935\\
\hline
\end{tabular}
\end{adjustbox}
\end{table}

\begin{table}[H]
\raggedright
\begin{adjustbox}{width=\textwidth}
\small
\begin{tabular}{ |c|c|c|c|c| }
\hline
\multicolumn{5}{|c|}{Cross Section $\delta\phi_{T}$, $0.2\,<\,\delta p_{T}\,<\,0.4\,GeV/\textit{c}$} \\
\hline
\hline
Bin \# & Low edge [deg] & High edge [deg] & Cross Section [$10^{-38}\frac{cm^{2}}{deg\,(GeV/\textit{c})\,^{40}Ar}$] & Uncertainty [$10^{-38}\frac{cm^{2}}{deg\,(GeV/\textit{c})\,^{40}Ar}$] \\
\hline
\hline
1 & 0 & 20 & 0.15290364 & 0.026143234\\
2 & 20 & 40 & 0.25120654 & 0.033752516\\
3 & 40 & 60 & 0.15996627 & 0.020905241\\
4 & 60 & 80 & 0.059050604 & 0.012096139\\
5 & 80 & 100 & 0.019140583 & 0.008853898\\
6 & 100 & 120 & 0.010670539 & 0.0075945702\\
7 & 120 & 140 & 0.0106064 & 0.007133161\\
8 & 140 & 160 & 0.0077663445 & 0.0058119678\\
9 & 160 & 180 & 0.004481286 & 0.005133954\\
\hline
\end{tabular}
\end{adjustbox}
\end{table}

\begin{table}[H]
\raggedright
\begin{adjustbox}{width=\textwidth}
\small
\begin{tabular}{ |c|c|c|c|c| }
\hline
\multicolumn{5}{|c|}{Cross Section $\delta\phi_{T}$, $\delta p_{T}\,>\,0.4\,GeV/\textit{c}$} \\
\hline
\hline
Bin \# & Low edge [deg] & High edge [deg] & Cross Section [$10^{-38}\frac{cm^{2}}{deg\,(GeV/\textit{c})\,^{40}Ar}$] & Uncertainty [$10^{-38}\frac{cm^{2}}{deg\,(GeV/\textit{c})\,^{40}Ar}$] \\
\hline
\hline
1 & 0 & 25 & 0.0019384644 & 0.0025388511\\
2 & 25 & 50 & 0.0099436444 & 0.0044065902\\
3 & 50 & 75 & 0.0239602 & 0.0063420662\\
4 & 75 & 105 & 0.023477874 & 0.0051465383\\
5 & 105 & 145 & 0.016670113 & 0.003744251\\
6 & 145 & 180 & 0.015435737 & 0.0041462476\\
\hline
\end{tabular}
\end{adjustbox}
\end{table}

\begin{table}[H]
\raggedright
\begin{adjustbox}{width=\textwidth}
\small
\begin{tabular}{ |c|c|c|c|c| }
\hline
\multicolumn{5}{|c|}{Cross Section $\delta p_{T,x}$, $All\,events$} \\
\hline
\hline
Bin \# & Low edge [GeV/$\textit{c}$] & High edge [GeV/$\textit{c}$] & Cross Section [$10^{-38}\frac{cm^{2}}{(GeV/\textit{c})\,^{40}Ar}$] & Uncertainty [$10^{-38}\frac{cm^{2}}{(GeV/\textit{c})\,^{40}Ar}$] \\
\hline
\hline
1 & -0.55 & -0.45 & 1.3017107 & 0.46826726\\
2 & -0.45 & -0.35 & 2.563368 & 0.59039691\\
3 & -0.35 & -0.25 & 5.5731445 & 0.89649636\\
4 & -0.25 & -0.15 & 11.707776 & 1.3957739\\
5 & -0.15 & -0.05 & 20.511688 & 2.3000876\\
6 & -0.05 & 0.05 & 26.020455 & 3.0236695\\
7 & 0.05 & 0.15 & 19.708245 & 2.2296454\\
8 & 0.15 & 0.25 & 11.103336 & 1.4077286\\
9 & 0.25 & 0.35 & 5.5393912 & 0.9180937\\
10 & 0.35 & 0.45 & 2.5182752 & 0.62433322\\
11 & 0.45 & 0.55 & 1.2015199 & 0.45891559\\
\hline
\end{tabular}
\end{adjustbox}
\end{table}

\begin{table}[H]
\raggedright
\begin{adjustbox}{width=\textwidth}
\small
\begin{tabular}{ |c|c|c|c|c| }
\hline
\multicolumn{5}{|c|}{Cross Section $\delta p_{T,x}$, $\delta p_{T,y}\,<\,-0.15\,GeV/\textit{c}$} \\
\hline
\hline
Bin \# & Low edge [GeV/$\textit{c}$] & High edge [GeV/$\textit{c}$] & Cross Section [$10^{-38}\frac{cm^{2}}{(GeV/\textit{c})^{2}\,^{40}Ar}$] & Uncertainty [$10^{-38}\frac{cm^{2}}{(GeV/\textit{c})^{2}\,^{40}Ar}$] \\
\hline
\hline
1 & -0.55 & -0.45 & 1.6208121 & 0.40811656\\
2 & -0.45 & -0.35 & 2.8270055 & 0.53216682\\
3 & -0.35 & -0.25 & 4.9291254 & 0.82899734\\
4 & -0.25 & -0.15 & 6.1606674 & 0.9663476\\
5 & -0.15 & -0.05 & 7.917514 & 1.0822931\\
6 & -0.05 & 0.05 & 8.7498076 & 1.1891885\\
7 & 0.05 & 0.15 & 8.5082736 & 1.2186225\\
8 & 0.15 & 0.25 & 7.2098121 & 1.0521901\\
9 & 0.25 & 0.35 & 5.6680756 & 0.864424\\
10 & 0.35 & 0.45 & 2.9003432 & 0.56377234\\
11 & 0.45 & 0.55 & 1.2978947 & 0.41077731\\
\hline
\end{tabular}
\end{adjustbox}
\end{table}

\begin{table}[H]
\raggedright
\begin{adjustbox}{width=\textwidth}
\small
\begin{tabular}{ |c|c|c|c|c| }
\hline
\multicolumn{5}{|c|}{Cross Section $\delta p_{T,x}$, $|\delta p_{T,y}|\,<\,0.15\,GeV/\textit{c}$} \\
\hline
\hline
Bin \# & Low edge [GeV/$\textit{c}$] & High edge [GeV/$\textit{c}$] & Cross Section [$10^{-38}\frac{cm^{2}}{(GeV/\textit{c})^{2}\,^{40}Ar}$] & Uncertainty [$10^{-38}\frac{cm^{2}}{(GeV/\textit{c})^{2}\,^{40}Ar}$] \\
\hline
\hline
1 & -0.55 & -0.45 & 1.2119704 & 0.61176744\\
2 & -0.45 & -0.35 & 2.6763573 & 0.82181016\\
3 & -0.35 & -0.25 & 7.3739087 & 1.055886\\
4 & -0.25 & -0.15 & 22.201232 & 2.2046742\\
5 & -0.15 & -0.05 & 46.579385 & 4.1454453\\
6 & -0.05 & 0.05 & 64.1127 & 5.923266\\
7 & 0.05 & 0.15 & 46.971297 & 4.3486902\\
8 & 0.15 & 0.25 & 21.927763 & 2.3914571\\
9 & 0.25 & 0.35 & 7.1648705 & 1.1876442\\
10 & 0.35 & 0.45 & 2.4895325 & 0.95614843\\
11 & 0.45 & 0.55 & 0.65303698 & 0.72564438\\
\hline
\end{tabular}
\end{adjustbox}
\end{table}

\begin{table}[H]
\raggedright
\begin{adjustbox}{width=\textwidth}
\small
\begin{tabular}{ |c|c|c|c|c| }
\hline
\multicolumn{5}{|c|}{Cross Section $\delta p_{T,x}$, $\delta p_{T,y}\,>\,0.15\,GeV/\textit{c}$} \\
\hline
\hline
Bin \# & Low edge [GeV/$\textit{c}$] & High edge [GeV/$\textit{c}$] & Cross Section [$10^{-38}\frac{cm^{2}}{(GeV/\textit{c})^{2}\,^{40}Ar}$] & Uncertainty [$10^{-38}\frac{cm^{2}}{(GeV/\textit{c})^{2}\,^{40}Ar}$] \\
\hline
\hline
1 & -0.55 & -0.35 & 0.23442021 & 0.22673131\\
2 & -0.35 & -0.25 & 1.0074053 & 0.55681324\\
3 & -0.25 & -0.15 & 2.7331277 & 0.88401443\\
4 & -0.15 & -0.05 & 5.7865516 & 1.3851403\\
5 & -0.05 & 0.05 & 7.4226631 & 1.5326132\\
6 & 0.05 & 0.15 & 5.8817533 & 1.2041397\\
7 & 0.15 & 0.25 & 3.1678851 & 0.75891207\\
8 & 0.25 & 0.35 & 1.4156176 & 0.51670928\\
9 & 0.35 & 0.55 & 0.46759544 & 0.27306014\\
\hline
\end{tabular}
\end{adjustbox}
\end{table}

\begin{table}[H]
\raggedright
\begin{adjustbox}{width=\textwidth}
\small
\begin{tabular}{ |c|c|c|c|c| }
\hline
\multicolumn{5}{|c|}{Cross Section $E^{Cal}$, $All\,events$} \\
\hline
\hline
Bin \# & Low edge [GeV] & High edge [GeV] & Cross Section [$10^{-38}\frac{cm^{2}}{GeV\,^{40}Ar}$] & Uncertainty [$10^{-38}\frac{cm^{2}}{GeV\,^{40}Ar}$] \\
\hline
\hline
1 & 0.2 & 0.35 & 2.4285916 & 0.79465249\\
2 & 0.35 & 0.5 & 8.5152669 & 1.2297922\\
3 & 0.5 & 0.65 & 13.117739 & 1.3413985\\
4 & 0.65 & 0.8 & 14.625952 & 1.4422233\\
5 & 0.8 & 0.95 & 13.511368 & 1.3447089\\
6 & 0.95 & 1.1 & 11.082323 & 1.2200843\\
7 & 1.1 & 1.25 & 7.7674502 & 1.1387079\\
8 & 1.25 & 1.4 & 4.8497867 & 1.0880583\\
9 & 1.4 & 1.6 & 1.8231661 & 0.55097001\\
\hline
\end{tabular}
\end{adjustbox}
\end{table}

\begin{table}[H]
\raggedright
\begin{adjustbox}{width=\textwidth}
\small
\begin{tabular}{ |c|c|c|c|c| }
\hline
\multicolumn{5}{|c|}{Cross Section $E^{Cal}$, $\delta p_{T}\,<\,0.2\,GeV/\textit{c}$} \\
\hline
\hline
Bin \# & Low edge [GeV] & High edge [GeV] & Cross Section [$10^{-38}\frac{cm^{2}}{(GeV^{2}/\textit{c})\,^{40}Ar}$] & Uncertainty [$10^{-38}\frac{cm^{2}}{(GeV^{2}/\textit{c})\,^{40}Ar}$] \\
\hline
\hline
1 & 0.2 & 0.34 & 7.4813194 & 1.6636418\\
2 & 0.34 & 0.48 & 23.725673 & 2.6926684\\
3 & 0.48 & 0.62 & 37.376454 & 2.9958466\\
4 & 0.62 & 0.76 & 42.373237 & 3.0989483\\
5 & 0.76 & 0.9 & 39.164318 & 3.122892\\
6 & 0.9 & 1.04 & 35.763087 & 3.2386372\\
7 & 1.04 & 1.18 & 30.738315 & 3.3618115\\
8 & 1.18 & 1.32 & 23.328121 & 3.1755891\\
9 & 1.32 & 1.6 & 8.3160489 & 1.3339053\\
\hline
\end{tabular}
\end{adjustbox}
\end{table}

\begin{table}[H]
\raggedright
\begin{adjustbox}{width=\textwidth}
\small
\begin{tabular}{ |c|c|c|c|c| }
\hline
\multicolumn{5}{|c|}{Cross Section $E^{Cal}$, $0.2\,<\,\delta p_{T}\,<\,0.4\,GeV/\textit{c}$} \\
\hline
\hline
Bin \# & Low edge [GeV] & High edge [GeV] & Cross Section [$10^{-38}\frac{cm^{2}}{(GeV^{2}/\textit{c})\,^{40}Ar}$] & Uncertainty [$10^{-38}\frac{cm^{2}}{(GeV^{2}/\textit{c})\,^{40}Ar}$] \\
\hline
\hline
1 & 0.2 & 0.34 & 8.5169003 & 1.3772991\\
2 & 0.34 & 0.48 & 17.371678 & 2.2343809\\
3 & 0.48 & 0.62 & 21.660624 & 2.2275417\\
4 & 0.62 & 0.76 & 23.069011 & 2.1799637\\
5 & 0.76 & 0.9 & 20.631725 & 2.3719543\\
6 & 0.9 & 1.04 & 15.887042 & 2.6183144\\
7 & 1.04 & 1.18 & 12.993749 & 2.6005735\\
8 & 1.18 & 1.32 & 9.975705 & 2.5421886\\
9 & 1.32 & 1.6 & 3.7075146 & 0.97376147\\
\hline
\end{tabular}
\end{adjustbox}
\end{table}

\begin{table}[H]
\raggedright
\begin{adjustbox}{width=\textwidth}
\small
\begin{tabular}{ |c|c|c|c|c| }
\hline
\multicolumn{5}{|c|}{Cross Section $E^{Cal}$, $\delta p_{T}\,>\,0.4\,GeV/\textit{c}$} \\
\hline
\hline
Bin \# & Low edge [GeV] & High edge [GeV] & Cross Section [$10^{-38}\frac{cm^{2}}{(GeV^{2}/\textit{c})\,^{40}Ar}$] & Uncertainty [$10^{-38}\frac{cm^{2}}{(GeV^{2}/\textit{c})\,^{40}Ar}$] \\
\hline
\hline
1 & 0.2 & 0.34 & 1.1820863 & 0.29856985\\
2 & 0.34 & 0.48 & 3.3702565 & 0.63180127\\
3 & 0.48 & 0.62 & 4.965046 & 0.68254502\\
4 & 0.62 & 0.76 & 5.3480289 & 0.72097161\\
5 & 0.76 & 0.9 & 4.6043633 & 0.73425928\\
6 & 0.9 & 1.04 & 3.8723977 & 0.72267063\\
7 & 1.04 & 1.18 & 2.8537449 & 0.64183313\\
8 & 1.18 & 1.32 & 2.1343443 & 0.64328401\\
9 & 1.32 & 1.6 & 0.5981865 & 0.23721122\\
\hline
\end{tabular}
\end{adjustbox}
\end{table}

\begin{table}[H]
\raggedright
\begin{adjustbox}{width=\textwidth}
\small
\begin{tabular}{ |c|c|c|c|c| }
\hline
\multicolumn{5}{|c|}{Cross Section $E^{Cal}$, $\delta\alpha_{T}\,<\,45^{o}$} \\
\hline
\hline
Bin \# & Low edge [GeV] & High edge [GeV] & Cross Section [$10^{-38}\frac{cm^{2}}{deg\,GeV\,^{40}Ar}$] & Uncertainty [$10^{-38}\frac{cm^{2}}{deg\,GeV\,^{40}Ar}$] \\
\hline
\hline
1 & 0.2 & 0.38 & 0.022951027 & 0.0049398235\\
2 & 0.38 & 0.55 & 0.046531518 & 0.0058663533\\
3 & 0.55 & 0.73 & 0.05373006 & 0.0053431075\\
4 & 0.73 & 0.9 & 0.057139142 & 0.0051453715\\
5 & 0.9 & 1.08 & 0.044232105 & 0.0050301203\\
6 & 1.08 & 1.23 & 0.031774025 & 0.0052320211\\
7 & 1.23 & 1.4 & 0.019600595 & 0.0049783453\\
8 & 1.4 & 1.6 & 0.0086225829 & 0.002714388\\
\hline
\end{tabular}
\end{adjustbox}
\end{table}

\begin{table}[H]
\raggedright
\begin{adjustbox}{width=\textwidth}
\small
\begin{tabular}{ |c|c|c|c|c| }
\hline
\multicolumn{5}{|c|}{Cross Section $E^{Cal}$, $45^{o}\,<\,\delta\alpha_{T}\,<\,90^{o}$} \\
\hline
\hline
Bin \# & Low edge [GeV] & High edge [GeV] & Cross Section [$10^{-38}\frac{cm^{2}}{deg\,GeV\,^{40}Ar}$] & Uncertainty [$10^{-38}\frac{cm^{2}}{deg\,GeV\,^{40}Ar}$] \\
\hline
\hline
1 & 0.2 & 0.34 & 0.01522215 & 0.0040685015\\
2 & 0.34 & 0.48 & 0.042879224 & 0.0060837195\\
3 & 0.48 & 0.62 & 0.061430253 & 0.0058403125\\
4 & 0.62 & 0.76 & 0.0678641 & 0.0062137779\\
5 & 0.76 & 0.9 & 0.061826648 & 0.0062462885\\
6 & 0.9 & 1.04 & 0.047889931 & 0.0080206275\\
7 & 1.04 & 1.18 & 0.029522517 & 0.0085618097\\
8 & 1.18 & 1.39 & 0.011720775 & 0.0064957436\\
9 & 1.39 & 1.6 & 0.0050754162 & 0.0035760336\\
\hline
\end{tabular}
\end{adjustbox}
\end{table}

\begin{table}[H]
\raggedright
\begin{adjustbox}{width=\textwidth}
\small
\begin{tabular}{ |c|c|c|c|c| }
\hline
\multicolumn{5}{|c|}{Cross Section $E^{Cal}$, $90^{o}\,<\,\delta\alpha_{T}\,<\,135^{o}$} \\
\hline
\hline
Bin \# & Low edge [GeV] & High edge [GeV] & Cross Section [$10^{-38}\frac{cm^{2}}{deg\,GeV\,^{40}Ar}$] & Uncertainty [$10^{-38}\frac{cm^{2}}{deg\,GeV\,^{40}Ar}$] \\
\hline
\hline
1 & 0.2 & 0.34 & 0.014022466 & 0.0051719568\\
2 & 0.34 & 0.48 & 0.054329958 & 0.0076302638\\
3 & 0.48 & 0.62 & 0.090182864 & 0.0077900063\\
4 & 0.62 & 0.76 & 0.09893367 & 0.0080747332\\
5 & 0.76 & 0.9 & 0.087007496 & 0.0097352415\\
6 & 0.9 & 1.04 & 0.06931574 & 0.011083051\\
7 & 1.04 & 1.18 & 0.056617758 & 0.010272729\\
8 & 1.18 & 1.39 & 0.033696028 & 0.0070340268\\
9 & 1.39 & 1.6 & 0.013876698 & 0.0031843772\\
\hline
\end{tabular}
\end{adjustbox}
\end{table}

\begin{table}[H]
\raggedright
\begin{adjustbox}{width=\textwidth}
\small
\begin{tabular}{ |c|c|c|c|c| }
\hline
\multicolumn{5}{|c|}{Cross Section $E^{Cal}$, $135^{o}\,<\,\delta\alpha_{T}\,<\,180^{o}$} \\
\hline
\hline
Bin \# & Low edge [GeV] & High edge [GeV] & Cross Section [$10^{-38}\frac{cm^{2}}{deg\,GeV\,^{40}Ar}$] & Uncertainty [$10^{-38}\frac{cm^{2}}{deg\,GeV\,^{40}Ar}$] \\
\hline
\hline
1 & 0.2 & 0.38 & 0.02738991 & 0.0047866583\\
2 & 0.38 & 0.55 & 0.084351701 & 0.009517808\\
3 & 0.55 & 0.73 & 0.11229756 & 0.011360333\\
4 & 0.73 & 0.9 & 0.10790313 & 0.010555327\\
5 & 0.9 & 1.08 & 0.080894441 & 0.010453967\\
6 & 1.08 & 1.25 & 0.056198812 & 0.012227031\\
7 & 1.25 & 1.43 & 0.029556203 & 0.010311689\\
8 & 1.43 & 1.6 & 0.011480862 & 0.0052568349\\
\hline
\end{tabular}
\end{adjustbox}
\end{table}

\begin{table}[H]
\raggedright
\begin{adjustbox}{width=\textwidth}
\small
\begin{tabular}{ |c|c|c|c|c| }
\hline
\multicolumn{5}{|c|}{Cross Section $E^{Cal}$, $\delta p_{T,y}\,<\,-0.15\,GeV/\textit{c}$} \\
\hline
\hline
Bin \# & Low edge [GeV] & High edge [GeV] & Cross Section [$10^{-38}\frac{cm^{2}}{(GeV^{2}/\textit{c})\,^{40}Ar}$] & Uncertainty [$10^{-38}\frac{cm^{2}}{(GeV^{2}/\textit{c})\,^{40}Ar}$] \\
\hline
\hline
1 & 0.2 & 0.34 & 1.9731427 & 0.44504811\\
2 & 0.34 & 0.48 & 5.6420022 & 0.8169704\\
3 & 0.48 & 0.62 & 8.1512474 & 0.90794023\\
4 & 0.62 & 0.76 & 8.7602012 & 0.93194497\\
5 & 0.76 & 0.9 & 7.704455 & 0.99275782\\
6 & 0.9 & 1.04 & 6.5634731 & 1.0697429\\
7 & 1.04 & 1.18 & 5.016838 & 1.0469953\\
8 & 1.18 & 1.32 & 3.2663924 & 0.94346708\\
9 & 1.32 & 1.6 & 0.84020999 & 0.27514394\\
\hline
\end{tabular}
\end{adjustbox}
\end{table}

\begin{table}[H]
\raggedright
\begin{adjustbox}{width=\textwidth}
\small
\begin{tabular}{ |c|c|c|c|c| }
\hline
\multicolumn{5}{|c|}{Cross Section $E^{Cal}$, $|\delta p_{T,y}|\,<\,0.15\,GeV/\textit{c}$} \\
\hline
\hline
Bin \# & Low edge [GeV] & High edge [GeV] & Cross Section [$10^{-38}\frac{cm^{2}}{(GeV^{2}/\textit{c})\,^{40}Ar}$] & Uncertainty [$10^{-38}\frac{cm^{2}}{(GeV^{2}/\textit{c})\,^{40}Ar}$] \\
\hline
\hline
1 & 0.2 & 0.34 & 5.7223658 & 1.212393\\
2 & 0.34 & 0.48 & 19.727209 & 1.9508156\\
3 & 0.48 & 0.62 & 27.37896 & 2.205255\\
4 & 0.62 & 0.76 & 32.178095 & 2.3526793\\
5 & 0.76 & 0.9 & 29.562168 & 2.4079725\\
6 & 0.9 & 1.04 & 25.428421 & 2.291771\\
7 & 1.04 & 1.18 & 20.135241 & 2.1749524\\
8 & 1.18 & 1.32 & 14.828573 & 2.1227496\\
9 & 1.32 & 1.6 & 4.9774309 & 0.90031019\\
\hline
\end{tabular}
\end{adjustbox}
\end{table}

\begin{table}[H]
\raggedright
\begin{adjustbox}{width=\textwidth}
\small
\begin{tabular}{ |c|c|c|c|c| }
\hline
\multicolumn{5}{|c|}{Cross Section $E^{Cal}$, $\delta p_{T,y}\,>\,0.15\,GeV/\textit{c}$} \\
\hline
\hline
Bin \# & Low edge [GeV] & High edge [GeV] & Cross Section [$10^{-38}\frac{cm^{2}}{(GeV^{2}/\textit{c})\,^{40}Ar}$] & Uncertainty [$10^{-38}\frac{cm^{2}}{(GeV^{2}/\textit{c})\,^{40}Ar}$] \\
\hline
\hline
1 & 0.2 & 0.34 & 2.380383 & 0.53739776\\
2 & 0.34 & 0.48 & 3.8496338 & 0.75841597\\
3 & 0.48 & 0.62 & 4.8834724 & 0.70558778\\
4 & 0.62 & 0.76 & 5.1299994 & 0.62286466\\
5 & 0.76 & 0.9 & 4.2400462 & 0.62877741\\
6 & 0.9 & 1.04 & 3.2096576 & 0.73787833\\
7 & 1.04 & 1.18 & 1.9033993 & 0.87637707\\
8 & 1.18 & 1.32 & 0.91623979 & 0.95294685\\
9 & 1.32 & 1.6 & 0.36490825 & 0.49299019\\
\hline
\end{tabular}
\end{adjustbox}
\end{table}


\clearpage
\section{Log-Likelihood Ratio Score}\label{LLRDecomp}

\begin{figure}[htb!]
\centering  
\includegraphics[width=0.49\linewidth]{\figures 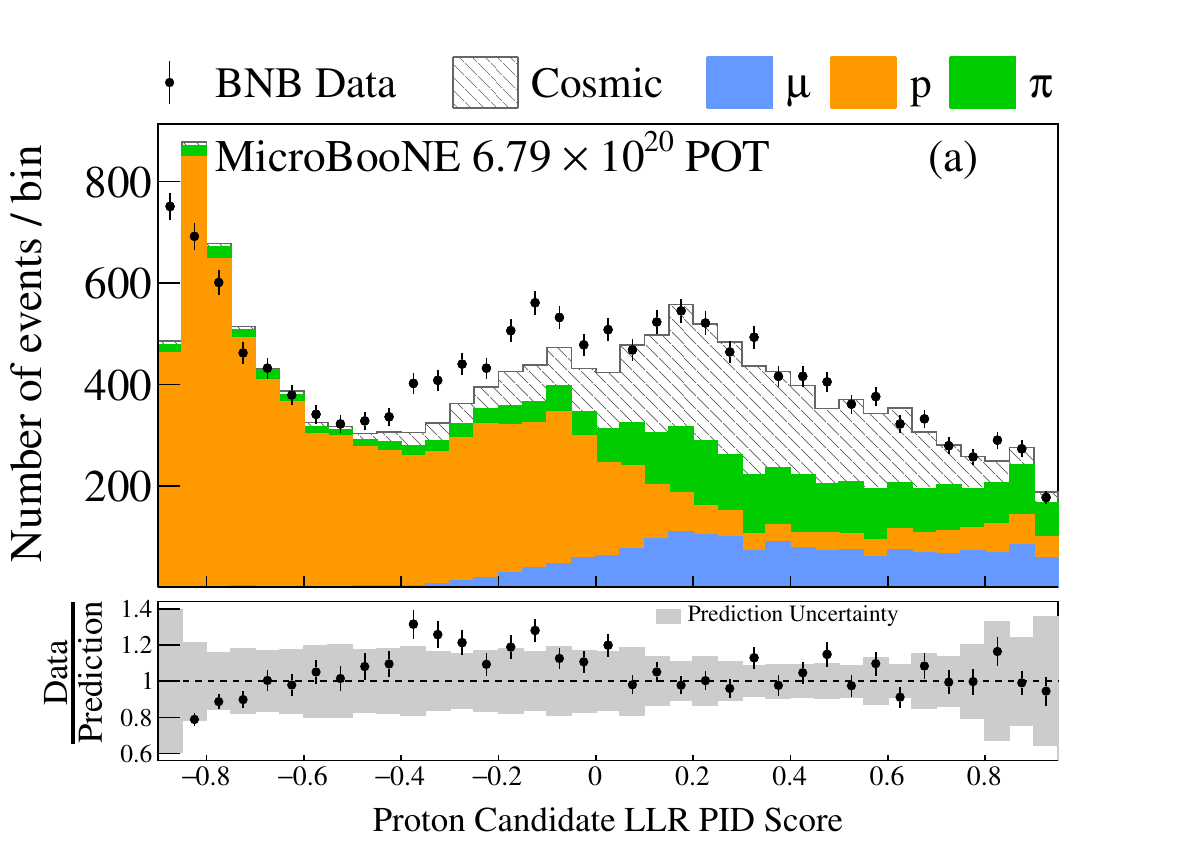}
\includegraphics[width=0.49\linewidth]{\figures 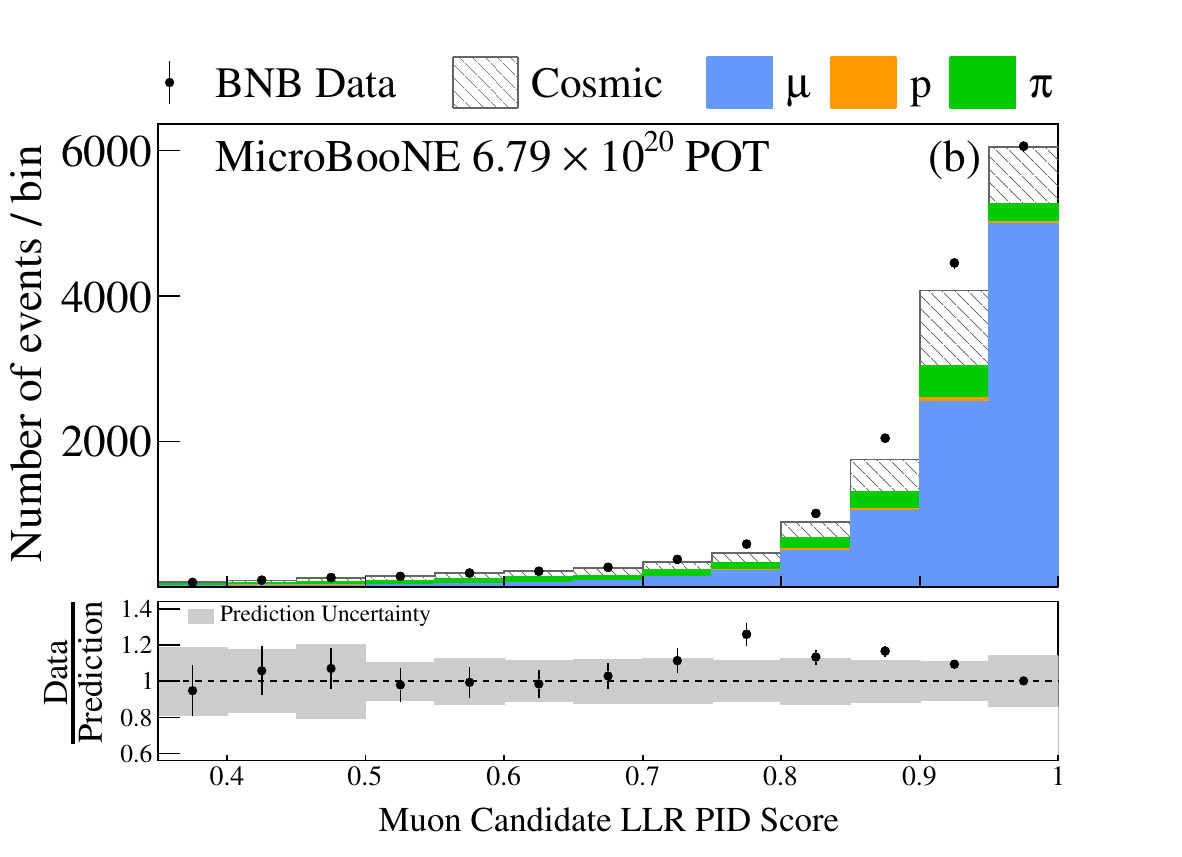}
\caption{
(Left) the proton candidate LLR PID score distribution, illustrating the particle composition of the variable.
(Right) the muon candidate LLR PID score distribution, illustrating a peak close to one.
Only statistical uncertainties are shown on the data.
The bottom panel shows the ratio of data to prediction.
}
\label{partpid}
\end{figure}

Candidate muon-proton pairs are isolated by requiring the existence of precisely two track-like and no shower-like objects, as classified by $\texttt{Pandora}$ using a track-score variable~\cite{VanDePontseele:2020tqz,PhysRevD.105.112004}.
The log-likelihood ratio (LLR) particle identification (PID) score~\cite{NicoloPID} is used to identify the muon and proton candidates.
Figure~\ref{partpid} shows the particle composition breakdown of the sample as a function of the LLR PID score.
Protons tend to have lower LLR PID score values than muons, thus the track with the lowest score is tagged as the candidate proton, as can be seen in Fig.~\ref{partpid}a.
Meanwhile, the track with the higher score is treated as the candidate muon, as shown in Fig.~\ref{partpid}b.

\section{Uncertainty Components}\label{UncComp}

Below we present the contribution of each one of the sources of systematic uncertainties for each of the variables of interest. 	
``Stat'' refers to the statistical uncertainty due to the data, cosmic, MC non-CC1p0$\pi$ background, and out-of-cryostat samples.
``LY'' refers to the systematic uncertainty due to light yield detector variations~\cite{WireMod}. 
``TPC'' refers to the systematic uncertainty due to wire modification TPC detector variations~\cite{WireMod}. 
``SCERecomb2'' refers to the systematic uncertainty due to the space charge (SCE) and recombination detector variations~\cite{WireMod}.
``XSec'' refers to the systematic uncertainty due to the cross-section related uncertainties~\cite{Andreopoulos:2009rq,Andreopoulos:2015wxa,GENIEKnobs}.	
``G4'' refers to the systematic uncertainty due to the $\texttt{GEANT4}$ reinteraction related uncertainties~\cite{Calcutt_2021}.
``Flux'' refers to the systematic uncertainty due to the flux related uncertainties~\cite{Aguilar-Arevalo:2013dva}. 	
``Dirt'' refers to the systematic uncertainty due to the out-of-cryostat strength related uncertainties.	
``MC\_Stat'' refers to the systematic uncertainty due to the MC CC1p0$\pi$ uncertainties obtained with a bootstrapping technique.
``POT'' refers to the proton-on-target uncertainty of our measurement.
``NTarget'' refers to the number-of-targets uncertainty of our measurement.		
``Total'' refers to all the previous uncertainties combined.

\begin{figure*}[htb!]
\centering 
\includegraphics[width=0.49\linewidth]{\figures 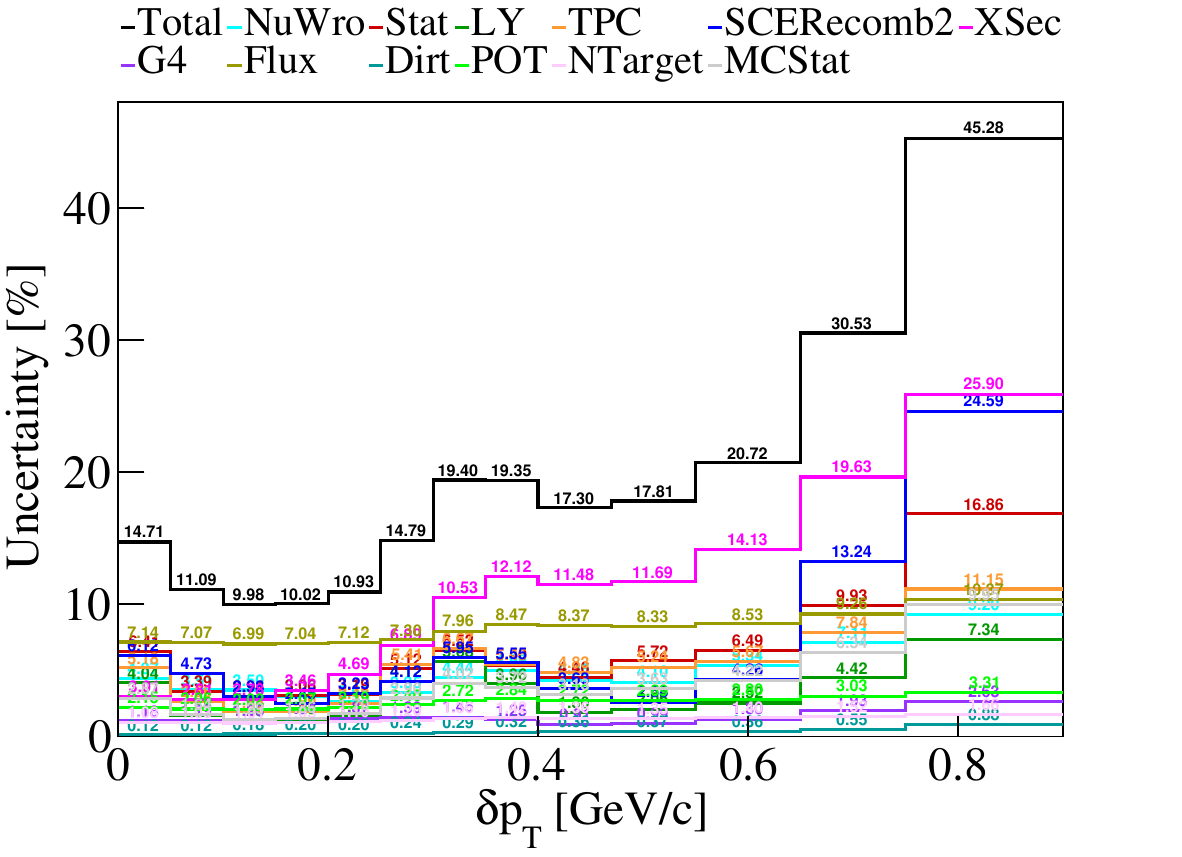}
\caption{
Uncertainty components for $\delta p_{T}$.
}
\label{SuppMatDelatPTUncComponents}
\end{figure*}

\begin{figure*}[htb!]
\centering 
\includegraphics[width=0.49\linewidth]{\figures 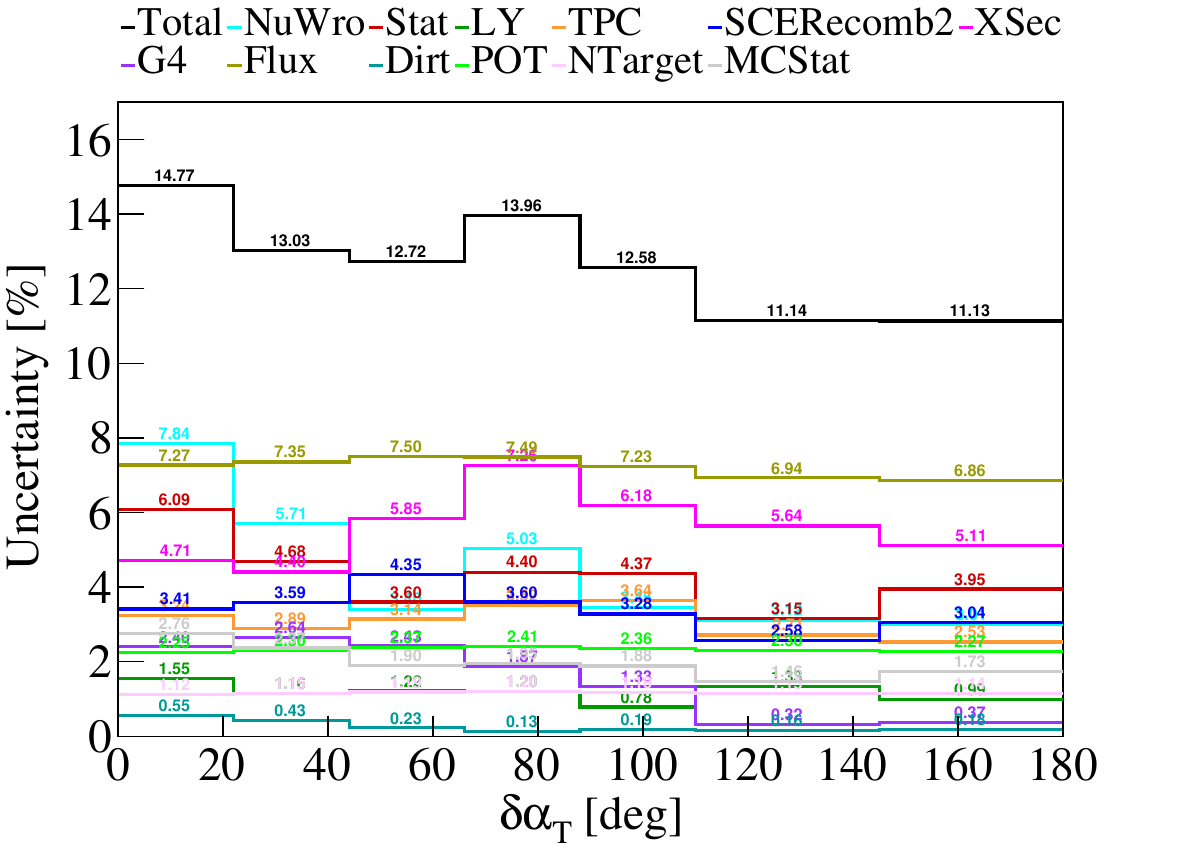}
\caption{
Uncertainty components for $\delta\alpha_{T}$.
}
\label{SuppMatDeltaAlphaTUncComponents}
\end{figure*}

\begin{figure*}[htb!]
\centering 
\includegraphics[width=0.49\linewidth]{\figures 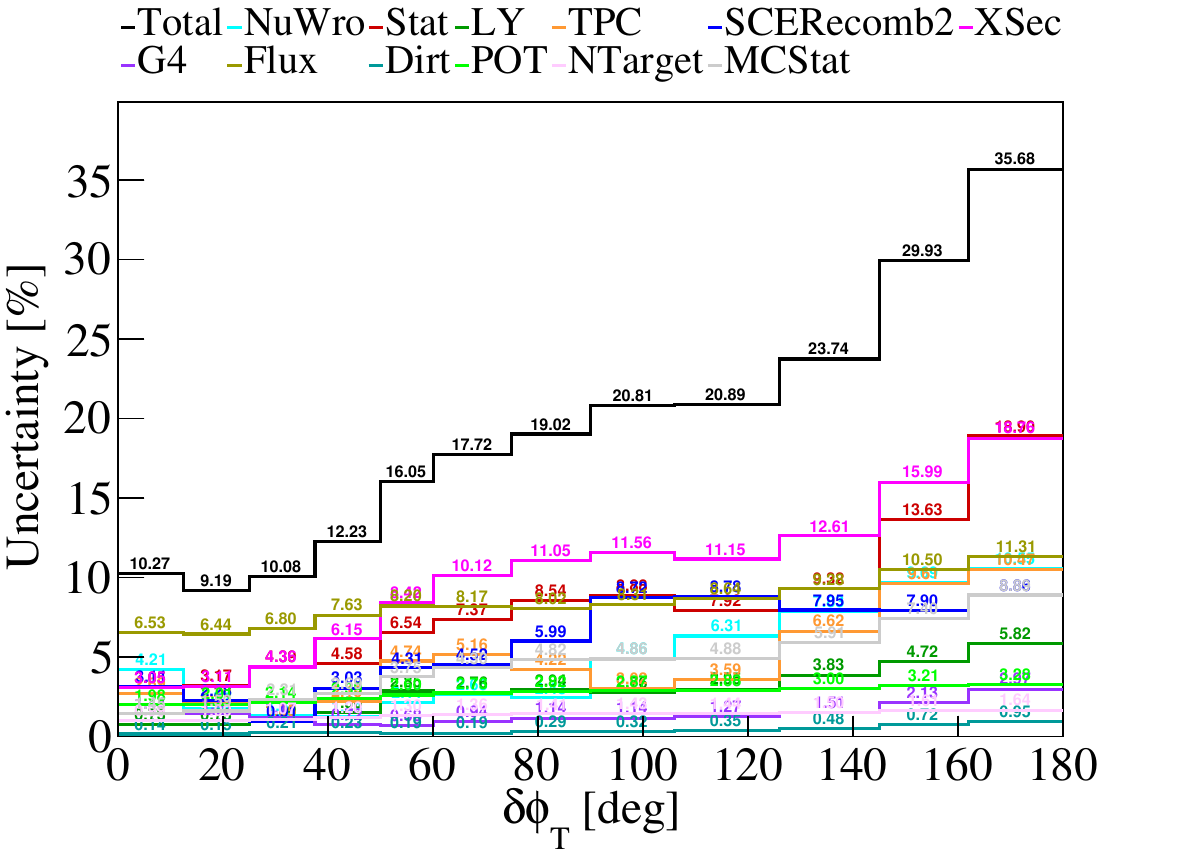}
\caption{
Uncertainty components for $\delta\phi_{T}$.
}
\label{SuppMatDeltaPhiTUncComponents}
\end{figure*}

\begin{figure*}[htb!]
\centering 
\includegraphics[width=0.49\linewidth]{\figures 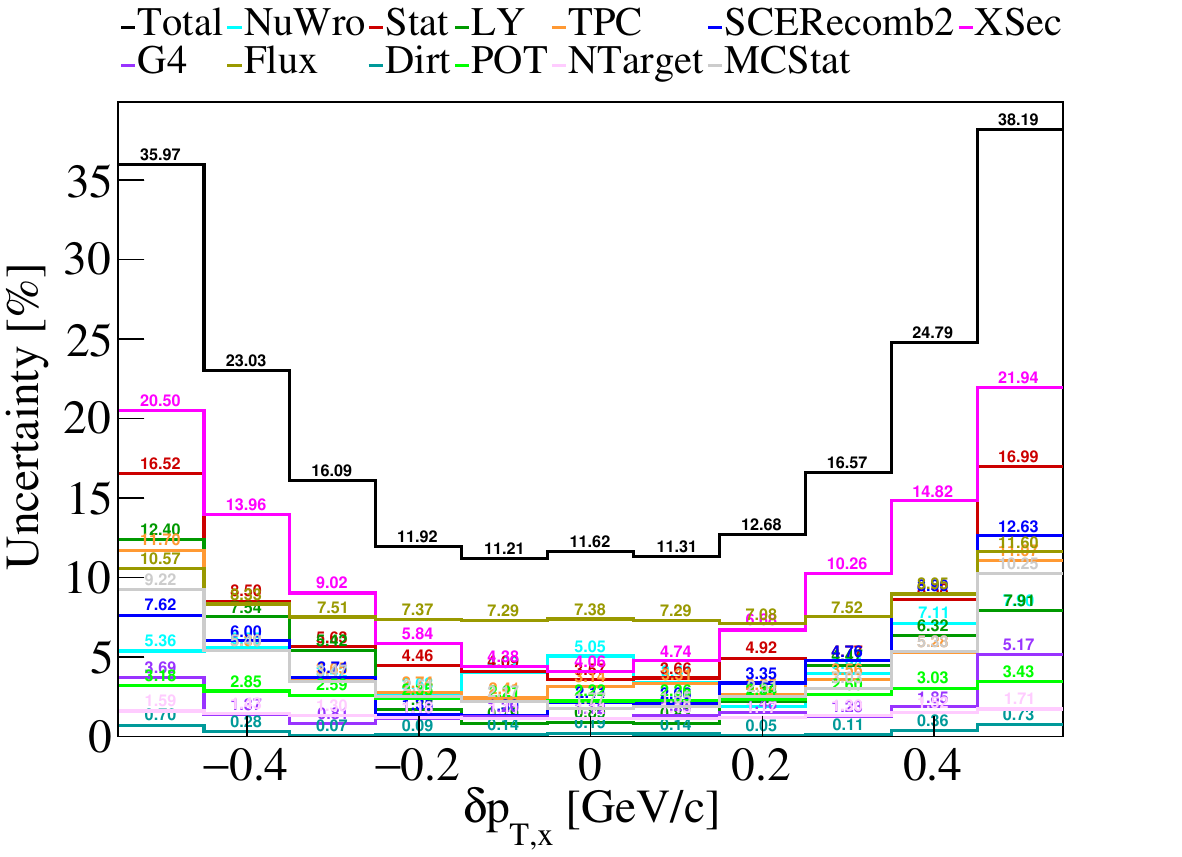}
\caption{
Uncertainty components for $\delta p_{T,x}$.
}
\label{SuppMatDeltaPtxUncComponents}
\end{figure*}

\begin{figure*}[htb!]
\centering 
\includegraphics[width=0.49\linewidth]{\figures 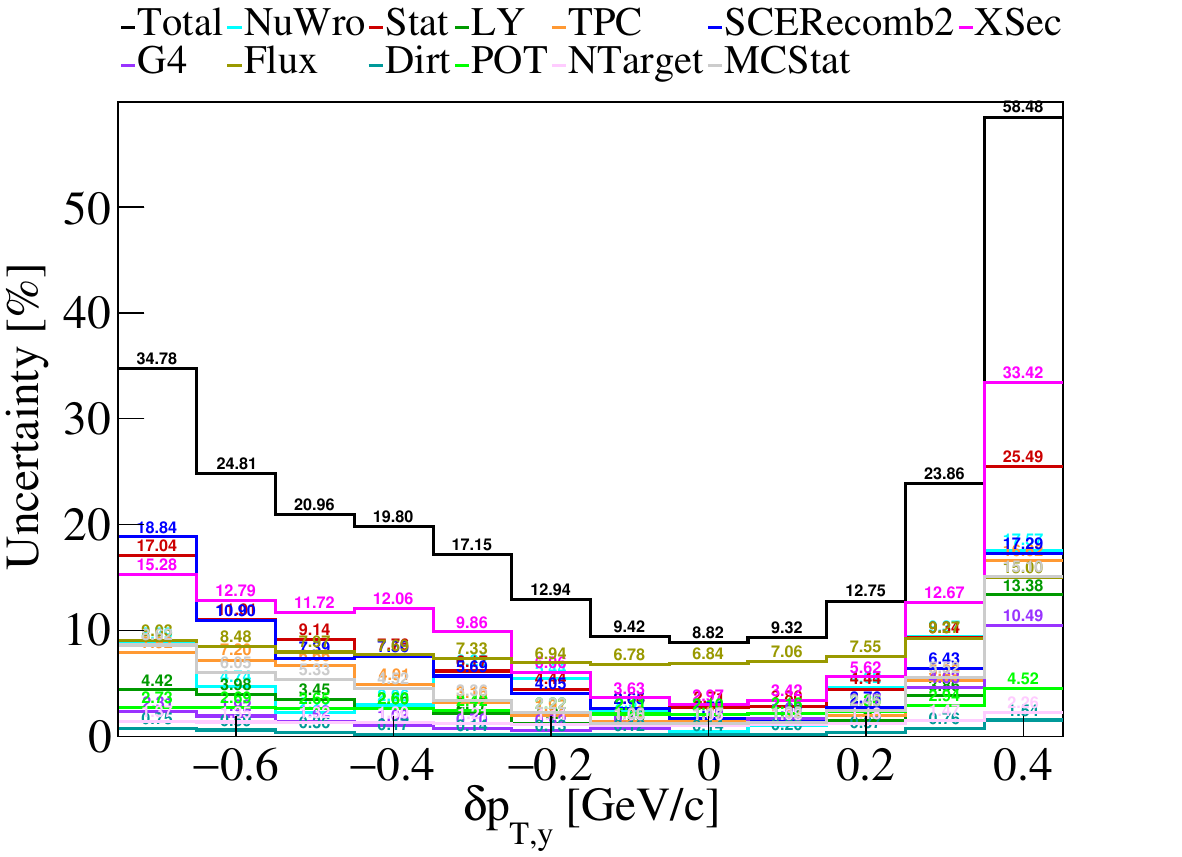}
\caption{
Uncertainty components for $\delta p_{T,y}$.
}
\label{SuppMatDeltaPtyUncComponents}
\end{figure*}

\begin{figure*}[htb!]
\centering 
\includegraphics[width=0.49\linewidth]{\figures 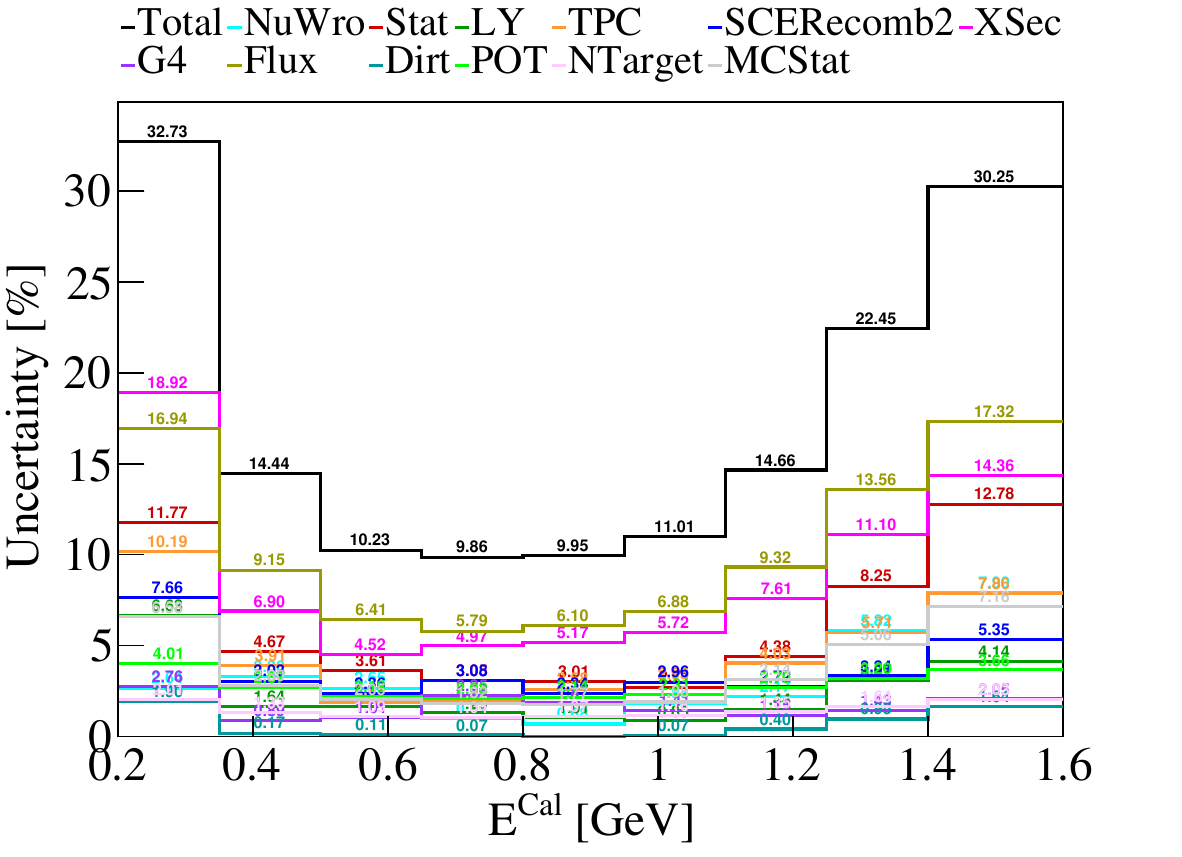}
\caption{
Uncertainty components for $E^{Cal}$.
}
\label{SuppMatECalUncComponents}
\end{figure*}


\clearpage
	\subsection{GENIE Cross Section Uncertainties}\label{genie}
	
    \justify	
    The systematic uncertainty due to the cross section modeling  with the $\texttt{GENIE}$ framework is calculated using the relevant EventWeight weights detailed in~\cite{GENIEKnobs}.
    The breakdown of the relevant uncertainties are shown in the tables below, both for the total as well as the signal (Sig) and background (Bkg) events. 

    \begin{table}[htbp]
    \centering
    \begin{adjustbox}{width=1\textwidth}
    \small
    \begin{tabular}{lllcc}
    \hline
    \textbf{Parameter}    & \textbf{Description}                                                                & \textbf{CV} & \textbf{$1\sigma$~Uncertainty} &  \textbf{\makecell{Contributing\\Uncertainty (\%)\\Total (Sig/Bkg)}}\\ 
    \hline
    \textcolor{blue}{\textbf{Quasi-Elastic Parameters}} & & & \\
    \texttt{MaCCQE}             &  CCQE axial mass                                                                      & 1.10~GeV     & $\pm$0.1~GeV   & 0.038 (0.256, 0.218)\\
    \texttt{RPA CCQE}             & Strength of the RPA correction                                                      & 0.151        & $\pm$0.4       &  2.094 (1.989, 0.104)\\    
    \texttt{MaNCEL}             &  Axial mass for NCEL                                                                  & 0.961242~GeV & $\pm$25\%      & 0.348 (0.010, 0.348)\\
    \texttt{EtaNCEL}            &  Empirical parameter used to account for sea quark contribution to NCEL form factor   & 0.12         & $\pm$30\%      & 0.010 (0.010, 0.010)\\
    \texttt{AxFFCCQEshape}        & Parametrisation of the nucleon axial form factor                                    & Dipole       & z-expansion    & 0.022 (0.016, 0.031) \\
    \texttt{VecFFCCQEshape}       & Parametrisation of the nucleon vector form factors                                  & BBA07        & Dipole         & 0.051 (0.186, 0.137)\\    
    \hline
    \textcolor{blue}{\textbf{MEC Parameters}} & & & &\\
    \texttt{NormCCMEC}          &  Energy-independent normalization for CCMEC                                           & 1.66         & $\pm$0.5       & 1.832 (0.514, 1.319)\\
    \texttt{NormNCMEC}          &  Energy-independent normalization for NCMEC                                           & 1            & $\pm$20\%      & 0.129 (0.010, 0.129)\\
    \texttt{FracPNCCMEC}        &  Fraction of initial nucleon pairs that are pn (0 = Valencia)                         & 0            & $\pm$20\%      & 0.041 (0.333, 0.293)\\
    \texttt{FracDeltaCCMEC}     &  Relative contribution of $\Delta$ diagrams to total MEC cross section (0 = Valencia) & 0            & $\pm$30\%      & 0.124 (0.124, 0.158)\\
    \texttt{XSecShape CCMEC}      & Changes shape of differential cross section                                         & 1.0          & 0.0            & 2.273 (1.762, 0.511)\\
    \texttt{DecayAngMEC}          & Changes angular distribution of nucleon cluster                                     & Isotropic    & $\cos^2\theta$ in rest frame  & 0.693 (0.290, 0.404) \\    
    \hline
    \textcolor{blue}{\textbf{Resonant Parameters}} & & & \\
    \texttt{MaCCRES}            &  CCRES axial mass                                                                     & 1.120~GeV    & $\pm0.2$       & 0.986 (0.119, 1.103)\\
    \texttt{MvCCRES}            &  Shape-only CCRES axial mass                                                          & 0.840~GeV    & $\pm0.1$       & 0.775 (0.121, 0.896)\\
    \texttt{MaNCRES}            &  NCRES axial mass                                                                     & 1.120~GeV    & $\pm0.2$       & 0.969 (0.010, 0.969)\\
    \texttt{MvNCRES}            &  NCRES vector mass.                                                                   & 0.840~GeV    & $\pm0.1$       & 0.395 (0.010, 0.395)\\
    \texttt{ThetaDelta2Npi}       & Interpolates angular distribution for $\Delta \rightarrow N + \pi$                  & Rein-Sehgal  & Isotropic      &  1.533 (0.142, 1.392)\\
    \texttt{ThetaDelta2NRad}      & Interpolates angular distribution for $\Delta \rightarrow N + \gamma$               & Rein-Sehgal  & $\cos^2\theta$ &  0.016 (0.016, 0.016)\\    
    \hline
    \end{tabular}
    \end{adjustbox}
    \label{tab:genie_all}
    \end{table}


    \begin{table}[htbp]
    \centering
    \begin{adjustbox}{width=1\textwidth}
    \small
    \begin{tabular}{lllcc}
    \hline
    \textbf{Parameter}          & \textbf{Description}                                                                & \textbf{CV} & \textbf{$1\sigma$~Uncertainty} & \textbf{\makecell{Contributing\\Uncertainty (\%)\\Total (Sig/Bkg)}} \\ 
    \hline
    \textcolor{blue}{\textbf{Non-Resonant Parameters}} & & & \\
    \texttt{NonRESBGvpNC1pi}    &  Non-resonant background normalization for $\nu$p NC1$\pi$                          & 0.1          & $\pm0.5$ & 0.041 (0.010, 0.041)\\
    \texttt{NonRESBGvpNC2pi}    &  Non-resonant background normalization for $\nu$p NC2$\pi$                          & 1            & $\pm0.5$ & 0.096 (0.010, 0.096)\\
    \texttt{NonRESBGvnNC1pi}    &  Non-resonant background normalization for $\nu$n NC1$\pi$                          & 0.3          & $\pm0.5$ & 0.390 (0.010, 0.390)\\
    \texttt{NonRESBGvnNC2pi}    &  Non-resonant background normalization for $\nu$n NC2$\pi$                          & 1            & $\pm0.5$ & 0.022 (0.010, 0.022)\\
    \texttt{NonRESBGvbarpNC1pi} &  Non-resonant background normalization for $\bar{\nu}$p NC1$\pi$                    & 0.3          & $\pm0.5$ & 0.010 (0.010, 0.010)\\
    \texttt{NonRESBGvbarpNC2pi} &  Non-resonant background normalization for $\bar{\nu}$p NC2$\pi$                    & 1            & $\pm0.5$ & 0.010 (0.010, 0.010)\\
    \texttt{NonRESBGvbarnNC1pi} &  Non-resonant background normalization for $\bar{\nu}$n NC1$\pi$                    & 0.1          & $\pm0.5$ & 0.010 (0.010, 0.010)\\
    \texttt{NonRESBGvbarnNC2pi} &  Non-resonant background normalization for $\bar{\nu}$n NC2$\pi$                    & 1            & $\pm0.5$ & 0.010 (0.010, 0.010)\\
    \texttt{NonRESBGvpCC1pi}    &  Non-resonant background normalization for $\nu$p CC1$\pi$                          & 0.007713     & $\pm0.5$ & 0.014 (0.010, 0.019)\\
    \texttt{NonRESBGvpCC2pi}    &  Non-resonant background normalization for $\nu$p CC2$\pi$                          & 0.787999     & $\pm0.5$ & 0.059 (0.010, 0.063)\\
    \texttt{NonRESBGvnCC1pi}    &  Non-resonant background normalization for $\nu$n CC1$\pi$                          & 0.127858     & $\pm0.5$ & 0.217 (0.034, 0.250)\\
    \texttt{NonRESBGvnCC2pi}    &  Non-resonant background normalization for $\nu$n CC2$\pi$                          & 2.11523      & $\pm0.5$ & 0.079 (0.010, 0.077)\\
    \texttt{NonRESBGvbarpCC1pi} &  Non-resonant background normalization for $\bar{\nu}$p CC1$\pi$                    & 0.127858     & $\pm0.5$ & 0.013 (0.010, 0.013)\\
    \texttt{NonRESBGvbarpCC2pi} &  Non-resonant background normalization for $\bar{\nu}$p CC2$\pi$                    & 2.11523      & $\pm0.5$ & 0.010 (0.010, 0.010)\\
    \texttt{NonRESBGvbarnCC1pi} &  Non-resonant background normalization for $\bar{\nu}$n CC1$\pi$                    & 0.007713     & $\pm0.5$ & 0.010 (0.010, 0.010)\\
    \texttt{NonRESBGvbarnCC2pi} &  Non-resonant background normalization for $\bar{\nu}$n CC2$\pi$                    & 0.787999     & $\pm0.5$ & 0.010 (0.010, 0.010)\\
    \texttt{AhtBY}              &  $A_{HT}$ higher-twist parameter in the Bodek-Yang model scaling variable $\xi_w$   & 0.538        & $\pm0.25$ & 0.010 (0.010, 0.010)\\
    \texttt{BhtBY}              &  BHT higher-twist parameter in the Bodek-Yang model scaling variable $\xi_w$        & 0.305        & $\pm0.25$ & 0.010 (0.010, 0.010)\\
    \texttt{CV1uBY}             &  CV1u valence GRV98 PDF correction parameter in the Bodek-Yang model                & 0.291        & $\pm0.3$ & 0.010 (0.010, 0.010)\\
    \texttt{CV2uBY}             &  CV2u valence GRV98 PDF correction parameter in the Bodek-Yang model                & 0.189        & $\pm0.4$ & 0.010 (0.010, 0.010)\\
    \hline
    \textcolor{blue}{\textbf{Hadronisation Parameters}} & & & & \\
    \texttt{AGKYxF1pi}          &  Hadronization parameter, applicable to true DIS interactions only                  & -0.385       & $\pm0.2$ & 0.108 (0.013, 0.120)\\
    \texttt{AGKYpT1pi}          &  Hadronization parameter, applicable to true DIS interactions only                  & 1/6.625      & $\pm0.03$ & 0.034 (0.014, 0.046)\\
    \hline
    \textcolor{blue}{\textbf{Final State Interaction Parameters}} & & & & \\
    \texttt{MFP$_{\pi}$}        &  $\pi$ mean free path                                                               & 0            & $\pm0.2$  & 0.032 (0.010, 0.028)\\
    \texttt{MFP$_{N}$}          &  Nucleon mean free path                                                             & 0            & $\pm0.2$  & 1.212 (1.067, 0.147)\\
    \texttt{FrCEx$_{\pi}$}      &  Fractional cross section for $\pi$ charge exchange                                 & 0            & $\pm0.5$  &  0.159 (0.025, 0.184)\\
    \texttt{FrInel$_{\pi}$}     &  Fractional cross section for $\pi$ inelastic scattering                            & 0            & $\pm0.4$  &  0.133 (0.042, 0.091)\\
    \texttt{FrAbs$_{\pi}$}      &  Fractional cross section for $\pi$ absorption                                      & 0            & $\pm0.3$  &  0.906 (0.066, 0.019)\\
    \texttt{FrCEx$_{N}$}        &  Fractional cross section for nucleon charge exchange                               & 0            & $\pm0.2$  &  0.953 (0.411, 0.542)\\
    \texttt{FrInel$_{N}$}       &  Fractional cross section for nucleon inelastic scattering                          & 0            & $\pm0.5$  &  0.289 (0.228, 0.061)\\
    \texttt{FrAbs$_{N}$}        &  Fractional cross section for nucleon absorption                                    & 0            & $\pm0.4$  &  0.906 (0.526, 0.380)\\
    \hline
    \textcolor{blue}{\textbf{Delta Resonant Decay Parameters}} & & & & \\
    \texttt{RDecBR1gamma}       &  Normalization for $\Delta \rightarrow \gamma$ decays                               & Nominal BR & $\pm0.5$ & 0.042 (0.017, 0.025)\\
    \texttt{RDecBR1eta}         &  Normalization for $\Delta \rightarrow \eta$ decays                                 & Nominal BR & $\pm0.5$ & 0.513 (0.198, 0.324)\\
    \hline
    \textcolor{blue}{\textbf{Coherent Parameters}} & & & & \\
    \texttt{NormCCCOH}            & Scaling factor for CCCOH $\pi$ production total cross section               & Nominal            & 100\% increase  &  0.027 (0.016, 0.027)\\
    \texttt{NormNCCOH}            & Scaling factor for NCCOH $\pi$ production total cross section               & Nominal            & 100\% increase  &  0.016 (0.016, 0.016)\\
    \hline
    \end{tabular}
    \end{adjustbox}
    \label{tab:genie_res}
    \end{table}   


\clearpage
\bibliography{main}
